%% file: ms.tex
\documentclass[twocolumn,trackchanges]{aastex62}
\input{Input_usepackage.tex}

\input{Input_newcommand.tex}

\usepackage{expl3}
\usepackage{varwidth}
\usepackage{tikz}
\usepackage{amsmath,amssymb}
\usepackage{bm}
\usetikzlibrary{arrows.meta}
\usetikzlibrary{shapes,arrows,chains}
\usetikzlibrary{calc}
\usetikzlibrary{positioning}
\input{Input_flow_chart_style.tex}
\input{Input_galaxy_flow_chart_style.tex}
\shorttitle{A$^3$COSMOS photometry \& catalogs}
\shortauthors{D. Liu et al.}

\begin{document}

\title{Automated Mining of the ALMA Archive in the COSMOS Field (A$^3$COSMOS): I. Robust ALMA Continuum Photometry Catalogs and Stellar Mass and Star Formation Properties for $\sim$700 Galaxies at $z=0.5$--6}

\email{dzliu@mpia.de}

\author[0000-0001-9773-7479]{Daizhong Liu}
\affiliation{Max-Planck-Institut f\"{u}r Astronomie, K\"{o}nigstuhl 17, D-69117 Heidelberg, Germany}

\author[0000-0002-5681-3575]{P. Lang}
\affiliation{Max-Planck-Institut f\"{u}r Astronomie, K\"{o}nigstuhl 17, D-69117 Heidelberg, Germany}

\author[0000-0002-6777-6490]{B. Magnelli}
\affiliation{Argelander-Institut f\"{u}r Astronomie, Universit\"{a}t Bonn, Auf dem H\"ugel 71, D-53121 Bonn, Germany}

\author[0000-0002-3933-7677]{E. Schinnerer}
\affiliation{Max-Planck-Institut f\"{u}r Astronomie, K\"{o}nigstuhl 17, D-69117 Heidelberg, Germany}

\author[0000-0002-4826-8642]{S. Leslie}
\affiliation{Max-Planck-Institut f\"{u}r Astronomie, K\"{o}nigstuhl 17, D-69117 Heidelberg, Germany}

\author[0000-0001-7440-8832]{Y. Fudamoto}
\affiliation{Department of Astronomy, Universit\'e de Gen\`eve, Chemin des Maillettes 51, 1290 Versoix, Switzerland}

\author{M. Bondi}
\affiliation{INAF--Istituto di Radioastronomia, Via Gobetti 101, I-40129, Bologna, Italy}

\author[0000-0002-9768-0246]{B. Groves}
\affiliation{Research School of Astronomy and Astrophysics, Australian National University, Canberra ACT, 2611, Australia}

\author[0000-0002-2640-5917]{E. Jim\'{e}nez-Andrade}
\affiliation{Argelander-Institut f\"{u}r Astronomie, Universit\"{a}t Bonn, Auf dem H\"ugel 71, D-53121 Bonn, Germany}

\author[0000-0001-5429-5762]{K. Harrington}
\affiliation{Argelander-Institut f\"{u}r Astronomie, Universit\"{a}t Bonn, Auf dem H\"ugel 71, D-53121 Bonn, Germany}

\author[0000-0002-8414-9579]{A. Karim}
\affiliation{Argelander-Institut f\"{u}r Astronomie, Universit\"{a}t Bonn, Auf dem H\"ugel 71, D-53121 Bonn, Germany}

\author[0000-0001-5851-6649]{P. A. Oesch}
\affiliation{Department of Astronomy, Universit\'e de Gen\`eve, Chemin des Maillettes 51, 1290 Versoix, Switzerland}

\author[0000-0003-1033-9684]{M. Sargent}
\affiliation{Astronomy Centre, Department of Physics and Astronomy, University of Sussex, Brighton BN1 9QH, UK}

\author[0000-0002-4437-1773]{E. Vardoulaki}
\affiliation{Argelander-Institut f\"{u}r Astronomie, Universit\"{a}t Bonn, Auf dem H\"ugel 71, D-53121 Bonn, Germany}

\author{T. B\v{a}descu}
\affiliation{Argelander-Institut f\"{u}r Astronomie, Universit\"{a}t Bonn, Auf dem H\"ugel 71, D-53121 Bonn, Germany}

\author{L. Moser}
\affiliation{Argelander-Institut f\"{u}r Astronomie, Universit\"{a}t Bonn, Auf dem H\"ugel 71, D-53121 Bonn, Germany}

\author[0000-0002-1707-1775]{F. Bertoldi}
\affiliation{Argelander-Institut f\"{u}r Astronomie, Universit\"{a}t Bonn, Auf dem H\"ugel 71, D-53121 Bonn, Germany}

\author[0000-0003-4569-2285]{A. Battisti}
\affiliation{Research School of Astronomy and Astrophysics, Australian National University, Canberra ACT, 2611, Australia}

\author[0000-0001-9759-4797]{E. da Cunha}
\affiliation{Research School of Astronomy and Astrophysics, Australian National University, Canberra ACT, 2611, Australia}

\author[0000-0002-7051-1100]{J. Zavala}
\affiliation{The University of Texas at Austin, 2515 Speedway Boulevard, Stop C1400, Austin, TX 78712, USA}

\author[0000-0002-6748-0577]{M. Vaccari}
\affiliation{Department of Physics and Astronomy, University of the Western Cape, Robert Sobukwe Road, 7535 Bellville, Cape Town, South Africa}
\affiliation{INAF--Istituto di Radioastronomia, Via Gobetti 101, I-40129, Bologna, Italy}

\author[0000-0002-2951-7519]{I. Davidzon}
\affiliation{IPAC, Mail Code 314-6, California Institute of Technology, 1200 East California Boulevard, Pasadena, CA 91125, USA}

\author[0000-0001-9585-1462]{D. Riechers}
\affiliation{Department of Astronomy, Cornell University, Space Sciences Building, Ithaca, NY 14853, USA}

\author[0000-0002-6290-3198]{M. Aravena}
\affiliation{N\'{u}cleo de Astronom\'{i}a, Facultad de Ingenier\'{i}a, Universidad Diego Portales, Av. Ej\'{e}rcito 441, Santiago, Chile}

\begin{abstract}
The rich information on (sub-)millimeter dust continuum emission from distant galaxies in the public Atacama Large Millimeter/submillimeter Array (ALMA) archive is contained in thousands of inhomogeneous observations from individual PI-led programs. To increase the usability of these data for studies deepening our understanding of galaxy evolution, we have developed automated mining pipelines for the ALMA archive in the COSMOS field (A3COSMOS) which efficiently exploit the available information for large numbers of galaxies across cosmic time, and keep the data products in sync with the increasing public ALMA archive: (a) a dedicated ALMA continuum imaging pipeline; (b) two complementary photometry pipelines for both blind source extraction and prior source fitting; (c) a counterpart association pipeline utilizing the multi-wavelength data available (including quality assessment based on machine-learning techniques); (d) an assessment of potential (sub-)mm line contribution to the measured ALMA continuum; and (e) extensive simulations to provide statistical corrections to biases and uncertainties in the ALMA continuum measurements. Application of these tools yields photometry catalogs with $\sim1000$ (sub-)mm detections (spurious fraction\,$\sim8-12\%$) from over 1500 individual ALMA continuum images. Combined with ancillary photometric and redshift catalogs and the above quality assessments, we provide robust information on redshift, stellar mass and star formation rate for $\sim$700 galaxies at redshifts 0.5-6 in the COSMOS field (with undetermined selection function). The ALMA photometric measurements and galaxy properties are released publicly within our blind-extraction, prior-fitting and galaxy property catalogs, plus the images. These products will be updated on a regular basis in the future. 
\end{abstract}


\section{Introduction}
\label{Section_Introduction}

The interstellar medium (ISM) is the raw material in galaxies out of which stars form. It plays a fundamental role when reconstructing the Universe through cosmological simulations. In galaxies harboring intensive star formation, cold neutral gas is the main component of the ISM dominating its mass; and a significant fraction of this cold gas is in the molecular phase (e.g., \citealt{Walter2008}; \citealt{Leroy2008}; \citealt{Bigiel2008}). Over the past four decades, molecular gas has been observed mainly via the Carbon Monoxide (CO) rotational transition lines in the rest-frame millimeter (mm) wavelengths (which are the most feasible observable tracers of molecular gas; e.g., see review by \citealt{Solomon2005}; \citealt{Carilli_and_Walter_2013}). 
However, CO observations at high redshift ($z>1$) mostly target the brightest sub-millimeter galaxies (SMGs; e.g., review by \citealt{Blain2002}) and quasi-stellar objects (QSOs). These objects are the most extreme cases and not representative of the more numerous, less starbursty galaxies, i.e., the star-forming galaxies that follow a tight main sequence (MS) in the stellar mass--SFR plane (e.g. \citealt{Brinchmann2004}; \citealt{Noeske2007}; \citealt{Elbaz2007}; \citealt{Daddi2007}). 
Observing CO in a large number (e.g., a few hundred) of main-sequence galaxies at $z>1$ (hence probing the ISM evolution) is in practise very time-consuming even with the most advanced facility, the Atacama Large Millimeter/submillimeter Array (ALMA), as firstly all galaxies are required to have a spectroscopic redshift in advance, secondly sufficient sensitivity is needed to detect the line within a small spectral bandwidth (typically $\sim300-500\;\mathrm{km/s}$), and thirdly the galaxy sample should cover enough parameter space in the main sequence plane. 

In recent years, a much more efficient approach --- carrying out broadband dust continuum observations to infer the ISM --- has been established. 
With the use of all the bandwidth of the receiver, usually much short integration time are needed for high-redshift galaxies than observing (sub)mm emission lines (see also \citealt{Carpenter2019_ALMA_Roadmap}). Then, the cold gas mass can be inferred either using the gas-to-dust mass ratio, $\deltaGDR$, which has been reasonably characterized as a function of gas phase metallicity (e.g., \citealt{Santini2010}; \citealt{Leroy2011GDR}; \citealt{Magdis2011SED,Magdis2012SED}; \citealt{Magnelli2012}; \citealt{Bolatto2013ARAA}; \citealt{RemyRuyer2014}; \citealt{Tan2014}; \citealt{Coogan2019}), or with the ratio between gas mass and dust continuum luminosity at Rayleigh-Jeans tail wavelengths (e.g., rest-frame 250--850\,$\mu$m), which has been calibrated with rich observations (e.g., \citealt{Scoville2014}; \citealt{Groves2015}; \citealt{Hughes2017}; \citealt{Bertemes2018}; \citealt{Saintonge2018}). 
The use of dust continuum observations to systematically survey the ISM content in hundreds of high-redshift galaxies is already proved to be fruitful (e.g., \citealt{Schinnerer2016}; \citealt{Scoville2016,Scoville2017}). 

Meanwhile, the continuously growing ALMA public archive offers a great opportunity of studying very large samples of high-redshift galaxies. The ALMA public archive consists of thousands of observations of high-redshift galaxies within deep fields led by individual Principle Investigator (PI) programs. Although ALMA has a small field of view, e.g., $\sim$0.5$'$ in diameter (primary beam FWHM) in Band~6, accumulating archival data compensates for this shortcoming and leads to several hundred arcmin$^2$ area. 
Comparing to contiguous deep field surveys with ALMA (e.g., \citealt{Hatsukade2011,Hatsukade2016,Hatsukade2018}; \citealt{Carniani2015}; \citealt{Walter2016}; \citealt{Aravena2016}; \citealt{Dunlop2017}; \citealt{Umehata2017,Umehata2018}; \citealt{Franco2018}), the discreteness of field of views makes the sample selection bias and cosmic comoving volume very unpredictable, but it also leads to a sample with large varieties in galaxy properties, which thus provides crucial constraints on galaxy ISM and star formation scaling relations and analytic evolution prescriptions (e.g., \citealt{Scoville2017}; \citealt{Tacconi2018}). Moreover, the ALMA archive also serves as a powerful test bed for automated pipelines as in this work and for future large facilities. 

Several recent studies have already been exploring the full ALMA archive \citep[e.g.][]{Scoville2017,Fujimoto2017,Zavala2018}, however, the photometric methods (including aperture photometry, peak pixel analysis, $uv$-plane fitting, etc.) and sample selection can significantly differ between different authors. 
Furthermore, none of these studies have statistically evaluated the effects of applying different photometric methods to ALMA images, especially for large numbers of ALMA images with widely varying sensitivity and synthesized beam properties.
Consequently, noticeable discrepancies on the cosmic evolution of the ISM are present among the aforementioned studies. In order to understand how potential biases of the photometric methods and gas mass calibration affect the outcome of ISM evolution studies, more dedicated efforts are required to exploit the public ALMA archive. 

In this work, we present automated pipelines for ``mining'' the public ALMA archive in the COSMOS field \citep{Scoville2007} (hereafter referred to as the ``A$^3$COSMOS'' (Automated mining of the ALMA Archive in COSMOS) project\,\footnote{\url{https://sites.google.com/view/a3cosmos}}.
This work provides the foundation for a systematic exploitation of the (sub-)mm continuum as a proxy for cold dust and gas for a diverse and large sample of high-redshift galaxies. The resulting catalog of galaxies with (sub-)mm continuum detections can be used to, e.g., study the cosmic evolution of the gas fraction and gas depletion time (Paper~\textsc{II}; D. Liu et al. 2019, submitted to ApJ.). 

\begin{figure*}[t!]
\centering%
\begin{minipage}[b]{\textwidth}
\resizebox{1.00\textwidth}{!}{\input{Input_flow_chart.tikz}}
\caption{%
Workflow of our automated mining of the ALMA archive in the COSMOS field (A$^3$COSMOS), which corresponds to Sects.~\ref{Section_Data}~to~\ref{Section_Monte_Carlo_Simulation_and_Correction} of this paper. 
\textbf{(Left branch:)} Starting from Sect.~\ref{Section_ALMA_Continuum_Images}, we create ALMA continuum images from raw ALMA data which are obtained from the full public ALMA archive data for COSMOS. Next two photometric pipelines are used: (a) blind source extraction (see Sect.~\ref{Section_Blind_Source_Extraction}) and (b) prior source fitting (see Sect.~\ref{Section_Prior_Source_Fitting}) utilizing the COSMOS master catalog compiled beforehand as described in Sect.~\ref{Section_Prior_Source_Catalogs} \textbf{(middle branch)}. 
\textbf{(Right branch:)} Two Monte Carlo simulation pipelines with largely different prior assumptions are employed to verify the two photometric methods, and to provide flux bias correction and flux error estimation for the two photometric catalogs (see Sect.~\ref{Section_Monte_Carlo_Simulation_and_Correction}, and  Appx.~\ref{Section_Appendix_MC_Sim} for details). 
After verification, the two photometric catalogs are combined to build a galaxy catalog with physical properties as described in Sect.~\ref{Section_Galaxy_Sample_and_Properties}. 
\label{Figure_flow_chart}
}%
\end{minipage}
\vspace{6mm}
\end{figure*}
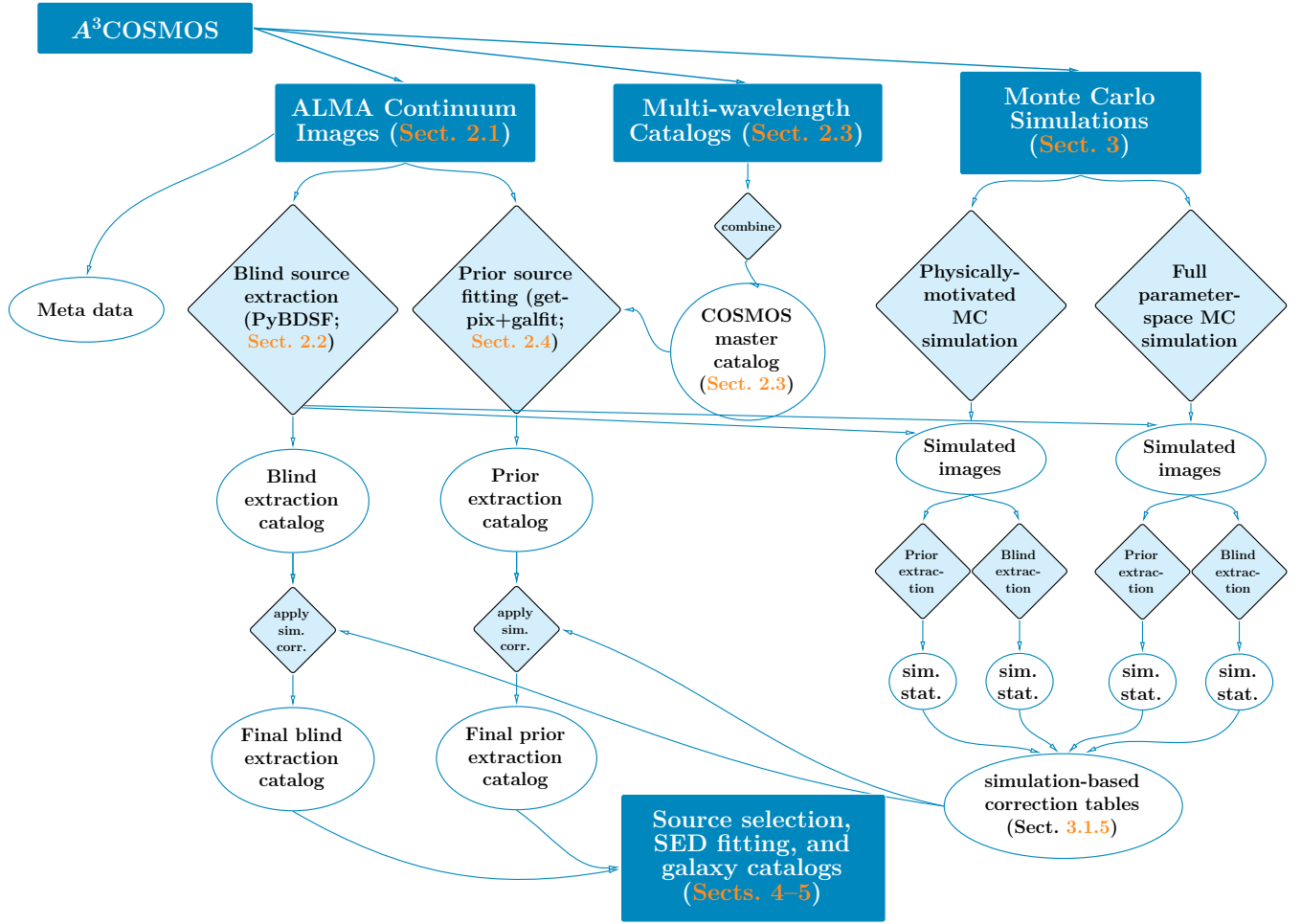

We present our workflow from the raw public ALMA data to the two robust photometric catalogs in Fig.~\ref{Figure_flow_chart}, which corresponds to Sects.~\ref{Section_Data}~to~\ref{Section_Monte_Carlo_Simulation_and_Correction} of the paper. 
We first describe our ALMA data reduction and continuum imaging procedures in Sect.~\ref{Section_ALMA_Continuum_Images}, and then we present the blind source extraction in Sect.~\ref{Section_Blind_Source_Extraction}, prior catalog compilation in Sect.~\ref{Section_Prior_Source_Catalogs}, and prior source fitting in Sect.~\ref{Section_Prior_Source_Fitting}.
Sect.~\ref{Section_Monte_Carlo_Simulation_and_Correction} includes our extensive Monte Carlo (MC) simulations and analyses to verify our photometry. 
Sect.~\ref{Section_Galaxy_Sample_and_Properties} describes how we combine the two photometry catalogs, remove spurious sources, and build a final well-characterized galaxy catalog (a workflow for these substeps is presented at the beginning of Sect.~\ref{Section_Galaxy_Sample_and_Properties}).
Finally, the resulting catalogs are described in Sect.~\ref{Section_Data_Delivery}, and we summarize the paper in Sect.~\ref{Section_Summary}.

Throughout the paper, we adopt a flat $\Lambda$CDM cosmology with $H_0=70\;\mathrm{km\,s^{-1}\,Mpc^{-1}}$, $\Omega_\mathrm{M}=0.3$, $\Omega_{0}=0.7$, and a \cite{Chabrier2003} initial mass function (IMF).

\vspace{0.5truecm}

\section{Data and Photometry}
\label{Section_Data}

\begin{figure*}[t!]
\centering%
\includegraphics[width=\textwidth, trim=12mm 5mm 20mm 0]{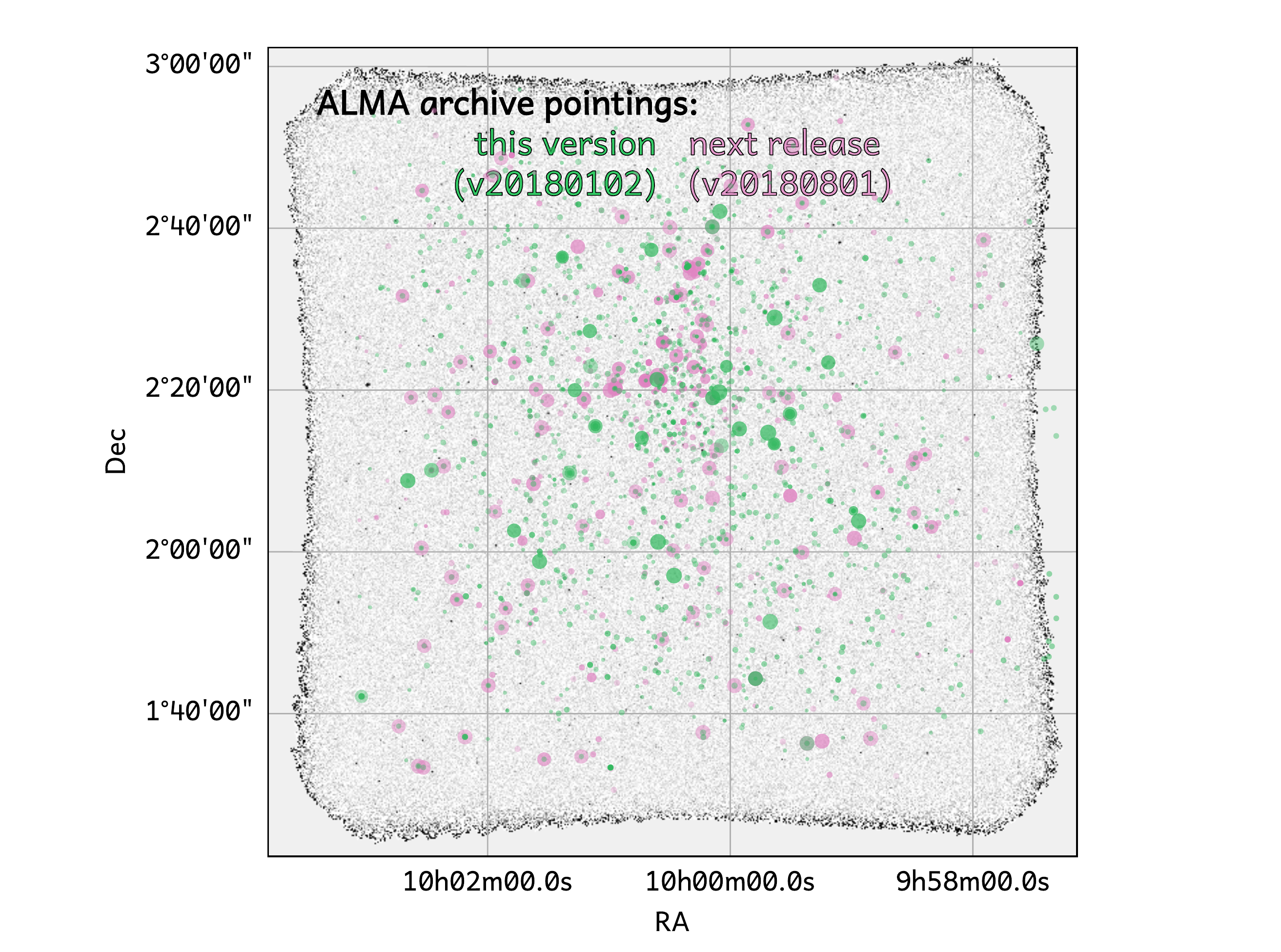}%
\caption{%
ALMA pointings in the COSMOS field that are publicly accessible. Green and magenta circles represent ALMA pointings that became public before \aaacosmosAlmaDataDate{} and \aaacosmosNextUpdateAlmaDataDate{} respectively. Circle sizes represent the FWHM of the ALMA 12m antennas' primary beam, and the shading reflects the on-source integration time (dark referring to longer integration times). 
The background image is the \textit{Herschel} PACS 100~$\mu$m data from the PACS Evolutionary Probe survey (PEP; \citealt{Lutz2011}).
\label{Figure_COSMOS_coverages}%
}
\vspace{3mm}
\end{figure*}

The public ALMA archive is growing rapidly through PI-led observations. 
These observations mainly focus on targeted scientific objectives (sources), which are usually at the phase center of each ALMA pointing. However, with ALMA's unprecedented sensitivity, and benefiting from the negative-$K$ correction at mm wavelengths (e.g., review by \citealt{Blain2002}; \citealt{Casey2014}), further (sub-)mm galaxies can serendipitously appear in any ALMA pointing. Such sources have a sizable chance for detection when their position falls within about twice the primary beam area\,\footnote{Here the primary beam area means the area enclosed in a circle with a radius equaling the primary beam's FWHM.} of the corresponding ALMA pointing (i.e., with a primary beam attenuation [PBA]~$\ge 0.2$).

Here, we conduct a systematic effort to exploit these observational data. We limit our selection to within the COSMOS field (R.A. = 10:00:28.6, decl. = +02:12:21.0, J2000; \citealt{Scoville2007}) because it is one of the deep fields with the richest, deepest multi-wavelength datasets, and there are numerous PI-led ALMA observations within its large area of 2\,deg$^2$ (compared to the Great Observatories Origins Deep Survey [GOODS] North and South fields with only 160\,arcmin$^2$ [0.044\,deg$^2$] each). We include all the available ALMA data in COSMOS regardless of the ALMA bands used (but excluded very long baseline data with a synthesized beam $<0.1''$; see Sect.~\ref{Section_ALMA_Continuum_Images}; and the only one mosaic project on the AzTEC-3 protocluster).

COSMOS has extensive imaging datasets covering all accessible wavelength ranges: 
X-ray (\citealt{Elvis2009}; \citealt{Civano2012,Civano2016}; \citealt{Marchesi2016}), 
UV (\citealt{Zamojski2007}), 
optical (\citealt{Leauthaud2007}; \citealt{Capak2007}; \citealt{Taniguchi2007,Taniguchi2015}), 
near-IR (\citealt{McCracken2010,McCracken2012}), 
mid-IR (\citealt{Sanders2007}, \citealt{LeFloch2009}), 
far-IR (\citealt{Lutz2011}; \citealt{Oliver2012}), 
sub-mm (\citealt{Geach_2016_SCUBA2}), 
mm (\citealt{Bertoldi2007}; \citealt{Aretxaga2011})
and radio (\citealt{Schinnerer2010}; \citealt{Smolcic2017a}). 
The depths of the X-ray, UV, optical and near-IR datasets are listed in \cite{Laigle2016}, and the depths of mid-to-far-IR, (sub-)mm and radio datasets are summarized in \cite{Jin2018}. 

Photometric redshifts have been obtained for $\sim 1.1 \times 10^{6}$ galaxies through optical to near-IR spectral energy distribution (SED) fitting by \cite{Muzzin2013}, \cite{Ilbert2013}, \cite{Laigle2016}, \cite{Davidzon2017} and \cite{Delvecchio2017}, and through optical to mm/radio SED fitting by \cite{Jin2018}. 

Spectroscopic redshifts also exist for $\sim\,7.1\,\times\,10^{4}$ galaxies, from the latest compilation by M. Salvato et al. (version Sept. 1st, 2017; available internally in the COSMOS collaboration), which includes almost all spectroscopic observations in the COSMOS field: 
\citet[][zCOSMOS Survey; with VLT/VIMOS]{Lilly2007,Lilly2009}; 
\citet[][with Spitzer/IRS]{FuHai2010Specz}; 
\citet[][with Keck II/DEIMOS]{Casey2012,Casey2017};
\citet[][with VLT/FORS2]{Comparat2015}; 
\cite{LeFevre2015} and \cite{Tasca2017} (VUDS Survey; with VLT/VMOS);
\citet[][with Keck II/DEIMOS]{Hasinger2018};
\citet[][MOSDEF Survey; with Keck I/MOSFIRE]{Kriek2015};
\citet[][with Keck II/NIRSPEC]{Marsan2017}; 
\citet[][with Keck II/DEIMOS]{Masters2017}; 
\citet[][with Keck I/MOSFIRE]{Nanayakkara2016};
\citet[][FMOS-COSMOS Survey; with Subaru/FMOS]{Silverman2015a};
\citet[][LEGA-C Survey; with VLT/VMOS]{vanderWel2016};
\citet[][with LMT/RSR]{Yun2015}
(listed only references whose spectroscopic redshifts are used in this work). 

We show the pointings of all public ALMA data for the COSMOS field as of \aaacosmosAlmaDataDate{} in Fig.~\ref{Figure_COSMOS_coverages}, overlaid on the \textit{Herschel Space Observatory} \citep{Pilbratt2010} Photodetector Array Camera and Spectrometer (PACS; \citealt{Poglitsch2010}) 100$\mu$m image. 
All the pointings processed for catalogs presented here are shown in green, and data that will be processed in our next release are shown in magenta, which includes ALMA data becoming public before August 1$^\mathrm{st}$, 2018. Circle size represents the primary beam\,\footnote{%
    Primary beam FWHMs are computed according to \url{https://www.iram.fr/IRAMFR/ARC/documents/cycle3/alma-technical-handbook.pdf}, Eq.\,(3.4).%
}. 
The sum of primary beam area of these observations reaches \aaacosmosAlmaPointingArea{}\,arcmin$^2$ as of \aaacosmosAlmaDataDate{}, and will reach \aaacosmosNextUpdateAlmaPointingArea{}\,arcmin$^2$ in our next release. 
Some pointings overlap because they are observed at different frequencies or with different spatial resolution. The overlapped area of all pointings is about 12\%. 
Thus, even considering the non-overlapped primary beam area, the current data already reach a spatial coverage similar to the area of the GOODS fields, and are much larger than any existing contiguous ALMA deep field survey (e.g., \citealt{Dunlop2017}, 4.5\,arcmin$^2$ with 1$\sigma\sim$35\,$\mu$Jy/beam; \citealt{Franco2018}, 69\,arcmin$^2$ with 1$\sigma\sim$0.18\,mJy/beam). 

\begin{figure}[t!]
\centering%
\includegraphics[width=\linewidth, trim=0 7mm 0 0]{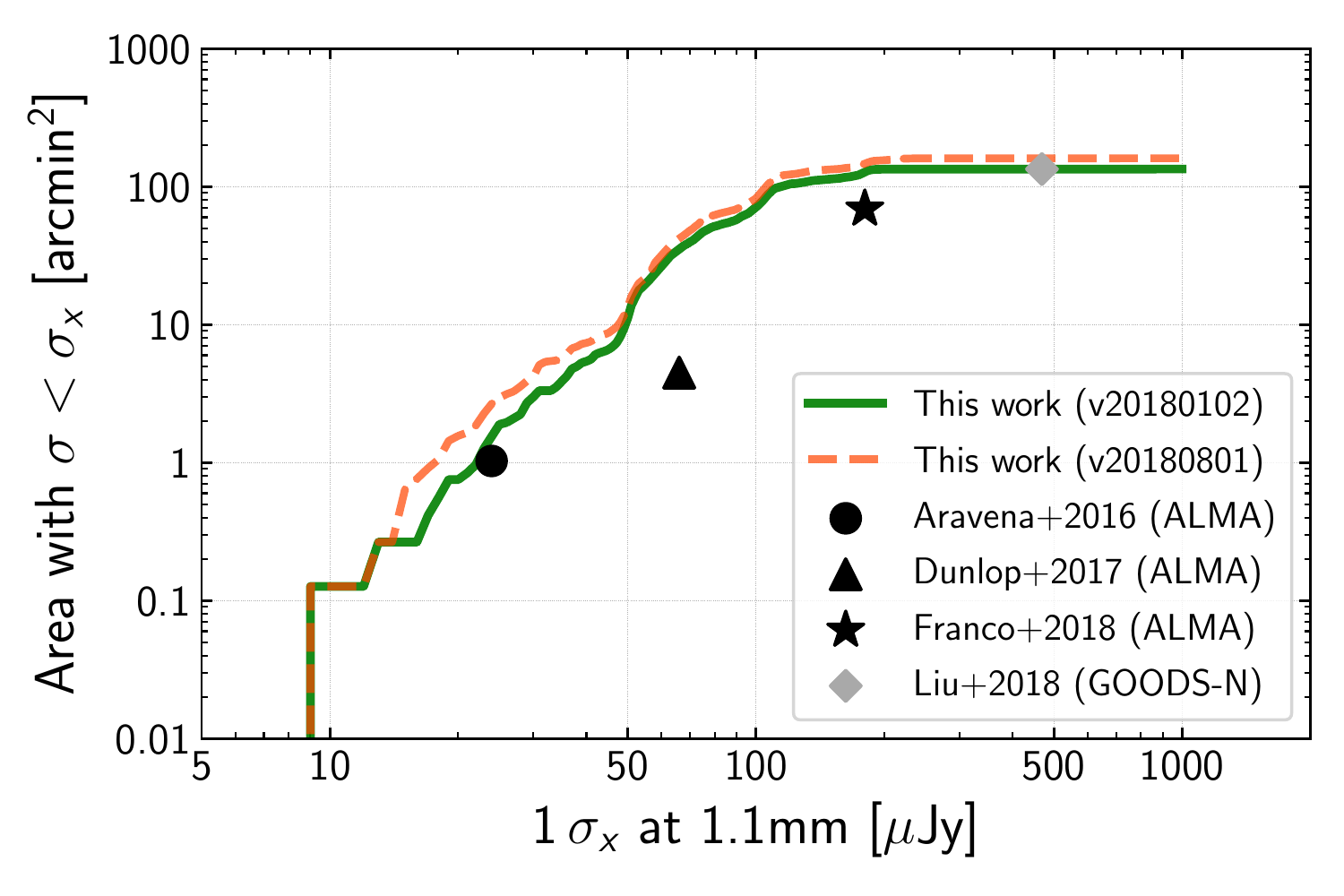}
\caption{%
Accumulated areal coverage of the public ALMA archival pointings at ALMA Band 6 and 7 used in this work as a function of the effective 1.1\,mm 1-$\sigma$ sensitivity (i.e., rms of pixel noise converted to the 1-$\sigma$ sensitivity at 1.1\,mm assuming a modified blackbody with $\beta=1.8$; pixel size varies and is about 0.2$\times$ the beam size of each ALMA cleaned image). 
The orange line represents our next data release (corresponding to the magenta circles in Fig.~\ref{Figure_COSMOS_coverages}). 
The area and sensitivity of three representative contiguous ALMA deep field surveys from \citet[][see also \citealt{Walter2016}]{Aravena2016}, \citet{Dunlop2017} and \citet{Franco2018} are shown as black symbols for comparison, as well as the 1-$\sigma$ rms of super-deblended SCUBA2 photometry in the GOODS-North deep field from \citet{Liudz2017} (gray diamond).
\label{Figure_A3COSMOS_Sensitivity_vs_Area}%
}
\end{figure}

In Fig.~\ref{Figure_A3COSMOS_Sensitivity_vs_Area}, we compare the depth and areal coverage of the ALMA archival data in COSMOS at ALMA Band 6 and 7 to the selected existing contiguous ALMA continuum deep fields: \citet[][see also \citealt{Walter2016}]{Aravena2016}; \citet{Dunlop2017}; and \citet[][PI: D. Elbaz]{Franco2018}. 
Other ALMA deep fields (e.g., \citealt{Hatsukade2016}; \citealt{Umehata2017}) have similar properties and are therefore not shown. 
We compute the depth of each ALMA image by converting its rms noise to an equivalent flux at observed-frame 1.1\,mm assuming a modified blackbody with $\beta=1.8$. The green (orange) curve represents the cumulative area of version \incode{20180102} (version \incode{20180801}) ALMA images reaching a given sensitivity. 
Given that the ALMA archival data in the COSMOS field alone cover a larger cumulative area at all sensitivities, a systematic mining of the ALMA archive is strongly motivated. 
Given the inhomogeneous science goals of the individual PI-led projects, the resulting catalogs will not have a well-characterized selection function and are not complete per se (see Sect.~\ref{Section_Final_Galaxy_Sample_and_Properties} for a discussion of the properties and completeness of the final galaxy catalog).

In the following sections, we describe the reduction and processing of ALMA raw data into image products (in Sect.~\ref{Section_ALMA_Continuum_Images}) and the photometric methods used (in Sects.~\ref{Section_Blind_Source_Extraction}~to~\ref{Section_Prior_Source_Fitting}). 
We employ two complementary photometric methods, blind source extraction and prior source fitting, to obtain source flux densities and sizes from the ALMA continuum images. 
A comparison of the two methods and further technical assessments are presented in Sects.~\ref{Section_Photometry_Quality_Check_1}~to~\ref{Section_Photometry_Quality_Check_3}.

\subsection{ALMA Continuum Images}
\label{Section_ALMA_Continuum_Images}

We start by querying the ALMA archive with the \incode{Python} package \incode{astroquery} \citealt{astroquery}, retrieving all projects publicly available within a search radius of 2 degrees centered on the COSMOS field. 
These datasets are calibrated with the Common Astronomy Software Applications (\CASA{}; \citealt{McMullin2007}) using the \incode{scriptForPI.py} scripts provided by the Joint ALMA Observatory together with the archived raw data. 

Calibrated visibilities are imaged and ``cleaned'' --- i.e., deconvolved with the ``dirty'' beam --- with the \CASA{} imaging pipeline version \incode{4.7.2}. 
With this systematic approach, we aim at obtaining data products as homogeneous as possible and also with maximized sensitivities. 
The pipeline is operated in ``continuum'' + ``automatic'' mode, leaving all but the weight parameters (set to ``Briggs'' with \incode{robust=2}) to their default values. 
In this mode, the spectral windows (SpWs) of each target are aggregated into a single continuum image calculated at the central frequencies of these SpWs using the multi-frequency synthesis (\incode{MFS}) algorithm with \incode{nterms=2}. 
The parameters controlling the deconvolution process (i.e., masked pixels, maximum number of iterations and stopping threshold) are automatically and homogeneously set by the pipeline based on the noise properties and dynamic range of the ``dirty'' images (i.e., before deconvolution). 
The output images sample the synthesized beam with 5 pixels and are masked where the PBA is $<0.2$. 
In case of obvious image artifacts in the cleaned images (as found by visual inspection, $<10\%$), we rerun the \CASA{} imaging pipeline flagging corrupted baselines and/or adopting \incode{robust=0.5}. 

The imaging pipeline uses masks to identify regions of bright emission prior to cleaning and the stopping criterion is set to a signal-to-noise ratio (S/N)~$=$~4. Given this approach, combined with the sparseness of high S/N~$>$~15 sub-mm sources in our catalog, we do not expect any overcleaning resulting in artificially low rms noise. Given the large number of antennas in the 12m array, the instantaneous dirty beam for a short integration ($\sim$30 seconds) as used for many programs has very low sidelobes, the presence of imaging artifacts is also minimized. The robustness of our cleaning process is also supported by our comparison of image-plane to $uv$-plane photometry (Sect.~\ref{Section_Photometry_Quality_Check_2}).

The \CASA{} imaging pipeline unfortunately could not be run for a few of our projects (mostly from Cycle 0) owing to backward compatibility issues. 
These projects are thus imaged with the \CASA{} task \incode{clean} with input parameters manually set using a similar imaging and ``cleaning'' strategy to the \CASA{} imaging pipeline, e.g., ``Briggs'' weighting with \incode{robust=2}, sampling of the synthesized beam by $\sim$5 pixels and masking based on the noise of the dirty image.

\begin{figure}[htb]
\centering%
\includegraphics[width=\linewidth, trim=0mm 0mm 0 0]{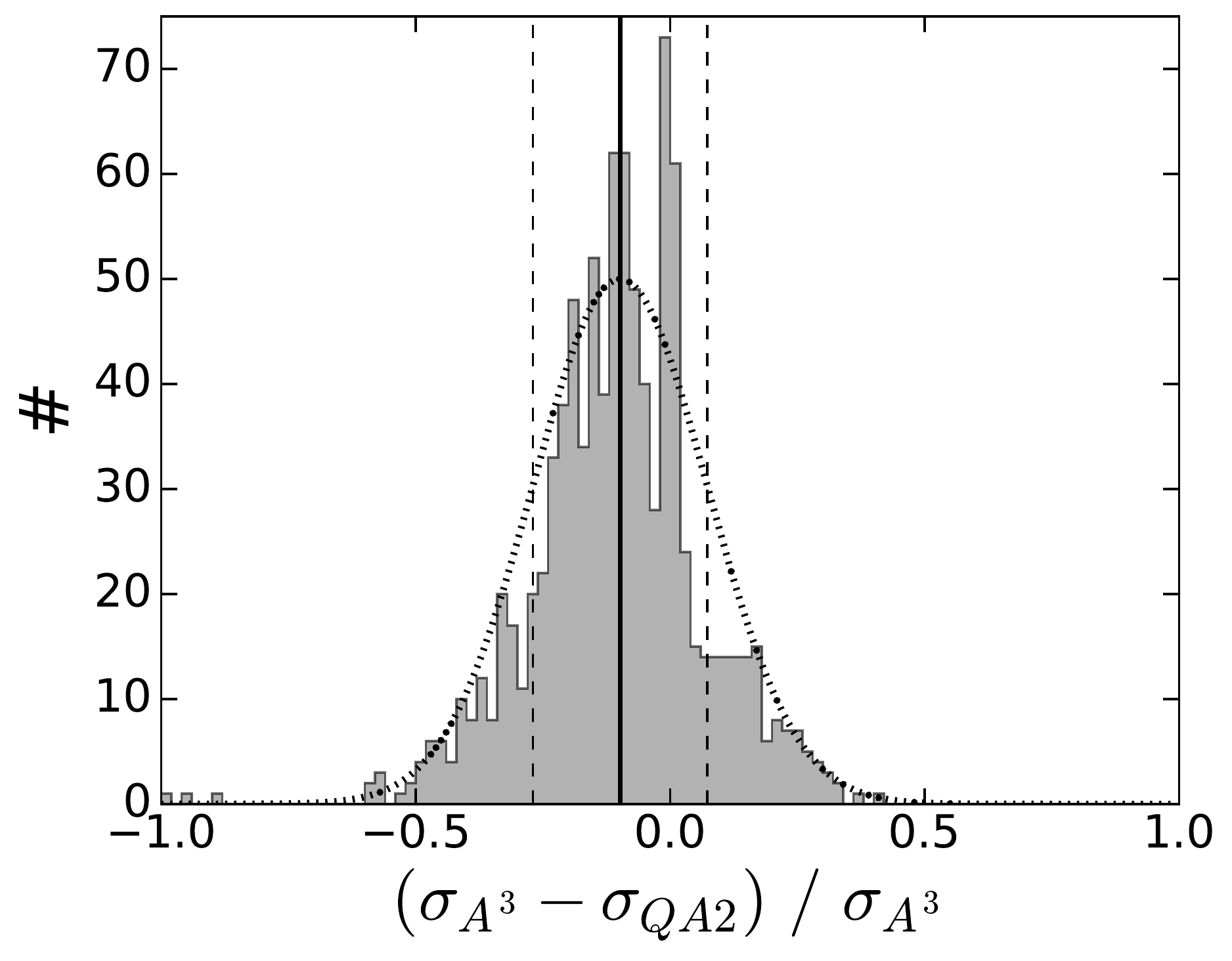}
\caption{%
Comparison of the 1$-\sigma$ sensitivities of our images (i.e., $\sigma_{\rm A^3}$) to those measured in continuum maps from the ALMA archive (i.e., $\sigma_{\rm QA2}$).
The dark histogram shows the $(\sigma_{\rm A^3} - \sigma_{\rm QA2}) / \sigma_{\rm A^3}$ distribution, while the vertical continuous and dashed lines represent its mean ($\sim-0.1$) and dispersion ($\sim0.17$).
A Gaussian distribution with similar characteristics is shown as a dotted line.
\label{Plot_QualityPipeline}
}
\end{figure}

As a test, we measure the 1$\,\sigma$ sensitivities (rms noise) of our images ($\sigma_{\rm A^3 }$) and compare with those measured in the continuum images available in the ALMA archive which are produced during the phase 2 of the ALMA Quality Assessment (QA2; $\sigma_{\rm QA2}$). 
Approximately 60\% of our images have QA2-based continuum images, while the remaining $\sim40\%$ were not imaged during the QA2 mostly because they are part of the scheduling blocks (SBs) with multiple targets, and the quality assessment was performed by imaging only a few of them. 
Consequently, for most projects, at least one QA2-based continuum image is available to perform our test. 
The $(\sigma_{\rm A^3 } - \sigma_{\rm QA2}) / \sigma_{\rm A^3 }$ follows a Gaussian distribution centered at $\sim-0.1$ with a dispersion of $\sim0.17$ (Fig.~\ref{Plot_QualityPipeline}). 
Our images have $\sim$10\% better sensitivities than those from the ALMA archive because they are produced with \incode{robust=2} -- i.e., favoring sensitivity over spatial resolution -- while most QA2 analyses are performed with \incode{robust=0.5}. 
We find no outliers with large positive values (e.g., $(\sigma_{\rm A^3} - \sigma_{\rm QA2}) / \sigma_{\rm A^3}>0.6$), as images with obvious artifacts were spotted by our visual inspection and already re-imaged. 
Finally, we find few outliers with $(\sigma_{\rm A^3COSMOS} - \sigma_{\rm QA2}) / \sigma_{\rm QA2}<-0.6$. 
We systematically checked these images and found that all of them correspond to projects in which the QA2-based analysis was performed with low \incode{robust} values (i.e., $<0$) and/or using only a fraction of the SpWs available. 
All these comparisons demonstrate the reliability of the ALMA imaging pipeline and thus of our image products.

\begin{figure}[htb]
\centering%
\includegraphics[width=\linewidth, trim=0 7mm 0 0]{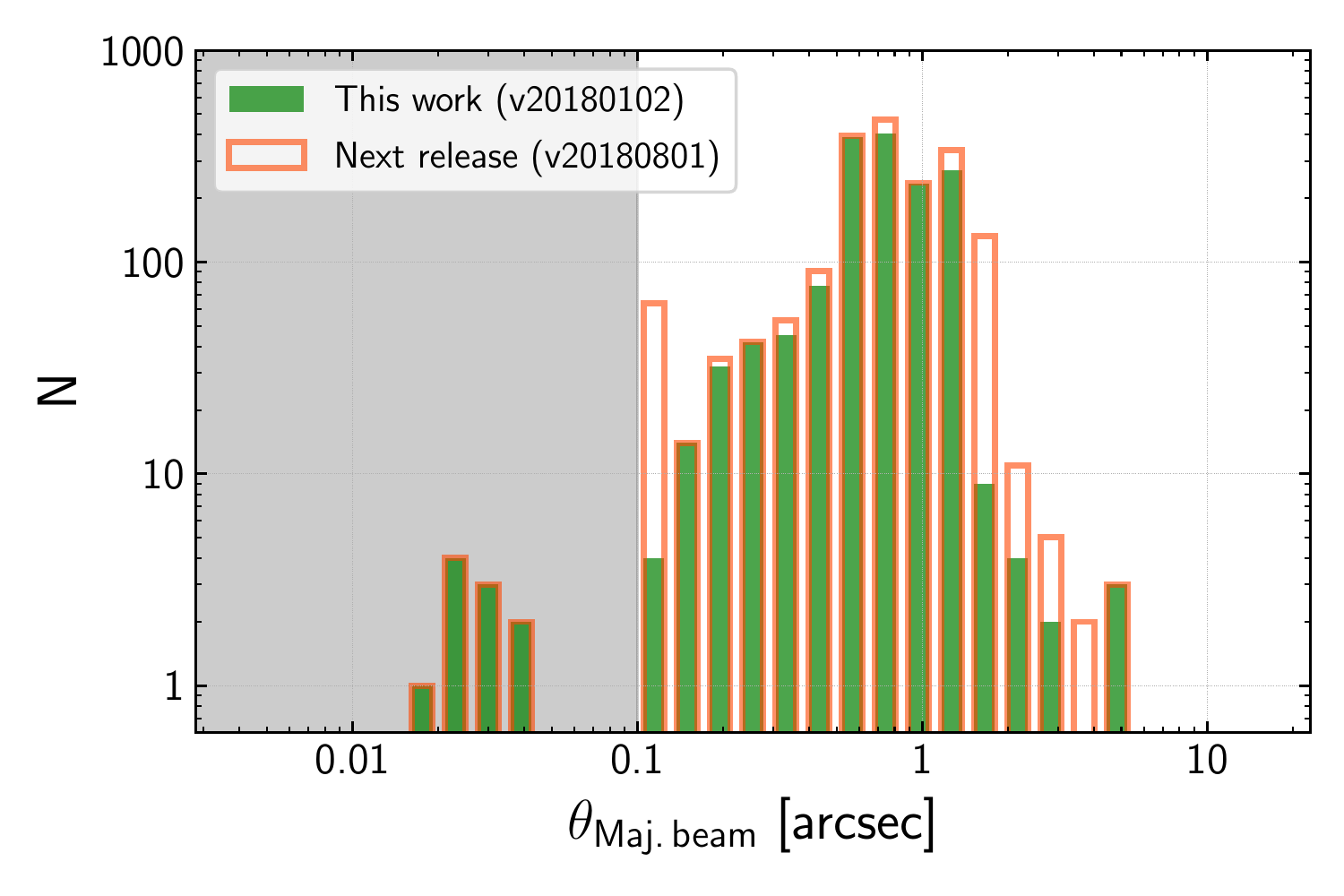}
\caption{%
Beam size distributions of two versions of A$^3$COSMOS data sets. The beam size is defined as the FWHM along the major axis of ALMA data's synthesized beam ($\theta_{\mathrm{Maj.,\,beam}}$) after the interferometric ``cleaning'' process. The gray shaded area indicates $\theta_{\mathrm{Maj.,\,beam}}<0.1''$, for which data were discarded owing to too high spatial resolution as discussed at the end of Sect.~\ref{Section_ALMA_Continuum_Images}. 
\label{Figure_A3COSMOS_Beamsize_Histogram}%
}
\end{figure}

The data products released here include all ``clean'' continuum images corrected and uncorrected for primary beam attenuation (PBA)\,\footnote{The PBA corrections are due to the non-uniform sensitivity within the Gaussian-approximated FWHM of the primary beam for each antenna.}. 
Note that although the aggregation of all SpWs available for a given target optimizes the sensitivities of our continuum images, it does not consider any possible line contamination. 
In Sect.~\ref{Section_Subtracting_Strong_Emission_Lines} we will further describe an effective approach to addressing potential line contamination. 

Furthermore, in Fig.~\ref{Figure_A3COSMOS_Beamsize_Histogram}, we show the distribution of the angular resolution of the ALMA data, as represented by the major-axis FWHM of ALMA data's synthesized beam $\theta_{\mathrm{Maj,\,{beam}}}$. 
Of current images, $\lesssim1\%$ have a very high angular resolution, i.e., $\theta_{\mathrm{Maj,\,{beam}}}<0.1''$. These images represent a more challenging case for our source extraction because our blind and prior source extraction methods are all optimized for only marginally resolved sources, while sources in the very high resolution images usually are significantly resolved (e.g., with a ratio of source to beam area $\gtrsim 10$). Also note that the large number of independent beams within these images ($\propto \mathrm{FoV}\,/\,\theta^2_{\mathrm{Maj,\,{beam}}}$) statistically translates into a significant contamination of ``spurious'' sources to our photometry catalog (even using a conservative $\SNR\sim5$ cut; see Sect.~\ref{Section_Spurious_Fraction}). Therefore, these $\lesssim1\%$ very high resolution ($\theta_{\mathrm{Maj,\,{beam}}}<0.1''$) images are excluded from our analysis. 

Currently, data for the same source taken at the same frequency arising from different projects are not combined.

The breakdown of the number of objects detected, the expected number of false objects, the area, median depth and resolution as a function of observing band are provided in Table~\ref{Table_per_band_info_table}.

\input{Table_per_band_info_table.tex}

\subsection{Blind Source Extraction}
\label{Section_Blind_Source_Extraction}

We perform the blind source extraction on our ``cleaned'' ALMA continuum images. We use the primary-beam-attenuation-uncorrected images because they have the advantage of a constant noise across the field of view, and thus source extraction can be run with uniform parameters across them. The primary beam attenuation corrections are applied after the photometry steps. 

We use the \textsc{Python Blob Detector and Source Finder} (aka \textsc{PyBDSM} or \textsc{PyBDSF}; hereafter \pybdsm{}; \citealt{PyBDSF})\,\footnote{\pybdsm{} documentation: \url{http://www.astron.nl/citt/pybdsf/index.html}; and its source code: \url{https://github.com/lofar-astron/PyBDSF}.}
to find sources blindly and extract their flux and size information. 
First, the code identifies ``islands'' of emission, i.e., with the peak pixel emission above 4 times the rms noise (\incode{thresh_pix}~$=4$), and surrounded by contiguous pixels with values all greater than 3 times the rms noise (\incode{thresh_isl}~$ = 3$). These thresholds are obtained from a series of tests by introducing mock sources into the ALMA images and recovering them with \pybdsm{}. The best performance was evaluated based on completeness and contamination (see Sects.~\ref{Section_Spurious_Fraction}~and~\ref{Section_MC_Sim_Completeness}). 
Next, \pybdsm{} fits multiple two-dimensional Gaussians to each ``island'' depending on the number of peaks identified within it. Thirdly, all Gaussians of the same ``island'' are grouped into one source, with the summed flux being the integrated source flux, the flux-weighted averaged position being the source position. The total intrinsic source size is obtained via a moment analysis\,\footnote{See details in \url{http://www.astron.nl/citt/pybdsm/process_image.html}.} on each individual Gaussian component's intrinsic size\,\footnote{Each Gaussian component's intrinsic size is their fitted Gaussian size deconvolved with the clean beam which is a two-dimensional Gaussian, and the deconvolution follows the Astronomical Image Processing System (AIPS; \citealt{AIPS}) \incode{DECONV.FOR} module (see also \citealt{Spreeuw2010PhDT}, Chapter~2).}. Finally, the errors of each fitted parameter (peak flux, total flux, and each size parameter) are computed using the formulae of Gaussian fitting errors calibrated by \citet{Condon1997}.

About 6\% of our ``islands'' are fitted with multiple Gaussians, while the vast majority (94\%) have a single Gaussian component. Most of these multi-Gaussian sources are isolated sources but exhibit non-smooth morphologies, either due to resolved spatial components and/or noise in the image. In a few cases, these multi-Gaussian sources might indeed be interacting galaxies. Utilizing information from the prior source catalog, we are able to reliably flag these sources in a later step of our analysis. Therefore, they are kept as a single source in the blind source catalog. 

Our final blind source catalog is obtained by correcting for flux bias and re-estimating flux errors (see Sect.~\ref{Section_Monte_Carlo_Simulation_and_Correction}), then applying a primary beam attenuation correction to the photometry of each source (i.e., to the peak flux, total flux, and associated uncertainties). Given that each source represents a high-redshift galaxy with a typical size of 0.5--2$''$, much smaller than the primary beam, using a single primary beam attenuation correction factor at the source's central position is reasonable.

\subsection{Prior Source Master Catalog}
\label{Section_Prior_Source_Catalogs}

\begin{figure}[htb]
\centering%
\includegraphics[width=\linewidth, trim=10mm 4mm 0 0]{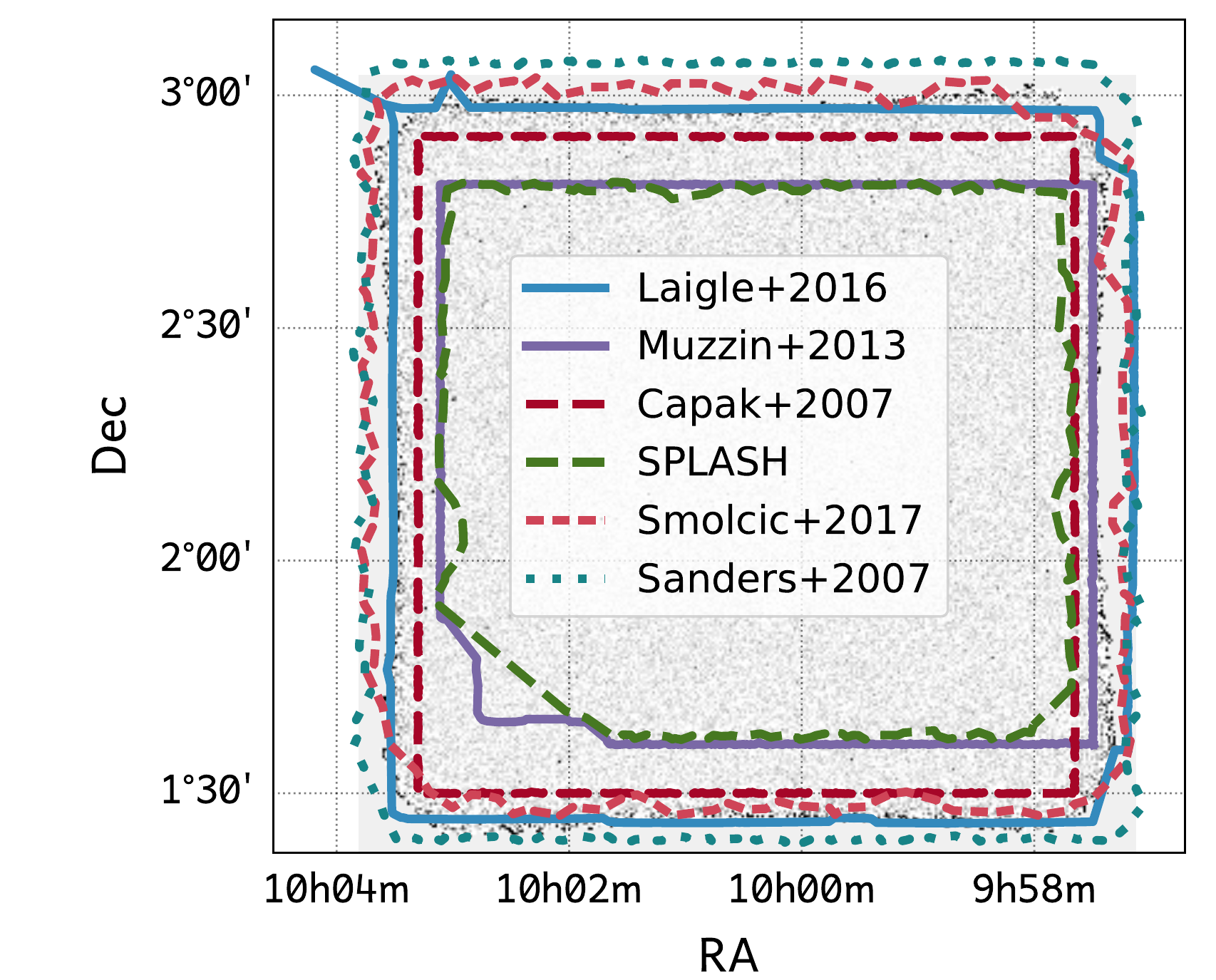}
\caption{%
Coverage of our prior catalogs for the COSMOS field as listed in Table~\ref{Table_prior_catalogs}. Background image is the \textit{Herschel} PACS 100\,$\mu$m image, same as in Fig.~\ref{Figure_COSMOS_coverages}. The colored lines encompass the area searched for sources in the respective prior catalog (see inset for catalog information). 
\label{Plot_catalog_sky_coverage}
}
\end{figure}

In addition to the blind source extraction, we utilize known source positions as a prior for the source fitting. This technique allows for deeper detection limits and lowers the spurious source fraction. 
Before starting the prior fitting, we compiled a ``COSMOS master catalog'' from a number of multi-wavelength catalogs for sources in the COSMOS field as listed in Table~\ref{Table_prior_catalogs}. The aim is to be as complete as possible in prior sources while ensuring that source duplication is solved among the various catalogs. 
Thus we loop over the prior catalogs in the order listed in Table~\ref{Table_prior_catalogs}. Their respective areal coverage is indicated in Fig.~\ref{Plot_catalog_sky_coverage}. 
To ensure that a given galaxy (which might be detected in multiple prior catalogs) has only one unique entry in the master catalog, 
we find out each uniquely-matched group among the prior catalogs (with matching radius 1$''$) and add into the master catalog only the source coming from the highest-quality (empirically sorted by angular resolution and relative depth) catalog, i.e. listed closest to the top in Table~\ref{Table_prior_catalogs}. 

Our $1''$ matching radius corresponds to a worst false-match probability of 13.3\% for other catalogs cross-matched to the \cite{Laigle2016} catalog based on Eq.~1 of \cite{Pope2006}\,\footnote{As an additional experiment, we estimated the false-match probability to be 9.9\% by first flipping the catalog to be cross-matched to the \cite{Laigle2016} catalog in R.A. positions, and then we did the cross-match.}. 
We emphasize that the false-match rate does not affect our photometric work because if a galaxy from another catalog in Table~\ref{Table_prior_catalogs} is falsely matched to the \citet{Laigle2016} catalog, we just use the prior position in the \citet{Laigle2016} catalog for our prior photometry. The source position is not forced to be at the exact prior position, as our photometry code will find the best fitting for position and flux (see next section). 
It may affect our galaxy property analysis via SED fitting in a later step because we use the redshift information from the literature as the prior. However, the influence is minimized by (a) collecting all possible prior redshift information in the literature, (b) verifying via photo-$z$ SED fitting (see Sect.~\ref{Section_Running_SED_Fitting}), and, in later steps, (c) only considering a source a robust galaxy if it passes all our quality assessments (Sect.~\ref{Section_Galaxy_Sample_and_Properties}). A falsely-matched source with a wrong prior redshift is unlikely to pass them as detailed in Sect.~\ref{Section_Galaxy_Sample_and_Properties}. 
Yet we cannot totally avoid false matches (which should be only a few out of a thousand in our final products), especially when the astrometry in optical/near-IR image data also affects our work (see next section and Appx.~\ref{Section_Astrometry}).

\input{Table_prior_catalogs_tabular.tex}

The combination of these prior catalogs results in the ``COSMOS master catalog'' with unique source IDs. In our current master catalog (version \incode{20170426}), because the COSMOS2015 catalog is our primary catalog, all 1,182,108 COSMOS2015 sources are in our ``COSMOS master catalog'' with the same IDs. A total of 443,688 (37.5\%) of them have counterparts in other catalogs. The remaining five catalogs contribute 110,768 new sources that are not in the COSMOS2015 catalog. The \cite{Muzzin2013} catalog contributes 18,536 sources, a fraction of which are from the COSMOS2015 masked regions close to bright stars. The $K_s$ data used for the \cite{Muzzin2013} catalog are shallower than those of the COSMOS2015 catalog (UltraVISTA DR2), so the reason for some new sources should be the different source extraction methods used: the COSMOS2015 catalog uses a $z,\,Y,\,J,\,H,\,K_s$-combined $\chi^2$ detection image while the \cite{Muzzin2013} catalog directly uses the $K_s$ image and therefore favors redder sources. 
The $i$-band-selected catalog contributes 31,159 sources, probably benefiting from its higher angular resolution detection image (see discussion in Sect.~4 of \citealt{Capak2007}). 
The IRAC catalog contributes another 4,685 sources. 
The radio catalog contributes 1,042 sources. 
Finally, the \cite{Sanders2007} catalog contributes 55,346 sources, but most of them are in an area outside the COSMOS2015 coverage (e.g., Fig.~\ref{Plot_catalog_sky_coverage}), while only 8,893 sources are new in the area covered by both catalogs. 

The total number of unique priors that fall in primary beam attenuation~$>0.2$ areas of our dataset version \incode{20180102} (\incode{20180801}) is 41,161 (73,387).

\subsection{Prior Source Fitting}
\label{Section_Prior_Source_Fitting}

Utilizing source positions from the COSMOS master catalog, we obtain the (sub-)mm photometry via prior source fitting of the ALMA continuum images. We implement two steps below to optimize the robustness of the fitting. Potential small astrometric inconsistencies between the prior source positions and the ALMA data are taken into account as follows before the full prior source fitting procedure is applied: we calculate the offsets between pre-run ALMA positions and the prior source positions directly from \citet{Laigle2016} and other prior catalogs, and then we derive a mean offset for each prior catalog and update all prior positions in our master catalog. Details of the astrometry analysis are given in Appx.~\ref{Section_Astrometry}.

As a first step, we identify potential candidate sources based on the S/N of their peak  (sub-)mm flux density ($\Speak$)
or integrated flux density ($\Stotal$). Following \cite{Scoville2014,Scoville2016,Scoville2017}, we measure
both flux densities in a series of apertures with radii from 0$\farcs$25 to 2$''$ in steps of 0$\farcs$25. 
We follow exactly the \cite{Scoville2016,Scoville2017} method so as to allow for a direct comparison.
Using the pixel rms noise calculated from Gaussian fitting to the pixel value distribution of each image, we obtain the S/N ratio for the peak flux density via $\SNRpeak \equiv \Speak/(\noise)$, and the one for integrated flux density $\SNRtotal$ by dividing $\Stotal$ by the integrated noise in each aperture (i.e., $\noise$ times the square root of pixel number in each aperture). We refer to this aperture photometry as the \getpix{} method hereafter (and compare its results with those from our other photometry methods in Sect.~\ref{Section_Photometry_Quality_Check_3}). 

This first \getpix{} step also provides guidance for the prior source fitting using \galfit{} \citep{Peng2002,Peng2010} in the next step. 
We select $\SNRtotal>2$ or $\SNRpeak>3.6$ sources (same as \citealt{Scoville2017}) as valid detections. This pre-selection of prior sources is important for applying \galfit{}, as it significantly reduces the required computational time for \galfit{} by avoiding the fitting of sources that mostly correspond to noise in the image. We have confirmed that this approach is sensible with our MC simulations (see Sect.~\ref{Section_Monte_Carlo_Simulation_and_Correction}). 
Also, our final catalogs are not sensitive to small changes of these thresholds, because in the end we apply a relatively high $\SNRpeak$ cut according to our MC simulation statistics. Note that in most ALMA images our priors do not have blending issues.

To optimize the \galfit{} fitting for source fluxes as well as sizes, an iterative approach is adopted: After the first-pass fitting with point-source models to all \galfit{} priors fixed at their original positions, we select sources with a fitted magnitude error of $<0.25$\,\footnote{\galfit{} fits magnitude instead of flux density} or $\Stotal>3\,\sigma$ ($\sigma$ being the pixel $\noise$) and allow their positions to vary by at most 0$\farcs$7\,\footnote{This is the 1\,$\sigma$ scatter of the spatial separations between our ALMA sources and their optical/near-infrared counterparts as we examined in Sect.~\ref{Section_Examining_counterpart_association}.} in the second-pass fitting. 
Then, in order to identify possible extended sources, we allow sources with fitted magnitude error $<0.20$ or $\Stotal$ above 3 times the rms noise to be fitted with circular Gaussian models (and in a next step S\'{e}rsic profiles) in the third-pass fitting. 
We note that our thresholds are very loose, and 98\% of the sources in our final prior photometry catalog (with a relatively high selection threshold, $\SNRpeak \gtrsim 5$, according to our MC simulation statistics, see Sect.~\ref{Section_Combining_two_photometry_catalogs}) are fitted with extended shapes.

For each fit, we ensure that the image background is zero (as already verified by the close-to-zero means of the distributions of the pixel values from the ALMA images). If a given \galfit{} iteration yields bad fits and/or non-convergence, the fitting is repeated with a higher limit for \galfit{} iteration\,\footnote{By default, \galfit{} iterates a maximum for a total of 100 times, and 10 times when converging to a local minimum. These numbers can be increased to, for example, 1000 total iterations and 255 iterations during convergence, e.g., \cite{Liudz2017}.}.

As the \galfit{} errors only consider the covariance matrix of the fitting, they do not reflect observational noise or correlated noise. Therefore, we estimate the error in $\Stotal$ ($\EStotal$) for Gaussian-fitted sources following \cite{Condon1997}. 
This error estimation determines $\EStotal$ purely from the rms noise, beam major and minor axes FWHM sizes ($\theta_{\mathrm{bmaj.}}$ and $\theta_{\mathrm{bmin.}}$ respectively), source major and minor axes FWHM sizes ($\theta_{\mathrm{maj.}}$ and $\theta_{\mathrm{min.}}$ respectively; fitted values and convolved with the beam) and source $\Speak$ and $\Stotal$. 
We further verify that this is in general consistent with our own MC simulations (see Sect.~\ref{Section_Monte_Carlo_Simulation_and_Correction}).

\subsection{Comparing Blind Extraction and Prior Fitting Results}
\label{Section_Photometry_Quality_Check_1}

\begin{figure}[htb]
\centering%
\includegraphics[width=\linewidth, trim=10mm 4mm 2mm 2mm]{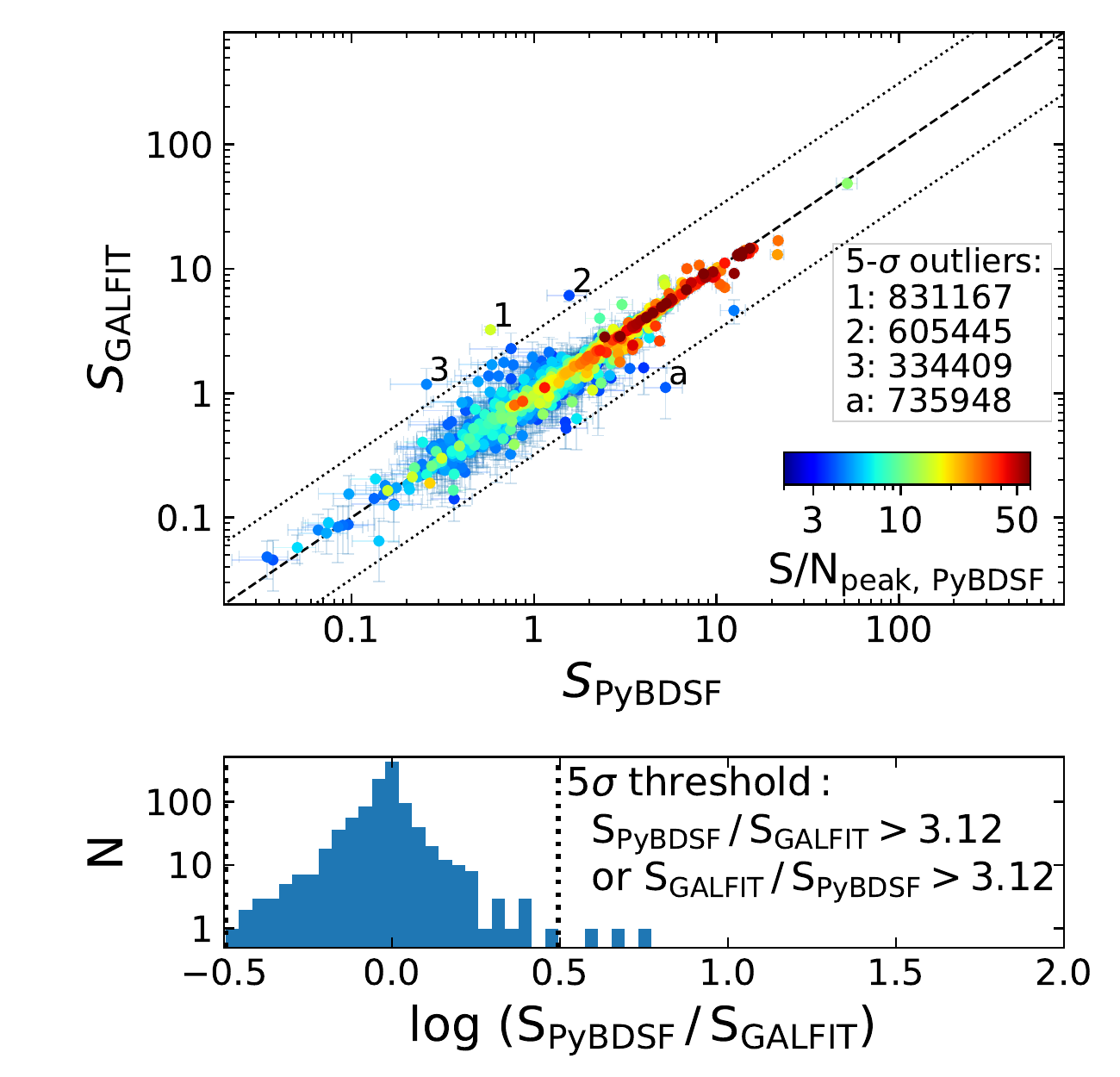}
\caption{%
\textbf{Top:} Comparison of total fluxes derived from the \pybdsm{} blind source extraction and the prior-based \galfit{} source fitting. Data points show sources matched within 1$''$ and measured on the same image. Color indicates their $\SNRpeak$, i.e., the ratio of source peak flux to rms noise of the image. The solid line shows the one-to-one correspondence and the two dashed lines indicate the $5\,\sigma$ range, where $\sigma$ is the scatter measured from the bottom panel. Outliers above $5\,\sigma$ are labeled and discussed in detail in Appx.~\ref{Section_Appendix_outliers_of_photometry}. 
\textbf{Bottom:} Histogram of the flux difference on a logarithmic scale, $\log_{10} (S_{\mathrm{PyBDSF}}/S_{\mathrm{GALFIT}})$. The mean is $-0.013$ with a standard deviation of 0.10\,dex. Dashed vertical lines indicate the same $5\,\sigma$ range as in the top panel. 
\label{Plot_catalog_comparison_a3cosmos_prior_vs_blind_total_flux}%
}
\end{figure}

As a quality check to both blind source extraction and prior source fitting, and to identify potential problem cases, we compare the total fluxes from \pybdsm{} to those from \galfit{} for common sources (within 1$\farcs$0 and using the same ALMA images) in Fig.~\ref{Plot_catalog_comparison_a3cosmos_prior_vs_blind_total_flux}. 96\% of sources have fluxes agreeing within 3\,$\sigma$.
Outliers with flux differences of $>5\,\sigma$ are labeled in the figure. 
Their \pybdsm{} and \galfit{} fitting models and residuals are further shown and discussed in Appx.~\ref{Section_Appendix_outliers_of_photometry}. 
The three outliers with a \galfit{} flux much larger than the \pybdsm{} flux are caused by poor fits of \pybdsm{} to their irregular morphologies. 
The one outlier with a much larger \pybdsm{} flux than the \galfit{} flux is due to a blending of prior sources
and given the complex morphology, both \galfit{} and \pybdsm{} could not provide an ideal fit.

With both \pybdsm{} and prior-based \galfit{} photometry, we not only obtain accurate independent fluxes which agree very well but also identify those few (0.5\%\,\footnote{We have 0.5\% such sources in our final photometry catalogs selected according to the threshold in Sect.~\ref{Section_Combining_two_photometry_catalogs}. This fraction goes up to only 2\% if we apply a threshold of $\SNRpeak \ge 3.0$ to both catalogs.}) sources which suffer from source multiplicity/blending issues. These sources need careful visual inspections as well as multi-wavelength diagnostics (e.g., SEDs) in order to fully deblend their ALMA flux, and thus will be analyzed in a future work. 

In our released two photometry catalogs, we flag sources 
for which the total fluxes from the two methods
disagree by more than a factor of $\sim$3.12 ($5\,\sigma$, where $\sigma$ is the scatter between \pybdsm{} and \galfit{} total fluxes, see Fig.~\ref{Plot_catalog_comparison_a3cosmos_prior_vs_blind_total_flux}) with a column \incode{Flag_inconsistent_flux} and exclude them in subsequent steps. In the next sections, we use the prior photometry flux for the SED fitting. But measurements from both photometry methods will be made public together with the final galaxy SED and property catalog (see Sect.~\ref{Section_Data_Delivery}).

\subsection{Comparison to $uv$-plane Source Fitting Results}
\label{Section_Photometry_Quality_Check_2}

Instead of measuring the source flux density in the image plane, it can also be directly measured in
the $uv$-plane by fitting source models to the visibilities. 
We use the \textsc{GILDAS}\,\footnote{\textsc{GILDAS} is an interferometry data reduction and analysis software developed by Institut de Radioastronomie Millim\'{e}trique (IRAM) and is available from \url{http://www.iram.fr/IRAMFR/GILDAS/}. The conversion of ALMA measurement sets to \textsc{GILDAS}/\textsc{MAPPING} $uv$ table data follows \url{https://www.iram.fr/IRAMFR/ARC/documents/filler/casa-gildas.pdf}.}
\incode{uv_fit} task to fit Gaussian and/or point-source models then compare the total flux with those measured from the image-plane \galfit{} and \pybdsm{} fitting. 
We verified that \textsc{GILDAS} \incode{uv_fit} gives similar results to the \CASA{} \incode{uvmodelfit} task for high-$\SNR$ sources (e.g., total flux $\SNR>10$). 

We run \textsc{GILDAS} \incode{uv_fit} in an iterative approach: first we fit point-source models, and next for high-$\SNR$ sources we fit extended Gaussian source models. We fit only for one source at the phase center and allow its position to vary freely by \incode{uv_fit}. In total we ran the $uv$-fitting for 
301 pointings from four representative ALMA projects: 2015.1.00137.S, 2013.1.00151.S, 2015.1.00379.S, and 2016.1.01208.S (these projects target the dust continuum for hundreds of galaxies from redshift 1 to 3; the PI of the first project is N. Scoville, and the PI for the other three is E. Schinnerer). 
The \incode{uv_fit} flux densities and the prior-based and blind (sub-)mm photometries agree very well. The difference between blind photometry and \incode{uv_fit} flux densities (on a logarithmic scale) has a median of 0.015\,dex and scatter of 0.08\,dex. The difference between prior photometry and \incode{uv_fit} flux densities has a median of -0.005\,dex and scatter of 0.13\,dex, showing a few more outliers (caused by blended priors, same as in Fig.~\ref{Plot_catalog_comparison_a3cosmos_prior_vs_blind_total_flux}).

\subsection{Comparison to Aperture Photometry Results}
\label{Section_Photometry_Quality_Check_3}

We further compare the fluxes from our \pybdsm{} and prior-based \galfit{} fitting with those derived from aperture photometry
\citep{Scoville2016,Scoville2017} (i.e. the \getpix{} method described in Sect.~\ref{Section_Prior_Source_Fitting}). 
For sources with \galfit{} $\SNRpeak\ge3$, the \getpix{} method provides flux densities consistent with the ones from \galfit{} (the mean of \getpix{} to \galfit{} flux ratio on a logarithmic scale is 0.004\,dex and the scatter is 0.18\,dex). Sources with $\SNRpeak>10$ are on average biased toward higher \getpix{} flux densities, but no more than 10\% (the mean value increases to 0.03\,dex and scatter 0.06\,dex; likely due to bright outlier sources that have non-Gaussian shapes).

The comparison between \getpix{} and \pybdsm{} flux densities yields similar results: for \pybdsm{} $\SNRpeak \ge 3$ sources the mean of $\log_{10} S_{\mathrm{GETPIX}}/S_{\mathrm{PyBDSF}}$ is 0.003\,dex with a scatter of 0.15\,dex; when considering only $\SNRpeak \ge 10$ sources the mean is still <0.01\,dex. 
For about 10 sources, we directly compared our flux densities to measurements from
\citet[][priv. comm.]{Scoville2016,Scoville2017}, finding similar results to those mentioned above.

\begin{figure}[thb]
\centering%
\includegraphics[width=0.46\textwidth, trim=5mm 13mm 15mm 0]{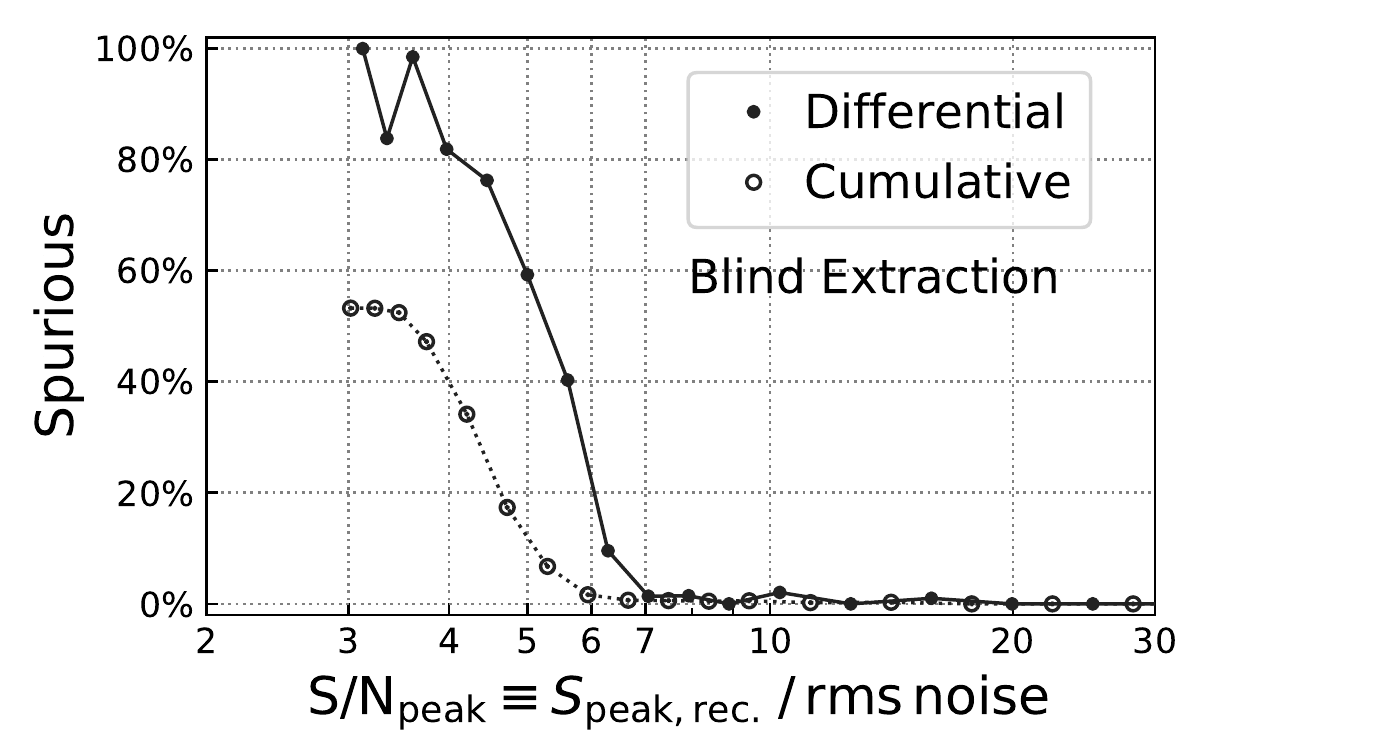}
\includegraphics[width=0.46\textwidth, trim=5mm 3.5mm 15mm 0]{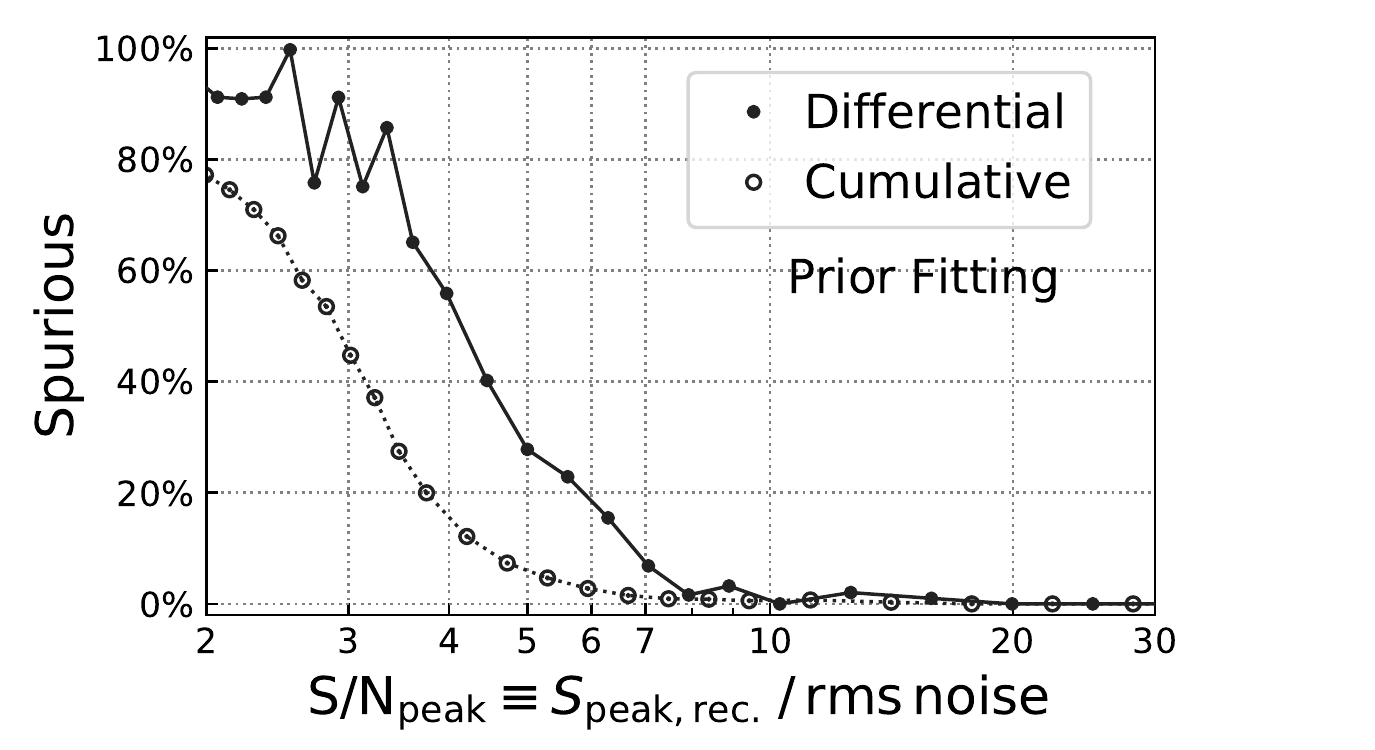}
\caption{%
    Fraction of spurious detection for the \pybdsm{}-based blind source extraction (upper panel) and \galfit{}-based prior source fitting (lower panel). The solid curves and filled data points represent the differential spurious fraction at each $\SNRpeak$ bin, while the dotted curves and open data points represent the cumulative values, i.e., for $\SNRpeak \ge$ the current bin's $\SNRpeak$. 
    \label{Plot_spurious_fraction}%
}
\end{figure}

\begin{figure}[thb]
\centering%
\includegraphics[width=0.46\textwidth, trim=5mm 3.5mm 15mm 0]{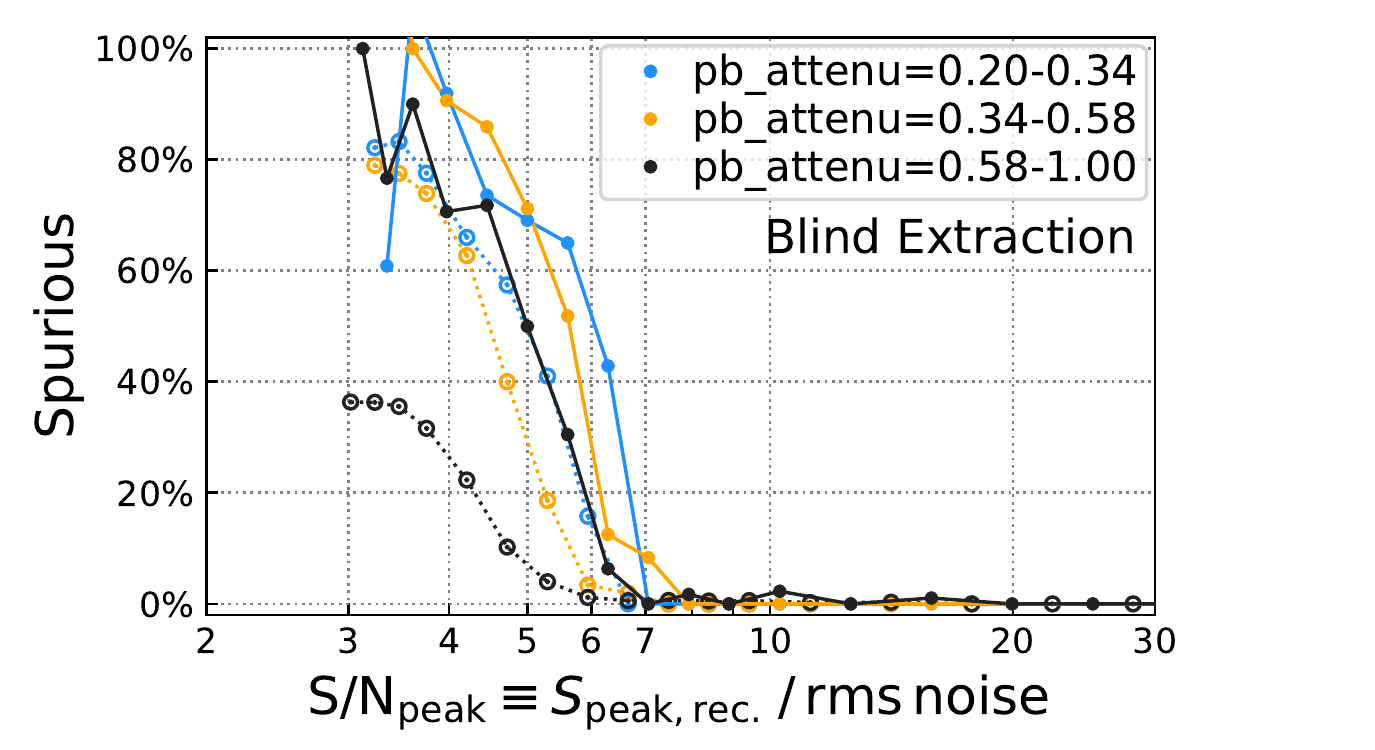}
\caption{%
    Spurious fraction for three bins of $\incode{pb_attenu}$: (0.2--0.34], (0.34--0.58] and (0.58--1.0], which are equally distributed in logarithm. The solid and dashed lines represent differential and cumulative curves, respectively (see Fig.~\ref{Plot_spurious_fraction} caption). We caution that the trend seen here suffers from a strong bias in statistics, because the lowest $\incode{pb_attenu}$ (farthest away from phase center) bin has only about 62 $\SNRpeak>5$ detections in original images and 27 in inverted images, while the numbers are 10 times larger in the innermost bin with $\incode{pb_attenu}\sim0.58-1.0$ (although they are equal in area). 
    \label{Plot_spurious_fraction_versus_pb_attenu}%
}
\end{figure}

\subsection{Inverted-image Fitting and the Fraction of Spurious Detection}
\label{Section_Spurious_Fraction}

We run our photometry tools (based on \pybdsm{} and \galfit{}) on the inverted images (i.e., the sign of each pixel value is inverted) to estimate the fraction (and probability) of spurious detections by comparing the number of sources detected in inverted images to that in original images. We define the spurious fraction as the number of sources detected in inverted images compared to the corresponding number in the original images as a function of $\SNRpeak$ (defined as $\Speak/\noise$ in Sect.~\ref{Section_Prior_Source_Fitting}), since this quantity does not depend on any fitted source size. 

Since prior fitting needs a prior catalog to proceed with, and because our prior catalog has a very high number density ($\sim700$ per arcmin$^2$) which acts like a random sampling in the image, we directly use our COSMOS master catalog as the prior catalog for the inverted-image \galfit{} photometry. The procedure is the same as described in Sect.~\ref{Section_Prior_Source_Fitting}, we first run the \getpix{} step then iteratively run \galfit{} source fitting. In addition, we checked that the spurious detection curve remains the same when shifting the positions of the entire prior catalog by $+/-2''$ in Right Ascension and/or Declination to avoid overlap with real galaxies.

Fig.~\ref{Plot_spurious_fraction} shows the derived spurious fraction curves as a function of $\SNRpeak$ for both \pybdsm{} (top) and \galfit{} (bottom) photometry. The differential curve (solid line) indicates the spurious fraction at each $\SNRpeak$. The cumulative curve (dotted line) provides the spurious fraction summed over all bins with $\SNRpeak$ greater than or equal to the current bin. 
As expected, spurious fractions are lower for the prior-based photometry compared to the blind source extraction due to the availability of information on the presence of a galaxy. Thus the prior-based photometry achieves deeper detection limits.

\begin{figure}[thb]
\centering%
\includegraphics[width=0.46\textwidth, trim=0 0 0 0]{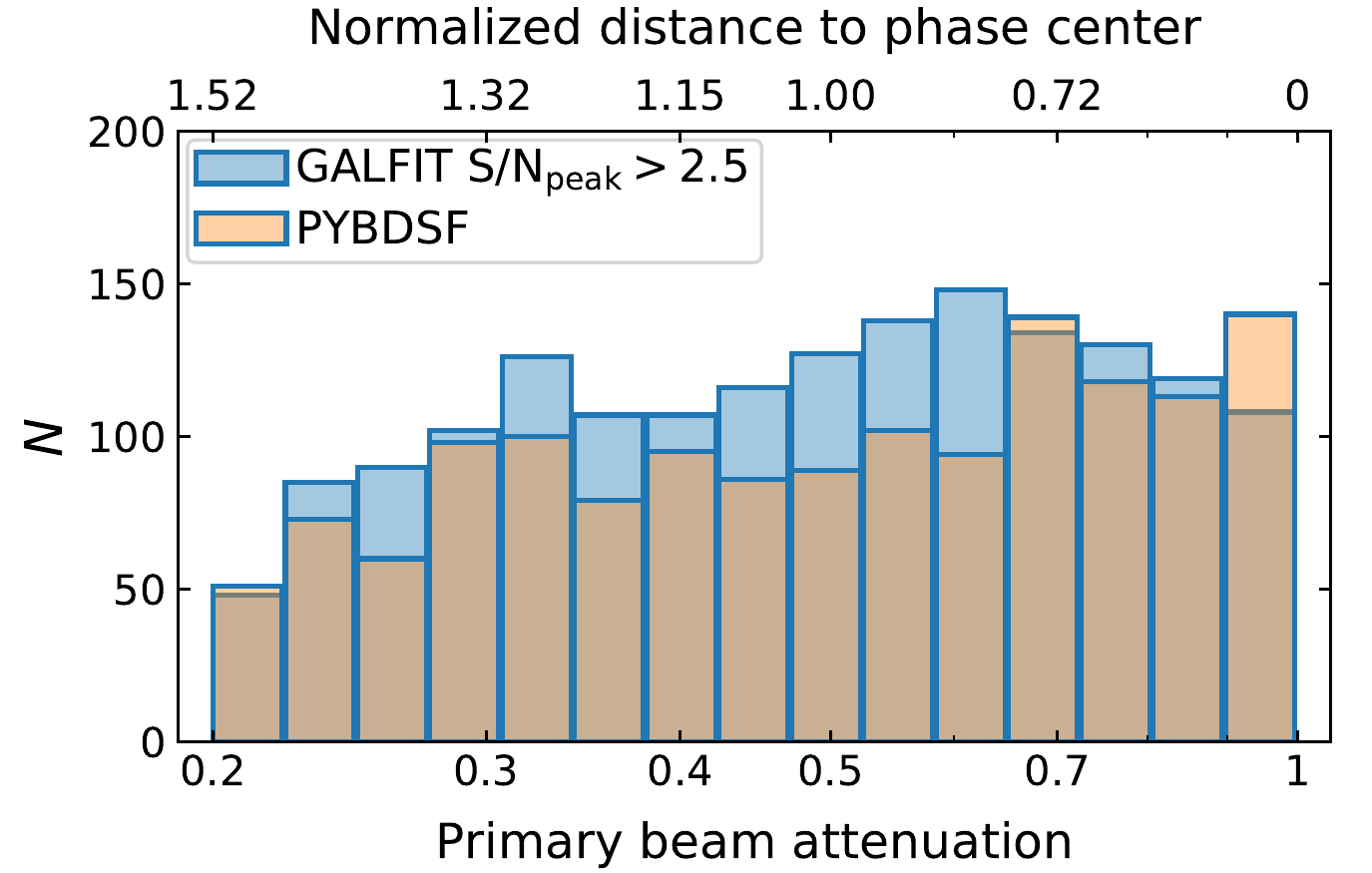}
\caption{%
    The radial distribution of spurious detections for the \pybdsm{}-based blind source extraction (orange) and \galfit{}-based prior source fitting (blue). The bottom $x$-axis is the primary beam attenuation, $\incode{pb_attenu}$, and the top axis is the normalized distance to the phase center, $\incode{dist.}/(0.5\times\mathrm{\incode{pb}})$, where $\incode{dist.}$ is the spatial distance to the phase center and $\incode{pb}$ is the FWHM of the primary beam. The histogram bins are equally distributed in $\ln (\incode{pb_attenu})$ (so that the area of each bin normalized by the primary beam area are equal). 
    \label{Plot_spurious_detection_radial_distribution}%
}
\end{figure}

To investigate whether the primary beam attenuation is affecting the false-positive detection, we have done two tests: one is dividing the spurious fraction curve in bins of primary beam attenuation ($\incode{pb_attenu}$) as shown in Fig.~\ref{Plot_spurious_fraction_versus_pb_attenu}, the other is plotting the radial distribution of all spurious detections from the inverted images in Fig.~\ref{Plot_spurious_detection_radial_distribution}. 
In the former test, we choose only three bins because of the low number of sources away from the phase center (low $\incode{pb_attenu}$). We bin in equal $\ln (\incode{pb_attenu})$ intervals which correspond to the same sky area, because $\incode{pb_attenu} \propto \exp(\incode{dist.}^2/\mathrm{\incode{pb}}^2)$, where $\incode{dist.}$ is the distance of the source to the phase center and $\incode{pb}$ is the FWHM of the primary beam. 
The spurious fraction decreases when $\incode{pb_attenu}$ becomes closer to 1.0, which is as expected. But we also caution that there is a strong bias in the statistics because the number of sources dramatically differs (see Fig.~\ref{Plot_spurious_fraction_versus_pb_attenu} caption). 
    
In Fig.~\ref{Plot_spurious_detection_radial_distribution}, we show the radial distribution of all sources detected in the inverted images with $\galfit{}$ $\SNRpeak>2.5$ or found by \pybdsm{}. Since the spurious fraction curve is slightly higher at larger radii, we might expect the spurious source density to be higher, however, the distribution remains fairly constant out to a $\incode{pb_attenu}$ of $\sim0.3$. We attribute the slight drop below $\sim0.3$ to the fact that instrumental systematics are likely becoming more prominent, namely (a) the approximation of the primary beam by a Gaussian might no longer be correct\,\footnote{E.g., see \url{https://help.almascience.org/index.php?/Knowledgebase/Article/View/234}.}, 
and (b) the frequency dependence of the primary beam across the frequency range sampled by the continuum (i.e., 16\,GHz between the upper and lower boundary of the spectral sidebands) will be more evident at large distances from the phase center. A more detailed investigation is beyond the scope of this paper.

In this work, we provide a photometry catalog out to a primary beam attenuation of 0.2 (i.e. covering the full area of the images that are made available) and provide the $\incode{pb_attenu}$ for each source in our catalog. Note that 91\% of our final selected sources lie within a primary beam attenuation of 0.5 and only 2\% beyond 0.3. Special care should be applied, e.g., considering a higher $\SNRpeak$ threshold as shown in Fig.~\ref{Plot_spurious_fraction_versus_pb_attenu} when studying sources below a $\incode{pb_attenu}$ of $\sim$0.5.

\vspace{1truecm}

\section{Monte Carlo Simulations}
\label{Section_Monte_Carlo_Simulation_and_Correction}

We run extensive Monte Carlo (MC) simulations to verify our two main photometry methods: \pybdsm{} and \galfit{}. The principle idea is to simulate model galaxies and recover them with the same analysis used to create our catalogs. The aims are (1) to test whether the recovered flux densities have a systematic offset to the simulated flux densities, which is hereafter referred to as ``flux bias'', and to understand its source and quantify it if it exists; 
(2) to quantify the overall uncertainty on the extracted flux densities and verify whether the aforementioned \cite{Condon1997} error estimates can statistically describe the uncertainty; 
(3) to quantify the fraction of sources being recovered from all sources simulated, which is hereafter referred to as ``completeness''; and (4) to verify whether the prior information used in the simulations will alter the output statistics or not.

In our simulations, we create artificial sources (of Gaussian shape), insert them into residual images (after blind extraction photometry), and recover them with our photometry pipelines. These steps are repeated several tens to hundreds of times for a large number of images with different properties (details are given in Appx.~\ref{Section_Appendix_MC_Sim}). Our artificial sources are created within a grid of input values of both flux density and size. We create two sets of simulations with quite different input distributions defining this grid: 
(1) We start with a full-parameter-space simulation (hereafter ``FULL'' simulation) in which the full parameter space of flux density and size is uniformly sampled: $\SNRpeak$ ranges from 2.5 to 100 in logarithmic intervals, and source major-axis size to beam major-axis size ratio ranges from 0.1 to 6. Each grid point with a given flux density and size contains the same number of simulated sources. (2) We create another physically-motivated MC simulation, hereafter ``PHYS'' simulation, where we simulate sources mimicking observed galaxy stellar mass functions (SMFs; e.g., \citealt{Davidzon2017}), star-forming MS relation (MS; e.g., \citealt{Sargent2014}) and starburst/MS classification (i.e., following the 2-star-formation model [2SFM] of \citealt{Sargent2012_2SFM, Sargent2014} and \citealt{Bethermin2012_2SFM,Bethermin2017Model}), as well as galaxies' size evolution (e.g., \citealt{vanderWel2014, Fujimoto2017}). 
The motivation for performing our ``PHYS'' simulation is that galaxies have non-uniform luminosity functions (or number counts) and size distributions. Fainter galaxies are much more numerous than brighter ones, and lower redshift galaxies are in general larger than higher redshift ones. 
Our comparison of the ``FULL'' and ``PHYS'' simulations tests whether the input distribution of the simulations influences the derived recovery statistics. 

Due to the large number (1500+) of individual ALMA imaging data, we select a subset (150+) of representative images for each Scheduling Block of each Science Goal in each ALMA project.
In this way we make sure that all different observing scenarios (frequencies, spatial resolutions, integration times, etc.) are covered.

For each selected image, we perform the ``FULL'' and ``PHYS'' type simulations 4225 and 273 times respectively, depending on the grid of simulation (see Appx.~\ref{Section_Appendix_MC_Sim}), resulting in 4225 and $\sim$3000--25000 simulated objects respectively. 
The number of sources in the ``PHYS'' simulation varies with the image field of view and the observing wavelength, and dominates with fainter sources due to the assumed galaxy SMFs and MS correlation as well as the SEDs. 
Details of the two simulations are presented in Appx.~\ref{Section_Appendix_MC_Sim}. 

We then recover the simulated objects with our \pybdsm{} and \galfit{} photometry pipelines respectively, using the identical settings as for the real ALMA data. 
Therefore, we have four sets of simulated-and-recovered data to analyze and compare: FULL-\pybdsm{}, FULL-\galfit{}, PHYS-\pybdsm{} and PHYS-\galfit{}. 

In the next sections, we discuss the flux bias and flux errors for each simulation set, and characterize them by two normalized parameters: 
the fitted source peak flux density normalized by the rms noise, 
\begin{equation}
\SNRpeak \equiv \Speak/\noise
\label{Equation_SNRpeak}
\end{equation}
and the fitted source area (convolved with the beam) normalized by the beam area, 
\begin{equation}
\Sbeam \equiv \sqrt{\mathrm{Area_{source,\,convol.}}/\mathrm{Area_{beam}}}
\label{Equation_Sbeam}
\end{equation}

Note that the different types of simulations yield clear differences in the parameters of interest, especially the flux bias correction,
as we will show in the following when comparing the results from all four simulated data sets.

\vspace{0.25truecm}

\subsection{Analyses of the ``FULL'' and ``PHYS'' Simulations}
\label{Section_MC_Sim_Statistics}

Although the simulated total source flux density, $\Ssim$, overall agrees well with the recovered total source flux density, $\Srec$, a substantial bias between $\Ssim$ and $\Srec$ becomes obvious when looking at the dependency on the flux $\SNR$. When normalizing the difference between $\Ssim$ and $\Srec$ by the measured flux error, the histogram distribution of $(\Ssim-\Srec)/\ESrec$ exhibits a non-zero mean and non-unity scatter (such histograms are illustrated later in Appendices~\ref{Section_Appendix_MC_Sim_FULL_Source_Recovery},~\ref{Section_Appendix_MC_Sim_PHYS_Source_Recovery}~and~\ref{Section_Appendix_MC_Sim_Final_Correction}). This indicates that the measured fluxes need to be corrected for flux biases, and the errors in the measured fluxes need to be re-estimated.

To analyze the flux bias and errors from our simulations, we bin all simulated and recovered sources in the 2D parameter space of $\SNRpeak$ and $\Sbeam$, and consider flux bias and error to be functions of these two parameters (\citealt{Condon1997}; \citealt{Bondi2003,Bondi2008}; \citealt{Schinnerer2010}; \citealt{JimenezAndrade2019}). 
Because $\SNRpeak$ and $\Sbeam$ are both normalized quantities, sources from different ALMA projects can be combined.

For each $\SNRpeak$ and $\Sbeam$ bin, we compute the mean and median of the relative flux density difference ($(\Ssim-\Srec)/\Ssim$). The flux bias is then defined as: 
\begin{equation}
\eta_{\textnormal{bias}} \equiv \langle (\Ssim-\Srec)/\Ssim \rangle
\label{Equation_eta_bias}
\end{equation}
which represents how the recovered flux density is biased relative to the simulated flux density. 
The corrected flux density can then be calculated as: 
\begin{equation}
\Scorr = \Srec / (1.0 - \eta_{\textnormal{bias}})
\label{Equation_eta_bias_application}
\end{equation}
We note that computing the flux bias using the noise-normalized flux density difference ($(\Ssim-\Srec)/\noise$) leads to no obvious difference. 

Then, we also compute the scatter of $(\Ssim-\Srec)/\noise$ (we computed the standard deviation and the lower and higher 68$^{\mathrm{th}}$ percentiles, see Sect.~\ref{Section_MC_Sim_Flux_Error}) and denote it as: 
\begin{equation}
\eta_{\textnormal{error}} \equiv \sigma_{\left[(\Ssim-\Scorr)/\noise\right]}
\label{Equation_eta_error}
\end{equation}
We do not use the relative difference ($(\Ssim-\Srec)/\Ssim$) because its scatter has an asymmetric distribution. 
The corrected flux density error can then be computed as:
\begin{equation}
\EScorr = \eta_{\textnormal{error}} \times \noise
\label{Equation_eta_error_application}
\end{equation}

\begin{figure*}[htb]
\centering%
\adjincludegraphics[width=0.90\textwidth, Clip={0.02\width} {0.04\height} {0.52\width} {0.07\height}]{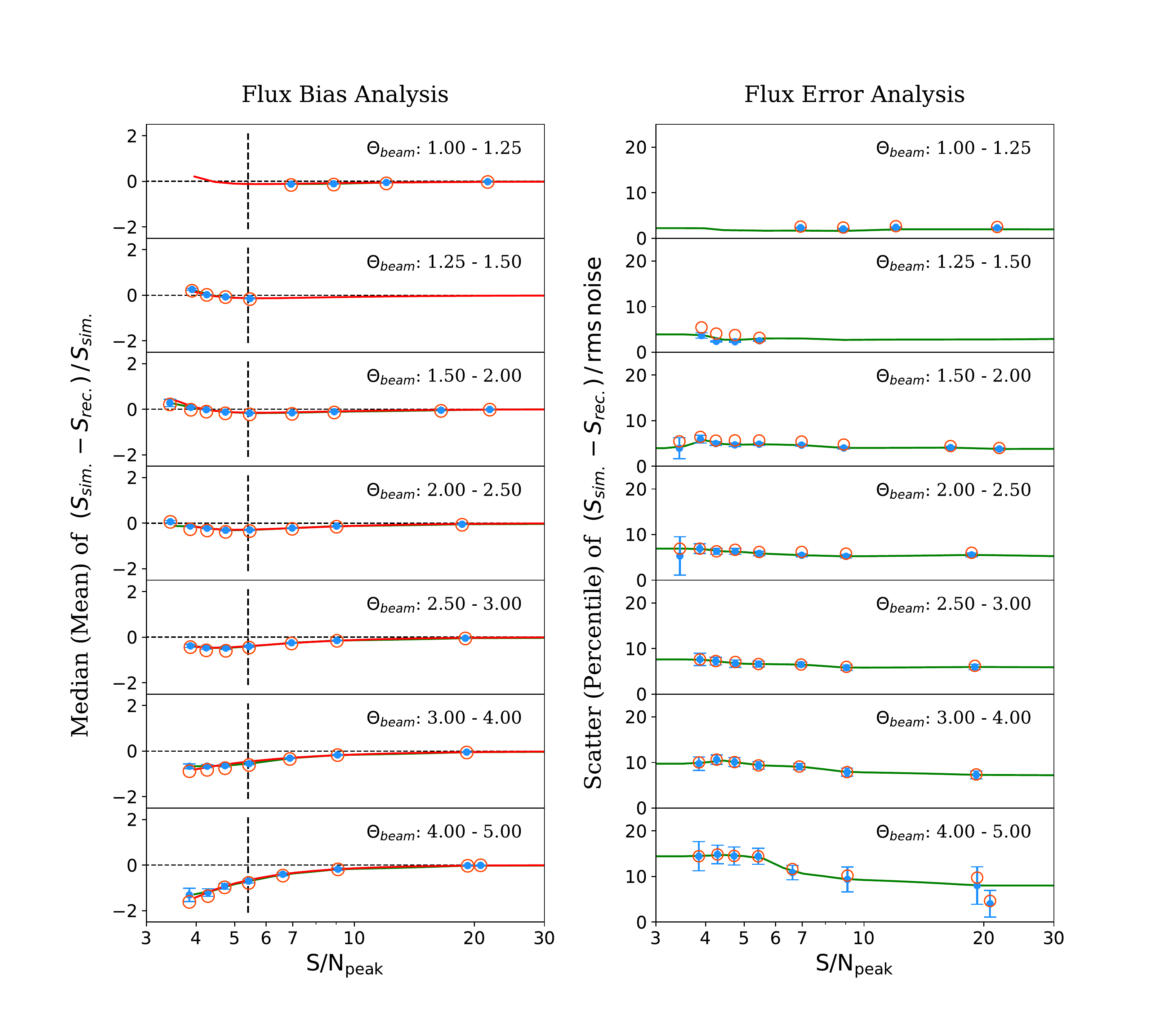}
\adjincludegraphics[width=0.90\textwidth, Clip={0.48\width} {0.04\height} {0.06\width} {0.07\height}]{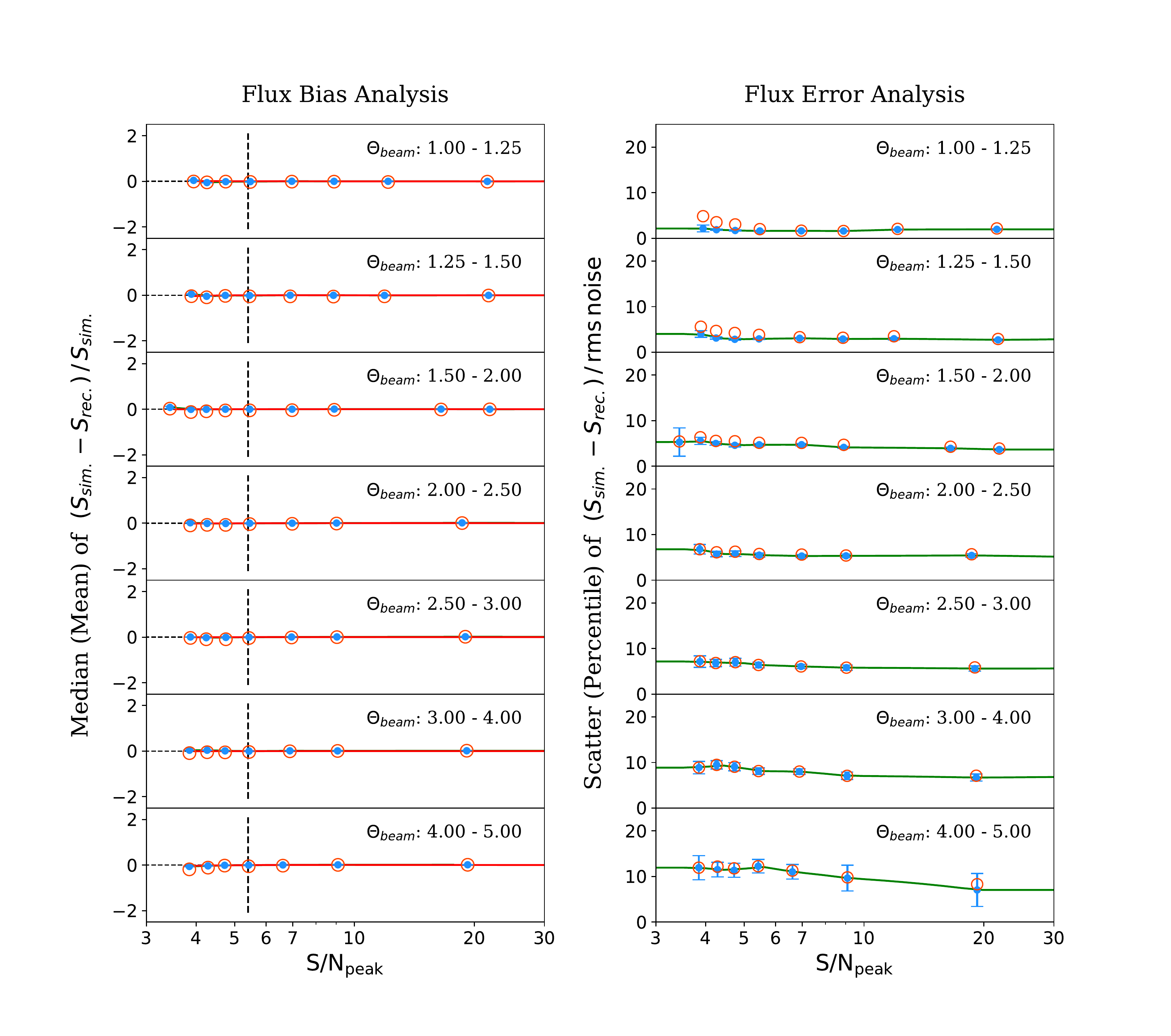}
\vspace{-2.5ex}
\caption{%
\pybdsm{} flux bias and flux error as functions of measured $\SNRpeak$ (Eq.~\ref{Equation_SNRpeak}) and $\Sbeam$ (Eq.~\ref{Equation_Sbeam}) in the left and right panels, respectively, from the ``FULL'' simulation. Sub-panels (from top to bottom) are bins with increasing $\Sbeam$ (as labeled). In the left panels, red and blue circles correspond to the mean and median of $(\Ssim-\Srec)/\Ssim$, respectively. Solid red lines are function fitting (with the form $a \, \SNRpeak^{\;m} + b \, \SNRpeak^{\;n}$) to the flux bias data points (but no feasible function form could be fitted for flux error), and solid green lines are interpolations or extrapolations (visible in the right panels). The vertical dashed line corresponds to our sample selection threshold as will be detailed in Sect.~\ref{Section_Combining_two_photometry_catalogs}. 
In the right panels, red and blue circles represent the scatter (standard deviation) and (the minimum of upper and lower) 68$^\mathrm{th}$ percentile of $(\Ssim-\Srec)/\noise$, respectively. See text in Sect.~\ref{Section_MC_Sim_Statistics}.
\label{Plot_MC_sim_simu_corr_FULL_PYBDSM}
}
\vspace{2.0ex}
\end{figure*}

\begin{figure*}[htb]
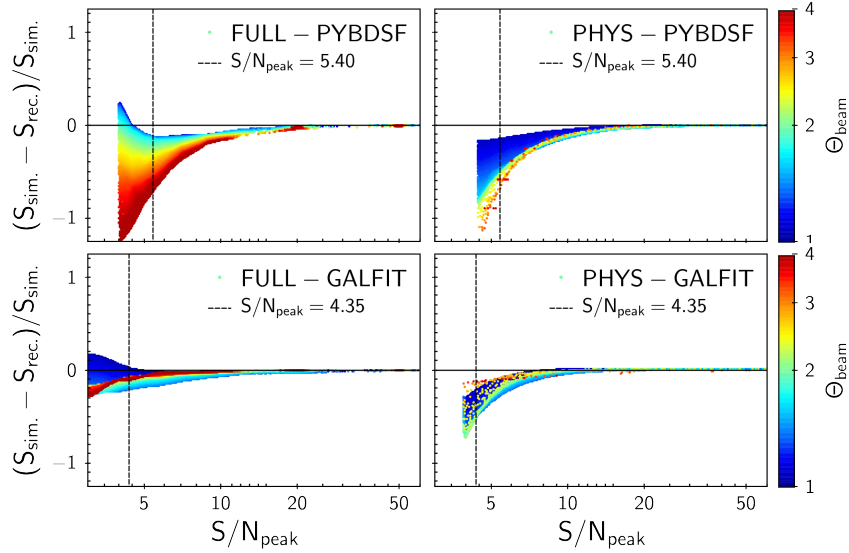

\centering%
\adjincludegraphics[scale=0.15, Clip={0.02\width} {0.23\height} {0.194\width} {0.00\height}]{{Plot_MC_Sim_FULL_PYBDSM_1_Plot_corrected_fbias}.png}
\adjincludegraphics[scale=0.15, Clip={0.194\width} {0.23\height} {0.00\width} {0.00\height}]{{Plot_MC_Sim_PHYS_PYBDSM_1_Plot_corrected_fbias}.png}\\
\adjincludegraphics[scale=0.15, Clip={0.02\width} {0.04\height} {0.194\width} {0.04\height}]{{Plot_MC_Sim_FULL_GALFIT_1_Plot_corrected_fbias}.png}
\adjincludegraphics[scale=0.15, Clip={0.194\width} {0.04\height} {0.00\width} {0.04\height}]{{Plot_MC_Sim_PHYS_GALFIT_1_Plot_corrected_fbias}.png}\\
\vspace{-1.5ex}%
\caption{%
Flux bias (as defined in Eq.~\ref{Equation_eta_bias}) vs. the measured $\SNRpeak$ (as defined in Eq.~\ref{Equation_SNRpeak}) statistics for \pybdsm{} ({\em top panels}) and \galfit{} ({\em bottom panels}) photometry, each based on our two types of Monte-Carlo simulations ({\em left} panels are ``FULL'' simulation and {\em right} panels are ``PHYS'' simulation; see the label in each panel). Color represents the geometric mean source-to-beam size ratio ($\Sbeam$, as defined in Eq.~\ref{Equation_Sbeam} in Sect.~\ref{Section_Blind_Source_Extraction}) and is the same in all four panels. Vertical lines are our $\SNRpeak$ thresholds for the sample selection in Sect.~\ref{Section_Combining_two_photometry_catalogs}. 
\label{Plot_MC_sim_fbias}
}
\vspace{-1.5ex}%
\end{figure*}

Combining all bins, we can measure $\eta_{\textnormal{bias}}$ and $\eta_{\textnormal{error}}$ as functions of $\SNRpeak$ and $\Sbeam$, which are illustrated in Fig.~\ref{Plot_MC_sim_simu_corr_FULL_PYBDSM} for the ``FULL'' simulation with \pybdsm{} recovery as the example (the other three simulation-recovery pairs are analyzed similarly, and the $\Sbeam$-collapsed figures can be seen in Fig.~\ref{Plot_MC_sim_fbias} and \ref{Plot_MC_sim_ecorr}). 
This figure demonstrates that the flux bias and error do strongly correlate with $\SNRpeak$ and $\Sbeam$.

\vspace{0.25truecm}

\subsubsection{Flux bias in the \textsl{\pybdsm{}} photometry}
\label{Section_MC_Sim_Flux_Bias}

The \pybdsm{} photometry measurement of a source always includes the intrinsic source flux plus a contribution from noise, thus it always fits positive source fluxes, and the measured fluxes are statistically boosted by a certain amount that we define as the flux bias. 

Based on our simulations, we characterize the flux bias correction factor ($\eta_{\mathrm{bias}}$, Eq.~\ref{Equation_eta_bias}) by the two measurable parameters $\SNRpeak$ and $\Sbeam$, and apply the flux bias correction to the measured/recovered flux with Eq.~\ref{Equation_eta_bias_application}. 
We find these two parameters to much more strongly affect the flux bias than other parameters, e.g., absolute source size or beam size. After the flux bias correction, the extracted total fluxes for simulated sources in maps of different spatial resolutions exhibit no obvious further bias from their simulated total fluxes. 
Here, we also found that the flux bias parameterization strongly depends on the input mock source populations of the MC simulation as demonstrated below. 

In Fig.~\ref{Plot_MC_sim_fbias}, we compare the flux bias of the \pybdsm{} photometry characterized from our ``FULL'' and ``PHYS'' simulations. $\SNRpeak$ is on the x-axes and $\Sbeam$ is indicated by the color. The flux bias is a strong function of both $\SNRpeak$ and $\Sbeam$. It rapidly becomes significant with  decreasing $\SNRpeak$. For example, $\vert\Ssim-\Srec\vert$ can be $>10\%$ of $\Ssim$ when $\SNRpeak\lesssim10$. Secondly, sources with larger sizes suffer a stronger flux bias: a source with a measured size 4 times the beam size can be boosted by $\sim$80\% of $\Ssim$ at an $\SNRpeak=5.77$ (where the spurious fraction at this $\SNRpeak$ is $\sim$40\%, see Fig.~\ref{Plot_spurious_fraction}); while an unresolved source is only boosted by $\sim$20\% at the same $\SNRpeak$. 

The flux bias functions derived from the two simulations are fully consistent at the bright end, e.g., $\SNRpeak \gtrsim 10-20$.
Discrepancies between the simulations become only obvious at the faint end of $\SNRpeak$ for sources with small $\Sbeam$. The flux bias in the ``FULL'' simulation is much smaller than compared to the ``PHYS'' simulation. This is due to the difference of the input populations of the two simulations. The effect of ``resolution bias'' is likely the main reason -- such a bias causes sources with low $\SNR$ and large \textit{simulated} sizes to have much smaller \textit{recovered} sizes or even be unresolved (or undetected) and also causes their fluxes to be underestimated instead of boosted by noise. This is common in radio photometry, where the spatial resolution is comparable to and even smaller than the sizes of galaxies at high redshift, e.g., as discussed in \cite{Bondi2003,Bondi2008}. The resolution bias is much more evident at the faint end of the ``FULL'' simulation than the ``PHYS'' simulation because of the higher number of large sources simulated in the former case. More discussion is presented in Appx.~\ref{Section_MC_Sim_FULL_Limitations}. 

In reality, the physical sizes of galaxies increase with cosmic time and scale with stellar masses \citep{vanderWel2014}, and their angular sizes (stellar component) increase quickly from $z\sim1$ to the present. This means that lower-redshift galaxies with high stellar masses tend to be largest. These galaxies can be bright at radio wavelengths but are in general much fainter and even undetectable at (sub-)mm wavelengths (due to the $K$-correction and the general drop in star formation activity). Therefore, in our ALMA (sub-)mm data, the real galaxy angular size distribution should be dominated by small sources, i.e., it is better described by the ``PHYS'' simulation rather than the ``FULL'' simulation. 
And thus we use ``PHYS'' simulation-based flux bias functions for the final correction of the photometry.

\begin{figure*}[htb]
\vspace{-1.5ex}%
\centering%
\adjincludegraphics[scale=0.15, Clip={0.02\width} {0.23\height} {0.194\width} {0.00\height}]{{Plot_MC_Sim_FULL_PYBDSM_2_Plot_corrected_ecorr}.png}
\adjincludegraphics[scale=0.15, Clip={0.194\width} {0.23\height} {0.00\width} {0.00\height}]{{Plot_MC_Sim_PHYS_PYBDSM_2_Plot_corrected_ecorr}.png}\\
\adjincludegraphics[scale=0.15, Clip={0.02\width} {0.04\height} {0.194\width} {0.04\height}]{{Plot_MC_Sim_FULL_GALFIT_2_Plot_corrected_ecorr}.png}
\adjincludegraphics[scale=0.15, Clip={0.194\width} {0.04\height} {0.00\width} {0.04\height}]{{Plot_MC_Sim_PHYS_GALFIT_2_Plot_corrected_ecorr}.png}\\
\vspace{-1.5ex}%
\caption{%
Similar to Fig.~\ref{Plot_MC_sim_fbias} but shows the flux error factor $\eta_{\mathrm{error}}$ (as defined in Eq.~\ref{Equation_eta_error}) versus the measured $\SNRpeak$ (as defined in Eq.~\ref{Equation_SNRpeak} in Sect.~\ref{Section_Blind_Source_Extraction}) for our simulated sources. Statistics for the two photometry methods  (\pybdsm: {\em top}; \galfit: {\em bottom}) based on our two types of simulation (``FULL'': {\em left}, ``PHYS'': {\em right}) are shown. Color represents the source-to-beam area ratio ($\Sbeam$, as defined in Eq.~\ref{Equation_Sbeam}) and is the same in all panels.
The horizontal colored lines show the expected flux errors using the \cite{Condon1997} prescription for $\Sbeam=1$, $2$ and $5$ (same color-coding as the data points). See text for further details. 
\label{Plot_MC_sim_ecorr}
}
\vspace{1.5ex}%
\end{figure*}

\begin{figure*}[htb]
\centering%
\adjincludegraphics[scale=0.15, Clip={0.00\width} {0.06\height} {0.19\width} {0.02\height}]{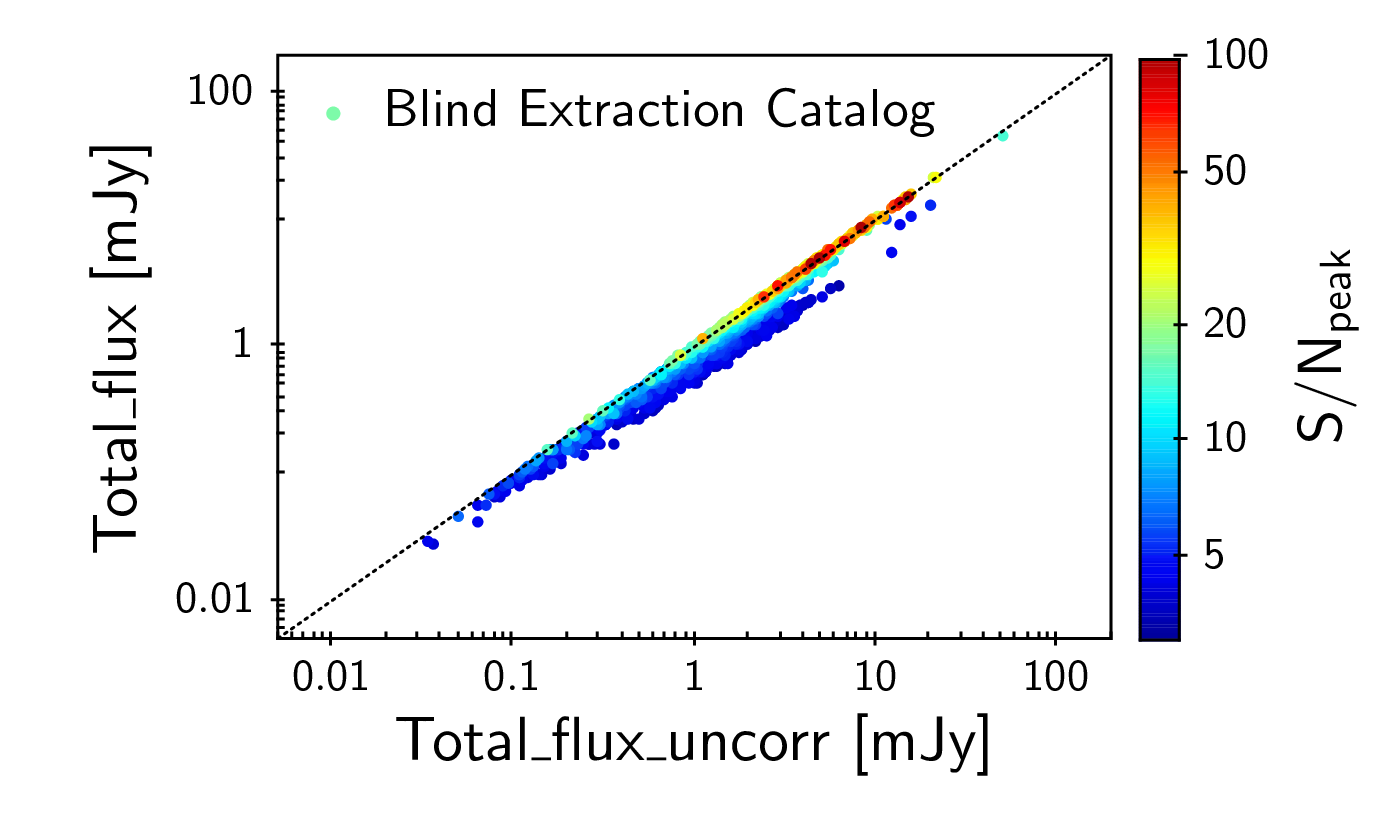}
\adjincludegraphics[scale=0.15, Clip={0.19\width} {0.06\height} {0.00\width} {0.02\height}]{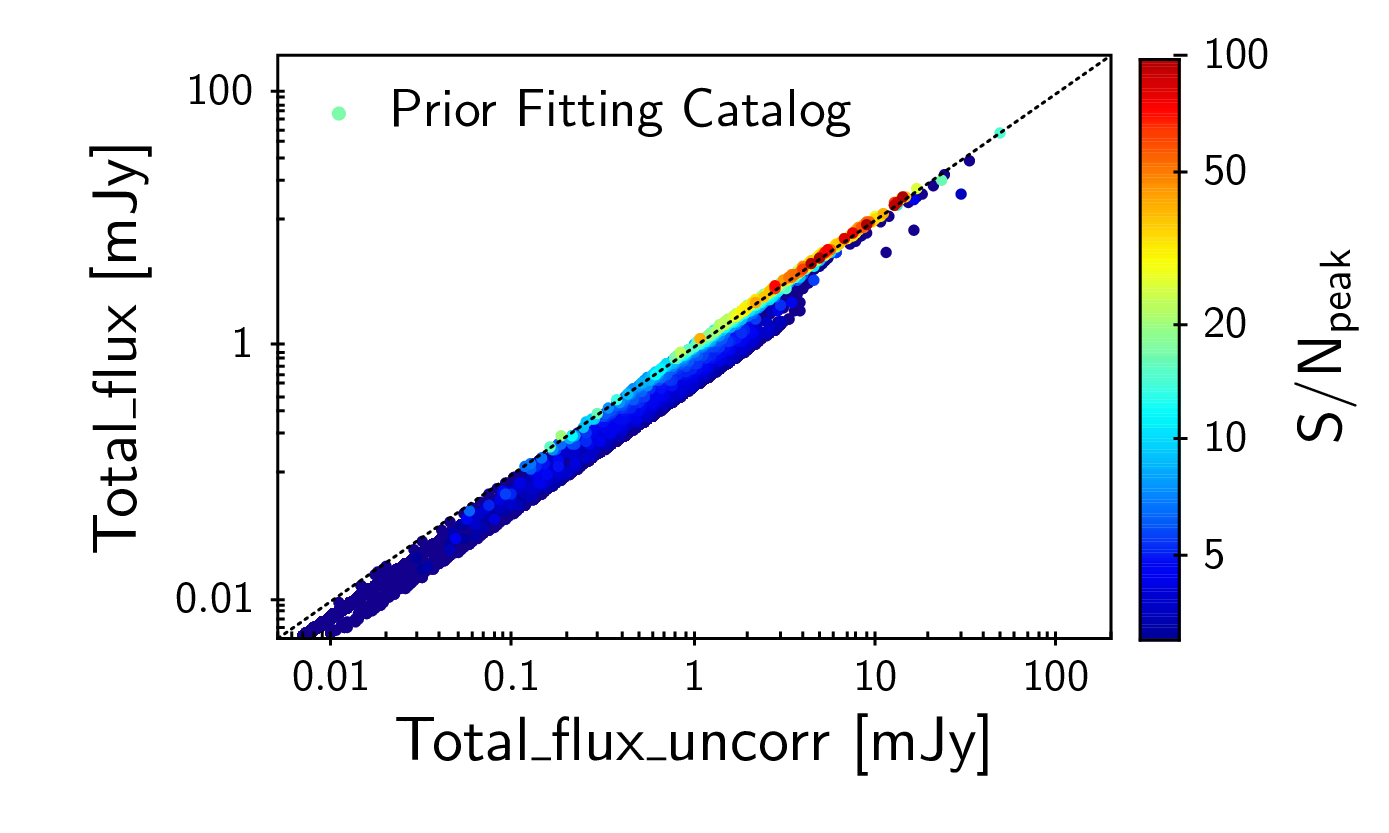}\\
\adjincludegraphics[scale=0.15, Clip={0.00\width} {0.08\height} {0.19\width} {0.02\height}]{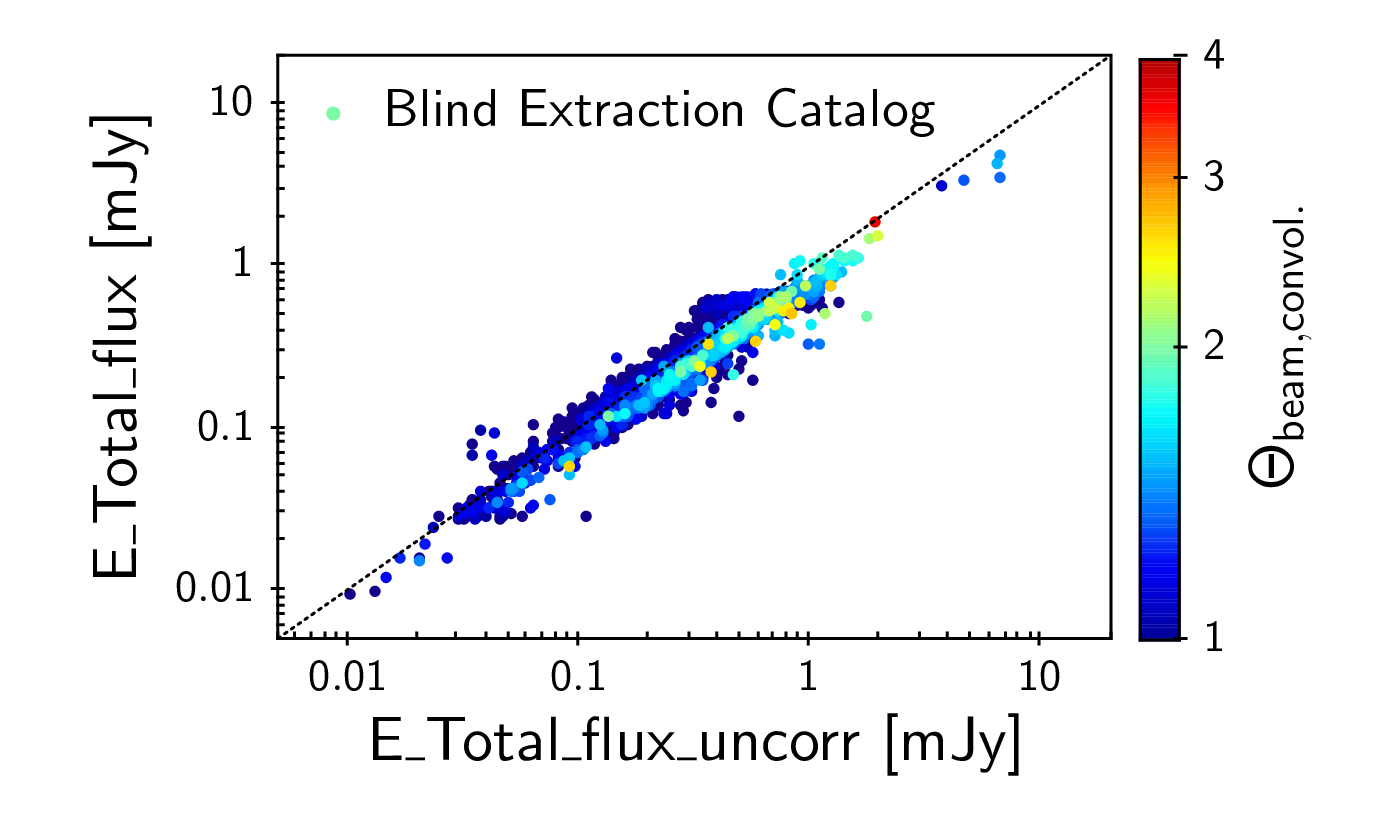}
\adjincludegraphics[scale=0.15, Clip={0.19\width} {0.08\height} {0.00\width} {0.02\height}]{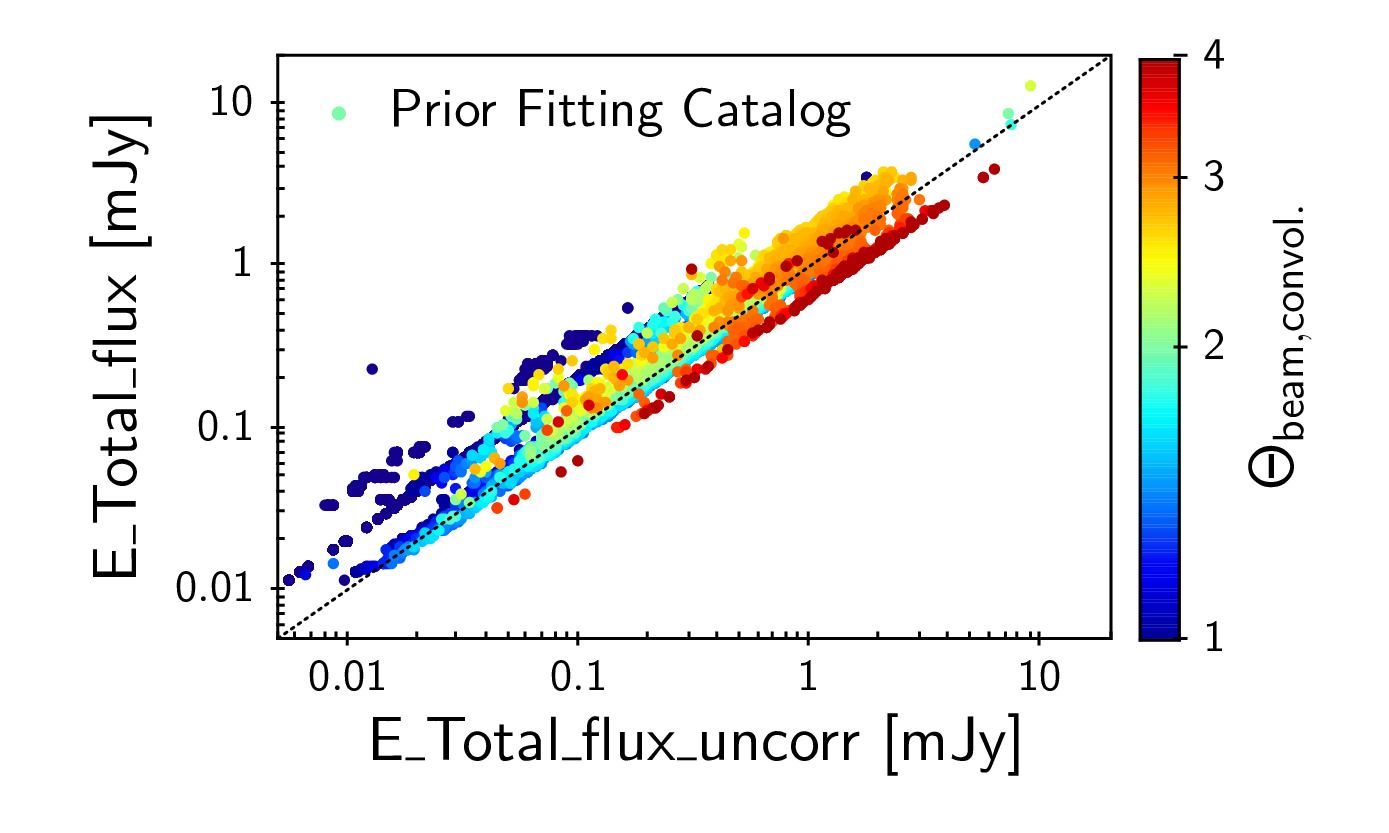}
\vspace{-1.0ex}%
\caption{%
Comparison of the final corrected and uncorrected fluxes (upper panels) and errors (lower panels) for real data's blind extraction and prior fitting photometry catalogs (left and right panels in each row respectively). The two upper panels have the same axes ranges and color bar indicating the measured $\SNRpeak$, and similar for the two lower panels but with the color bar indicating the measured $\Sbeam$. 
\label{Plot_final_corrected_flux_pybdsm}
}
\vspace{1.5ex}
\end{figure*}

\vspace{0.25truecm}

\subsubsection{Flux bias in the \textsl{\galfit{}} photometry}
\label{Section_MC_Sim_Flux_Bias_for_GALFIT}

In Fig.~\ref{Plot_MC_sim_fbias}, we show the flux bias parameterizations derived for the \galfit{} photometry based on both simulations. Similar to the \pybdsm{} photometry, the \galfit{} photometry also shows both flux boosting due to noise and flux underestimation due to the ``resolution bias''. 

The \galfit{} photometry has a smaller flux bias, which is likely due to the use of known prior position information for the photometry and the optimized iterative photometry approach (Sect.~\ref{Section_Prior_Source_Fitting}). 
It even achieves a better accuracy for sources with largest measured sizes ($\Sbeam\sim5$) than those with slightly smaller measured sizes ($\Sbeam\sim3$), if their $\SNRpeak$ are above 20 or so.

\vspace{0.25truecm}

\subsubsection{Flux error estimation for \textsl{\pybdsm{}} photometry}
\label{Section_MC_Sim_Flux_Error}

With Eqs.~\ref{Equation_eta_error}~and~\ref{Equation_eta_error_application}, we estimate the flux error factor ($\eta_{\mathrm{error}}$) from our simulation bins and parameterize it by $\SNRpeak$ and $\Sbeam$ (after the correction for flux bias). We compute $\eta_{\mathrm{error}}$ in a given bin by computing both the standard deviation and the upper and lower 68$^\mathrm{th}$ percentiles. Because the data do not usually follow a normal distribution in $(\Ssim-\Srec)$, both of these error estimates do not always agree with each other. This can be seen in the right panels of Fig.~\ref{Plot_MC_sim_simu_corr_FULL_PYBDSM}, especially for low-$\SNRpeak$ data points, where the standard deviation is usually larger than the one derived from the percentiles. And we find that the minor value of the upper and lower 68$^\mathrm{th}$ percentiles can better represent the underlying scatter (which are shown in later figures). 

\cite{Condon1997} proposed a mathematical recipe for estimating the errors of a six-parameter Gaussian fit with correlated noise. As shown by their Eqs.~32, 41 and 42, the total flux error can be characterized by the following parameters: the convolved source size parameters (major and minor axes FWHM sizes, denoted as $\theta_{\mathrm{maj.}}$ and $\theta_{\mathrm{min.}}$ respectively, corresponding to $\theta_{\mathrm{M}}$ and $\theta_{\mathrm{m}}$ respectively in \citealt{Condon1997}), the beam size parameters (major and minor axes FWHM sizes, denoted as $\theta_{\mathrm{bmaj.}}$ and $\theta_{\mathrm{bmin.}}$ respectively, corresponding to $\theta_{\mathrm{N}}$ and $\theta_{\mathrm{n}}$ respectively in \citealt{Condon1997}), and the measured total flux ($\Stotal$). Such a recipe has later been adopted in \cite{Bondi2003,Bondi2008}, \cite{Schinnerer2010} and \cite{Smolcic2017a} for the VLA source fitting photometry. 

In this work, because our ALMA data have different beam sizes, we express these size quantities in the normalized form: the geometric mean of the source size normalized by the beam size, $\Sbeam$ as defined in Eq.~\ref{Equation_Sbeam} which equals $\Sbeam^2 \equiv (\theta_{\mathrm{maj.}} \theta_{\mathrm{min.}})/(\theta_{\mathrm{bmaj.}} \theta_{\mathrm{bmin.}})$; 
the size of the source major axis normalized by the beam, $\Theta_{\mathrm{maj.}} \equiv \theta_{\mathrm{maj.}}/\theta_{\mathrm{bmaj.}}$; and the size of the source minor axis normalized by the beam, $\Theta_{\mathrm{min.}} \equiv \theta_{\mathrm{min.}}/\theta_{\mathrm{bmin.}}$.

Because the total flux is the product of peak flux and source area, we can write: 
\begin{equation}
\Stotal = \Speak \times \frac{(\theta_{\mathrm{maj.}} \theta_{\mathrm{min.}})}{(\theta_{\mathrm{bmaj.}} \theta_{\mathrm{bmin.}})} \equiv \Speak \times \Sbeam^2
\end{equation}

Therefore, the \citet[][C97]{Condon1997} recipe can be rewritten as:
\begin{equation}
\begin{split}
& \frac{\sigma_{S_{\mathrm{total}},\,\mathrm{C97}}^2}{{\mathrm{[rms\,noise]}}^2} \ = \ \\[0.3ex]
& \qquad \qquad \bigg( \Sbeam^2 \times \frac{8}{[1+\frac{1}{\Theta_{\mathrm{maj.}}^2}]^{1.5}[1+\frac{1}{\Theta_{\mathrm{min.}}^2}]^{1.5}} \bigg) \\
& \qquad \qquad \qquad \qquad + \frac{8}{[1+\frac{1}{\Theta_{\mathrm{maj.}}^2}]^{2.5}[1+\frac{1}{\Theta_{\mathrm{min.}}^2}]^{0.5}} \\
& \qquad \qquad \qquad \qquad +  \frac{8}{[1+\frac{1}{\Theta_{\mathrm{maj.}}^2}]^{0.5}[1+\frac{1}{\Theta_{\mathrm{min.}}^2}]^{2.5}} \\
\label{Equation_Condon1997_errors}
\end{split}
\end{equation}

\cite{Condon1997} validated the coefficients/indices in their equations using $\sim$3000 simulations. Because our ALMA photometry is more diverse than their simulations in both data complexity (the variety of beam size, rms noise) and photometry method (e.g., involving iterations), we need to verify that the \cite{Condon1997} recipe is still appropriate for our analysis.

In Fig.~\ref{Plot_MC_sim_ecorr}, we present how our estimated $\eta_{\mathrm{error}}$ changes with $\SNRpeak$ and $\Sbeam$, and compared with the \cite{Condon1997} errors (horizontal lines). The four panels show the same diagram for our two photometry methods and the two simulations. 

According to Eq.~\ref{Equation_Condon1997_errors}, the flux error normalized by the $\noise$ should be independent of $\SNRpeak$ but strongly dependent on $\Sbeam$. Fig.~\ref{Plot_MC_sim_ecorr} indeed shows a strong dependency on $\Sbeam$ but also indicates a weak dependency on $\SNRpeak$. For sources with small sizes (relative to the beam), the flux error becomes larger for larger $\SNRpeak$ (by about 15\% within the range indicated in the figure). However, for sources with large sizes (relative to the beam), it becomes smaller for larger $\SNRpeak$ (by about 40\% within the range of the figure).

The expected \cite{Condon1997} errors for $\Sbeam=1$, $2$ and $\Sbeam=5$ cases are shown as horizontal lines in Fig.~\ref{Plot_MC_sim_ecorr}, computed using Eq.~\ref{Equation_Condon1997_errors} and assuming a minor/major axis ratio of 1. Note that a smaller axis ratio will lead to a smaller \cite{Condon1997} error value (by about 15\% for $\Sbeam=5$ when reducing the axis ratio from 1 to 0.1.
Our simulation-derived errors (colored data points) are consistent with \cite{Condon1997} errors (colored lines) at the low-$\SNRpeak$ end and at smallest and largest sizes (represented by the colors). However, the ``FULL'' simulation panel indicates that \cite{Condon1997} errors are overestimated by about 40\% for large, high-$\SNRpeak$ sources; while the ``PHYS'' simulation panel indicates that \cite{Condon1997} errors are underestimated by about 15\% for small, high-$\SNRpeak$ sources. 
Both simulations show that the \cite{Condon1997} errors are slightly overestimated at $\SNRpeak\sim5-10$ for small and intermediate-sized sources. 

In our final catalog, we provide both our simulation-derived total flux errors and those given by our photometry pipelines which are based on \citet{Condon1997}\,\footnote{Note that in \pybdsm{}, if a source is fitted with a single-Gaussian component, then its total flux error is based on \citet{Condon1997}, but if it is fitted with multiple Gaussian components, then the error is propagated.}.

\vspace{0.25truecm}

\subsubsection{Flux error estimation for \textsl{\galfit{}} photometry}
\label{Section_MC_Sim_Flux_Error_for_GALFIT}

The flux errors are analyzed in a similar way for \galfit{} photometry. The same diagnostic plots are shown in the bottom panels of Fig.~\ref{Plot_MC_sim_ecorr}. The trends for the \galfit{} photometry is very similar to those for \pybdsm{}. The \galfit{} photometry has even smaller flux errors for large size sources than \pybdsm{} photometry. Both methods involve multiple iteration or multi-source fitting (rather than one-time simple 2D Gaussian fits), and thus the reason for these trends is not very clear. Yet the different inputs for the two types of simulations do not have a sizeable impact here.

\begin{figure*}[htb]
\centering%
\includegraphics[width=0.85\textwidth]{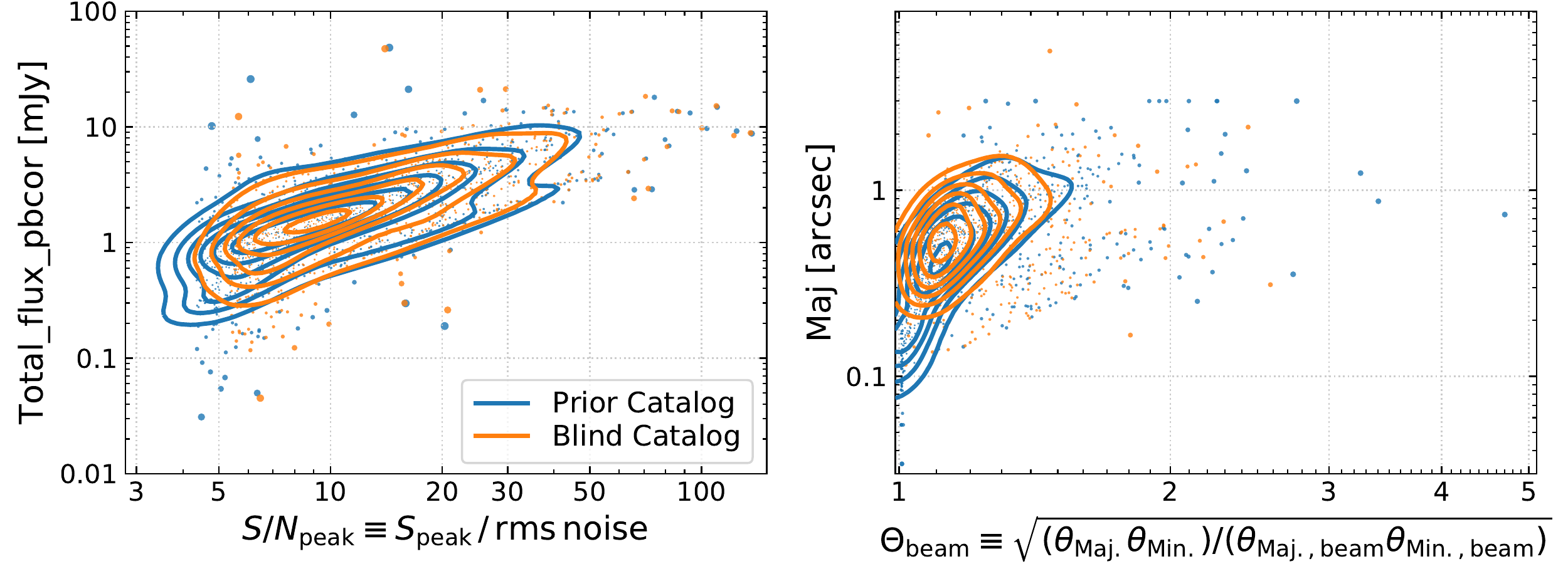}
\vspace{-1.5ex}%
\caption{%
Total flux (primary beam corrected) vs. source peak flux to rms noise ratio $\SNRpeak$ (see Eq.~\ref{Equation_SNRpeak}) (\textit{left panel}) and the fitted intrinsic source major-axis FWHM versus the beam-normalized source size $\Sbeam$ (see Eq.~\ref{Equation_Sbeam}) (\textit{right panel}) for the ALMA detections with $\SNRpeak>4.35$ and 5.40 in our two prior- and blind-photometry catalogs respectively (the $\SNRpeak$ thresholds are determined in Sect.~\ref{Section_Combining_two_photometry_catalogs}). 
Contours are the density of the data points. The size of a data point scales inversely with the density for illustration purposes. 
\label{Plot_A3COSMOS_photometry_catalog_statistics_2}
}
\vspace{2ex}
\end{figure*}

\begin{figure*}[htb]
\centering%
\includegraphics[width=0.85\textwidth]{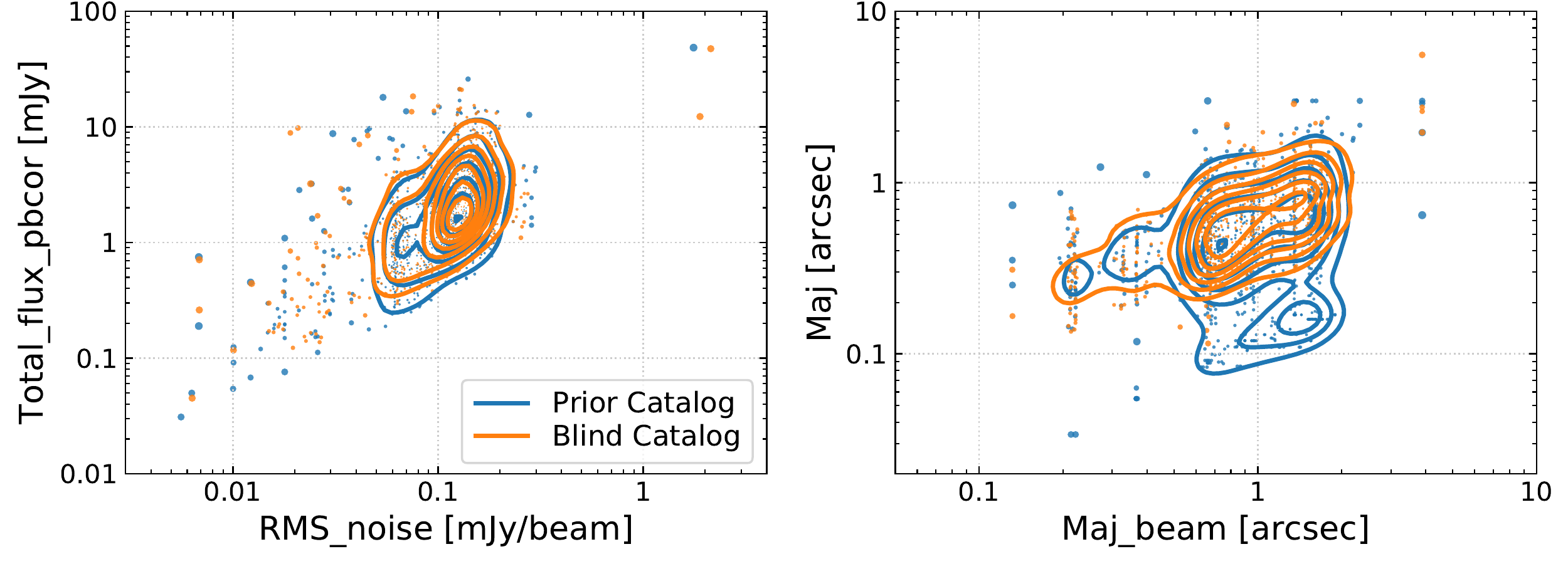}
\vspace{-2.0ex}%
\caption{%
Total flux (primary beam corrected) vs. rms noise (\textit{left panel}) and the fitted intrinsic source major-axis FWHM versus beam major-axis FWHM (\textit{right panel}). 
Contours and data points are in the same style as in Fig.~\ref{Plot_A3COSMOS_photometry_catalog_statistics_2}. 
\label{Plot_A3COSMOS_photometry_catalog_statistics_3}
}
\vspace{1ex}
\end{figure*}

\vspace{0.25truecm}

\subsubsection{Final Corrections}
\label{Section_MC_Sim_Final_Correction}

We finally correct flux biases and re-estimate flux errors for both the simulation catalogs and the real data's blind extraction and prior fitting catalogs, based on our aforementioned recipes (as functions of $\SNRpeak$ and $\Sbeam$; Eqs.~\ref{Equation_eta_bias_application}~and~\ref{Equation_eta_error_application}, respectively). 
We choose the ``PHYS'' simulation for the final correction, considering the discussion in the previous sections, i.e., ``PHYS'' simulation is more representative of our real data. Note that using ``FULL'' simulation would underestimate the flux bias correction and hence lead to larger fluxes especially, for large sources. 

The comparison of corrected and uncorrected fluxes and errors for real catalogs are shown in Fig.~\ref{Plot_final_corrected_flux_pybdsm}. Based on which, we find that our corrected fluxes and errors follow well-behaved statistics (see details in Appx.~\ref{Section_Appendix_MC_Sim_Final_Correction}), which means flux biases (e.g., flux boosting) are fully removed and flux errors can fully reflect the scatters of photometry measurements introduced by the noise in the data.

Further, in Figs.~\ref{Plot_A3COSMOS_photometry_catalog_statistics_2}~and~\ref{Plot_A3COSMOS_photometry_catalog_statistics_3} we present the distributions of primary-beam-corrected total flux and fitted intrinsic size versus source peak-to-rms noise $\SNRpeak$ (see Eq.~\ref{Equation_SNRpeak}), beam-normalized source size $\Sbeam$ (see Eq.~\ref{Equation_Sbeam}), and the rms noise and beam major-axis FWHM of the ALMA data. These figures show that our detections span a large range in flux and size. Note that the continuum wavelengths of the ALMA detections also vary: about 44\% of the data are at $\sim870\,\mu\mathrm{m}$, about 49\% at $\sim1.0-1.5\,\mathrm{mm}$ (mostly $1.25\,\mathrm{mm}$), $\sim1\%$ at $\sim1.9-2.3\,\mathrm{mm}$ and $\sim6\%$ at $\sim2.5-3.4\,\mathrm{mm}$. 
Thus the sensitivity shown cannot straightforwardly be compared to single-band ALMA continuum surveys.  
From these figures, good consistency between the two photometry catalogs is also evident. The prior catalog extends to a slightly fainter regime and only a minor fraction of sources are fitted with smaller sizes. 
As the aim here is to obtain good continuum photometry catalogs, the study of the uncertainty on source sizes is the topic of future work.

\vspace{0.25truecm}

\subsection{Completeness}
\label{Section_MC_Sim_Completeness}

In this section, we analyze the completeness of our photometry by examining the fraction of simulated sources that are successfully recovered to the total simulated number. 
The photometry is incomplete for several reasons: 
(1) some faint sources are undetected due to noise fluctuation; 
(2) \pybdsm{} groups blend multiple sources into one source; 
(3) \galfit{} might give wrong best-fit results in case of severely clustered priors; 
and (4) \pybdsm{} has certain flagging criteria to filter out nonphysical sources\,\footnote{According to the \pybdsm{} documentation \url{http://www.astron.nl/citt/pybdsf/process_image.html\#flagging-opts}, \pybdsm{} flags apparently nonphysical sources. See more details therein.}. 
To assess the contribution of these effects, we calculate the completeness curves as a function of $\SNR$ and source sizes (normalized by the beams).

We use both PHYS and FULL simulations to verify the completeness. Note that the two simulations have very different source flux and spatial distributions. Sources are isolated and have flat flux distribution in the FULL simulation, whereas in the PHYS simulations sources have instead realistic spatial distribution, as well as a flux distribution that fully agrees with the observed mm number counts (see Appx.~\ref{Section_Appendix_MC_Sim_PHYS}).

We cross-match the \pybdsm{} source recovery catalog to the simulated catalog for each image by coordinate using a search radius of 1.5$''$\,\footnote{%
    This corresponds to a false-match probability of $\le 1.3$\% for PHYS simulations according to Eq.~1 of \cite{Pope2006}. 
}, 
and we match by ID for the \galfit{} recovery catalog. 
We measure the completeness as the ratio of the number of sources in the cross-matched catalog and those in the simulated catalog for each bin of $\SNRpeak$ and $\Sbeam$.
We confirmed that the wide range in rms noise and beam size does not affect the completeness estimates by splitting our simulations in random half.
Using a smaller search radius has a very minor effect, as only 4\% (10\%) of sources have recovered position shifted by more than 1.0$''$ (0.6$''$) from the simulated position. 

Moreover, 
the completeness is associated to certain detection criteria. Within \pybdsm{}, the detection is defined as an extracted source that passes \incode{thresh_pix}, \incode{thresh_isl} and other flagging criteria. Therefore, the remaining discussion within this section is focused on the \pybdsm{} setups (Sect.~\ref{Section_Blind_Source_Extraction}). In \galfit{}, a detection is slightly more complex to define, as \galfit{} always fits a positive flux density for each prior. Thus, we apply an $\SNR$ cut to the \galfit{} catalog before computing the completeness (without such an $\SNR$ cut, the recovery rate would be 100\%, as every prior is fitted with a flux density). 

\begin{figure*}[ht]
\centering%
\includegraphics[scale=0.6, trim=3mm 10.1mm -4mm 0, clip]{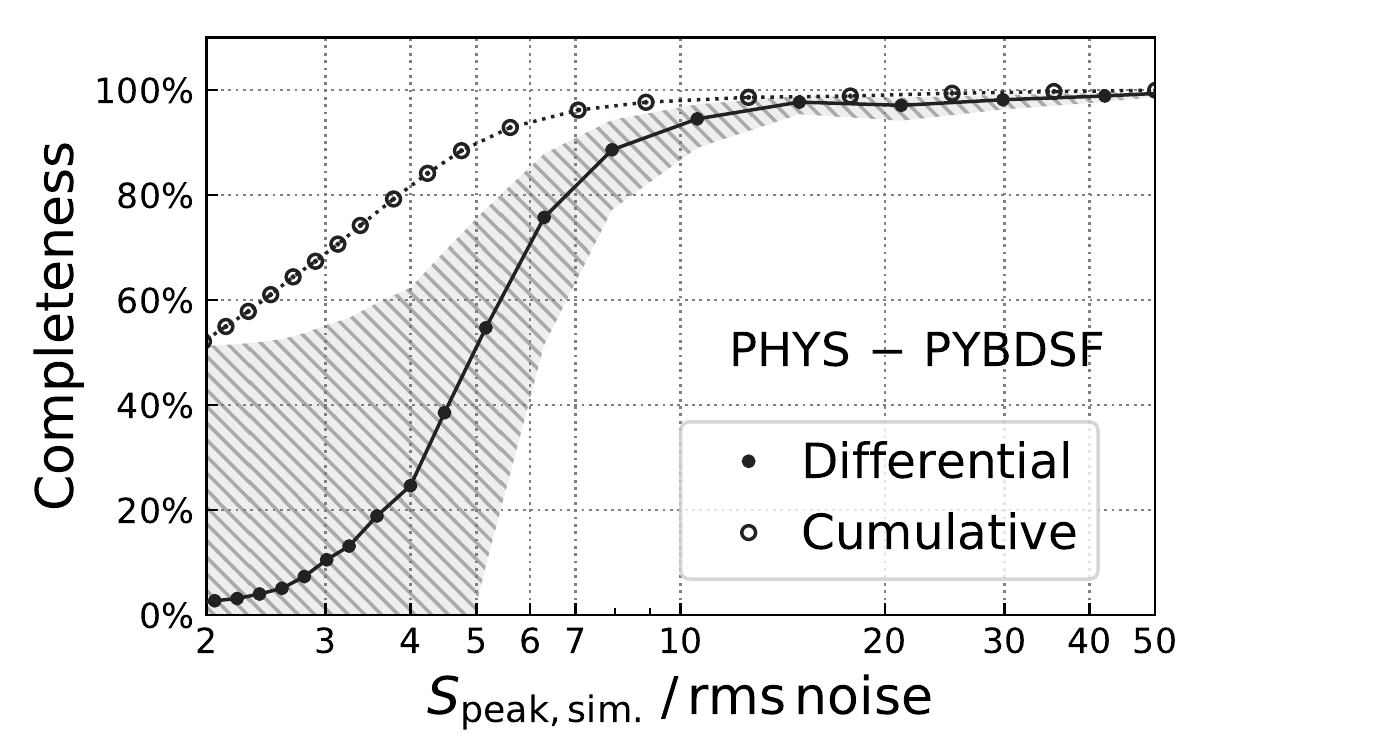}
\includegraphics[scale=0.6, trim=8.6mm 10.1mm 0 0, clip]{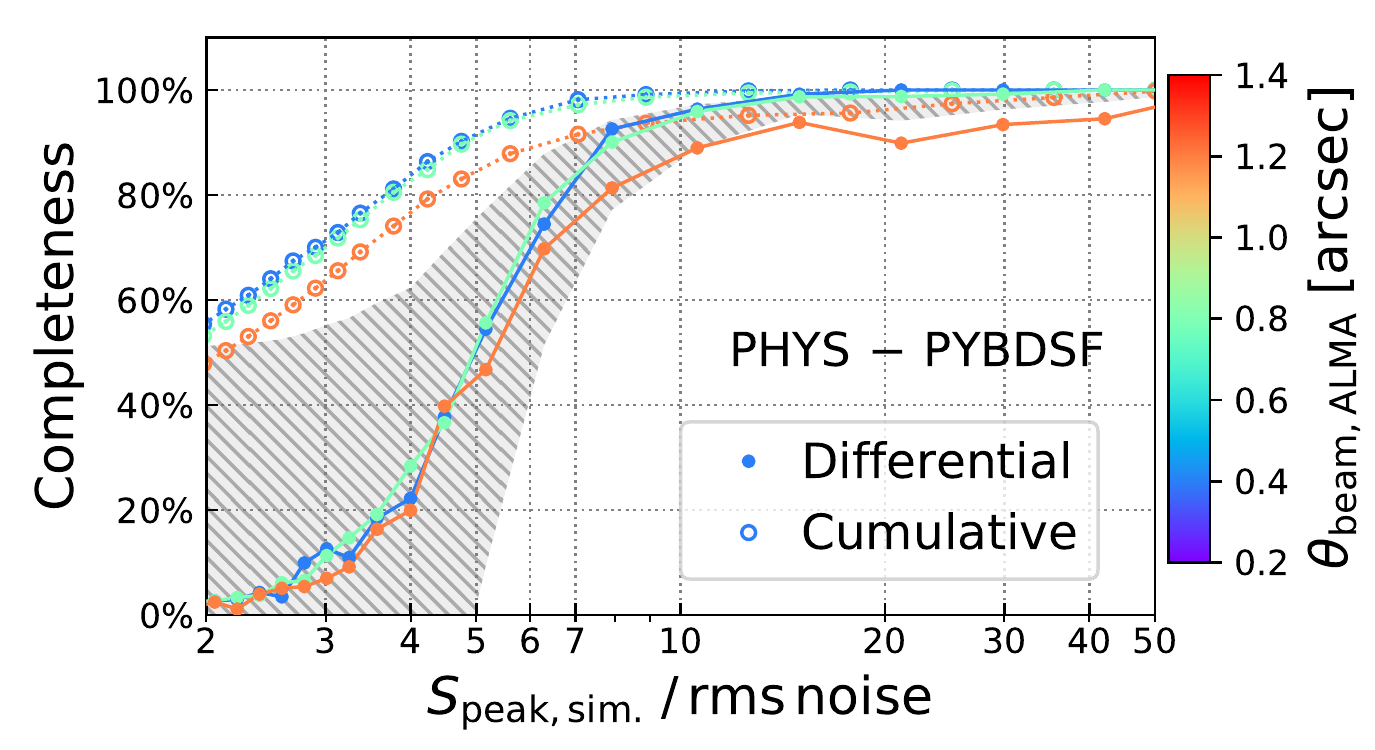}
\includegraphics[scale=0.6, trim=3mm 0 -4mm 0, clip]{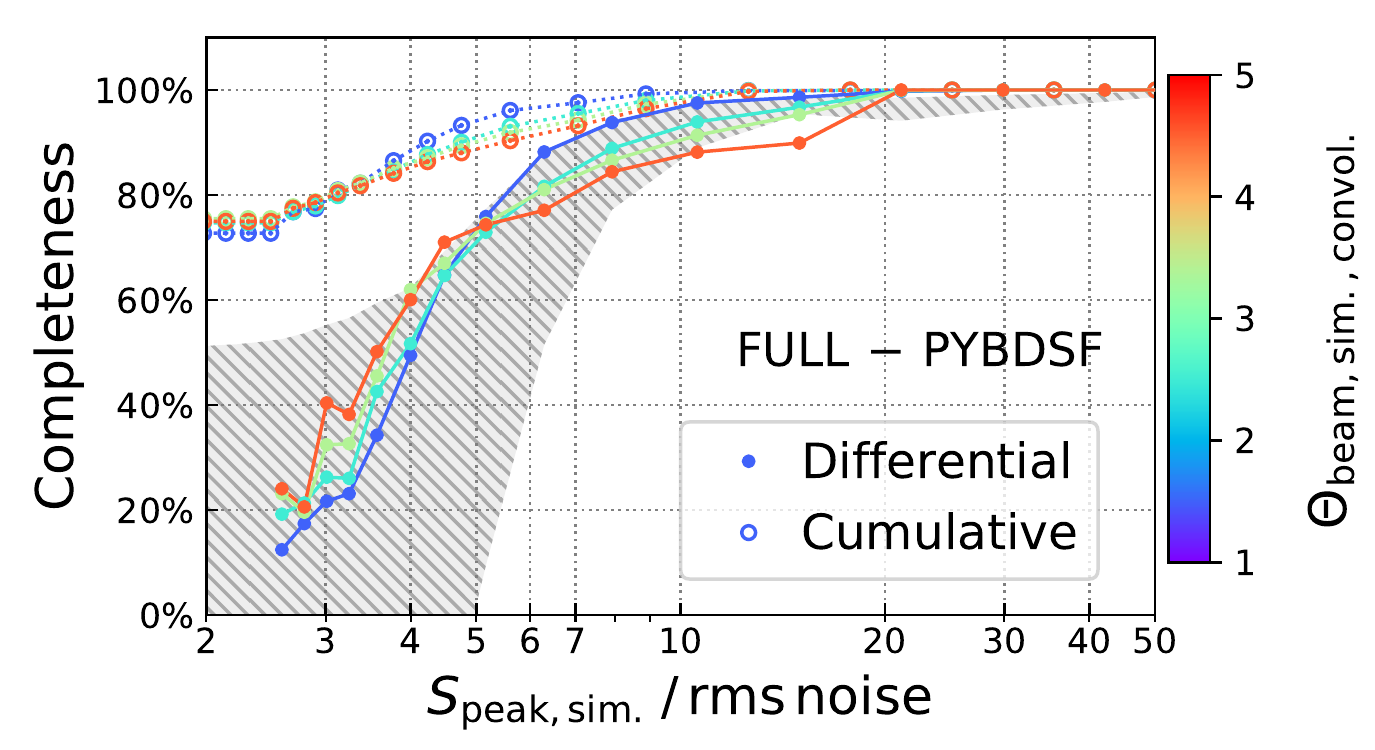}
\includegraphics[scale=0.6, trim=8.6mm 0 0 0, clip]{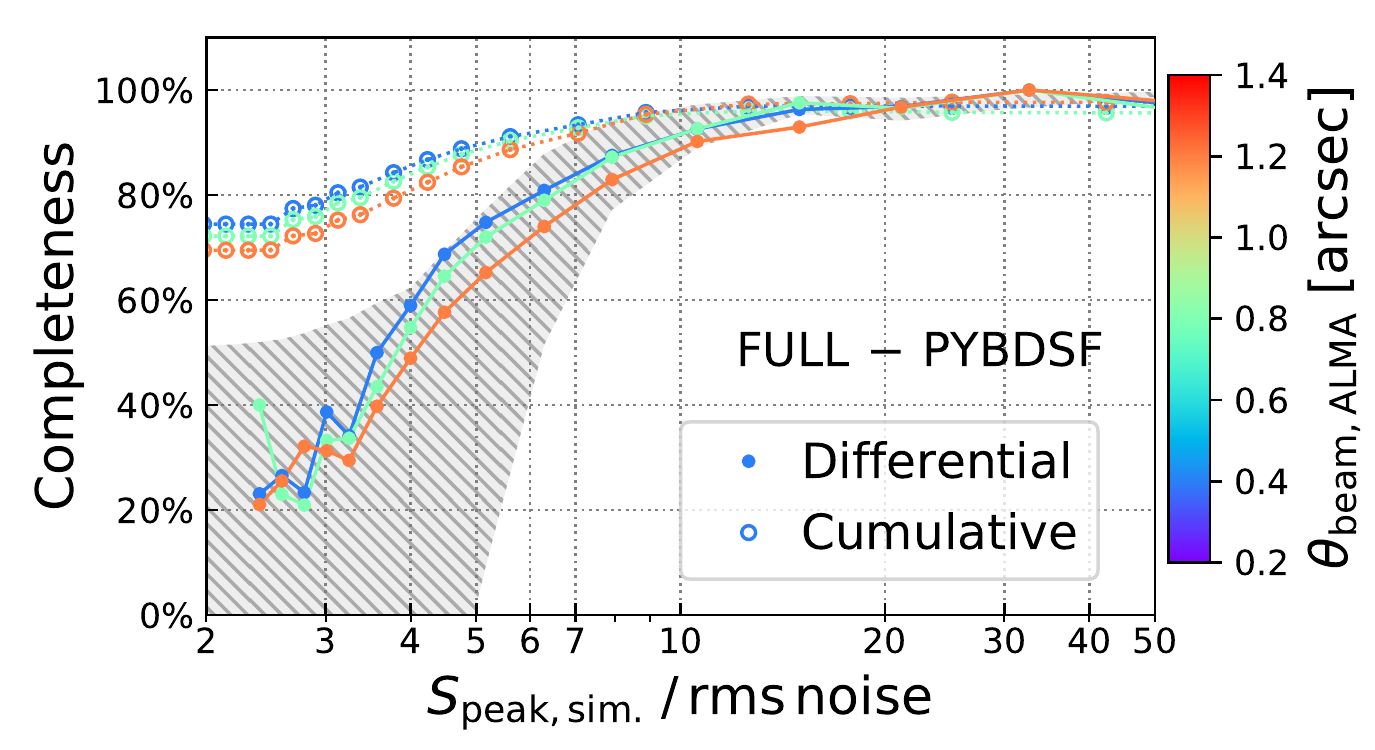}
\caption{%
    Completeness of the \pybdsm{} source extraction as a function of $\SNRpeak$ based on our simulations (``PHYS'': {\em top panels}; ``FULL'': {\em bottom panels}). 
    Color in the {\em bottom left panel} represents simulated source size (convolved, normalized by the ALMA beam, i.e., $\Sbeam$ as defined in Eq.~\ref{Equation_Sbeam}) and color in the {\em top and bottom right panels} are both absolute ALMA beam sizes ($\theta$ in units of $''$).
    Differential completeness at a given $\SNRpeak$ is marked by solid symbols, while the cumulative completeness for the range above a given $\SNRpeak$ is shown by open symbols. The shaded areas indicate a factor of two uncertainty in the incompleteness for the ``PHYS'' simulations in the first panel and is repeated in other three panels for comparison. 
    \label{Plot_MC_sim_completeness}
}
\vspace{1ex}
\end{figure*}

In the left panels of Fig.~\ref{Plot_MC_sim_completeness}, we show the completeness curves for the \pybdsm{} photometry as a function of $\SNRpeak$. As sources tend to be small relative to the beam size (with a median [mean] observed size of $\ConvSimSbeam\sim1.2$ [$1.6$]) in the ``PHYS'' simulation (top-left panel), we do not distinguish between source sizes. 
The ``FULL'' simulations (bottom-left panel) have sufficient statistics to study the effect of source sizes, thus we
show completeness curves for different simulated source sizes in the bottom-left panel. Here, we consider simulated size instead of recovered size, as the latter is unavailable for undetected sources.
Large sources are slightly more complete than small sources at very low $\SNRpeak\sim2-4$. This trend reverses at a higher $\SNRpeak$ (up to $\SNRpeak\sim20$) above which the completeness for sources reaches $\sim$100\%. 
In principle, at a given $\SNRpeak$, sources with larger recovered size should have higher completeness. We speculate that the previously discussed resolution bias, spatial noise fluctuation and the ``island'' feature of \pybdsm{} all play a role in the low- to intermediate-$\SNRpeak$ regime --- a larger simulated source is easier to detect due to a higher number of pixels above the threshold, but at the same time it has a chance of being recovered with a smaller size or even as an (or multiple) unresolved source(s) by \pybdsm{} (especially for the largest simulated sizes). Thus, these effects lead to a lower completeness for the largest simulated sources even at $\SNRpeak\sim10-20$. 
While fine-tuning the \pybdsm{} parameters can achieve better detection for large sources, this would require more dedicated effort beyond our systematic approach, which is tailored to the bulk of source properties expected. Moreover, our prior photometry is fitting well large sources ($<3''$), thus such cases will be identified when we cross-match the prior- and blind-photometry catalog (see Sect.~\ref{Section_Combining_two_photometry_catalogs}), and currently no such source is found in our dataset as we excluded beam~$<0.1''$ ALMA data. 

The shaded areas in Fig.~\ref{Plot_MC_sim_completeness} indicate an uncertainty of a factor of two in the estimated incompleteness in ``PHYS''--\pybdsm{}, and are the same in all other panels. 
Comparison between the completeness for the smallest sources in the ``FULL'' simulation and the one from the ``PHYS'' simulation gives a $\sim3\%$ lower completeness at $\SNRpeak\sim20-40$. This difference is caused by source blending and exactly corresponds to the 3.5\% multi-Gaussian sources detected in our data set. As described in Sect.~\ref{Section_Blind_Source_Extraction}, when several sources are blended, \pybdsm{} fits multiple Gaussians and groups them as one island which is then output as a single source.

In the right panels of Fig.~\ref{Plot_MC_sim_completeness}, we show how different ALMA beam sizes (absolute values in units of $''$) would impact the completeness. We find that as long as the ALMA beam is between $0.2''-1''$, the completeness is not obviously affected. For ALMA beams larger than $1''$, completeness drops by $\sim5-10\%$ even for a high $\SNRpeak\sim20-50$ source in our PHYS simulation, which is likely because sources are clustered and large ALMA beam starts to cause blending effect, and also because \pybdsm{} has the ``island''-grouping feature (Sect.~\ref{Section_Blind_Source_Extraction}).

In addition, in Fig.~\ref{Plot_MC_sim_completeness_2D} in Appendix~\ref{Section_Appendix_MC_Sim_Completeness}, we show the completeness as 2D functions of both $\Speak/\noise$ and $\Stotal/\noise$ and $\Sbeam$. We find a good agreement between our completeness analysis and similar work by \cite{JimenezAndrade2019} for \pybdsm{} photometry in their COSMOS VLA data as well as by \cite{Franco2018} for \textsc{Blobcat} \citep{Hales2012} photometry on their ALMA deep field data. Further, we discuss 
the comparisons of our completeness to other (sub-)mm/radio photometry works (\citealt{Karim2013}; \citealt{Ono2014}; \citealt{Aravena2016}; \citealt{Hatsukade2016}; \citealt{Franco2018}), which confirm that more realistic simulations are required to better recover the statistics. 

Given our finding that completeness shows an obvious dependency on source sizes, if selecting a sample with a total flux $\SNR$ threshold, the sample will have different completeness for different sizes. But when selecting with a constant $\SNRpeak$ threshold, the sample will have a homogeneous completeness. Thus we use $\SNRpeak$ to select our final sample (see the next section). 
Furthermore, we confirm that the spurious fractions derived from the simulations are consistent with those based on inverted-image fitting in Sect.~\ref{Section_Spurious_Fraction}. 
The robust estimates of the fractions of completeness and spurious sources provide us with a good handle of the performance of our photometry methods. For a given $\SNRpeak$-selection threshold, we know how many real sources are missed and how many could be spurious. While there is no way to improve on the non-detections, there are a number of automated examinations that can significantly reduce the number of spurious sources in our final galaxy catalog (see next sections).

\vspace{1truecm}

\section{Galaxy Catalog and Properties}
\label{Section_Galaxy_Sample_and_Properties}

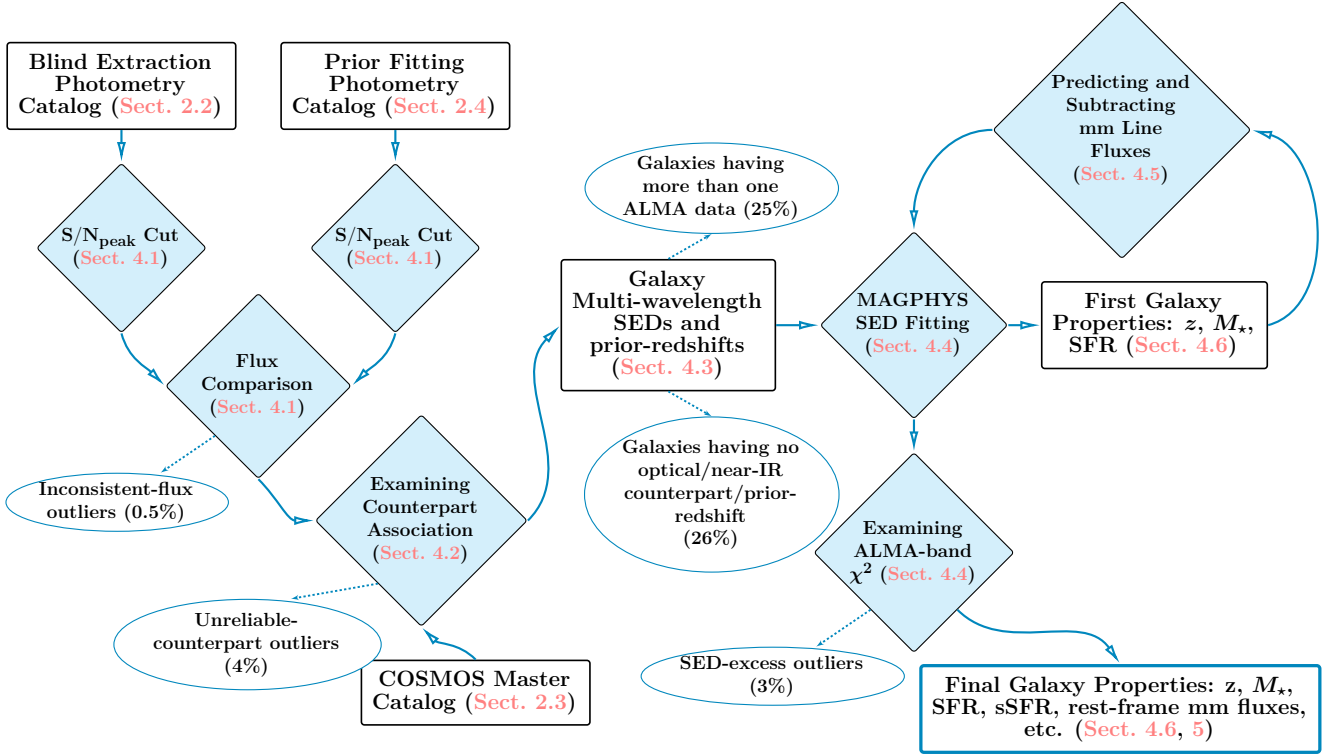
\begin{figure*}[ht]
\centering
\resizebox{\textwidth}{!}{\input{Input_galaxy_flow_chart.tikz}}
\caption{%
    Workflow for the selection of a reliable galaxy sample and the determination of its properties (see Section\,\ref{Section_Galaxy_Sample_and_Properties}). We first apply an $\SNRpeak$ cut to our two photometry catalogs and then apply a counterpart association code (based on machine learning) to construct our galaxy multi-wavelength catalog. Next, we run SED fitting to identify outliers that are due to either spurious ALMA sources or ``suspicious'' (inconsistent) redshifts in the literature. Finally, after discarding spurious sources and refinement of inconsistent prior redshifts, SED fitting is repeated to obtain physical properties of our galaxies. The corresponding subsections in the text are provided in parentheses. 
    \label{Figure_galaxy_flow_chart}
}
\end{figure*}

In this section, we discuss the selection of reliable ALMA detections from the two photometry catalogs and the construction of our `galaxy catalog'. 
Given the extensive information on galaxies in the COSMOS field that is available in the literature, we have devised
rigorous inspections to ensure that our galaxy sample and its SEDs are reliable.
These inspections include the identification of spurious sources and galaxies with inconsistent photometric and/or spectroscopic redshifts in the literature. We further discuss how galaxy properties are obtained via multiple SED fitting techniques including consistency and reliability checks. 
The workflow of this analysis step (including Sects.~\ref{Section_Combining_two_photometry_catalogs}~to~\ref{Section_Obtaining_Galaxy_Properties})
is illustrated in Fig.~\ref{Figure_galaxy_flow_chart}. 

\vspace{0.25truecm}

\subsection{Combining the two photometric catalogs}
\label{Section_Combining_two_photometry_catalogs}

\begin{figure}[htb]
\centering%
\includegraphics[width=0.48\textwidth, trim=5mm 5mm 0 0]{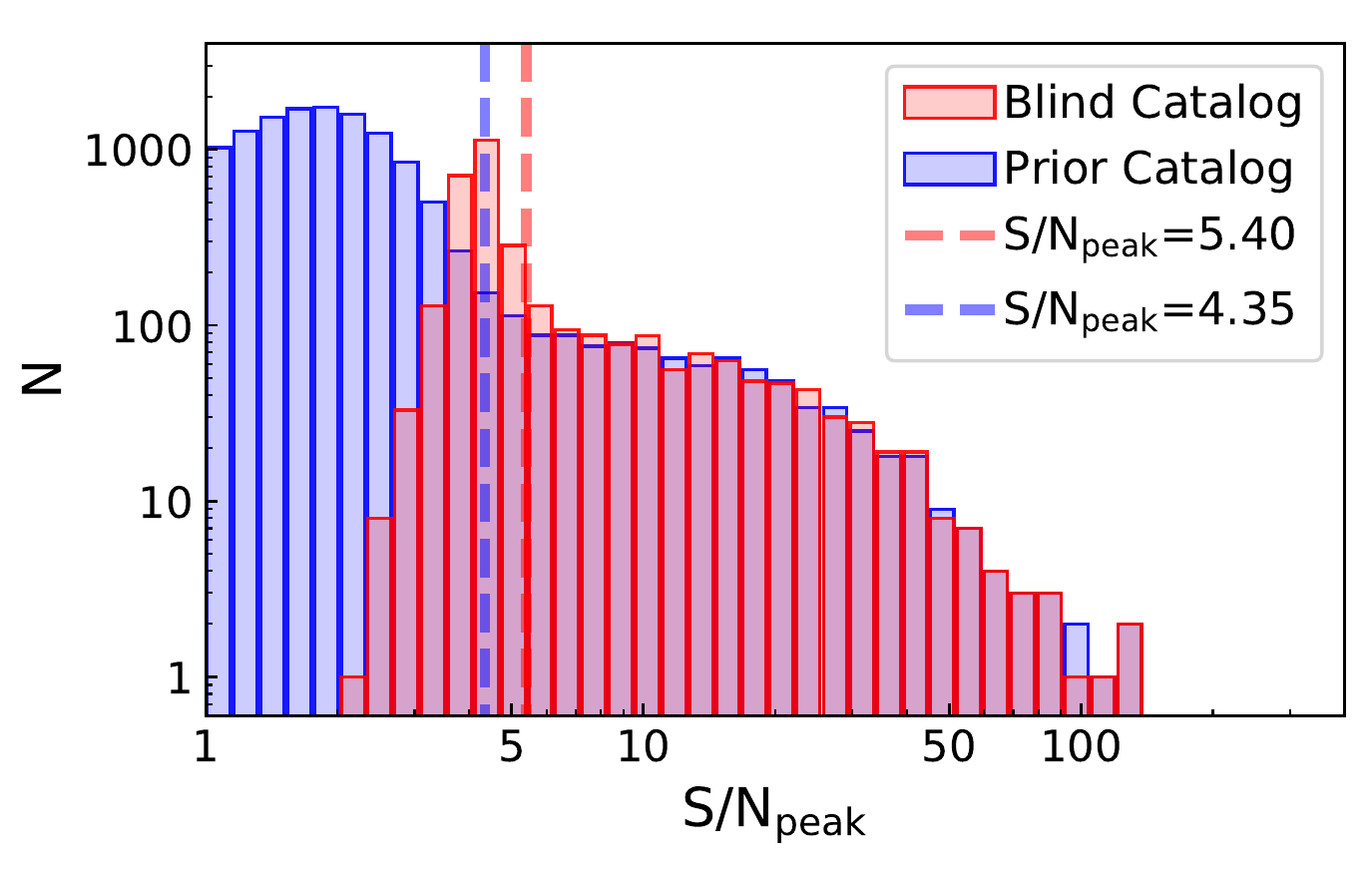}
\caption{%
$\SNRpeak$ histograms of our blind-extraction and prior-fitting catalogs. The blue and red vertical dashed lines indicate the $\SNRpeak$ thresholds we applied to select our sample from the prior and blind catalogs, respectively. 
\label{Plot_sample_selection_SNR_peak_histogram_before_selection}
}
\end{figure}

We apply an $\SNRpeak$ cut at 5.40 to our blind source extraction catalog (Sect.~\ref{Section_Blind_Source_Extraction}) and an $\SNRpeak$ cut at 4.35 to our prior source fitting catalog (Sect.~\ref{Section_Prior_Source_Fitting}). These thresholds are selected such that the differential spurious fractions are both 50\%
at the applied $\SNRpeak$ cut level, and the cumulative spurious fractions are $<$8\% and $<$12\% for the blind- and prior-selected samples, respectively (see Sect.~\ref{Section_Spurious_Fraction} and Fig.~\ref{Plot_spurious_fraction}). The corresponding differential completenesses at those thresholds are 57\% and 98\%, and the cumulative ones are as high as $>$92\% and $>$99\%, respectively (see Sect.~\ref{Section_MC_Sim_Completeness} and Figs.~\ref{Plot_MC_sim_completeness},~\ref{Plot_MC_sim_completeness_2D}). 
In Fig.~\ref{Plot_sample_selection_SNR_peak_histogram_before_selection}, we show the $\SNRpeak$ histograms of the blind and prior catalogs and the applied thresholds. 

\begin{figure}[htb]
\centering%
\includegraphics[width=0.48\textwidth, trim=5mm 5mm 0 0]{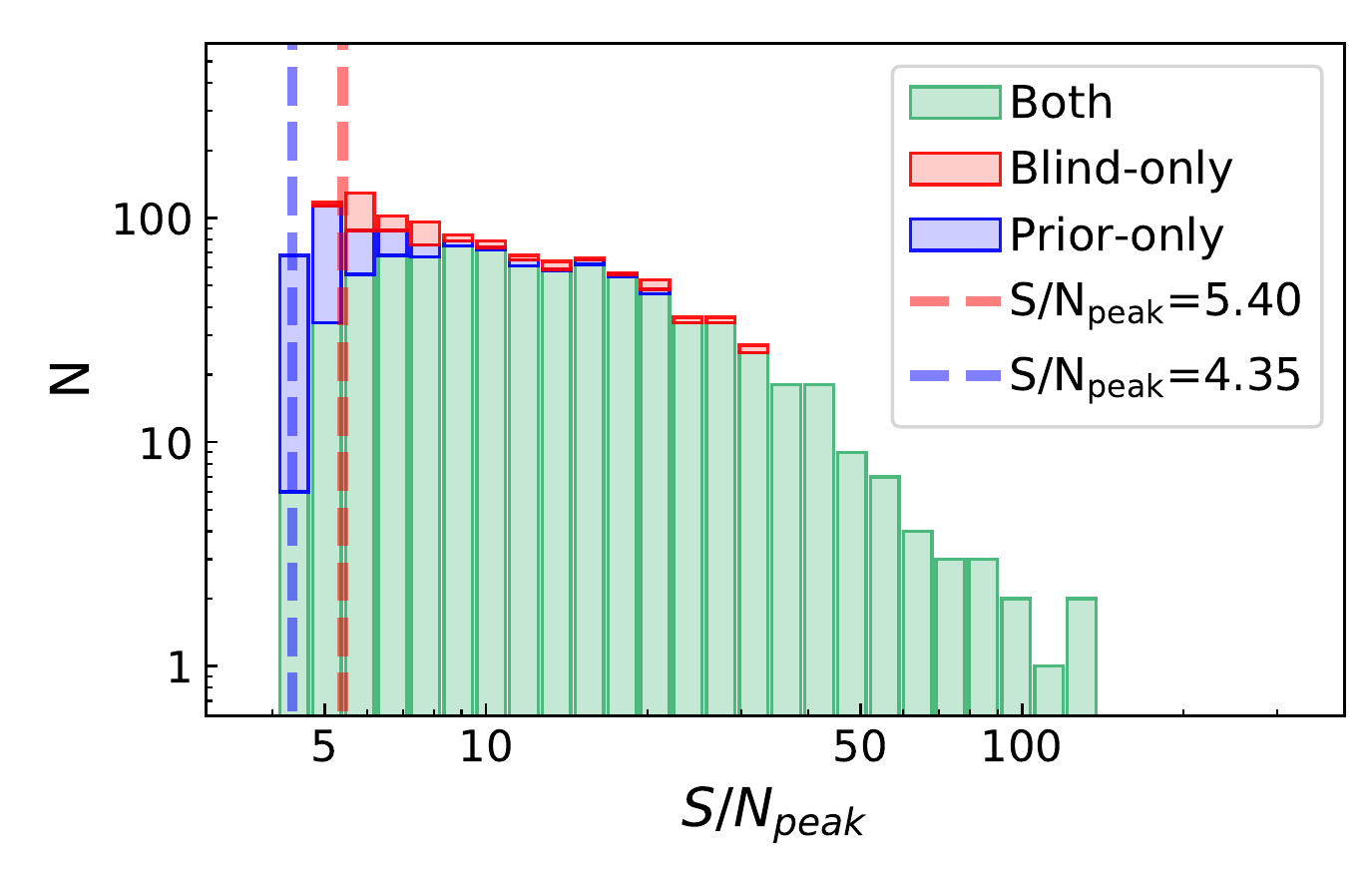}
\caption{%
Vertically-stacked $\SNRpeak$ histograms of our selected sample from the blind-extraction and prior-fitting catalogs. Sources in both catalogs are indicated by green bars and shown with the prior-catalog $\SNRpeak$, while those in the blind-extraction (prior-fitting) catalog with $\SNRpeak$ above the labeled threshold are shown with red (blue) bars. The height of each stacked bar indicates the relative number, and the total height of the histogram represents our selected sample size. 
(Comparing to Fig.~\ref{Plot_sample_selection_SNR_peak_histogram_before_selection}, the difference in the third-highest bin is due to different $\SNRpeak$ between prior- and blind-photometry as detailed in Sect.~\ref{Section_Photometry_Quality_Check_1}.)
\label{Plot_sample_selection_SNR_peak_histogram_after_selection}
}
\end{figure}

To merge the two photometric catalogs, we spatially cross-match their sources with a radius of 1.0$''$ (false-match probability 0.5\% applying Eq.~1 of \citealt{Pope2006}; see also further discussion of the counterparts association in the next section), 
and we find \aaacosmosPriorAndBlindDetectionNumber{} sources in common. Another \aaacosmosPriorXorBlindDetectionNumber{} sources are only present in one catalog (\aaacosmosPriorOnlyNumber{} sources in the prior catalog and \aaacosmosBlindOnlyNumber{} sources in the blind catalog).
The $\SNRpeak$ histograms of those sources (Fig.~\ref{Plot_sample_selection_SNR_peak_histogram_after_selection}) show that the sources only present in the prior catalog (prior-only sources) mostly lie at the lowest-$\SNRpeak$ end, where the spurious fraction is 50\%. The few prior-only sources at high $\SNRpeak$ are blends with nearby prior sources, such that only one source is cross-matched to the corresponding \pybdsm{} counterpart. The sources only present in the blind catalog could be spurious (if at low ALMA $\SNR$) or, if at high ALMA $\SNR$, real dusty, high-redshift galaxies whose optical/near-IR/radio emission are too faint to be detected in the prior catalogs. However, as there is currently no optical/near-IR information available for these blind-only sources, we exclude them from the analysis in the rest of this paper.

After accounting for \aaacosmosGalaxyWithMultiplePhotometryFraction{} of galaxies having more than one ALMA observations, due to either different wavelengths or spatial resolutions, we have \aaacosmosGalaxyNumber{} unique galaxies (with dataset version \incode{20180201}). 
The ALMA flux densities and their errors are then corrected for the PBA. 
As \aaacosmosFractionOfNoPriorz{} of these galaxies do not have sufficient optical/near-IR data, i.e., not in the \cite{Laigle2016} catalog, it is not possible to obtain reliable stellar masses for them. While some of these sources emit weakly in the deeper IRAC $3.6$ and $4.5\,\mu\mathrm{m}$ data from the \textit{Spitzer} Large Area Survey with Hyper-Suprime-Cam (SPLASH) survey (PI: P. Capak; I. Davidzon, priv. comm.) and are also 
present in the IRAC catalogs from the \textit{Spitzer} Matching Survey of the UltraVISTA Ultra-deep Stripes (SMUVS; \citealt{Ashby2018}), their stellar masses and photometric redshifts have large uncertainties due to the lack of shorter-wavelength information. 
We therefore omit these sources from our galaxy catalog (see the ``no optical/near-IR prior-redshift galaxies'' entry in Fig.~\ref{Figure_galaxy_flow_chart}; they are kept in the ALMA photometry catalogs, e.g., those with IRAC/radio priors). We plan to update our galaxy catalog when deeper optical-to-$K$-band data become available, e.g., from the UltraVISTA Data Release 4. 

In the next sections, we further exclude some outliers from the ALMA photometry catalogs to construct our final galaxy catalog. We list the numbers and fractions of sources excluded at each step in Table~\ref{Table_number_of_sources_excluded}.

\input{Table_number_of_sources_excluded.tex}

\vspace{0.25truecm}

\subsection{Examining counterpart association}
\label{Section_Examining_counterpart_association}

Our ALMA data set has excellent spatial resolution ($\sim1''$) compared to data from single-dish (sub-)mm telescopes ($>10''$), and for most sources a unique counterpart at optical/near-IR/radio wavelengths can be easily identified by examining the spatial separation. However, a small number of ambiguous cases remain for both prior fitting and blind extraction photometry. Note that we have already corrected for the known astrometry offsets between prior and ALMA positions before our final run of prior fitting (for more details on astrometry, see Appx.~\ref{Section_Astrometry}). 

During our prior fitting photometry, we allow the source position to vary if the source has high $\SNR$ (see Sect.~\ref{Section_Prior_Source_Fitting}). This implies that any ALMA source not in our prior master catalog close to a prior position will be wrongly attributed to that prior. 
In these cases, they are more likely to have a certain spatial offset. 
But this scenario needs to be distinguished from the case where the prior source is an extended galaxy and its dust emission peak is offset from its optical position (e.g., \citealt{Hodge2016}; \citealt{ChenCC2017}).

Besides, spurious sources caused by noise boosting ($\sim$10\% spurious sources are expected from our statistical analysis with our selection thresholds in Sect.~\ref{Section_Combining_two_photometry_catalogs}) can also exhibit larger offsets, as the signal boosted by noise is randomly spatially distributed. Thus, by examining the counterpart association, we can identify most of these outliers ($\sim$4\% in this step; or in total $\sim8.4\%$ including the steps in the next sections) and reduce the number of spurious sources in our final catalog\,\footnote{%
    Note that examining the counterpart association is not helpful in identifying line-of-sight boosting by noise or blending by background source. Therefore the outlier fraction found in this step is only $\sim$4\%, about half of our expected spurious fraction $\sim8-12\%$ (Sect.~\ref{Section_Combining_two_photometry_catalogs}). However, as shown in the next section, SED fitting is a powerful tool to exclude $\sim3\%$ of sources line-of-sight outliers and further reduce the spurious source fraction in our final catalog. 
    In total, after \incode{Flag_inconsistent_flux} (Sect.~\ref{Section_Photometry_Quality_Check_1}), \incode{Flag_outliers_CPA} (Sect.~\ref{Section_Examining_counterpart_association}) and \incode{Flag_outliers_SED} (Sect.~\ref{Section_Blind_Source_Extraction}), we excluded 
    \aaacosmosNumberOfQADiscardedGalaxies{} 
    sources as spurious for 
    \aaacosmosNumberOfQAedGalaxies{} 
    quality-assessed galaxies. 
    This is basically in agreement with our statistics ($8-12\%$). 
    \label{Footnote_flagging_spurious}
}.

In order to correctly identify such ambiguous cases in an automated fashion, 
we quantify the counterpart association process by several measurable parameters as follows:
\begin{itemize}[topsep=0em, leftmargin=1.5em, noitemsep]
    \item The projected separation between the positions of the ALMA and counterpart source, normalized by the projected ALMA source radius (denoted as $\mathbf{Sep.}$);
    \item The ALMA total flux $\SNR$ (denoted as $\mathbf{S/N_{ALMA}}$);
    \item The $\SNR$ of the aperture-integrated flux in optical/near-IR/radio images, measured with an aperture centered at the ALMA source position ($\mathbf{S/N_{S.}}$) and at the reference counterpart positions ($\mathbf{S/N_{Ref.}}$), as well as their respective ratio (denoted as $\mathbf{{S./Ref.}}$). The aperture size is determined via measurements with a series of concentric apertures where the aperture with the maximum $\SNR$ is taken; 
     \item An extension parameter $\mathbf{{Ext.}}$ that traces the amount of extended optical/near-IR/radio emission within the location between the ALMA and counterpart positions. This is quantified by deriving the optical/near-IR/radio surface brightness level within a series of fixed-size apertures (equal to the fitted ALMA source size) centered along the connecting line between the ALMA position and the reference counterpart position. The linear slope of the relation between surface brightness and increasing (linear) distance from the ALMA position is adopted as $\mathbf{{Ext.}}$: if the source is an extended galaxy and the optical emission is attenuated by dust at the ALMA position, then $\mathbf{{Ext.}}$ is around or slightly larger than 1. However if the ALMA source is a dusty galaxy with non-detectable optical emission and is wrongly associated to a counterpart in optical catalog at some distances away, $\mathbf{{Ext.}}$ will be very large or even not measurable in the counterpart optical image (as we require $\mathbf{S/N_{S.}}>3$ in the apertures to measure the $\mathbf{{Ext.}}$ parameter). \\
\end{itemize}

These parameters have been defined to best describe the counterpart association process, and are best suited to distinguish between those considered true by visual classification from those cases where the visual classification suggests that the ALMA source is unrelated to the counterpart source. 
These parameters are then measured for each ALMA detection (Sect.~\ref{Section_Combining_two_photometry_catalogs}) and its master catalog counterpart (Sect.~\ref{Section_Prior_Source_Catalogs}) in four counterpart images: \textit{Hubble Space Telescope} (\textit{HST}) ACS $i$-band image from \cite{Capak2007}; UltraVISTA $K_s$-band image from \cite{McCracken2010,McCracken2012}; \textit{Spitzer} IRAC 3.6\,$\mu$m image from the SPLASH survey (PI: P. Capak); and VLA 3\,GHz image from \cite{Smolcic2017a}. Other images have worse spatial resolution and/or sensitivity and therefore are less helpful in distinguishing the quality of counterpart associations. 

Empirically, we find counterparts with larger $\mathbf{Sep.}$ and lower $\mathbf{S/N_{ALMA}}$ are less reliable (i.e., less confident to say that the ALMA emission belongs to the counterpart galaxy, based on our visual identification). However, those could be more reliable if we see extended emission between the ALMA and counterpart position (i.e. $\mathbf{{Ext.}} \sim 1$), which could be the aforementioned case where the galaxy's dust emission is offset from its optical emission and has a smooth transition in-between. We show an example of our counterpart association diagnostic in Appx.~\ref{Section_Counterpart_association_examples}.

With these parameters, we proceed with machine learning techniques to establish the linkage between these parameters and the confidence of a counterpart association. 
To build up a training data set, three team members visually classified all the 1000+ ALMA detections individually. We visually inspected ALMA contours overlaid on ACS $i$-band, UltraVISTA $K_s$-band, IRAC 3.6\,$\mu$m, and 3\,GHz images and assigned each source a classification of 1 (robust) or 0 (spurious or incorrect association). 
We adopt the median classification from the three sets as truth. In order to automate this classification for future data releases, we use the results from visual inspection to train an algorithm that takes as input the parameters described above ($\mathbf{Sep.}$, $\mathbf{S/N_{ALMA}}$, $\mathbf{S/N_{Ref.}}$, $\mathbf{{S./Ref.}}$ and $\mathbf{{Ext.}}$) calculated for the ACS, Ks, IRAC\,3.6$\mu$m, and 3GHz cutouts. In addition, we include a flag for $\mathbf{crowdedness}$ (defined as the density of master catalog sources weighted by a 2D Gaussian with an FWHM of PSF size; see \citealt{Liudz2017} Eq.\,1) and $\mathbf{clean}$ parameter (defined as the number of master catalog sources within 3$''$ radius; \citealt{Elbaz2011}), as they are helpful in identifying extremely blended cases. 

For this supervised machine learning task, we use the \textsc{Python} \textsc{scikit-learn} package \citep{scikitlearn}. 
For sources with missing parameters, we replace the missing values with the mean of that parameter from the entire sample. 
Then, we randomly select 60\% of the sample with visual classifications for training, leaving the final 40\% for model validation. After testing a number of different classifiers available in \textsc{scikit-learn}, we decide to use the Gaussian Process (GP) classifier, which implements Gaussian Processes for probabilistic classification. Running our trained model on the validation sample gave an accuracy of $\sim96.5$\%.  For the total sample of 1027 analyzed sources, we find that 94\% (965) of sources are classified as robust by both the visual and GP classifications. 
3\% of the sources (32) are classified as not-robust/spurious by both visual and GP classifications (bringing the overall accuracy to 97\%). Only 1\% of the sources (7) are classified as robust visually but missed by the GP classification. 2\% of the sources were classified as not-robust visually but assigned a robust classification by the GP classifier. Reassuringly, the cases where the visual and GP classifications disagree are all borderline cases where the three visual inspectors are also not in full agreement. 
The model was saved and can be re-used to predict the robustness of counterpart associations for future A$^3$COSMOS runs without the need for visual classification, provided that our current training sample is representative of future datasets. 

After this automated counterpart association step, \aaacosmosNumberOfOutlierCPA{} sources are flagged as spurious sources (they could potentially be noise-boosted or a co-aligned real dusty galaxy). We flag them by the \incode{Flag_outliers_CPA} column in our final galaxy catalog, and discard them for our further analysis in this paper.

\vspace{0.25truecm}

\subsection{Combining multi-wavelength photometry and prior redshifts in the literature}
\label{Section_Combining_prior_redshifts}

To combine the multi-wavelength photometric and spectroscopic information for our prior catalog, we adopt the optical/near-IR photometry from the \cite{Laigle2016} catalog, and use the 3$''$ diameter aperture fluxes to be consistent with \cite{Laigle2016}\,\footnote{\cite{Laigle2016} found that the 3$''$ aperture fluxes lead to better photometric redshift determination and are less affected by uncertainties in the
astrometry. See their Sect.~4.1.}.

Further, we adopt the far-IR/(sub-)mm/radio photometry from \cite{Jin2018}. The authors use detailed ``super-deblended'' procedures following \cite{Liudz2017} to overcome the severe source confusion in their far-IR/(sub-)mm data, which is due to the large beam sizes of the \textit{Herschel} and ground-based single-dish far-IR/(sub-)mm telescopes. Their photometry is prior-based, with the prior catalog constructed by combining the \cite{Laigle2016}, \cite{Muzzin2013} and \cite{Smolcic2017a} catalogs, 
all of which are also in our master catalog. 
The ``super-deblended'' photometry uses the prior information of galaxies' photometric redshifts and SEDs to ``freeze'' low-redshift sources, and includes the step of blindly extracting sources in the residual images and re-fitting together with initial priors. 
Therefore, sources not in the prior catalog or even co-aligned sources at a significantly higher redshift than the prior source have already been reasonably well accounted for (e.g., if prior redshift $<1$, its SED will predict a too low far-IR flux and it gets ``frozen'' during fitting; see details in \citealt{Liudz2017} and \citealt{Jin2018}). 
More complex situations arise if an unknown far-IR source is blending with a prior source whose SED is not constrained well. However, the ALMA data have typically the spatial resolution and sensitivity to distinguish them. 
In this work, we do find about a hundred ALMA sources not in the prior catalog used by \cite{Jin2018}, of which only about 10 are blended with a \cite{Jin2018} prior source (within $1''$), and their ALMA $\sim1$\,mm fluxes ($<1$\,mJy) indicate that they are undetectable by \textit{Herschel} and SCUBA-2. Therefore, using the \cite{Jin2018} catalog for far-IR photometry seems appropriate, especially for those with common priors.

For the SED fitting in this work, we first consider a prior spectroscopic or optical/near-IR photometric redshift if available in the literature. Using photometric redshift is motivated by the sufficiently good agreement between photometric and spectroscopic redshifts as demonstrated by \cite{Laigle2016}.

\begin{figure}[t]
\centering%
\includegraphics[width=0.48\textwidth, trim=5mm 7mm 0 2mm]{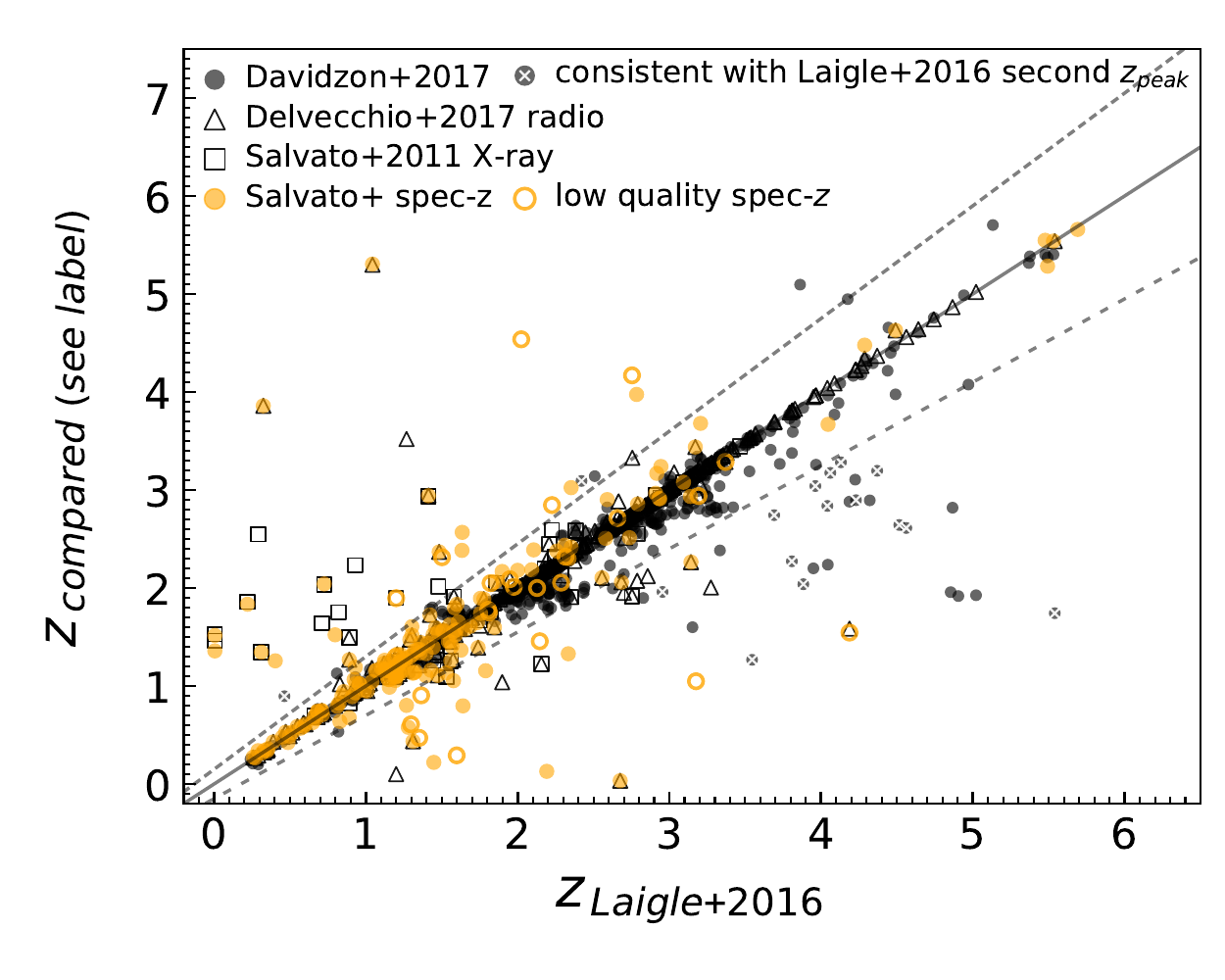}
\caption{%
Comparison between literature photometric and spectroscopic redshifts available for our galaxy sample. Redshifts from various studies in the literature: \cite{Davidzon2017}; \cite{Delvecchio2017}; \cite{Salvato2011} and the M. Salvato et al. compilation catalog of spectroscopic redshifts are plotted against the photometric redshifts from \cite{Laigle2016}.
Each data point represents a master catalog source that has a counterpart in the second respective catalog (see Sect.~\ref{Section_Combining_prior_redshifts} footnote~\ref{Footnote_backward_cross_matching} for the cross-matching). 
Solid orange circles indicate sources with robust spectroscopic redshifts ($\ge2$ spectral features); low quality spectroscopic redshifts with only one spectral feature are shown as open orange circles. 
From the set of sources with photometric redshifts in \cite{Davidzon2017}, we highlight those that have a consistent second probability peak in redshift in the \cite{Laigle2016} catalog by a white cross inside the black circle (mostly around redshift 4 in \cite{Laigle2016} catalog). 
The solid line presents the one-to-one relation, and the dashed gray lines indicate $\pm0.15\times(1+z)$ catastrophic errors (e.g., \citealt{Laigle2016}, Sect.~4.3). 
\label{Plot_z_spec}
\vspace{0.5cm}
}
\end{figure}

In this work, we examine all the spectroscopic and photometric redshifts in the literature listed in Sect.~\ref{Section_Data}. 
We show the comparison of these redshifts (hereafter prior redshift, or ``prior-$z$'') in Fig.~\ref{Plot_z_spec}, where each data point represents a galaxy in our galaxy catalog 
and has prior-$z$ from both the \cite{Laigle2016} catalog and other catalogs\,%
\footnote{%
    To make sure we select common sources in these catalogs, 
    we first do a backward cross-matching from each compared catalog to our full COSMOS master catalog (Sect.~\ref{Section_Prior_Source_Catalogs}; with $1''$ radius). Then we identify common sources by matching the exact master catalog ID. This avoids linking of different sources in the different catalogs which are closer than our cross-matching radius of $1''$.
    While the nominal false-match probability with this matching radius is 11\% (applying Eq.~1 of \citealt{Pope2006}), 
    we note that the it is only indicative of the likelihood of spurious cross matches between catalogs in a statistical sense, based on the number density of sources and distance between counterparts, but does not include physical information about these matches. Since we have a priori information about whether catalog matches are physically realistic, the actual value of the ``false-match probability'' will be lower than the listed values in this manuscript. 
    \label{Footnote_backward_cross_matching}
}:
the M. Salvato et al. spectroscopic redshift catalog; the \cite{Davidzon2017} photometric catalog for the same UltraVISTA galaxies as \cite{Laigle2016} but with optimized SED fitting for $z>2.5$ sources; the \cite{Delvecchio2017} photometric catalog for radio-detected galaxies; and the \cite{Salvato2011} photometric catalog for X-ray-detected active galactic nuclei (AGNs). 

The majority of our sample galaxies show good consistency among all available prior redshifts. However, we do find several types of outliers: 
(1) About 14 X-ray-detected AGNs have higher redshifts in the \cite{Salvato2011} than in the \cite{Laigle2016} catalog (see open squares in Fig.~\ref{Plot_z_spec}), but about half (6) of them have spectroscopic redshifts in good agreement with the \cite{Salvato2011} values (see overlap between open squares and yellow circles in Fig.~\ref{Plot_z_spec}). 
(2) About 25 $z>3$ galaxies have lower redshifts in \cite{Davidzon2017} than in \cite{Laigle2016}, as indicated by the black solid circles in Fig.~\ref{Plot_z_spec}, but about half (14) of them have consistent second redshift peaks in \cite{Laigle2016} (see the black solid circles with white cross in Fig.~\ref{Plot_z_spec}). 
(3) About 10 low quality spectroscopic redshifts (i.e., with two or less detected spectral features to determine the respective redshift) disagree with \cite{Laigle2016}, yet both could have large uncertainties (see yellow circles outside the area enclosed by dashed lines in Fig.~\ref{Plot_z_spec}). 

In our next step, we will run SED fitting to obtain galaxies' stellar mass and SFR properties, but with redshift fixed to a prior-$z$\,\footnote{We have also run another set of SED fitting without a prior-$z$, which is presented later in the last paragraph of Sect.~\ref{Section_Running_SED_Fitting}.}. For galaxies with consistent prior-$z$ or a single prior-$z$ from the above catalogs, we directly use it for the SED fitting. But for galaxies with inconsistent prior-$z$ ($\Delta{z}>0.15\times(1+z)$) from the above catalogs, we run SED fitting for each inconsistent prior-$z$ and take the one with minimum-$\chi^2$ at the ALMA bands as our best fit. The details are presented in the next section.

\vspace{0.25truecm}

\subsection{SED fitting}
\label{Section_Running_SED_Fitting}

We use \textsc{MAGPHYS} \citep{daCunha2008,daCunha2015}\,\footnote{\url{http://www.iap.fr/magphys/}} for the SED fitting, as it has rich stellar SED libraries and has been widely tested on local and high-redshift galaxies (e.g., \citealt{Smith2012}; \citealt{Berta2013}; \citealt{Rowlands2014a,Rowlands2014b}; \citealt{Smith2015}; \citealt{Hayward2015}; \citealt{Smolcic2015}; \citealt{Miettinen2017a,Miettinen2017b}; \citealt{Delvecchio2017}; \citealt{Hunt2019}). 
It assumes an energy balance between the energy attenuated by dust in the UV/optical and that radiated by dust at IR/mm wavelengths. 
As it is debated whether this energy balance is still robust for very dusty galaxies (e.g., \citealt{Simpson2017};
\citealt{Casey2017}), 
we provide some supporting evidence for the assumption of energy balance in our whole galaxy sample (see below in this section). 

Due to the large number of templates being fitted, \textsc{MAGPHYS} per default fits the SED at a fixed prior-$z$ (which can be either photo-$z$ or spec-$z$ from the literature). A wrong prior-$z$ can easily lead to a poor fit with a large residual at the wavelengths of the ALMA bands, which is measured by the reduced chi-square:
\begin{equation}
\chi^2_{\mathrm{ALMA}} \equiv \sum\limits_{\mathrm{ALMA}}
\frac{(|S_{\mathrm{SED}}-S_{\mathrm{OBS}}|)^2}{\sigma_{S_{\mathrm{OBS}}}^2}/N_{\mathrm{ALMA}}
\end{equation}
where $\sigma_{S_{\mathrm{OBS}}}$ is the flux error and $N_{\mathrm{ALMA}}$ is the number of ALMA data points. 
Therefore, 
we consider all possible prior-$z$'s for a given galaxy and fit each of them before choosing the fit with the lowest $\chi^2_{\mathrm{ALMA}}$ as the final best fit.\,\footnote{We treat spec-$z$'s the same as photo-$z$'s, except that only when the $\chi^2_{\mathrm{ALMA}}$ of a fitting at a spec-$z$ is at least a factor of 1.5 worse than that fitted at a photo-$z$ do we discard the spec-$z$ fitting.}

\begin{figure}[t!]
\centering%
\includegraphics[width=0.40\textwidth, trim=10mm 3mm 0 0]{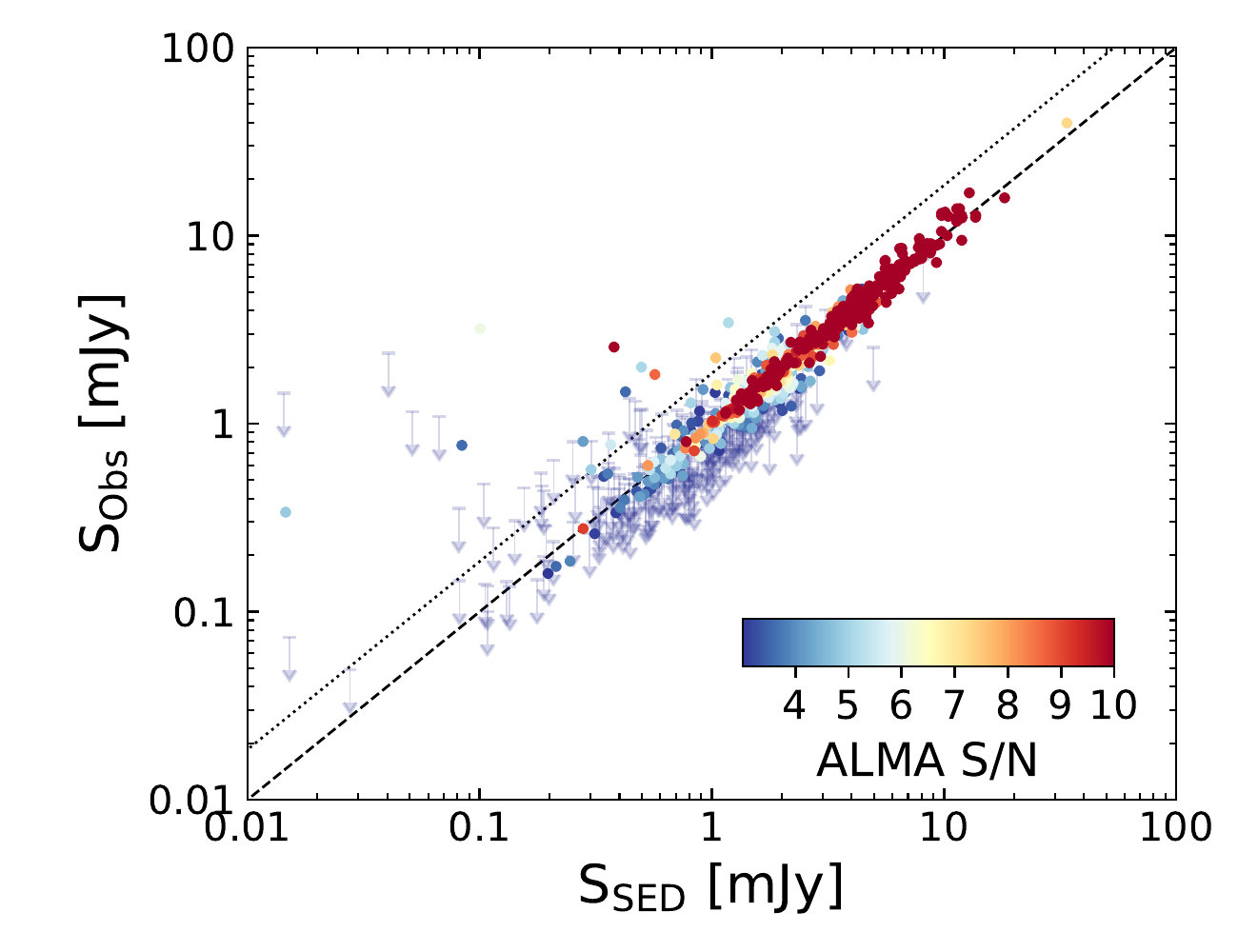}
\includegraphics[width=0.40\textwidth, trim=0 3mm 0 0]{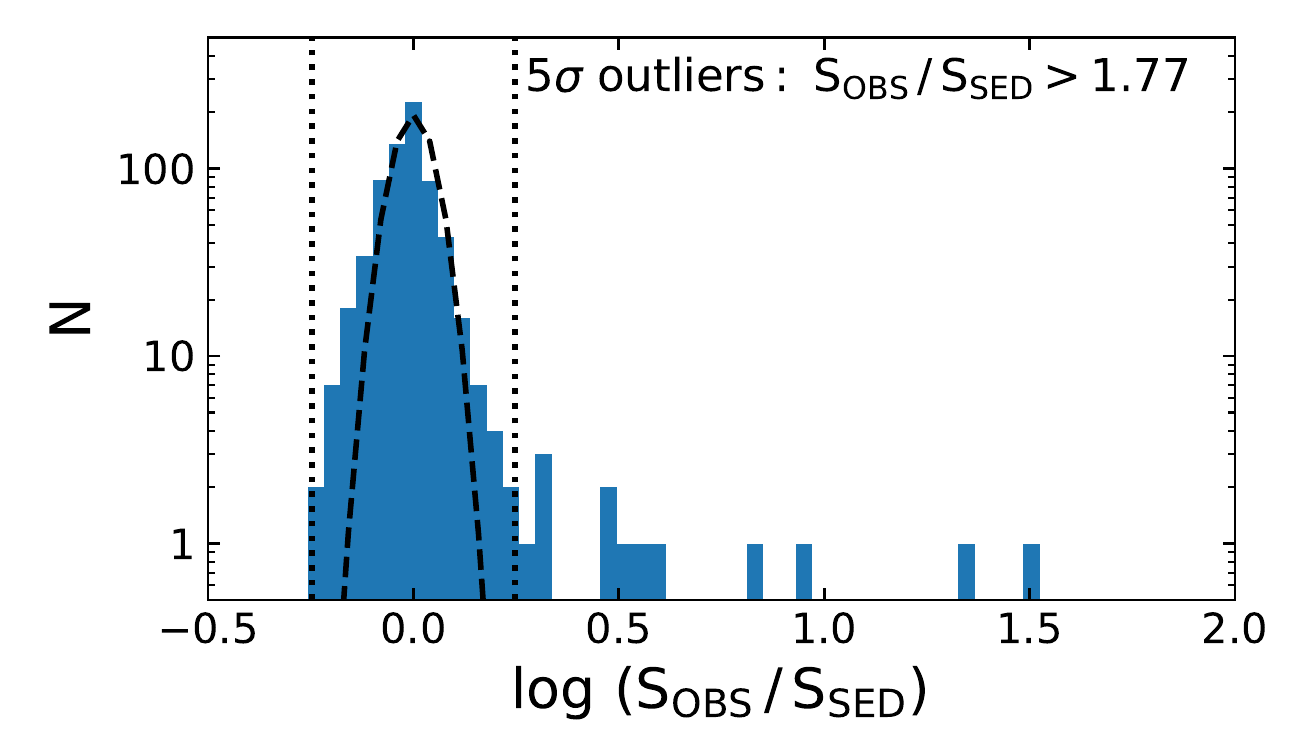}
\caption{%
\textbf{Upper panel:} 
Comparison of \textsc{MAGPHYS} SED-predicted ($S_{\mathrm{SED}}$) and observed fluxes ($S_{\mathrm{OBS}}$; already corrected for flux bias and error based on our simulation in Sect.~\ref{Section_MC_Sim_Final_Correction}) at all available ALMA photometric bands for each galaxy in our sample (Sect.~\ref{Section_Combining_two_photometry_catalogs}; removed spurious sources/outliers in Sect.~\ref{Section_Examining_counterpart_association}). 
Color indicates the $\SNR$ of the measured total ALMA flux. Arrows are the 3$\sigma$ upper limits for sources with total flux $\SNR<3$. 
The dashed line shows the one-to-one relation and the dotted line indicates the 5-$\sigma$ threshold derived from the histogram in the lower panel. 
\textbf{Lower panel:} Histogram of $\log_{10}(S_{\mathrm{OBS}}/S_{\mathrm{SED}})$ for sources with a total flux of $\SNR\ge3$ (i.e., excluding upper limits). 
The dashed curve shows the best-fit 1D Gaussian. The vertical dotted lines indicate the 5$\sigma$ range. We identify sources outside the 5$\sigma$ range (i.e., $S_{\mathrm{OBS}}/S_{\mathrm{SED}}>1.77$) as ``SED-excess'' outliers (see text for details). 
\label{Plot_SED_fitting_magphys_fObs_fSED}
}
\end{figure}

The final values of $\chi^2_{\mathrm{ALMA}}$ are generally well behaved. In Fig.~\ref{Plot_SED_fitting_magphys_fObs_fSED}, we compare the difference between $S_{\mathrm{SED}}$ and $S_{\mathrm{OBS}}$ at all available ALMA bands for each galaxy.
The median of $S_{\mathrm{SED}}-S_{\mathrm{OBS}}$ for all total flux $\SNRtotal>3$ ALMA photometry is consistent with being zero, suggesting that \textsc{MAGPHYS} fitting has no
obvious systematic over or underestimation of the flux. There are about 25\% of data points with $\SNRtotal<3$ (but $\SNRpeak$ meets our sample selection criterion) which are shown as 3-$\sigma$ upper limits, and 60\% of them are consistent with the SED flux (being above the one-to-one line). 
The histogram of $\log_{10}(S_{\mathrm{OBS}}/S_{\mathrm{SED}})$ in the lower panel is fitted with an 1D Gaussian with $\mu=-0.01$ and $\sigma=0.05$. Its upper 5-$\sigma$ envelope corresponds to $S_{\mathrm{OBS}}/S_{\mathrm{SED}}=1.77$, above which we do find 3\%
outliers. Most of these ``SED-excess'' outliers have low total flux $\SNR$ (i.e., $\SNRtotal<4-5$ as indicated by the color-coding in Fig.~\ref{Plot_SED_fitting_magphys_fObs_fSED}). 

We speculate that the outliers are most likely spurious sources boosted by noise which by chance align with their optical/near-IR counterparts and are thus not removed by our earlier counterpart association step. Since their $\SNRpeak$ pass our previous sample selection criterion, they tend to be large in angular size. 
And this number is actually supported by the statistics: we expect $\lesssim12\%$ ($\lesssim$140) spurious sources due to our $\SNRpeak$ selection in Sect.~\ref{Section_Combining_two_photometry_catalogs}, 
which is then reduced by $\sim4\%$ by our counterpart association examination in Sect.~\ref{Section_Examining_counterpart_association}. 
Meanwhile, we have $\sim$130,000 master catalog sources within the current dataset totaling 946~arcmin$^2$ regardless of primary beam areas ($\sim$23,000 within primary beam areas, which sum up to \aaacosmosAlmaPointingArea{}~arcmin$^2$); 
so we expect a false-match probability of 3\% with a matching radius of $0.5''$ (Eq.~1 of \citealt{Pope2006}), i.e., only $\sim$4 spurious sources to coincide with some prior sources by chance alignment. 

However, we note that there is also a chance that there is an unidentified ALMA source at the same line-of-sight as the foreground prior source thereby boosting the ALMA flux to much higher than what SED could fit. These SED-excess outliers are rare but do exist, e.g., the $z\sim5.7$ background ALMA source ``CRLE'' found by \cite{Pavesi2018}, which is not in any optical/near-IR/radio catalog but is at the same line-of-sight with a foreground $z\sim0.3$ galaxy in the \cite{Laigle2016} catalog. 

Similar to the counterpart association flagging, we flag \aaacosmosNumberOfOutlierSED{} sources as SED-excess outliers. They are indicated by the \incode{Flag_outlier_SED} column in our final galaxy catalog, and will no longer be considered in our further scientific analysis. 

Furthermore, in order to verify whether doing a completely blind photometric redshift scan could lead to better fits (smaller $\chi^2$) or not, we adopt the recently developed photo-$z$ version of the \textsc{MAGPHYS} code (\textsc{MAGPHYS+photo-z}; A. Battisti et al. in prep.). It considers redshift as a free parameter between $z$=0 and 8 and generates identical libraries to the original version of \textsc{MAGPHYS} for each redshift. The output of this step is a probability distribution function (PDF) of the photometric redshift. We perform this photo-$z$ fitting for all our sources and compare the best-fit redshifts (derived as the median of the PDF) to available spectroscopic redshifts, finding no obvious systematic offset (an 1D Gaussian fitting to the distribution of $(z_{\mathrm{photo.}}-z_{\mathrm{spec.}})/(1+z_{\mathrm{spec.}})$ gives $\mu=-0.015$ and $\sigma=0.045$). The comparison with all prior-$z$ also shows no obvious systematic offset (an 1D Gaussian fitting to the distribution of $(z_{\mathrm{photo.}}-z_{\mathrm{prior.}})/(1+z_{\mathrm{prior.}})$ gives $\mu=0.000$ and $\sigma=0.076$).

Comparing the physical properties obtained from the two SED fitting for common sources, we find a median difference (scatter) of 0.0\,dex (0.05\,dex) and 0.0\,dex (0.04\,dex) for $\log\Mstar$ and $\log\SFR$, respectively. However, we do note that the uncertainties in $\log\Mstar$ and $\log\SFR$ are systematically larger in photo-$z$ SED fitting when the uncertainties in redshift are included. (The histogram of the difference in uncertainty has a median of 0.0\,dex but has a second peak at 0.2\,dex and extends to 0.4\,dex.) Therefore, for $\log\Mstar$ and $\log\SFR$ in our final galaxy catalog, we take the uncertainties from the photo-$z$ SED fitting which includes the redshift uncertainty, while keep the best-fit values still from the best prior-$z$ fit.

In this photo-$z$ experiment, we also tested the photo-$z$ of the SED-excess outliers, finding that for 7 of them the photo-$z$ are between $z$=2--4 whereas the prior-$z$'s are below $z$=1, while the remaining 14 have photo-$z$ and prior-$z$ consistent with $z$=0--2. Note that the \textsc{MAGPHYS} photo-$z$ fitting places more weight on the stellar SED when the optical/near-IR bands have more data points than the FIR/mm bands. Thus these SED-excess outliers will still show an excess in their observed ALMA fluxes relative to the SED predicted flux. 
Given their unreliable photo-$z$'s, such sources will benefit from a better FIR/mm coverage as will be available from future submm/mm surveys like JCMT/SCUBA2 S2COSMOS (at 850\,$\mu$m; PI: I. Smail), STUDIES (at 450\,$\mu$m; PI: W. Wang), IRAM~30m/NIKA2 Cosmology Legacy Survey (N2CLS; at 1\,\&\,2\,mm; PI: G. Lagache), and the LMT/TolTEC Ultra-Deep Galaxy Survey (at 1\,\&\,2\,mm).

\vspace{0.25truecm}

\subsection{Correcting significant contribution from emission lines}
\label{Section_Subtracting_Strong_Emission_Lines}

In sensitive (sub-)mm observations like the majority of the ALMA observations in the COSMOS field, strong (sub-)mm spectral lines like \CII_{}, \NII_{} and high-$J$ CO emission from high-redshift galaxies can strongly bias the dust continuum measurement if they are bright enough and fall in the bandwidth of the spectral setup\,\footnote{For example, ALMA can detect \CII_{} from a $\SFR\lesssim50\,\Msyr$, $z\sim5$ galaxy with $\lesssim30$\,min on-source time (\citealt{Capak2015}; or only $\sim2$\,min if $\SFR\sim1000\,\Msyr$; \citealt{Swinbank2012}; \citealt{Cooke2018}), or high-$J$ CO lines from a $\SFR\sim500\,\Msyr$, $z\sim1.5$ galaxy with $\lesssim30$\,min on-source time (\citealt{Silverman2015b}).}. In special cases, these lines will dominate the emission from the whole bandwidth, e.g., mostly $\sim$8~GHz of the current ALMA receiver. This is more significant in the lower frequency 3mm observations, and will become more critical in the future with even deeper observations from ALMA and for the large surveys mentioned in the previous section. It is therefore necessary to consider strong submm/mm line emission together with the dust continuum in photometry pipelines. Although when the observation is not intended for line detection, the chance of a strong emission line being in the bandwidth is very low (e.g., $\sim1.6\%$, from the blind \CII_{} line search work by \citealt{Cooke2018} who found 10 line emitters out of 695 ALMA continuum sources), but when the number of sources becomes large as in this and future works (with automated pipelines), the line emitters must be systematically corrected for. 

As our continuum images are obtained by directly collapsing all channels of all spectral windows ignoring whether the PI intended a line detection or not, a strong (sub-)mm emission line could potentially ``contaminate'' the measured continuum flux. Therefore we developed a pipeline to automatically identify such cases and to apply a first rough correction for these lines. Direct blanking of channels affected by line emission before construction of the continuum image would require either a good a-priori knowledge of the redshift or dedicated line searches (that are not part of this project) as well as special treatment of each source present in a single pointing. Both aspects not only result in a significant
increase in data volume and analysis time required but also in an inhomogeneous dataset. Given the small fraction of potentially affected sources of 7\% (see below), our adopted approach is sufficient for our purpose. 

Our pipeline uses the redshift and SFR (and IR color, e.g., rest-frame $S_{70\,\mu\mathrm{m}}/S_{160\,\mu\mathrm{m}}$ from SEDs, when necessary) to predict for each source the low- to high-$J$ CO (upper level quantum number $2 \le J_{\mathrm{upper}} \le 10$), \CI_{} $^3P_2\to^3P_1$ and $^3P_1\to^3P_0$ (at rest-frame 370 and 609$\,\mu$m respectively), \NII_{} $^3P_2\to^3P_1$ and $^3P_1\to^3P_0$ (at rest-frame 122$\,\mu$m and 205$\,\mu$m respectively) and \CII_{} $^2P_{3/2}\to^2P_{1/2}$ (at rest-frame 158$\,\mu$m). We do not account for other lines in this work because those are predicted to fall outside the frequency range or are generally much weaker. The line prediction follows empirical luminosity--luminosity correlations:  \CII_{}--$L_{\mathrm{IR}}$ correlation from \cite{DeLooze2011}, with a \CII_{} deficit roughly proportional to $L_{\mathrm{IR}}^{-0.335}$ when $L_{\mathrm{IR}}>10^{10}\;\mathrm{L_{\odot}}$ which fits the data best; \NII_{}--$L_{\mathrm{IR}}$ correlation from \cite{ZhaoYinghe2013,ZhaoYinghe2016}; CO(1-0)--$L_{\mathrm{IR}}$ correlation from \cite{Sargent2014}; high-$J$ ($J_{\mathrm{upper}}\ge4$) CO--$L_{\mathrm{IR}}$ correlation from \cite{Liudz2015}; and \CI_{}--$L_{\mathrm{IR}}$ correlation based on the data sets in \cite{Liudz2015} and \cite{Valentino2018}. For CO $2 \le J_{\mathrm{upper}} \le 3$ lines, we interpolate the line luminosity using the the CO(1-0)--$L_{\mathrm{IR}}$ and CO(4-3)--$L_{\mathrm{IR}}$ correlations.

\begin{figure}[t!]
\centering%
\includegraphics[width=\linewidth, trim=5mm 5mm 5mm 2mm]{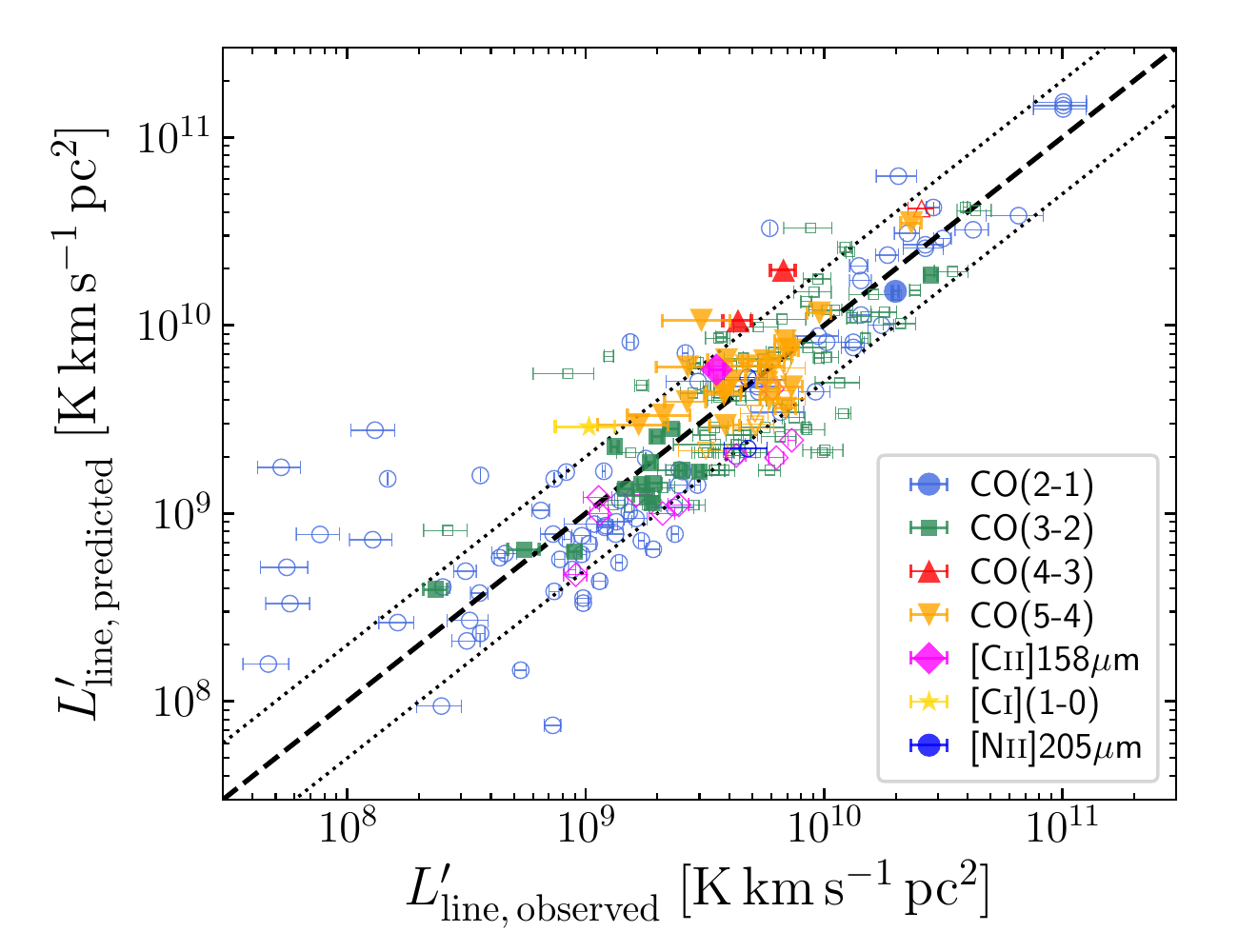}
\caption{%
Comparison of predicted and observed (sub-)mm molecular/atomic line luminosities for a large sample of galaxies with available CO, \CI_{}, \CII_{} or \NII_{} luminosity and SFR or IR luminosity in the literature and from this work. This figure verifies our line prediction pipeline which corrects the measured ALMA continuum flux for the emission line ``contamination'' (see description in Sect.~\ref{Section_Subtracting_Strong_Emission_Lines}). 
Color and symbol indicate different emission lines. Solid symbols are $\sim$50 sources which have (sub-)mm lines contributing to their measured continuum flux by more than 20\% by our prediction. We inspected their data cubes and extracted their (sub-)mm lines and therefore compared to the prediction. Open symbols are 234 galaxies with CO, \CI_{}, \CII_{} or \NII_{} detections with $\SNR>3$ in the literature (see references in Sect.~\ref{Section_Subtracting_Strong_Emission_Lines}). The dashed line is a one-to-one line, and the thin dotted lines indicate a factor of 2 scatter. 
\label{Plot_line_prediction}
}
\end{figure}

Meanwhile, we obtain the exact frequency setups for each ALMA observation from the ALMA archive, and identify the predicted strong (sub-)mm lines within the frequency setups. We estimate the line contribution to the measured continuum by dividing the predicted line flux by the total bandwidth and compare that to the measured continuum. Our prediction suggests that $\sim$50 ($\sim7\%$) sources have (sub-)mm lines contributing more than 20\% to the measured continuum. We looked into their data cubes and found that most of them do have line emission as predicted, as all except four have accurate redshift from the M. Salvato spectroscopic redshift compilation. A strong emission line is predicted but not found to be present for only three sources with spectroscopic redshift (A3COSMOS master catalog IDs~1236908, 350733 and 418763)
and two with photometric redshift (IDs~339509 and 1236904). 
Interestingly, two sources (IDs~990180 and 861198) without spectroscopic redshifts from the M. Salvato compilation do show a line detection, 
and their spectroscopic redshifts are also reported in the literature (\citealt{Lee2017}; Cassata et al. in prep.). More details of the A3COSMOS line search work will be presented in future papers. Here we have measured those (sub-)mm lines\,\footnote{%
The line search is done in the $uv$-plane adapting the methodology of \cite{Silverman2015b} and D. Liu et al. (2019, in preparation), with \CASA{} and \textsc{GILDAS}.}
to verify our prediction, and the comparison is presented in Fig.~\ref{Plot_line_prediction} where solid symbols are these A$^3$COSMOS sources. Their measured line luminosity ($x$-axis) and predicted line luminosity ($y$-axis) show good agreement (the dashed lines indicate a factor of 2 range). 
The pipeline also predicts $<20\%$ line contributions for more sources, but as these lines could not be measured at sufficient $\SNR$ in the data cube, they are omitted from the figure. 

In Fig.~\ref{Plot_line_prediction}, we added 234 line detections with $\SNR>3$ for CO, \CI_{}, \CII_{} or \NII_{} from the literature as follows: 
\cite{%
Albrecht2007,
Baan2008,
Bauermeister2013,
Bertemes2018,
Capak2015,
Carilli_and_Walter_2013,
Daddi2015,
Lee2017,
Magdis2017,
Magnelli2012,
Pavesi2018,
Saintonge2017,
Silverman2015,
Spilker2018,
Tacconi2013,
Tan2014,
Yao2003}. 
SFRs from these works and in addition from \cite{Sanders2003} and \cite{Brinchmann2004} are used for our line prediction. 
The distribution of $\log_{10} (L^{\prime}_{\mathrm{line,\,observed}}/L^{\prime}_{\mathrm{line,\,predicted}})$ has a mean of 0.07 and scatter of 0.27. 
Some disagreement can be found at the lowest end where line luminosity $L^{\prime}_{\mathrm{line,\,observed}}\sim10^{8}\;\mathrm{K\,km\,s^{-1}\,pc^{2}}$. As our current data do not cover this faint regime, improvement is postponed to a future work.

After the correction for strong (sub-)mm line ``contamination'', we reiterate over the SED fitting step. Note that in Fig.~\ref{Plot_SED_fitting_magphys_fObs_fSED} the data points represent already the final continuum fluxes corrected for line contamination.

\vspace{0.25truecm}

\subsection{Obtaining galaxy properties from SED fitting}
\label{Section_Obtaining_Galaxy_Properties}

From \textsc{MAGPHYS} SED fitting, we obtain the following galaxy properties: stellar mass ($\Mstar$), mass-weighted stellar age, $V$-band attenuation $A_V$, star formation history (SFH) integrated $\mathrm{SFR_{SFH}}$, and total IR luminosity $L_{\mathrm{IR}}$ (integrated over 8--1000$\,\mu$m). 
For each property, \textsc{MAGPHYS} gives a minimum-$\chi^2$ (i.e., best-fit) value, as well as the median and the lower and upper 68$^{\mathrm{th}}$ percentiles of the PDF. 

Our final SFRs are computed from the IR luminosity with the \cite{Kennicutt1998SFL} calibration and assuming a \cite{Chabrier2003} IMF:
\begin{equation} 
\frac{\mathrm{SFR_{IR}}}{[\mathrm{M_{\odot} \; yr^{-1}}]} = \frac{L_{\mathrm{IR,\,8\mathrm{-}1000{\mu}m}}}{[\mathrm{L_{\odot}}]} \times 10^{-10}
\end{equation}
By comparing $\mathrm{SFR_{SFH}}$ and $\mathrm{SFR_{IR}}$, we find that the distribution of $\log_{10}(\mathrm{SFR_{IR}}/\mathrm{SFR_{SFH}})$ has more pronounced wings than an 1D Gaussian, with a mean of 0.14 and a standard deviation of 0.15. As mentioned in \cite{Kennicutt1998SFL}, the calibration of $\mathrm{SFR_{IR}}$ is based on the starburst synthesis models of \cite{Leitherer1995} assuming a constant SFH with a young age of 10--100\,Myr (in which time the bolometric luminosity-to-SFR ratio is relatively constant), and assuming that dust re-radiates all the bolometric luminosity. The difference between $\mathrm{SFR_{IR}}$ and $\mathrm{SFR_{SFH}}$ could thus come from either the actual fitted SFHs, 
the fraction of bolometric luminosity re-radiated by dust, the variation of bolometric luminosity-to-SFR ratio with stellar population ages, or other additional effects. In the following analysis, we will use $\mathrm{SFR_{IR}}$ (and hereafter SFR) because the simple \cite{Kennicutt1998SFL} calibration is widely used in studies focused on the dusty galaxy population at high redshift and given the fact that our sample is biased toward massive, dusty galaxies at high redshift. 

Through a detailed simulation and recovery study, \cite{Hayward2015} tested the accuracy of \textsc{MAGPHYS} in recovering galaxies' physical properties. They found that for isolated disk galaxies, \textsc{MAGPHYS} recovers well the physical properties above mentioned. However, for galaxy mergers, there might be some bias in the determined dust masses (\cite{Hayward2015} found that \textsc{MAGPHYS} underestimates by 0.1--0.2\,dex [and up to 0.6\,dex] the dust mass during the post-starburst phase of a galaxy merger). 
Therefore, we do not provide dust masses in our final catalog and defer this to later work (our Paper~\textsc{II}).

\begin{figure}[t!]
\centering
\includegraphics[width=\linewidth, trim=12mm 5mm 0 0]{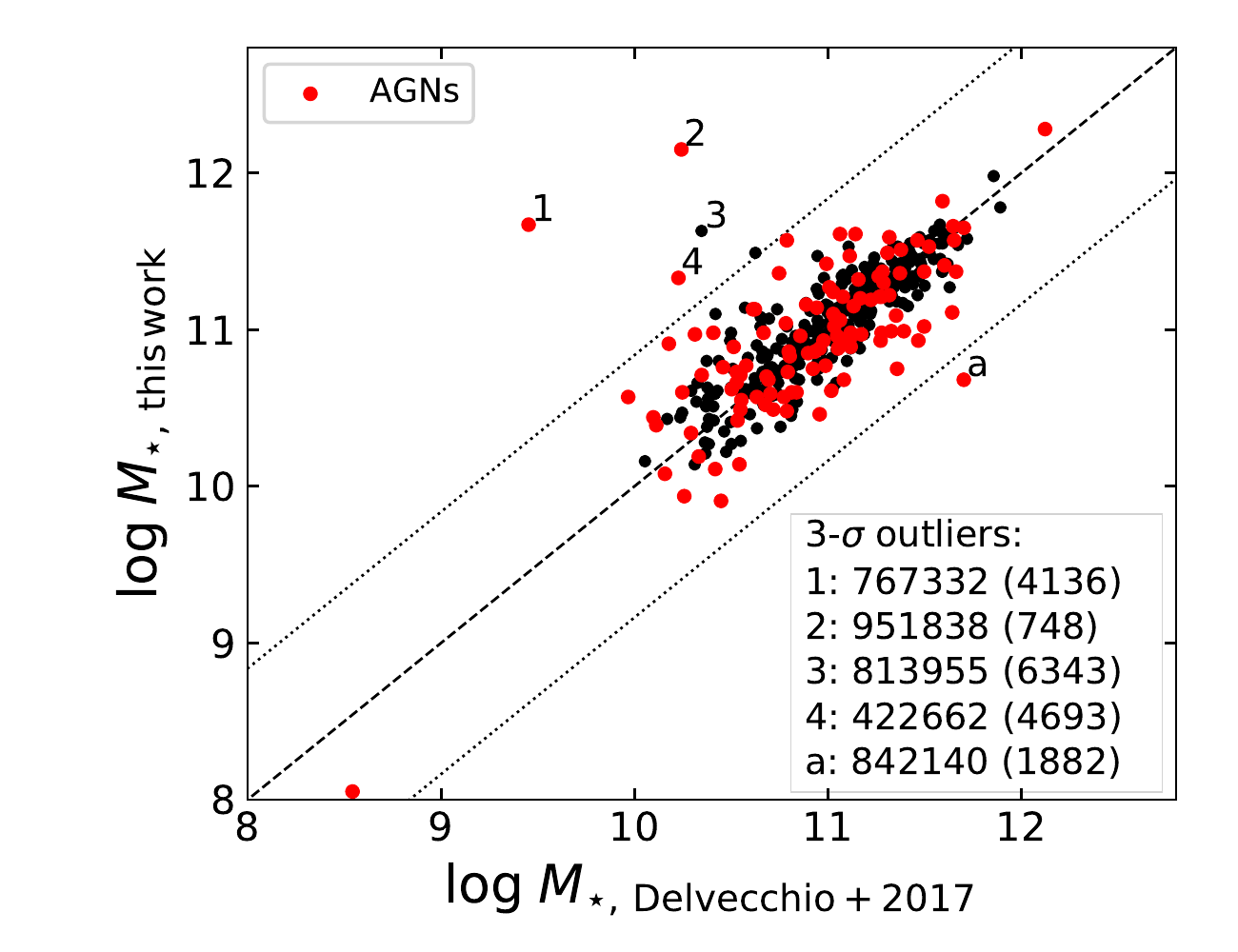}
\caption{%
    Comparison of our final $\Mstar$ from the multi-run \textsc{MAGPHYS} SED fitting and those of \cite{Delvecchio2017} who used \textsc{SED3FIT} to account for the mid-IR AGN component. AGNs classified by \cite{Delvecchio2017} are highlighted in red. Four sources
    have highly overestimated $\Mstar$ by our method (labeled \incode{1}, \incode{2}, \incode{3} and \incode{4}), while one sources exhibits a significantly underestimated $\Mstar$ (labeled \incode{a}). Their corresponding A$^3$COSMOS master catalog ID (and \citealt{Delvecchio2017} \incode{ID_VLA3} in brackets) are listed. They are discussed in Sect.~\ref{Section_Obtaining_Galaxy_Properties}. 
    The dashed line is the one-to-one relation and the dotted lines indicate the $\pm$3$\sigma$ range (with $\sigma$ being the standard deviation). 
    \label{Plot_comparison_with_Delvecchio2017_for_Mstar}
}
\end{figure}

\cite{Hayward2015} also found that for AGN host galaxies, when the AGN does not significantly contribute to the UV--mm luminosity (e.g., $<25\%$), the absence of a mid-IR AGN component in \textsc{MAGPHYS} is not significantly affecting the best-fit results. 
However, stronger mid-IR AGNs can lead to an overestimation of stellar mass and SFR. 
In our final sample (after removing outliers in Sect.~\ref{Section_Examining_counterpart_association}~and~\ref{Section_Running_SED_Fitting}), 34 galaxies are AGN hosts in the \cite{Salvato2011} \textit{XMM-Newton} catalog and 48 are in the \cite{Salvato2011} \textit{Chandra} catalog. Meanwhile, 112 are classified as AGNs via SED fitting with an AGN component using \textsc{SED3FIT} (\citealt{Berta2013}) by \citet{Delvecchio2017}. 
These catalogs have overlaps, thus the final number of AGNs is 158 ($\sim$23\%). 

We try to assess the mid-IR AGN problem by running \textsc{MAGPHYS} twice, one time including and the other time excluding the mid-IR 24$\,\mu$m flux information. Then we adopt the fit with the smaller $\chi^2$ as our final best fit. The fitting excluding the 24$\,\mu$m data usually leads to a better $\chi^2$. 
The overall difference between the derived IR luminosity is very small: 
the distribution of the difference in $\log_{10} L_{\mathrm{IR}}$ between the two SED fitting results has a median of 0.0~dex and sigma of 0.17~dex. This distribution is slightly broadened to a sigma of 0.25~dex for the AGN subsample, but the median is still close to zero. About 20 sources are 3-sigma outliers, but for most cases the difference is caused by low $\SNR$ data at the FIR/mm wavelengths. Only 4 of them are AGNs according to the \citet{Delvecchio2017} classification. 

In Fig.~\ref{Plot_comparison_with_Delvecchio2017_for_Mstar} we further compare our final $\Mstar$ to the \cite{Delvecchio2017} \textsc{SED3FIT} fitted $\Mstar$ for 396 sources in common (with consistent redshifts and coordinates). AGNs are highlighted in red. This demonstrates a good agreement (within 3$\sigma$). We find five outliers (labeled with \incode{1}-\incode{4} if our $\Mstar$ larger and ``\incode{a}'' if our $\Mstar$ is smaller) exceeding the 3$\sigma$ envelope of the distribution. Their corresponding A$^3$COSMOS master catalog IDs and \cite{Delvecchio2017} IDs are listed in the figure. 
Through detailed inspection, we find that the difference is mainly caused by including the ALMA data in the SED fitting, which leads to a higher dust attenuation and thus higher stellar mass.

In addition, the source shown with the highest stellar mass of $\sim10^{12}\;\Msun$ (ID\,223951)
in Fig.~\ref{Plot_comparison_with_Delvecchio2017_for_Mstar}
is the strong AGN XID2028 at $z = 1.593$ studied by \cite{Brusa2015}, \cite{Cresci2015}, \cite{Perna2015a} and \cite{Brusa2018}.  \cite{Brusa2018} estimated a stellar mass of $\log_{10}(\Mstar/\Msun) = 11.65^{+0.35}_{-0.35}$ via optical-to-mm SED fitting including an AGN component. For comparison, we obtain $\log_{10}(\Mstar/\Msun) = 12.28 \pm 0.07$, almost consistent with their upper boundary. 
Interestingly, the reduced-$\chi^2$ at the stellar wavelengths of our \textsc{MAGPHYS} SED fitting is as poor as for the outliers \incode{2} and \incode{3} with a \incode{rchi2_star}~$\sim 6.8$ (top $\sim$10\% of the worst fits). 
\cite{Delvecchio2017} accounting for mid-IR AGN contamination obtain $\log_{10}(\Mstar/\Msun) = 12.12$ (with an uncertainty of the order of 0.1~dex, see their Sect.~6.1). This indicates that our estimate is still acceptable for such an extreme case (although they should be treated with caution in individual studies). 

To summarize our detailed comparison of the robustness of the derived parameters for AGNs, we find: 
\begin{enumerate}[label=(\arabic*), topsep=0pt, noitemsep]
    \item Our current multi-run, iterative \textsc{MAGPHYS} SED fitting, although without an AGN component, achieves in general good agreement with SED fitting that includes an AGN component. The agreement is valid even for the AGN population identified in \cite{Delvecchio2017} and is within the uncertainties even for the most extreme AGNs, e.g., reported in \cite{Brusa2018}. 
    
    \item 
    For very few (5 out of 396) sources our stellar masses lie outside the 3$\sigma$ range when comparing to the \cite{Delvecchio2017} stellar masses. Three of them exhibit strong mid-IR AGN emission contaminating near-IR IRAC and even optical bands. Thus their stellar SEDs are poorly fitted, with \incode{rchi2_star}~$\gtrsim 10$. 
    These extreme outliers are further discussed in Appx.~\ref{Section_Appendix_outliers_of_SED_fitting_AGNs}. Their stellar mass estimates in this work should be treated with caution when used in individual studies. 
    
    \item The inclusion of ALMA (and far-IR/(sub-)mm) data points is crucial for codes like \textsc{MAGPHYS} which assumes energy balance. If the energy balance is valid for these dusty, ALMA-detected sources studied here, our stellar masses and IR luminosities are more reliable than optical-only estimates. 
\end{enumerate}

\begin{figure*}[htb]
\centering%
\includegraphics[width=0.49\textwidth, trim=0 5mm 0 0]{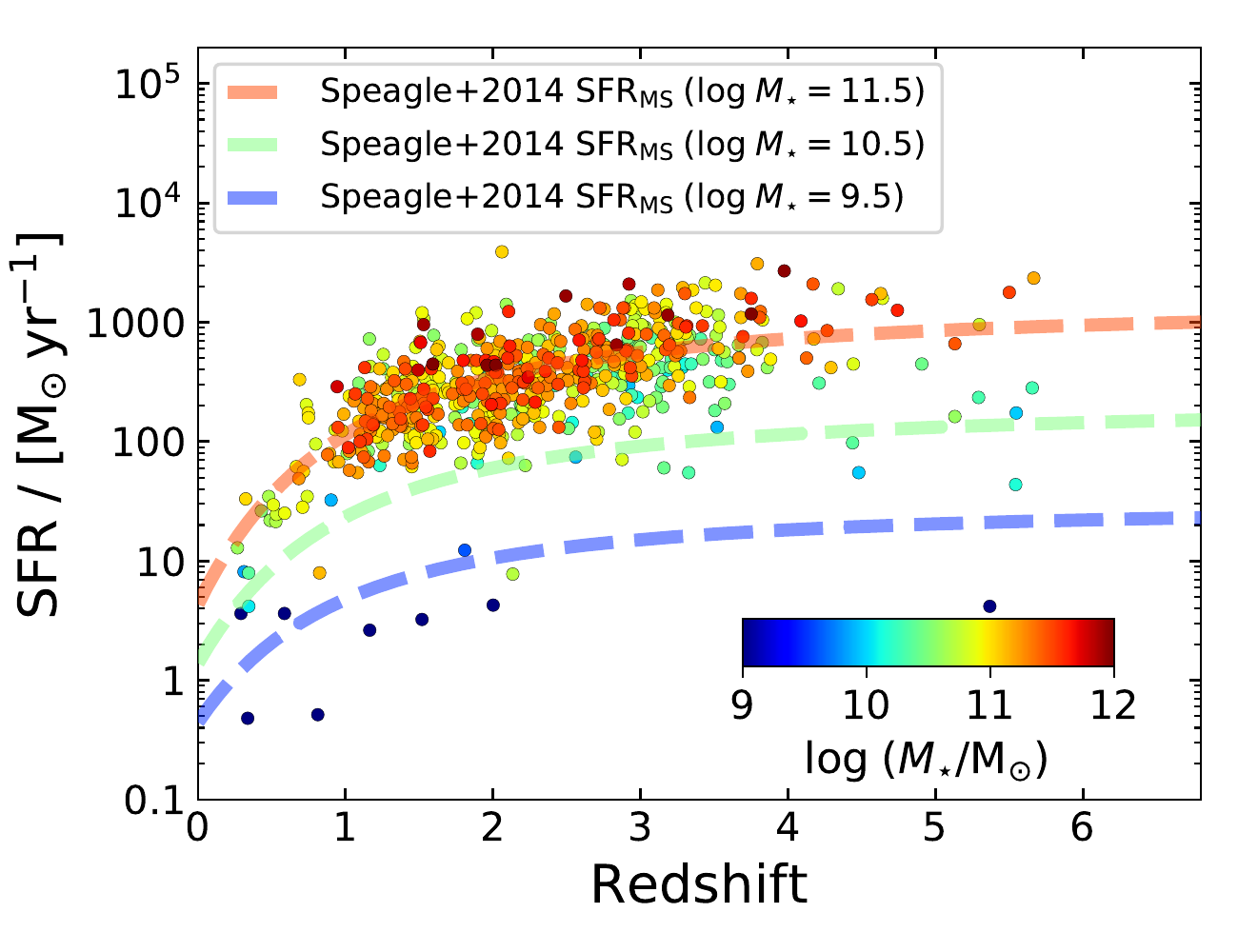}
\includegraphics[width=0.49\textwidth, trim=0 5mm 0 0]{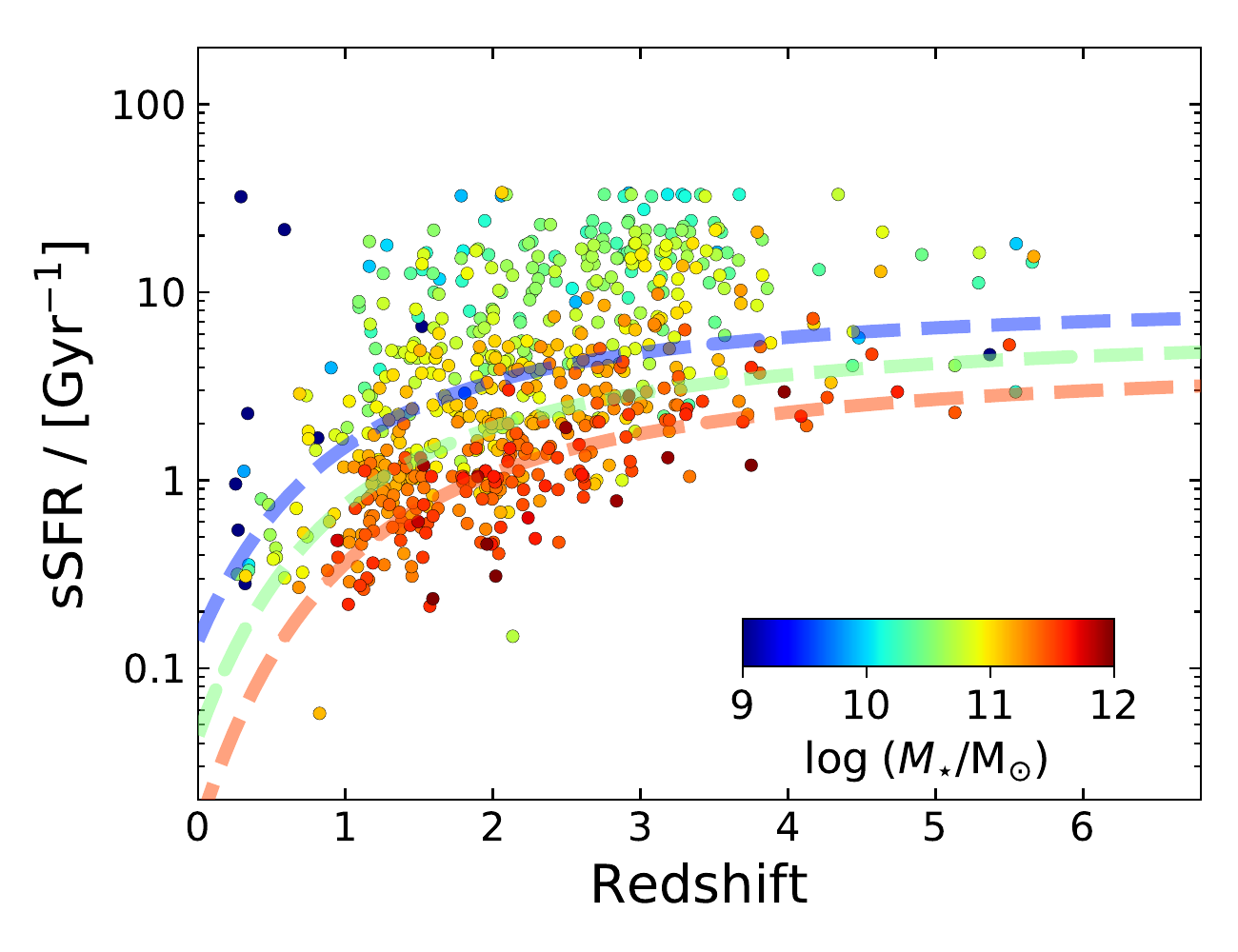}
\caption{%
Distribution of SFR ({\em left}) and specific-SFR (sSFR) ({\em right}) versus redshift. 
Empirical evolution curves of the star-forming main sequence at different stellar masses ($\log_{10} (\Mstar/\mathrm{M_{\odot}})=9.5$, 10.5 and 11.5, respectively) are shown as blue, green and red dashed lines, which are computed as a function of redshift and $\Mstar$ following \cite{Speagle2014} (the \#49 model in their Table~7). The color bar indicates $\log_{10}\Mstar$ in both panels.
\label{Plot_z_SFR}%
}
\vspace{0.5ex}
\end{figure*}

\begin{figure*}[htb]
\centering%
\includegraphics[width=0.49\textwidth, trim=0 5mm 0 0]{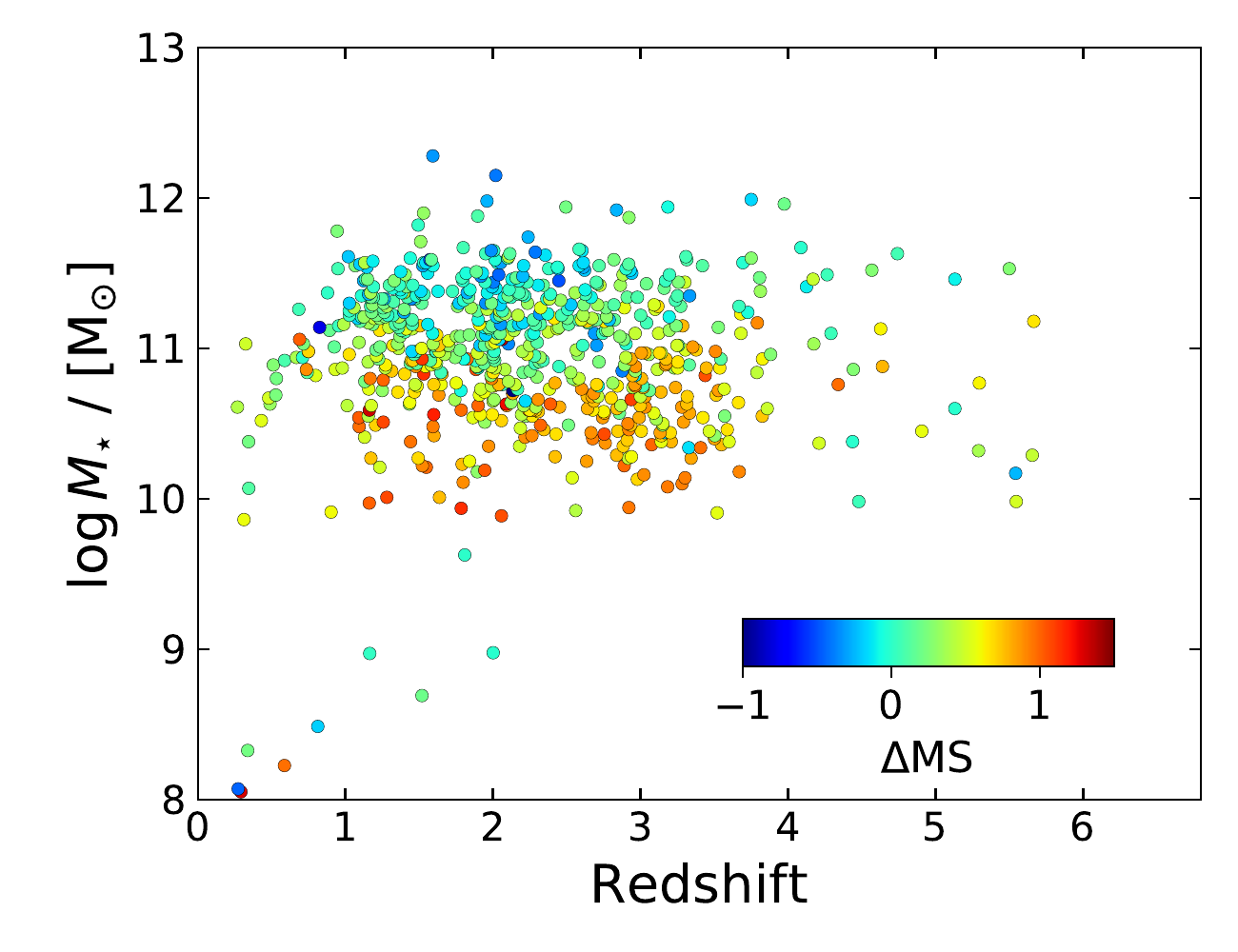}
\includegraphics[width=0.49\textwidth, trim=0 5mm 0 0]{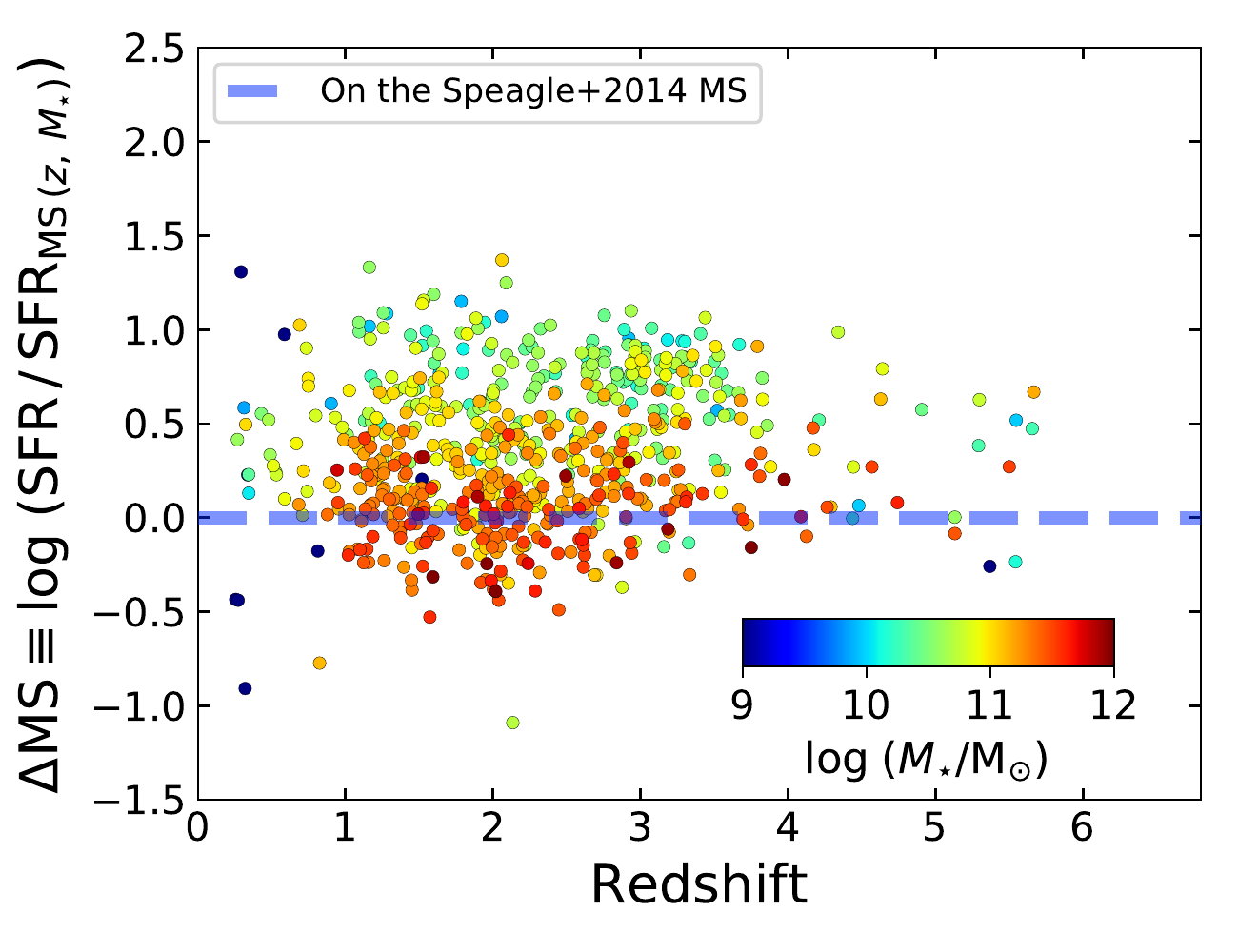}
\caption{%
Stellar mass $\Mstar$ ({\em left}) and main sequence offset $\DeltaMS\equiv\log_{10}(\mathrm{SFR}/\mathrm{SFR_{MS}})$ ({\em right}) versus redshift, where the main sequence SFR $\mathrm{SFR_{MS}}$ is computed as a function of redshift and $\Mstar$ for each source following \cite{Speagle2014} (the \#49 model in their Table~7). The color bars indicate $\DeltaMS$ and $\log_{10}\Mstar$ in the left and right panels, respectively. 
The dashed blue line represents $\Delta\mathrm{MS}=0$ in the right panel. 
\label{Plot_z_DeltaMS}%
}
\vspace{0.5ex}
\end{figure*}

\begin{figure*}[htb]
\centering%
\includegraphics[width=0.96\linewidth, trim=2mm 2mm 0 0]{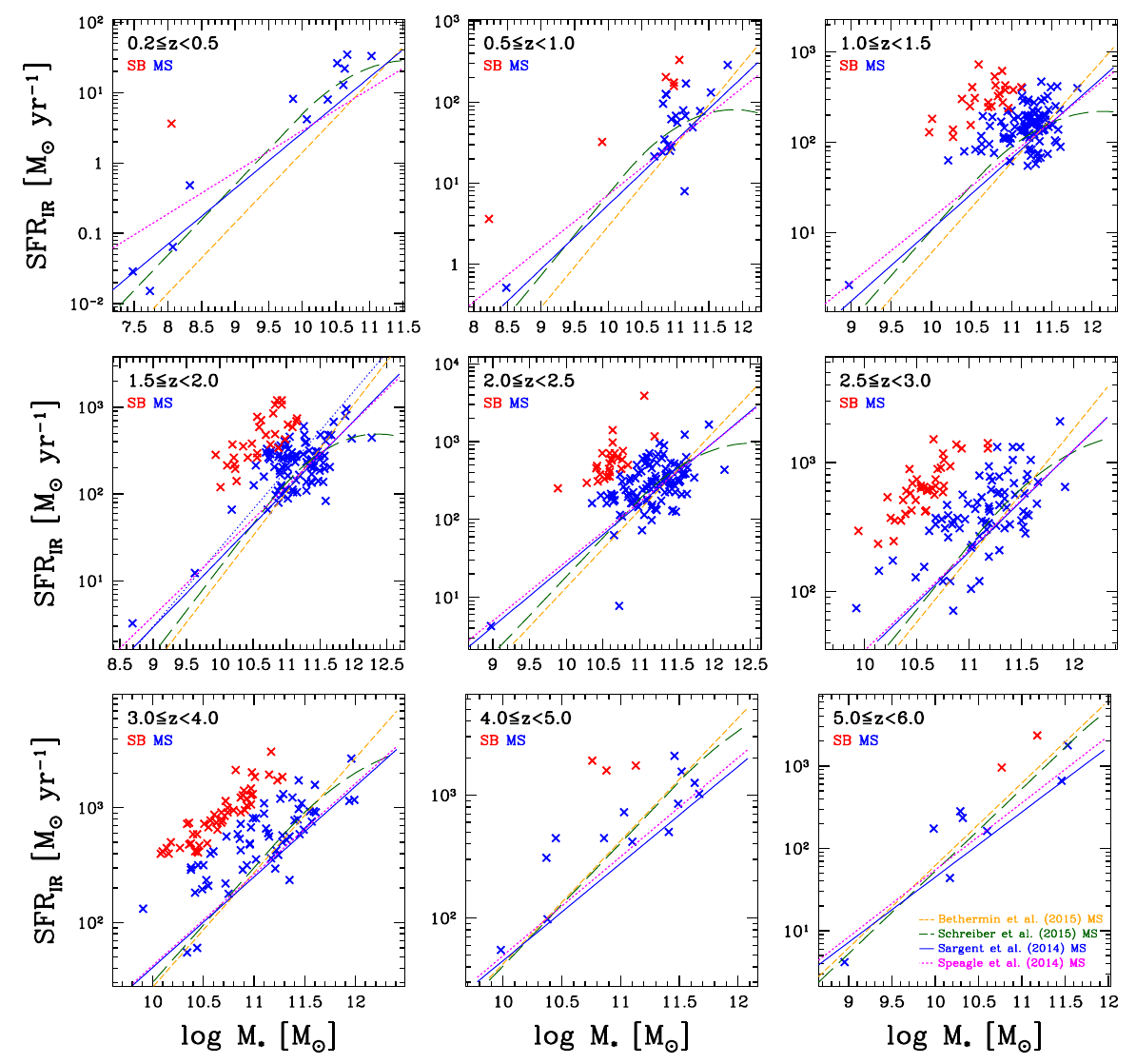}
\caption{%
SFR vs. $\Mstar$, i.e., the star-forming MS diagram, in nine redshift bins from $z\sim0.35-5.5$. Galaxies on the MS and 
starbursts whose SFR is 0.6\,dex above the empirical MS of \cite{Speagle2014} are shown as blue and red crosses, respectively. Several empirical MSs from \cite{Speagle2014}, \cite{Sargent2014}, \cite{Schreiber2015} and \cite{Bethermin2015}
are shown for reference (see bottom right panel for information on the color coding). 
\label{Plot_Mstar_SFR}%
}
\vspace{1.0ex}
\end{figure*}

\begin{figure*}[htb]
\centering
\includegraphics[width=0.96\linewidth, trim=2mm 2mm 0 0]{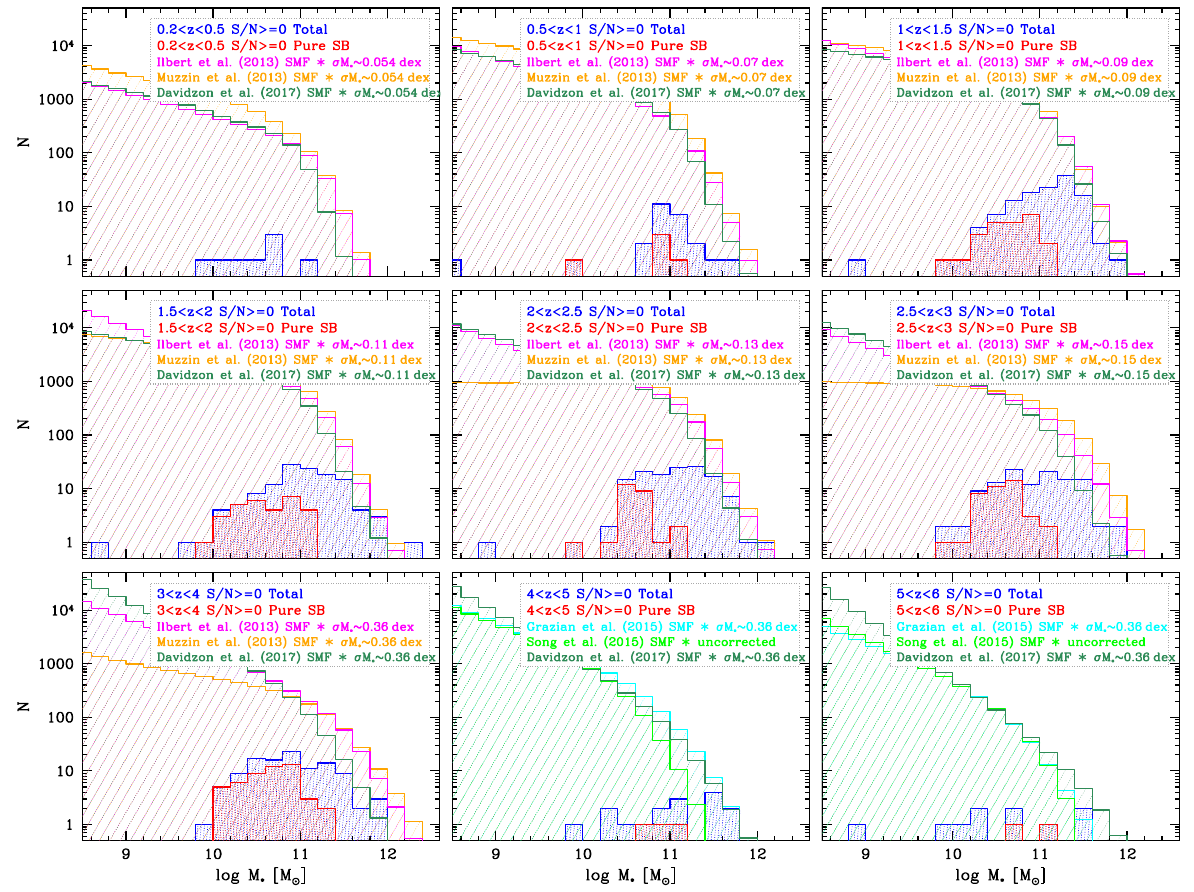}
\caption{%
Stellar mass histograms of our A$^3$COSMOS sources compared to expected stellar mass distributions in nine redshift bins from $z\sim0.35-5.5$. The A$^3$COSMOS sources
are selected within a \aaacosmosAlmaPointingArea{}\,arcmin$^2$ area covered by \aaacosmosAlmaPointingNumber{} discrete ALMA pointings. 
A$^3$COSMOS star-forming main sequence galaxies and starbursts as distinguished in Fig.~\ref{Plot_Mstar_SFR} are shown by blue and red histograms, respectively. 
The colored lines (with line shading) are the stellar mass distributions of all star-forming galaxies within the COSMOS 2\,deg$^2$ area for each redshift bin, computed with empirical stellar mass functions (SMFs) from \cite[][magenta]{Ilbert2013}, \cite[][yellow]{Muzzin2013}, \cite[][dark green]{Davidzon2017},  \cite[][cyan]{Grazian2015} and \cite[][light green]{Song2016}. These SMFs are derived for star-forming galaxies; they have been expressed for a Chabrier IMF when necessary, and convolved with typical stellar mass uncertainties ($\sigma_{\Mstar}$) as indicated by the labels in each panel, following Appx.~A of \cite[][except for \citealt{Song2016}, which is the directly observed SMF]{Ilbert2013}. 
The SMFs of \cite{Ilbert2013} and \cite{Muzzin2013} are measured up to $z\sim4$, while the SMFs of \cite{Grazian2015} and \cite{Song2016} are measured at $z\sim4-6$. \cite{Davidzon2017} SMF probes $z\sim0.2-5.5$ but does not fully cover the highest redshift bin, so we applied a linear extrapolation. 
\label{Plot_Mstar_Histogram}%
}
\vspace{1.0ex}
\end{figure*}

\vspace{0.25truecm}

\subsection{Final Galaxy Catalog and Properties}
\label{Section_Final_Galaxy_Sample_and_Properties}

Our final ``robust galaxy catalog'' contains \aaacosmosGoodGalaxyNumber{} galaxies with reliable stellar mass and SFR properties, from a parent sample \aaacosmosGalaxyNumber{} galaxies with at least one ALMA detection in \aaacosmosAlmaPointingNumber{} ALMA archive images (from 142 ALMA projects) available for the COSMOS field (version \incode{v20180201}). 
In this catalog, 56\% of the galaxies have a primary beam correction factor $<1.01$ (corresponding to a 2'' offset from the phase center at 230\,GHz), i.e., they are the primary targets of the PI-led observations.
We caution that due to the selection functions of the PI-led ALMA observations in the archive, our sample is not complete in any quantity, e.g., cosmic comoving volume, stellar mass, and SFR.
This bias exists even for sources away from the phase center because galaxies suffer from clustering effects, and also the input coordinates for single-dish-selected sub-mm galaxies are uncertain (possibly resulting in a few arcsec offset from the phase center).
Bearing these limitations in mind, we show the redshift, stellar mass and SFR properties of our galaxy catalog in this section and compare with known galaxy correlations and population properties in the literature.

In Fig.~\ref{Plot_z_SFR} we show the distributions of their SFRs and specific SFRs (hereafter sSFR, which is defined as $\mathrm{sSFR}\equiv\mathrm{SFR}/\Mstar$, in units of $\mathrm{Gyr^{-1}}$) versus redshift. Our sample spans a large range of SFR from $\sim$1 to $\sim$2000, but the main portion of the sample is SFR-limited at $z>1$ with SFR$\,\sim100-1000$. 

The low number of $z<1$ galaxies is mainly due to the selection function, the quick drop of flux density at the Rayleigh-Jeans tail in a galaxy's redshifted SED, and the rapid decline of the cosmic SFR density at $z<2$ (e.g., \cite{Madau2014a}; \citealt{Liudz2017}). 
Furthermore, given the smaller volume sampled at low redshift, lower source density as well as cosmic variance could play a role as well. 
Therefore at $z<1$, our sample is very different from far-IR selected samples (e.g., see a \textit{Herschel} sample in \citealt{Liudz2017} Fig.~23; see also \citealt{Bethermin2015}; to name a few). Several $z<1$ galaxies in our sample are strongly biased toward less massive systems (e.g., $\log_{10} \Mstar/\Msun < 10.5$) 
but also relatively low SFR, and they all have low ALMA $\SNR$s (total flux $\SNR\sim4-5$). Although they passed our rigorous spurious source examinations, they could statistically still be spurious. Should they be real, they are of interest in their own right. However, given these uncertainties, we recommend treating these galaxies with caution, especially for individual studies. 

In Fig.~\ref{Plot_z_DeltaMS}, we show for our A$^3$COSMOS galaxies the distribution of their stellar masses and their SFR offsets to the $\mathrm{SFR_{MS}}$ expected for star-forming MS galaxies ($\DeltaMS\equiv\log_{10}(\mathrm{SFR}/\mathrm{SFR_{MS}})$), where the MS $\mathrm{SFR_{MS}}$ is defined as a function of redshift and $\Mstar$ and is empirically measured by a number of works from $z\sim0$ (e.g., \citealt{Brinchmann2004}; \citealt{Chang2015}) to $\sim4$ (e.g., \citealt{Speagle2014}; \citealt{Sargent2014}; \citealt{Schreiber2015}; \citealt{Bethermin2015}; \citealt{Pearson2018}). Here we adopt the \cite{Speagle2014} MS (the \#49 model in their Table~7). 

The majority of our sample lies on the MS (i.e., their sSFRs are within a factor of 4, or equivalently $\pm$0.6\,dex, of the sSFR$_\mathrm{MS}$; \citealt{Rodighiero2011}). 
However, less massive galaxies tend to be above the MS. This strong anti-correlation between $\Mstar$ and $\DeltaMS$ is primarily an effect of detection limit/sample selection 
and is more evident in Fig.~\ref{Plot_Mstar_SFR}, where the $\Mstar$ versus SFR are plotted for A$^3$COSMOS galaxies in nine redshift bins ranging from $z=0.35-5.5$, overlaid with four empirical MS parameterizations from \cite{Speagle2014}; \cite{Sargent2014}; \cite{Schreiber2015} and \cite{Bethermin2015}. 
The differences between these MSs are relatively small (but see S. Leslie et al. 2019, submitted, for a detailed comparison) 

The fraction of sources classified as starbursts indicates that our ALMA catalog is biased toward starbursts. It is roughly constant at $\sim20\%$ in our catalog at each redshift in Fig.~\ref{Plot_Mstar_SFR}. But this is a factor of 2--5 higher than that from a \textit{Herschel}-selected sample, e.g., \cite{Liudz2017}, and much higher than that from a mass-complete sample. For example, \cite{Rodighiero2011} find with a starburst fraction of 2\%--3\% for a sample complete down to $\Mstar = 10^{10}\,\Msun$ at $1.5<z<2.5$; and \cite{Schreiber2015} report 2--4\% for a sample complete down to $\Mstar = 2\times10^{10}\,\Msun$ and is constant up to $z=4$. 
Fig.~\ref{Plot_Mstar_SFR} further shows that it is mainly the less-massive range within which our catalog is dominated by starbursts (e.g., $\Mstar < 10^{10.5}\,\Msun$). 

Finally, in Fig.~\ref{Plot_Mstar_Histogram}, we compare the $\Mstar$ histogram of our sample to the SMFs of star-forming galaxies \citep{Ilbert2013,Muzzin2013,Davidzon2017,Grazian2015,Song2016}, corresponding to the area of full 2~deg$^2$ COSMOS field. The completeness of our sample to the full star-forming galaxy population, which can be considered as the fraction of the $\Mstar$ histogram to the SMFs, strongly depends on redshift and stellar mass. Although the area covered by all the ALMA archival pointings is only about \aaacosmosAlmaPointingArea{}\,arcmin$^2$ (or only 4.2\% of the full 2~deg$^2$ area of COSMOS), our sample at $z\sim1-3$ probes 
a significant fraction ($\sim10-100\%$ depending on the used SMF and redshift bin) of all
very massive ($\log_{10}\Mstar>11.5$) star-forming galaxies present in the full 2~deg$^2$ area. 

Due to the large variety of ALMA programs contributing to out dataset, we find no obvious differences between sources at the phase center and in the outer area, even out to a PBA of 0.2. The sample selection bias is dominated by the range of sensitivities of the ALMA data rather than the PIs' targeted sources.

\vspace{0.25truecm}

\subsection{Properties of the source not included in the final robust galaxy catalog}
\label{Section_Discareded_Sources}

As listed in Table~\ref{Table_number_of_sources_excluded}, a significant number ($\sim26\%$) of ALMA detections are not included in our final galaxy catalog (see Table~\ref{Table_number_of_sources_excluded} caption $b$). 
Half of them come from the prior-photometry catalog with most having only IRAC 3.6 and 4.5\,$\mu$m and/or VLA 3\,GHz priors without optical/near-IR (up to Ks-band) counterparts (hence they do not have a photo-$z$ as their prior-$z$). These ``Ks-dropouts'' are potential very dusty $z\sim3-4$ galaxies or less dusty sources at even higher redshifts, i.e. similar to the sample of \citet{WangTao2016} and the \textit{HST}-dark sample of \citet{Franco2018}. This is in particular true for the sources with significant detections well above our threshold. The remaining half comes from the blind-photometry catalog and has typically low significance, implying that they could be spurious, 
as the differential spurious fraction strongly depends on the actual $\SNR$ (see Fig.~\ref{Plot_spurious_fraction}). 

As we do not have high spatial resolution optical/near-IR imaging nor accurate photometric redshifts for these sources, it is not possible to do similar counterpart association or SED fitting quality assessments to better identify spurious ones. 
If we assume that the fraction of spurious source is the same (about 10\% based on our quality assessment) for the sources we have done the quality assessment (74\% of the total ALMA detections) and for those unable to perform a quality assessment (26\% of the total ALMA detections), then we also expect a small number of spurious source from the latter sources.  
Adding the two together gives a total spurious fraction of $\sim10.1\%$, in good agreement with the statistics ($\sim8-12\%$). 

Further discussion of these interesting sources is not the focus of this work. 
Future deeper optical/near-IR (up to $K_s$-band) observations, e.g., the new data release of the UltraVISTA survey, will enable an analysis similar to the one done here, so that they could be included in the robust galaxy catalog in the future.

\vspace{0.25truecm}

\subsection{The effect of galaxy-galaxy gravitational lensing}

The galaxy-galaxy gravitational lensing has been found to be significant in several ALMA follow-up studies of brightest sub-mm galaxies over large areas, e.g., \cite{Negrello2010}, \cite{Bussmann2013,Bussmann2015} and \cite{Spilker2016}. Empirically, the strong-lensing cases (magnification $\mu>2$) therein exhibit the following common features: (1) Very bright observed sub-mm flux, e.g., $S_{870\,\mu\mathrm{m}} \gtrsim 15 \,\mathrm{mJy}$ for all the $\mu>2$ galaxies in \cite{Bussmann2013,Bussmann2015} and \cite{Spilker2016}. (Although we note that lensing is not just limited to the very brightest submillimeter objects but happens at all flux levels, see also below.)
(2) Bright optical emission within 1--2'' which belongs to a low-redshift (usually $z<1$) massive galaxy. 
(3) Usually two or more sub-mm components at each side of the optical emission 
or roughly distributed as an Einstein ring with $\sim$2'' size.

We estimate the number of strongly lensed ($\mu>2$) cases among our sub-mm galaxies to be very low as follows. 

First, given the flux distribution of our photometry catalogs, we only find 0.2\% sources with equivalent $S_{870\,\mu\mathrm{m}} \gtrsim 15 \,\mathrm{mJy}$, i.e., four sources in current data set (\incode{v20180102}). 
Three of them have only very weak or no optical emission in their 1--$2''$ vicinity, while the fourth one ID~180903 has a low-redshift ($z=0.347$) optically bright galaxy within $0.5''$ and has already been studied in detail by \cite{Pavesi2018}. ID~180903 does not exhibit multiple images as expected for strong lenses, fully consistent with its magnification factor of only 1.09 \citealt{Pavesi2018}. 

Second, considering the second feature of a close distance to a low-$z$ galaxy, our prior source fitting and SED excess assessment can test for this: if the ALMA flux coming from our prior catalog is originating from a lensed higher redshift galaxy, the SED fitting with a much lower redshift as the prior-$z$ will not be able to fit the ALMA data and therefore be classified as a SED-excess outliers (Sect.~\ref{Section_Running_SED_Fitting}). Among the 21 SED-excess outliers listed in Table~\ref{Table_number_of_sources_excluded}, we searched for multiple sub-mm images or distorted feature but found no obvious lensed candidates, except for one case ID~650923 where there are three optical components ($z=0.568$ in \citealt{Laigle2016}) surrounding the East and South sides of the ALMA emission at a distance of $\sim1''$ (although the ALMA data has a beam of $1.5''\times1.0''$). 
    
Third, there are no multiple sub-mm sources within 1-2'' or sources being part of an Einstein ring. This is based on visual identification. 
In addition, this is confirmed through the comparison between prior- and blind-extraction photometry, which can in principle identify sources with irregular multi-component morphology.

Lastly, our low number of strongly lensed sources is consistent with the analytic galaxy modeling of \citet[][see also \citealt{Bethermin2012}]{Bethermin2017Model}. In their modeling, 1.5 million galaxies are simulated from redshift 0 to 10 within a light cone of 2\,deg$^2$, the same area as the COSMOS field. Their modeled galaxies follow the clustering effect matched to dark matter halos, and strong- and weak-lensing effects are modeled following \cite{Hezaveh2011} and \cite{Hilbert2007}, respectively. 
According to our galaxy sample properties, we down-selected 3176 of their galaxies with the criteria $1<z<6$, $\Mstar>2 \times 10^{10}\;\Msun$ and $\SFR>200\;\Msyr$ over the full 2\,deg$^2$. Among this subsample, only 16 have $\mu>2$. Scaling to our galaxy catalog source number of \aaacosmosGalaxyNumber{}, 
only three strongly ($\mu>2$) lensed sources are expected. Note that as discussed in \cite{Hezaveh2011}, there remains significant uncertainty in the estimation of the probability of lensed sources, e.g., the assumed mass model for the lensing halos, the ellipticity of lenses, etc.

Therefore, we conclude that strong lensing is not affecting the properties of most of our galaxies.

\vspace{1truecm}

\section{Data Products}
\label{Section_Data_Delivery}

As the result of this work, we produce three public catalogs: two photometric ones (blind-extraction and prior-fitting) and one galaxy catalog (with SED-derived properties). We describe the columns in the first two catalogs in Table~\ref{Table_photometry_catalog_columns}, and those in the third catalog in Table~\ref{Table_galaxy_catalog_columns}. 

The two photometric catalogs have most columns in common, except that the prior photometry catalog has information on the prior source (\incode{ID}, \incode{ID_PriorCat} and \incode{Ref_ID_PriorCat} columns), and some \incode{Flag_*} columns differ. 

The \incode{ID} column lists the IDs in our A3COSMOS master catalog which is a combination of 6+ prior catalogs after solving source cross-matching (Sect.~\ref{Section_Prior_Source_Catalogs}). The \incode{ID} equals to the COSMOS2015 (\citealt{Laigle2016}) catalog ID when \incode{ID}~$\le 1182108$. The \incode{ID_PriorCat} column lists the original IDs in those prior catalogs, so that users of our prior photometric catalog can trace back into the prior catalogs. The \incode{Ref_ID_PriorCat} column lists the reference rank number of the prior catalog in Table~\ref{Table_prior_catalogs}, e.g., the COSMOS2015 (\citealt{Laigle2016}) catalog has \incode{Ref_ID_PriorCat}~$=1$, \cite{Smolcic2017a} catalog has \incode{Ref_ID_PriorCat}~$=5$, etc. Note that this reference number indicates in which catalog the source is first included, i.e., has no counterpart in all previous catalogs with smaller \incode{Ref_ID_PriorCat}. Thus our catalog does not contain the information of whether a source with \incode{Ref_ID_PriorCat}~$=1$ has a counterpart in \incode{Ref_ID_PriorCat}~$>1$ catalogs (but this information is in our master catalog upon request). Also note that our master catalog will be updated in the future with more deeper prior catalogs; thus, we caution that the source ID will be different when a future updated master catalog is used. 

The \incode{Flag_*} columns carry important information for quality assessment and should be taken into account when using the catalog for specific science applications. 
For the blind photometry catalog, \incode{Flag_multi} indicates whether the source is fitted with multiple Gaussian components or a single-Gaussian model by \pybdsm{}. \incode{Flag_inconsistent_flux} indicates whether the source has inconsistent fluxes between prior-fitting and blind-extraction photometry catalog (see Sect.~\ref{Section_Photometry_Quality_Check_1} and Fig.~\ref{Plot_catalog_comparison_a3cosmos_prior_vs_blind_total_flux})
When \incode{Flag_multi}~$=1$ and \incode{Flag_inconsistent_flux}~$=1$, it is likely that the source is a merger system or has a close companion in the ALMA image, as shown in Appx.~\ref{Section_Appendix_outliers_of_photometry}. For the prior photometry catalog, \incode{Galfit_reduced_chi_square} indicates the quality of the final \galfit{} source fitting. \incode{Flag_size_upper_boundary} indicates whether the fitted source size reaches the upper boundary of 3.0 arcsec we set in the photometry, in which case the source is either blended or dominated by noise and should be used with caution. 

The final galaxy property catalog also has two important flags: \incode{Flag_outlier_CPA} which indicates the outliers from our counterpart association examination (Sect.~\ref{Section_Examining_counterpart_association}), and \incode{Flag_outlier_SED} which indicates the outliers with SED-excess from our SED fitting (Sect.~\ref{Section_Running_SED_Fitting}). We recommend to only use galaxies with both \incode{Flag}~$=0$ for scientific analysis. 

The catalogs are available from the COSMOS archive at IPAC/IRSA and in electronic form from the journal. Further, the ALMA continuum images are also provided via the COSMOS archive.

\vspace{1truecm}

\section{Summary}
\label{Section_Summary}

The growing information in the ALMA archive is ideal for systematic exploitations of specific scientific questions, such as the number and properties of high-redshift galaxies detected in their (sub-)mm continuum emission in selected cosmological deep fields. Given the large number of observations already available in the archive, e.g. for the COSMOS field \aaacosmosAlmaPointingNumber{} pointings covering an area of \aaacosmosAlmaPointingArea{}~arcmin$^2$ have been publicly available since \aaacosmosAlmaDataDate{}, we have developed a highly automatic approach towards mining these data (A$^3$COSMOS --- Automated ALMA Archive mining in the COSMOS field). Here we summarize our workflows (Figs.~\ref{Figure_flow_chart}~and~\ref{Figure_galaxy_flow_chart}) implemented to obtain quality controlled (sub-)mm source catalogs based on two different identification approaches as well as a catalog of galaxies with (sub-)mm detections and reliable properties. 

We present two (sub-)mm continuum source catalogs from public ALMA archival data. For the source identification, the calibrated archival data were homogenously imaged to provide a single continuum image with best sensitivity (i.e. using all available bandwidth and natural weighting). The first catalog is based on a blind extraction using \pybdsm{} on the continuum images, and the second catalog used prior positions from a master catalog that combines sources detected in the optical, IR and radio. Extensive simulations using two mock samples with highly different distributions in (sub-)mm source properties provide robust information on the completeness limits, spurious source fraction, flux boosting factors as well as uncertainties on the measured quantities in both catalogs. In particular, we used these simulations to refine the widely used \cite{Condon1997} prescription for error estimation of radio continuum sources. 
After further quality control steps, the final catalogs (version \incode{v20180201}) contain \aaacosmosBlindDetectionNumber{} sources above a peak flux $\SNRpeak=5.40$ for the blindly detected sources (with a cumulative spurious fraction of $\sim8\%$) and \aaacosmosPriorDetectionNumber{} sources above a peak flux $\SNRpeak=4.35$ for the prior selected sources (with a cumulative spurious fraction of $\sim12\%$).

We combine the two (sub-)mm continuum source catalogs to remove inconsistent-flux outliers and use the prior catalog to produce a single sample of high-redshift galaxies with robust (sub-)mm detections by ALMA (\aaacosmosGalaxyWithMultiplePhotometryFraction{} having more than one ALMA photometric measurement usually at different wavelengths) and mostly homogeneously determined galaxy properties (stellar mass, SFR). 
The construction included the development of a sophisticated method to automatically qualify counterpart associations with the (sub-)mm continuum sources taking into account astrometric uncertainties (both absolute and relative) as well as complex, differing source structure across wavelength. 
Further steps were applied to remove spurious (sub-)mm continuum detections and/or sources with highly uncertain redshift information based on SED fitting results. 
The final galaxy catalog (version \incode{20180201}) contains \aaacosmosGoodGalaxyNumber{} galaxies in the range of $z = 0.25 - 5.67$, $\Mstar = 3\times10^{7} - 1.9\times10^{12} \; \Msun$ and $\SFR = 0.02 - 4000 \;\Msyr$. (Despite the vast number of star-forming galaxies presented in this work, we caution that this catalog is not complete in cosmic comoving volume, stellar mass or SFR.)

The latest versions of our catalogs are available from the COSMOS archive at IPAC/IRSA\,\footnote{\url{https://irsa.ipac.caltech.edu/data/COSMOS/images/a3cosmos/}} and in electronic form.

\rule{0pt}{28pt}

\acknowledgments

D.L., P.L., and E.S. acknowledge support and funding from the European Research Council (ERC) under the European Union's Horizon 2020 research and innovation programme (grant agreement No. 694343). 
S.L. acknowledges funding from Deutsche Forschungsgemeinschaft (DFG) Grant SCH 536/9-1.
B.G. acknowledges the support of the Australian Research Council as the recipient of a Future Fellowship (FT140101202).
Part of this research was carried out within the Collaborative Research Centre 956, sub-project A1, funded by the Deutsche Forschungsgemeinschaft (DFG) -- project ID 184018867. 
We thank Annalisa Pillepich and the Max Planck Computing \& Data Facility for very helpful computing cluster resources. 

This paper makes use of the following ALMA data: 
\path{ADS/JAO.ALMA#2011.0.00064.S}, 
\path{ADS/JAO.ALMA#2011.0.00097.S}, 
\path{ADS/JAO.ALMA#2011.0.00539.S}, 
\path{ADS/JAO.ALMA#2011.0.00742.S}, 
\path{ADS/JAO.ALMA#2012.1.00076.S}, 
\path{ADS/JAO.ALMA#2012.1.00323.S}, 
\path{ADS/JAO.ALMA#2012.1.00523.S}, 
\path{ADS/JAO.ALMA#2012.1.00536.S}, 
\path{ADS/JAO.ALMA#2012.1.00919.S}, 
\path{ADS/JAO.ALMA#2012.1.00952.S}, 
\path{ADS/JAO.ALMA#2012.1.00978.S}, 
\path{ADS/JAO.ALMA#2013.1.00034.S}, 
\path{ADS/JAO.ALMA#2013.1.00092.S}, 
\path{ADS/JAO.ALMA#2013.1.00118.S}, 
\path{ADS/JAO.ALMA#2013.1.00151.S}, 
\path{ADS/JAO.ALMA#2013.1.00171.S}, 
\path{ADS/JAO.ALMA#2013.1.00208.S}, 
\path{ADS/JAO.ALMA#2013.1.00276.S}, 
\path{ADS/JAO.ALMA#2013.1.00668.S}, 
\path{ADS/JAO.ALMA#2013.1.00815.S}, 
\path{ADS/JAO.ALMA#2013.1.00884.S}, 
\path{ADS/JAO.ALMA#2013.1.00914.S}, 
\path{ADS/JAO.ALMA#2013.1.01258.S}, 
\path{ADS/JAO.ALMA#2013.1.01292.S}, 
\path{ADS/JAO.ALMA#2015.1.00026.S}, 
\path{ADS/JAO.ALMA#2015.1.00055.S}, 
\path{ADS/JAO.ALMA#2015.1.00122.S}, 
\path{ADS/JAO.ALMA#2015.1.00137.S}, 
\path{ADS/JAO.ALMA#2015.1.00260.S}, 
\path{ADS/JAO.ALMA#2015.1.00299.S}, 
\path{ADS/JAO.ALMA#2015.1.00379.S}, 
\path{ADS/JAO.ALMA#2015.1.00388.S}, 
\path{ADS/JAO.ALMA#2015.1.00540.S}, 
\path{ADS/JAO.ALMA#2015.1.00568.S}, 
\path{ADS/JAO.ALMA#2015.1.00664.S}, 
\path{ADS/JAO.ALMA#2015.1.00704.S}, 
\path{ADS/JAO.ALMA#2015.1.00853.S}, 
\path{ADS/JAO.ALMA#2015.1.00861.S}, 
\path{ADS/JAO.ALMA#2015.1.00862.S}, 
\path{ADS/JAO.ALMA#2015.1.00928.S}, 
\path{ADS/JAO.ALMA#2015.1.01074.S}, 
\path{ADS/JAO.ALMA#2015.1.01105.S}, 
\path{ADS/JAO.ALMA#2015.1.01111.S}, 
\path{ADS/JAO.ALMA#2015.1.01171.S}, 
\path{ADS/JAO.ALMA#2015.1.01212.S}, 
\path{ADS/JAO.ALMA#2015.1.01495.S}, 
\path{ADS/JAO.ALMA#2015.1.01590.S}, 
\path{ADS/JAO.ALMA#2015.A.00026.S}, 
\path{ADS/JAO.ALMA#2016.1.00478.S}, 
\path{ADS/JAO.ALMA#2016.1.00624.S}, 
\path{ADS/JAO.ALMA#2016.1.00735.S}. 
ALMA is a partnership of ESO (representing its member states), NSF (USA) and NINS (Japan), together with NRC (Canada), MOST and ASIAA (Taiwan), and KASI (Republic of Korea), in cooperation with the Republic of Chile. The Joint ALMA Observatory is operated by ESO, AUI/NRAO and NAOJ.

\facility{ALMA}

\input{Table_photometry_catalog_columns.tex}

\input{Table_galaxy_catalog_columns.tex}

\makeatletter\onecolumngrid@push\makeatother
\FloatBarrier
\makeatletter\onecolumngrid@pop\makeatother
\clearpage

\appendix
\counterwithin{figure}{section}
\counterwithin{table}{section}
\FloatBarrier

\section{Astrometry accuracy between prior catalogs}
\label{Section_Astrometry}

Variations in the absolute astrometric calibration between different catalogs can cause small, but noticeable offsets between source positions at different wavelengths. As we use prior positions from sources selected from catalogs covering the optical to radio regime, it is important to verify that potential offsets are small.
Here we report astrometric offsets between the prior catalogs used from the literature (Sect.~\ref{Section_Prior_Source_Catalogs}) and our ALMA prior fitting photometry catalog. 
These astrometric offsets between the prior positions and the fitted ALMA positions are small ($<0.1''$). 

\begin{figure}[ht!]
\centering
\includegraphics[width=0.45\textwidth]{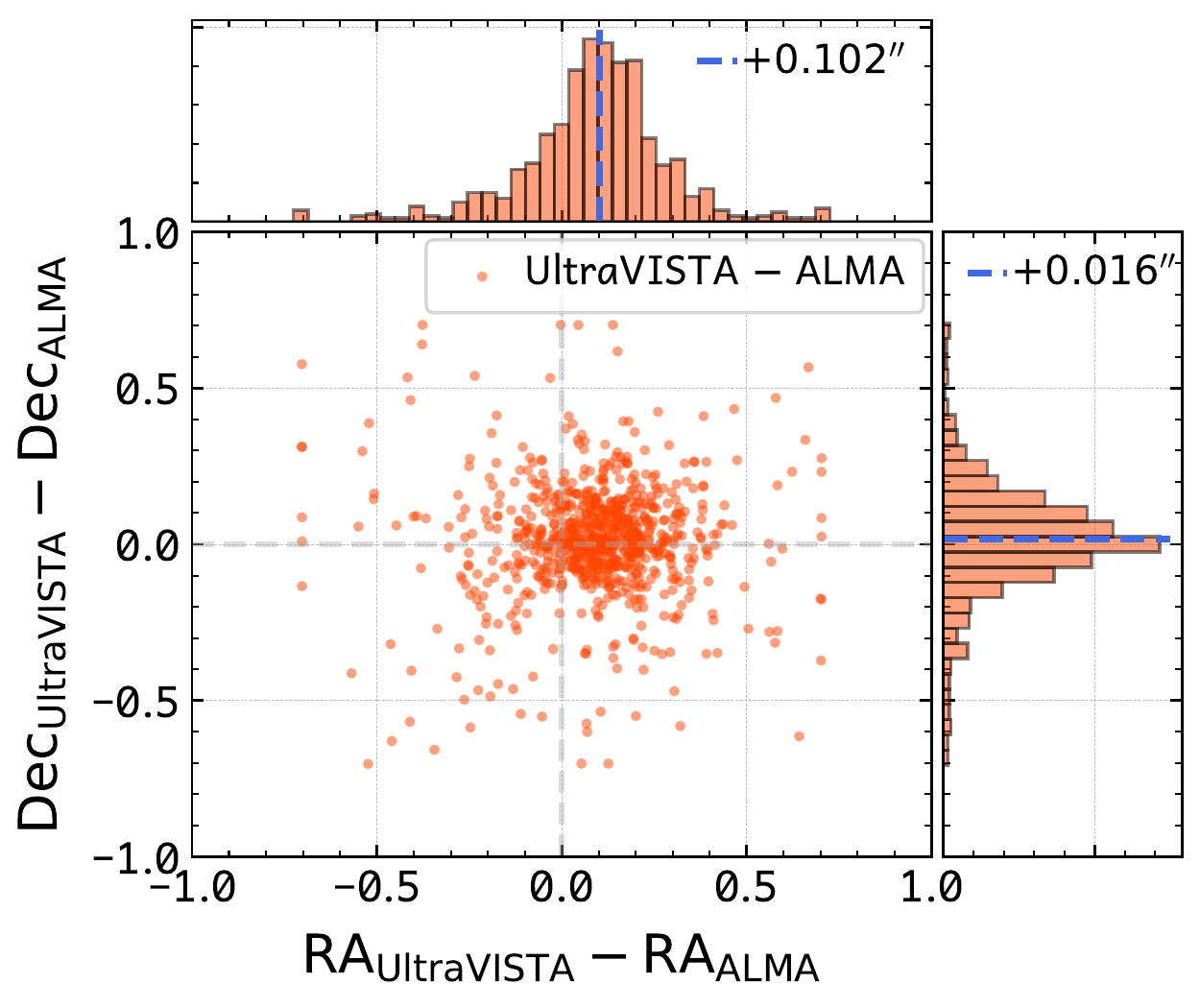}
\includegraphics[width=0.45\textwidth]{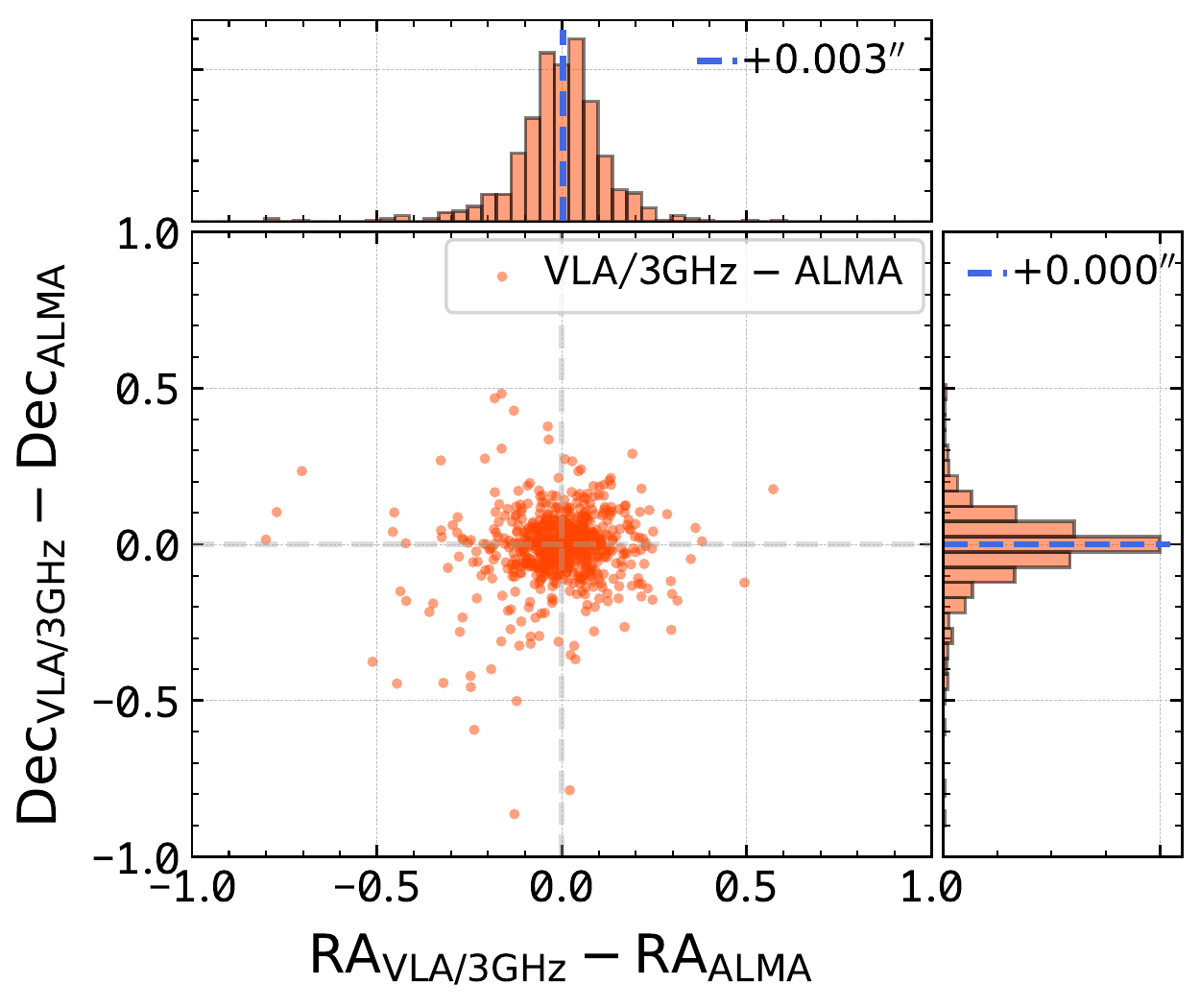}
\vspace{1ex}
\includegraphics[width=0.45\textwidth]{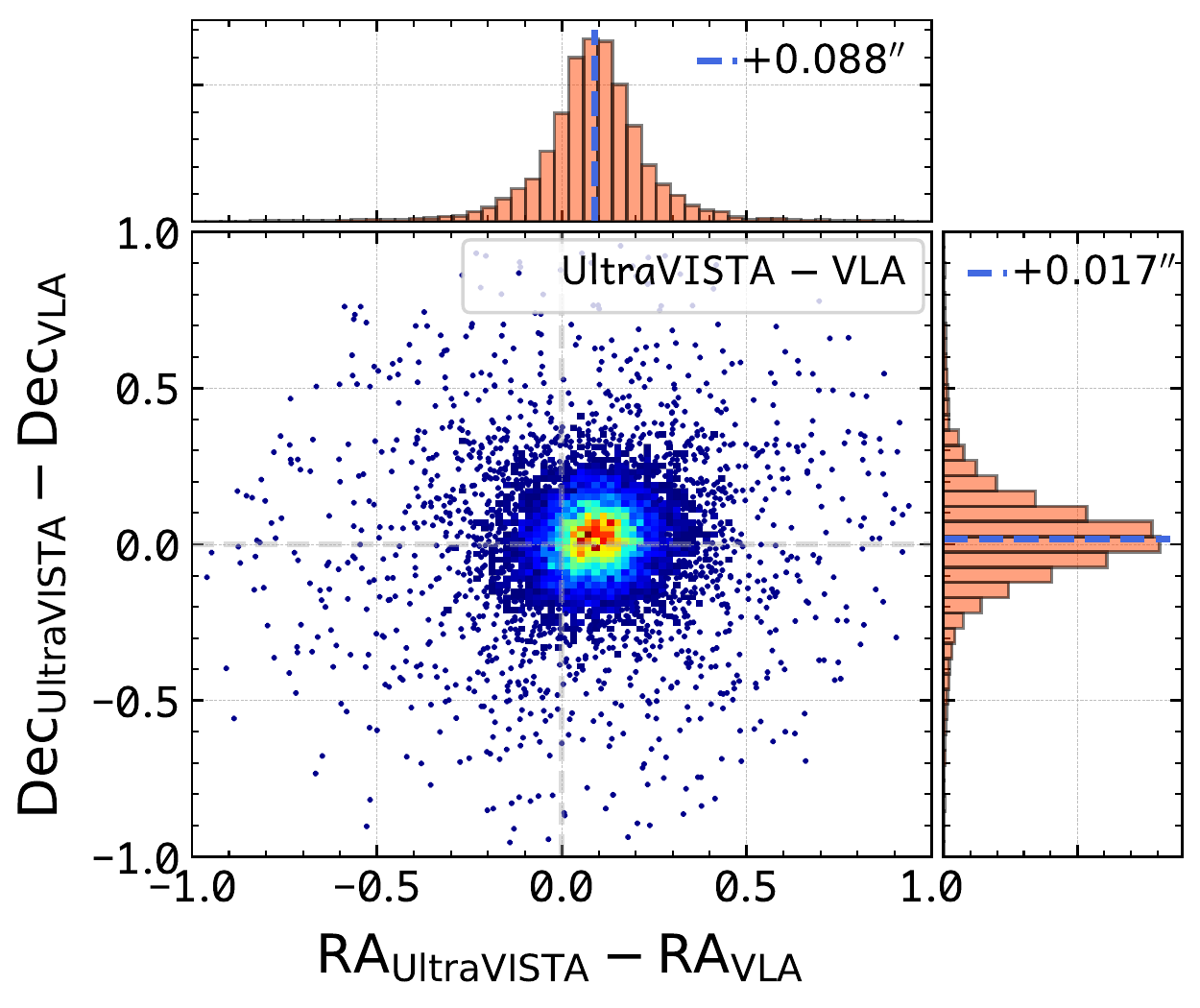}
\includegraphics[width=0.45\textwidth]{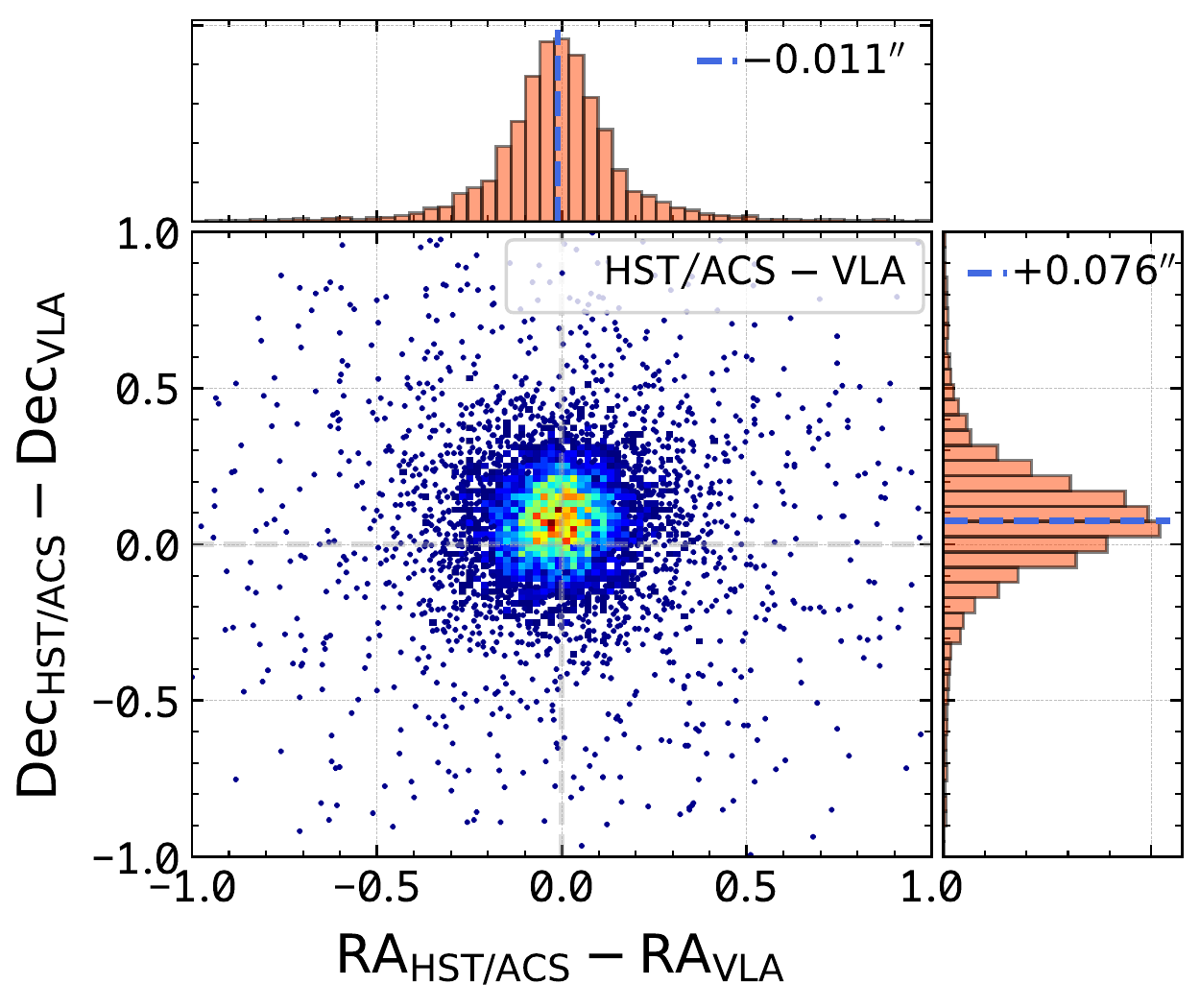}
\caption{
Astrometric offsets between source positions in the prior catalogs used from the literature (Sect.~\ref{Section_Prior_Source_Catalogs}; \citealt{Laigle2016}; \citealt{Smolcic2017a}; \citealt{Capak2007}) and positions from our ALMA prior-based detections. Each panel shows the R.A. and Dec. offset distribution for sources common in two catalogs as labeled. Histograms of the R.A. and Dec. offsets are shown at each top and right axis, respectively. Median values of the R.A. and Dec. offsets are indicated by the dashed blue line and the text therein. See Appx.~\ref{Section_Astrometry} for the details. 
\label{Plot_astrometry_1}}
\end{figure}

In Fig.~\ref{Plot_astrometry_1}, we plot the offsets in R.A. and Dec. for sources common in two catalogs using the UltraVISTA/COSMOS2015 catalog \citep{Laigle2016}, the VLA-COSMOS 3GHz catalog \cite{Smolcic2017a}, the HST/ACS $i$-band \citep{Capak2007} and the fitted positions of the (sub-)mm sources in our prior-based ALMA catalog (see Sect.\ref{Section_Prior_Source_Fitting}).
First we confirm that the ALMA astrometry is indeed excellent (as expected for a (sub-)mm interferometer at the angular resolutions and frequencies of our observations) by comparison to the positions of 699 VLA-COSMOS 3GHz sources ({\em top right panel}). 
Comparison between the UltraVISTA and our ALMA positions for 827 sources ({\em top left panel})
yields a a relatively large offset of $+0.10''$ in R.A., but the offset in Dec. is very small ($+0.01''$). We confirm this astrometric offset of the UltraVISTA catalog by examining positions for 9373 sources in common with the VLA-COSMOS 3GHz catalog ({\em bottom left panel}). Given the order of magnitude larger number of sources the offset of $+0.088''$ in R.A. is statistically meaningful and consistent with the offset seen between UltraVISTA and ALMA source positions.
Finally, comparison between positions of 7369 sources in common in the HST/ACS $i$-band and VLA catalogs ({\em bottom right panel}) yields a lower significantly offset in R.A. but a more substantial offset in Dec. 

Since we allow the source position to vary by a relatively large amount ($\sim0.7''$, see Sect.~\ref{Section_Prior_Source_Fitting}) during the prior-based detection of (sub-)mm continuum source, it is not necessary to repeat the initial detection step. However, we have applied a correction to take the small offsets into account during our counterpart association process (Sect.~\ref{Section_Examining_counterpart_association}). 
We note that \cite{Smolcic2017a} report astrometric offsets of similar size between VLA-COSMOS 3GHz and UltraVISTA source positions using a more complex analysis identifying variations in the astrometry across the full COSMOS field (see their Appx.~A.1 and Fig.~A.1). As the numbers of sources analyzed per R.A. and Dec. bin are only a few hundred, we prefer to apply only a single value when correcting for the astrometric offset of the UltraVISTA sources.

We note that a new COSMOS photometry catalog is under construction using the UltraVISTA DR4 data which are astrometrically corrected using GAIA data, providing a much better astrometry of a few milli-arcsecs (see \url{https://calet.org/}). 
Our next A$^3$COSMOS updates will use it when available.

\vspace{1truecm}
\FloatBarrier

\section{Sources with inconsistent (sub-)mm continuum photometry in our two catalogs}
\label{Section_Appendix_outliers_of_photometry}

\begin{figure}[h!]
\centering%
\includegraphics[width=0.46\textwidth, trim=0 0 0 0, clip]{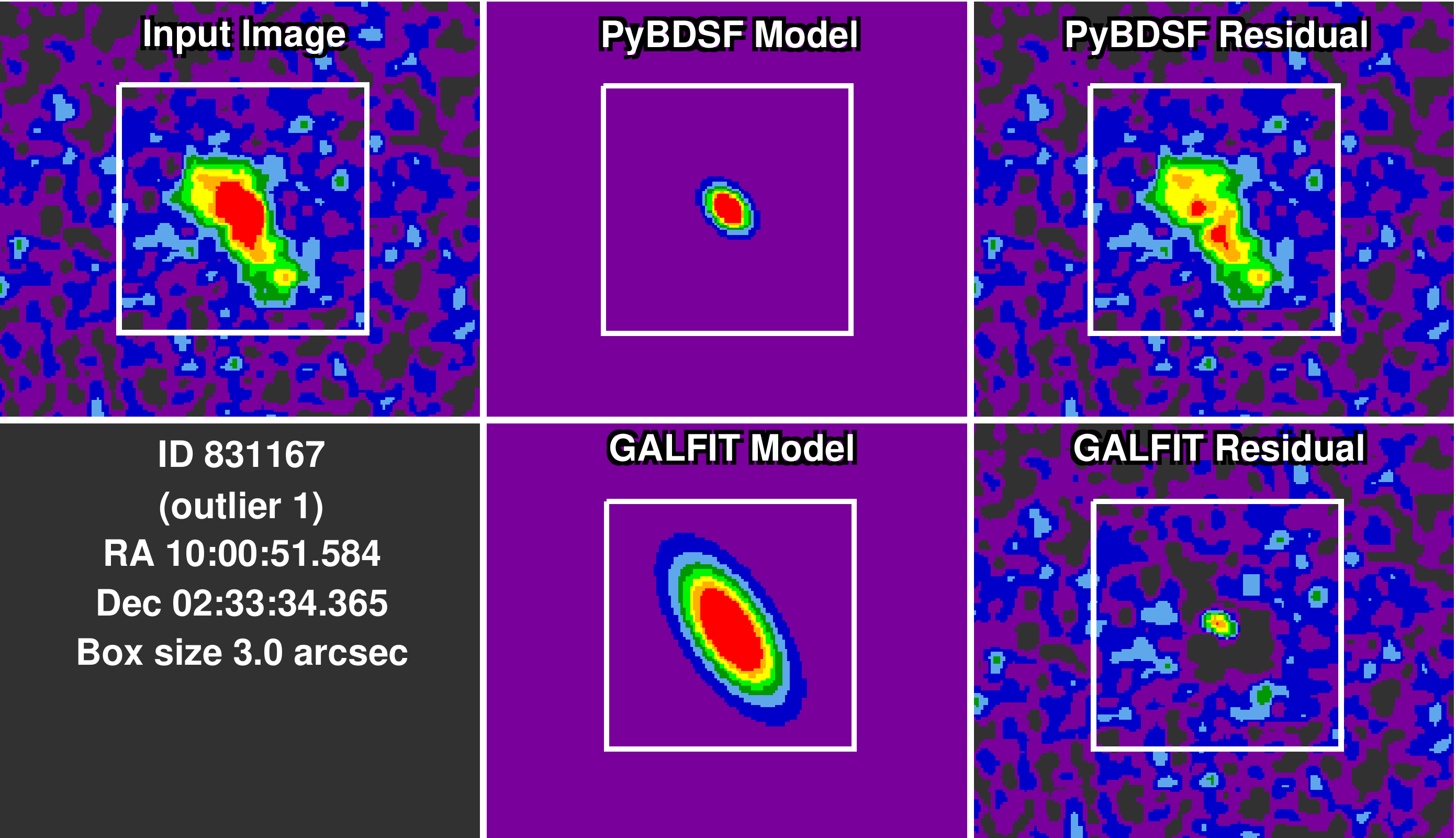}
\hspace{2.2mm}
\includegraphics[width=0.46\textwidth, trim=0 0 0 0, clip]{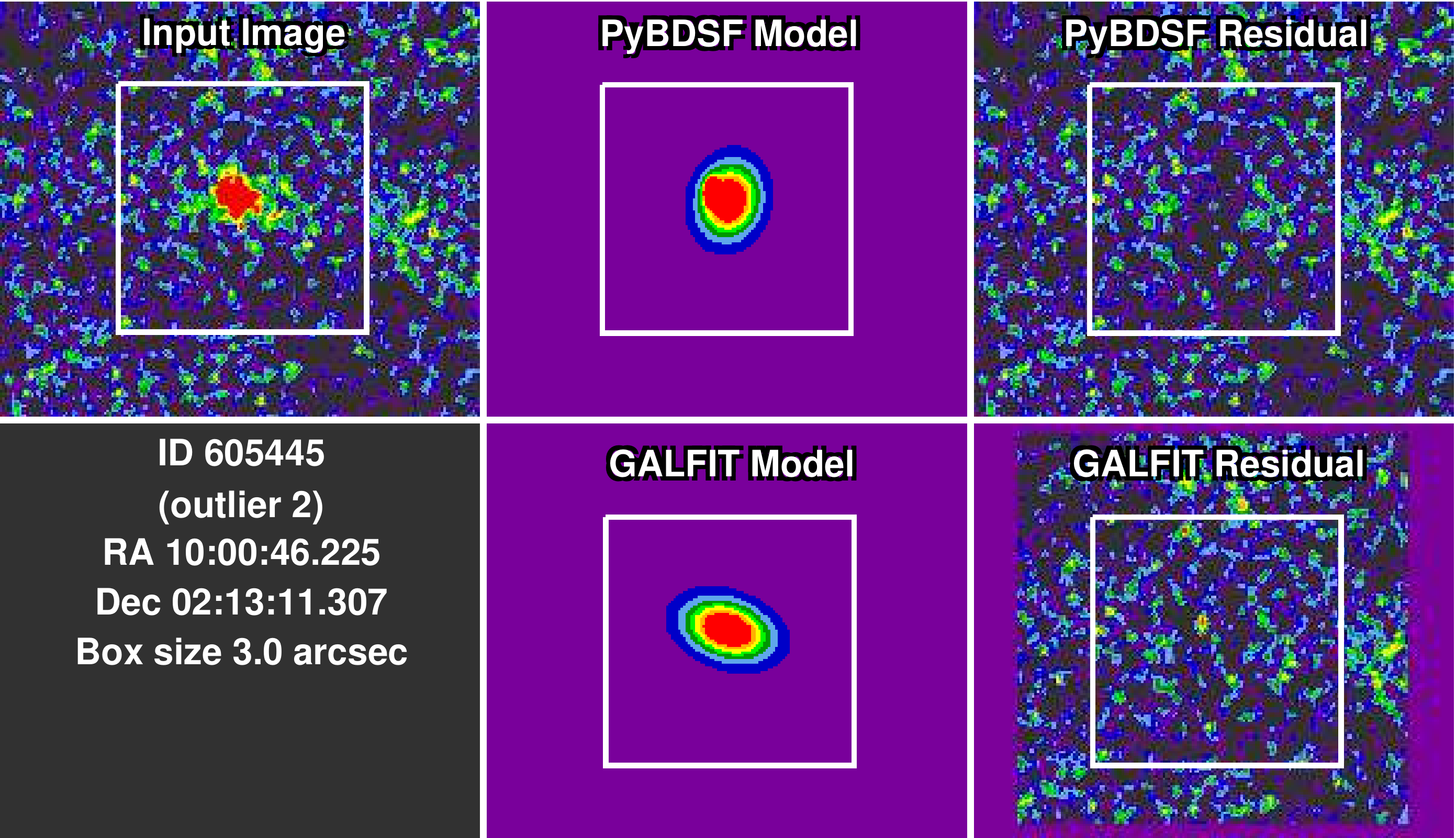}\\
\vspace{2.2mm}
\includegraphics[width=0.46\textwidth, trim=0 0 0 0, clip]{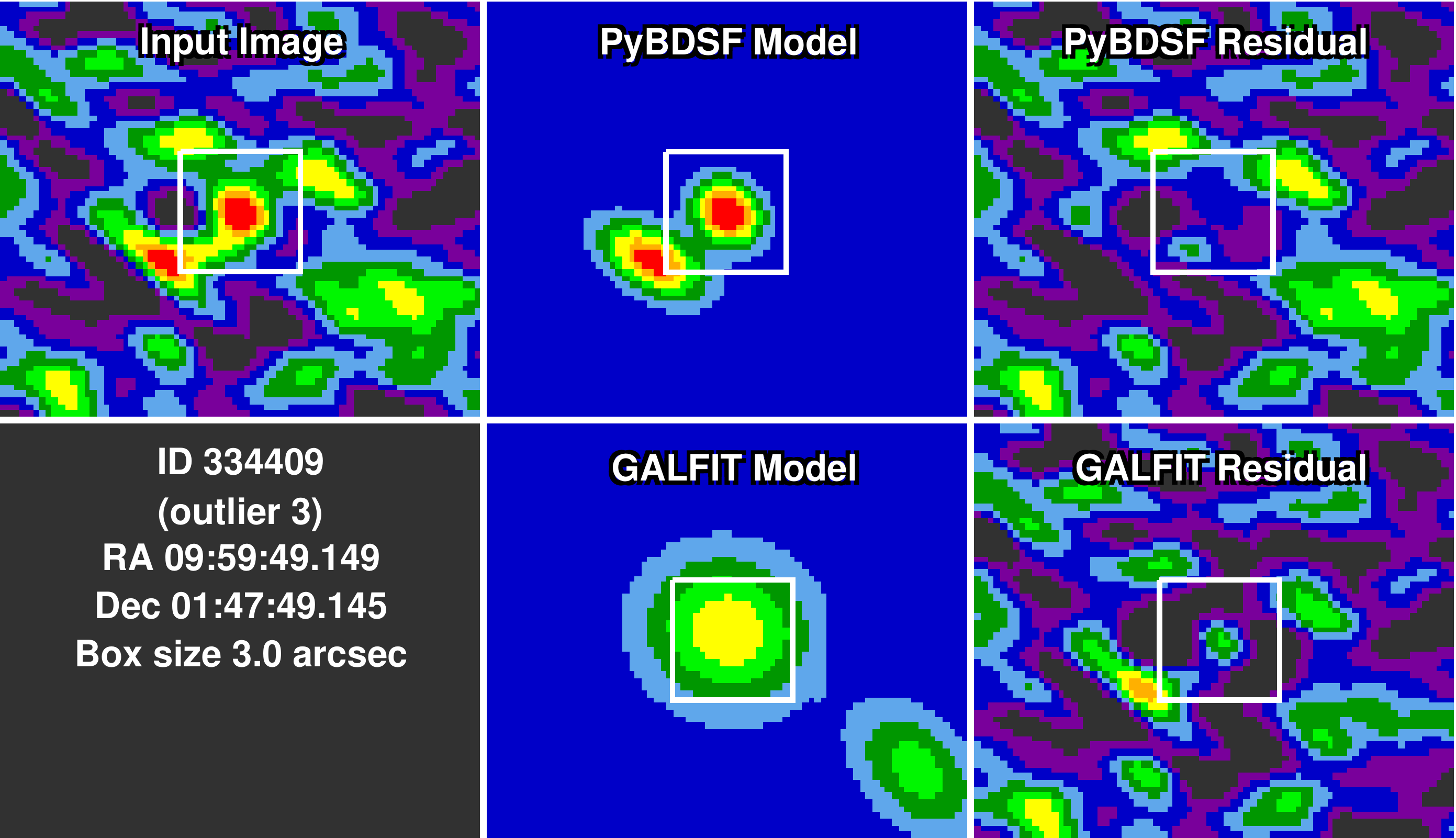}
\hspace{2.5mm}
\includegraphics[width=0.46\textwidth, trim=0 0 0 0, clip]{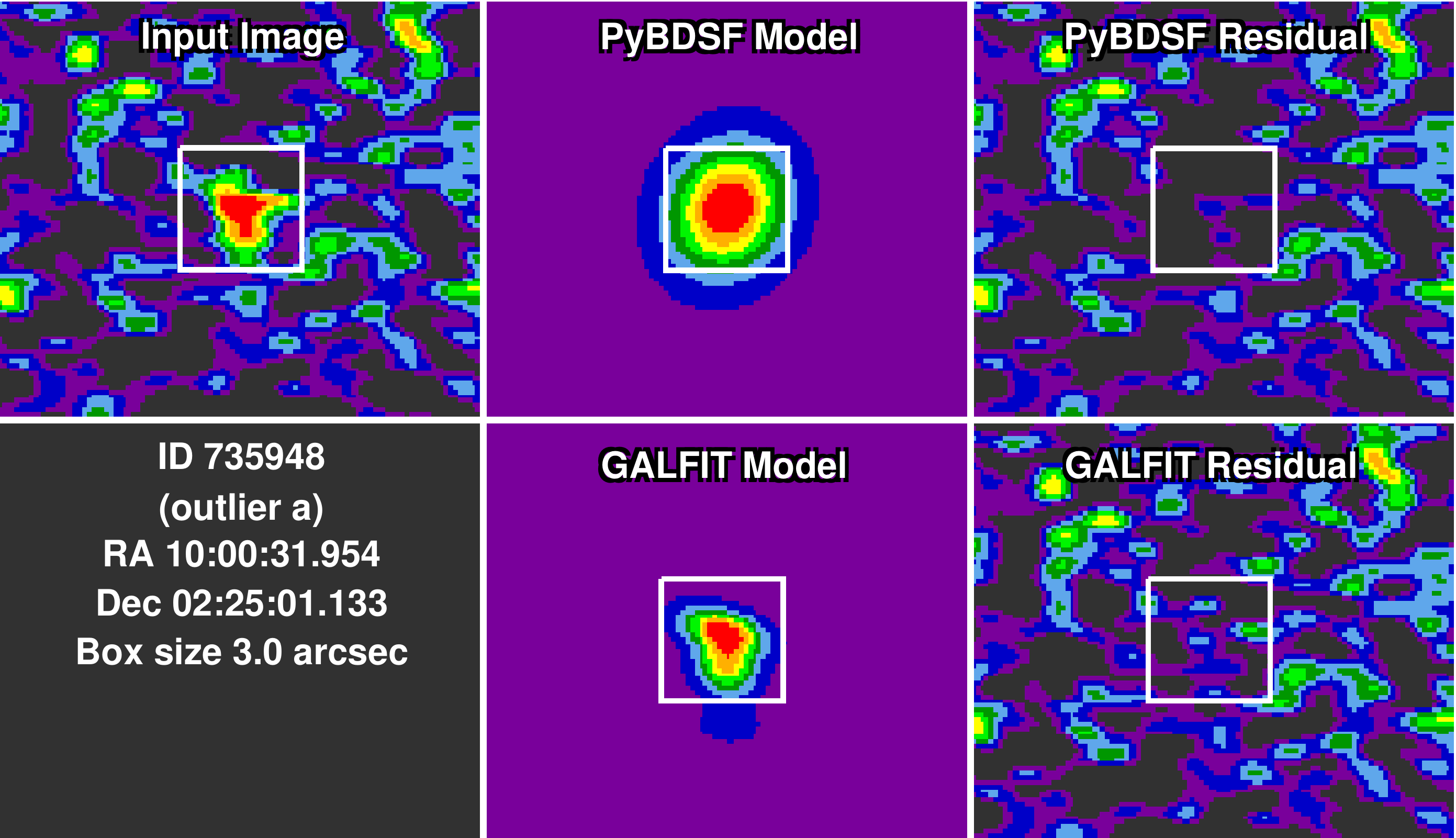}\\
\vspace{-1.5mm}
\caption{%
\pybdsm{} and \galfit{} fitting images for the four outliers labeled as \incode{1}, \incode{2}, \incode{3} and \incode{a} in Fig.~\ref{Plot_catalog_comparison_a3cosmos_prior_vs_blind_total_flux}
where the measured (sub-)mm continuum
fluxes from our \pybdsm{} and \galfit{} photometry differ significantly ($>5\sigma$. Their IDs in our master catalog (being the same as in the \cite{Laigle2016} catalog) are 831167, 605445, 334409 and 735948, respectively. For
each source six sub-panels are shown, namely input image, \pybdsm{} model image, \pybdsm{} residual image ({\em top row, left to right}), source information, \galfit{} model image and \galfit{} residual image ({\em bottom row, left to right}). The color scale is the same for all sub-panels for each source, but varies from source to source. A white box with a size of 3$''$ is shown in each sub-panel for reference. 
\label{Figure_flux_comp_galfit_vs_pybdsm_outlier_a}%
}
\end{figure}

Here we present the ALMA images of the (sub-)mm continuum sources that
have inconsistent 
total fluxes in the prior-based and blindly extracted catalogs and are labeled in Fig.~\ref{Plot_catalog_comparison_a3cosmos_prior_vs_blind_total_flux}.
Three outliers with labels \incode{1} to \incode{3} have $>5\sigma$ higher \galfit{} fluxes than \pybdsm{} fluxes, and one outlier with label \incode{a} has the opposite situation. Their ALMA images, prior fitting and blind extraction model images and residual images are shown in Fig.~\ref{Figure_flux_comp_galfit_vs_pybdsm_outlier_a} (each outlier has six sub-panels; see caption for details). 

In general the \galfit{} source models provide better fits to the original ALMA images, with less residual emission in the residual images, except for outlier \incode{3} which seems to 
be composed of two ALMA sources while our COSMOS master catalog contains only one prior source. 

For the fourth outlier with label \incode{a} in Fig.~\ref{Plot_catalog_comparison_a3cosmos_prior_vs_blind_total_flux} and shown in the bottom-right of Fig.~\ref{Figure_flux_comp_galfit_vs_pybdsm_outlier_a}, the \galfit{} model is more complex than the simple Gaussian shaped \pybdsm{} model because multiple prior sources are fitted. Therefore, as long as we have a good knowledge of prior sources in the ALMA field of view, i.e., from our compiled COSMOS master catalog, the \galfit{} fitting typically provides very good photometry results (with a small enough reduced-$\chi^2$ in the residual image).

\vspace{1truecm}
\FloatBarrier

\section{Detailed Description of our Monte Carlo Simulations}
\label{Section_Appendix_MC_Sim}

We briefly introduced our two sets of Monte Carlo simulations in Sect.~\ref{Section_Monte_Carlo_Simulation_and_Correction} --- the full-parameter-space (``FULL'') simulation and the physically-motivated (``PHYS'') simulation. Below we provide in-depth details of how we model the artificial sources (Appx.~\ref{Section_Appendix_MC_Sim_FULL_Source_Simulation}~and~\ref{Section_Appendix_MC_Sim_PHYS_Source_Simulation}), inject them into ALMA residual images (after blind extraction photometry) (Appx.~\ref{Section_Appendix_MC_Sim_FULL_Source_Simulation}~and~\ref{Section_Appendix_MC_Sim_PHYS_Source_Placement}), recover the sources with our two types of photometry pipelines (Appx.~\ref{Section_Appendix_MC_Sim_FULL_Source_Recovery}~and~\ref{Section_Appendix_MC_Sim_PHYS_Source_Recovery}), and analyze the statistics (Appx.~\ref{Section_Appendix_MC_Sim_Final_Correction}). We also discuss the limitations of each simulation in Appx.~\ref{Section_MC_Sim_FULL_Limitations}~and~\ref{Section_Appendix_MC_Sim_PHYS_Limitations}. Note that both simulations have limitations which could bias our final flux and error estimations. Only by doing both simulations and comparing them with each other as done here, these limitations can be understood and the least biased way to implement corrections to obtain final photometry results can be identified. 

\subsection{Full parameter space (``FULL'') MC Simulation}
\label{Section_MC_Sim_FULL}

\subsubsection{Source simulation and injection}
\label{Section_Appendix_MC_Sim_FULL_Source_Simulation}

In the ``FULL'' simulation, we simulate one source at a time for each of the $\sim$150 representative ALMA continuum images (Sect.~\ref{Section_Monte_Carlo_Simulation_and_Correction}), with source peak flux density ranging from 3.0 to 100 times the rms noise (the ratio is denoted as $\SNRpeak$; see Eq.~\ref{Equation_SNRpeak}), and size (Gaussian major-axis FWHM, convolved with the beam) ranging from 0.1 to 6.0 times the synthesized beam size (clean beam, Gaussian major-axis FWHM) (the ratio is denoted as $\Sbeam$; see Eq.~\ref{Equation_Sbeam}). 

There are 13 grid points in the first parameter ($\SNRpeak$) and also 13 in the second parameter ($\Sbeam$). For each grid point, we generate 25 mock sources by randomizing the injecting position. 

The simulated source is assumed to be of Gaussian shape (the minor-axis FWHM is generated with an axis ratio randomly picked between 0.2 to 1.0). Then, the source is convolved with the clean beam and injected into the residual image derived from \pybdsm{} where sources were already blindly extracted and removed. We randomly cut a box area around the source with a size of 8 times the intrinsic source size to ensure enough empty sky area for source extraction. 

In total, we have $\sim$4225 ``FULL'' simulations per ALMA image, and repeating this for $\sim$150 representative ALMA images (one for each independent ALMA scheduling block) we have $\sim$3750 sources per grid point in the two-dimensional parameter space.

\subsubsection{Source recovery}
\label{Section_Appendix_MC_Sim_FULL_Source_Recovery}

We run our \pybdsm{} and \galfit{} photometry tools to recover those simulated sources one by one. For \pybdsm{}, we keep the exact same conditions as for the real catalog, i.e., setting the background to zero and the rms noise to the values we measured from the previous photometry run (from fitting the pixel histograms; Sect.~\ref{Section_Blind_Source_Extraction}), and using the same thresholds as for the original ALMA images (Sect.~\ref{Section_Blind_Source_Extraction}). 
For \galfit{}, the only difference is the input prior catalog. We assume no source blending issue and only fit the simulated source. 

\begin{figure}[ht!]
\centering%
\includegraphics[width=0.34\textwidth, trim=30mm 15mm 0 0]{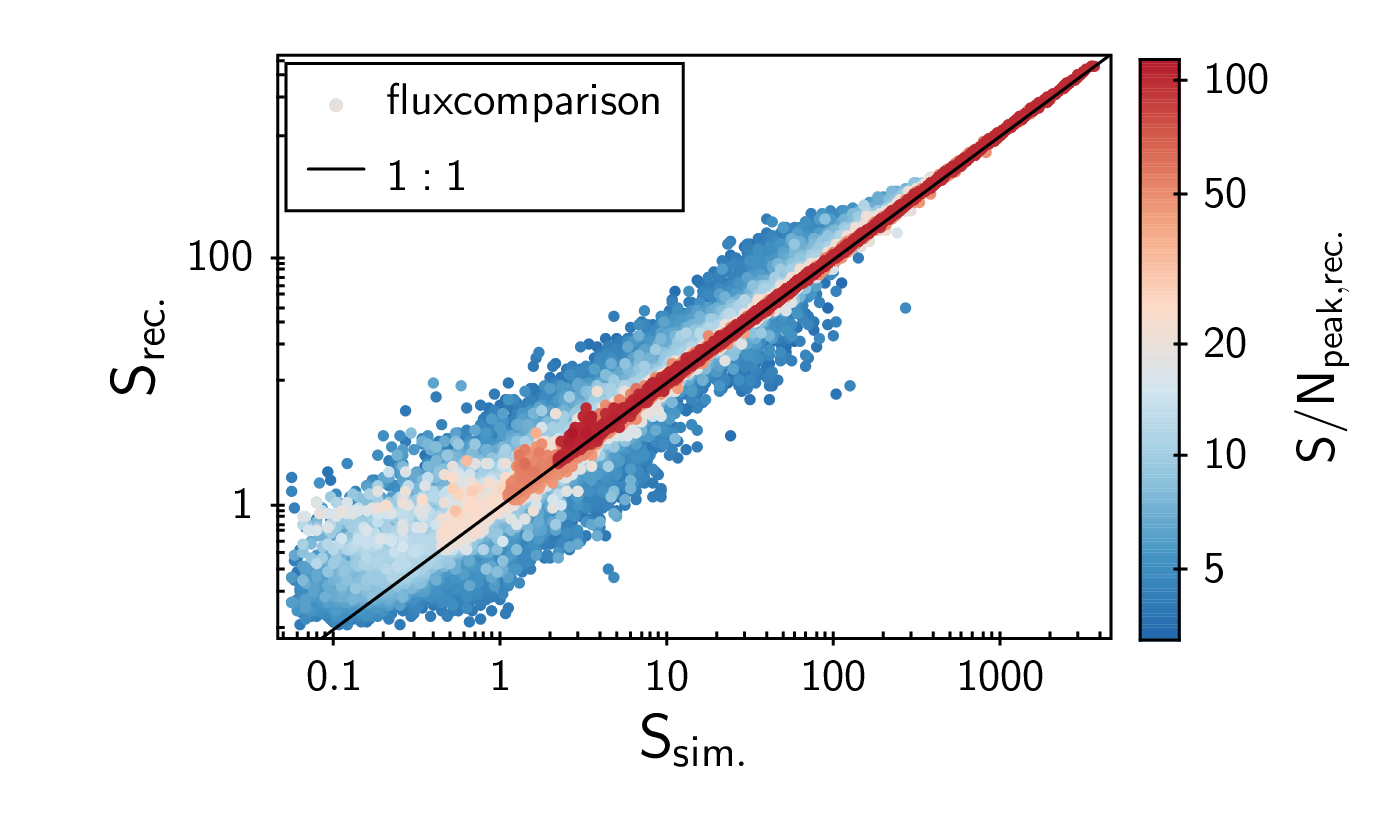}
\includegraphics[width=0.34\textwidth, trim=22mm 15mm 8mm 0]{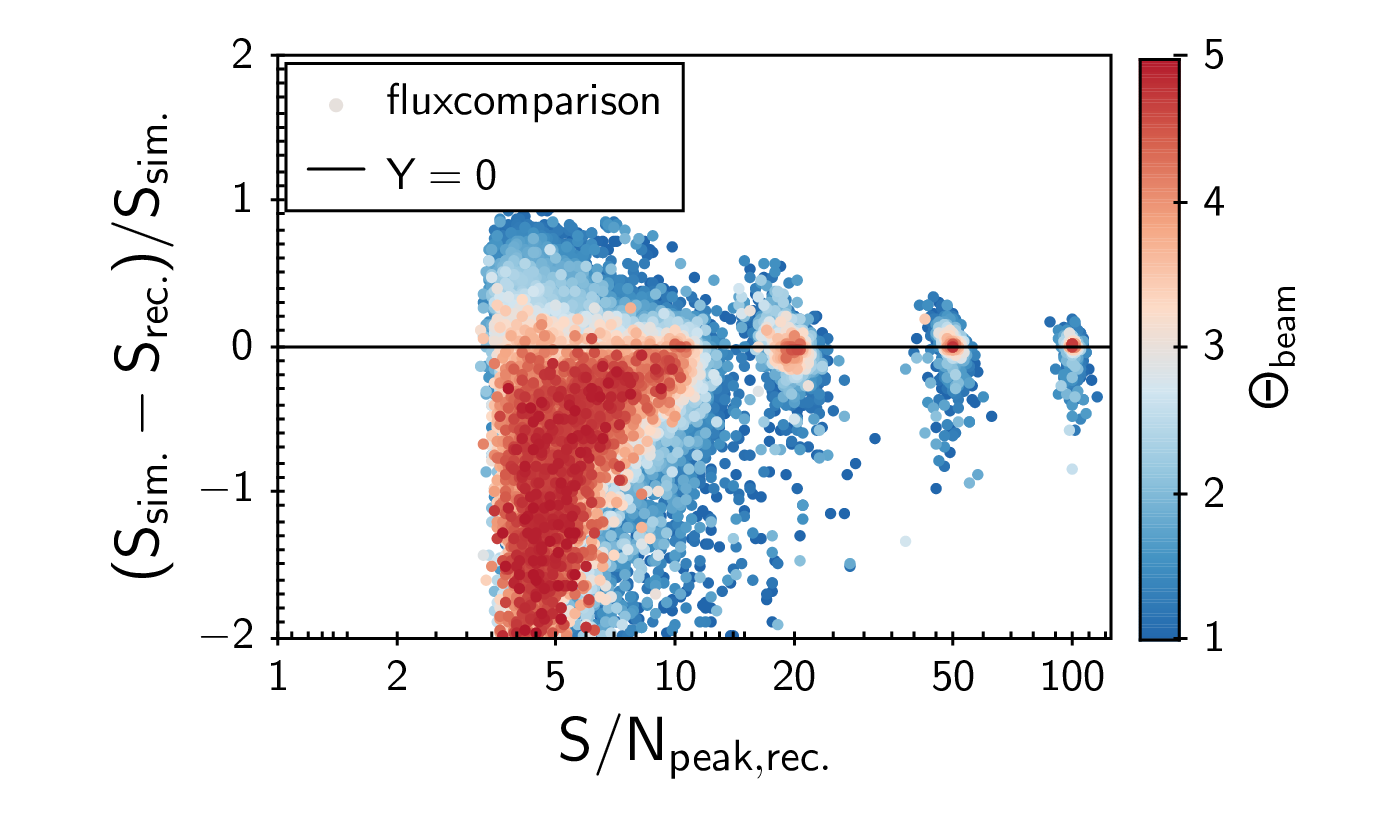}
\includegraphics[width=0.30\textwidth, trim=15mm 0 5mm 0]{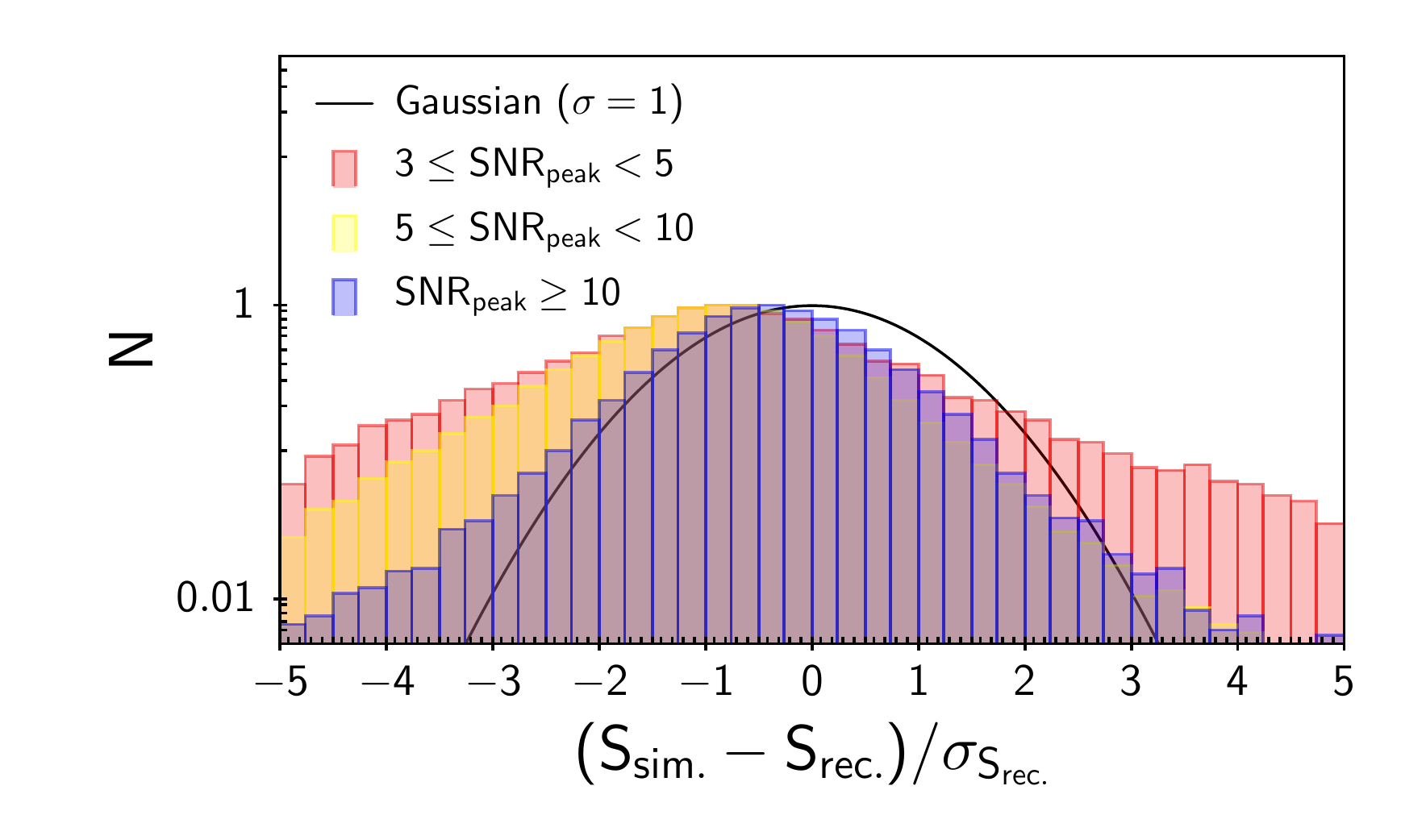}
\includegraphics[width=0.34\textwidth, trim=30mm 15mm 0 0]{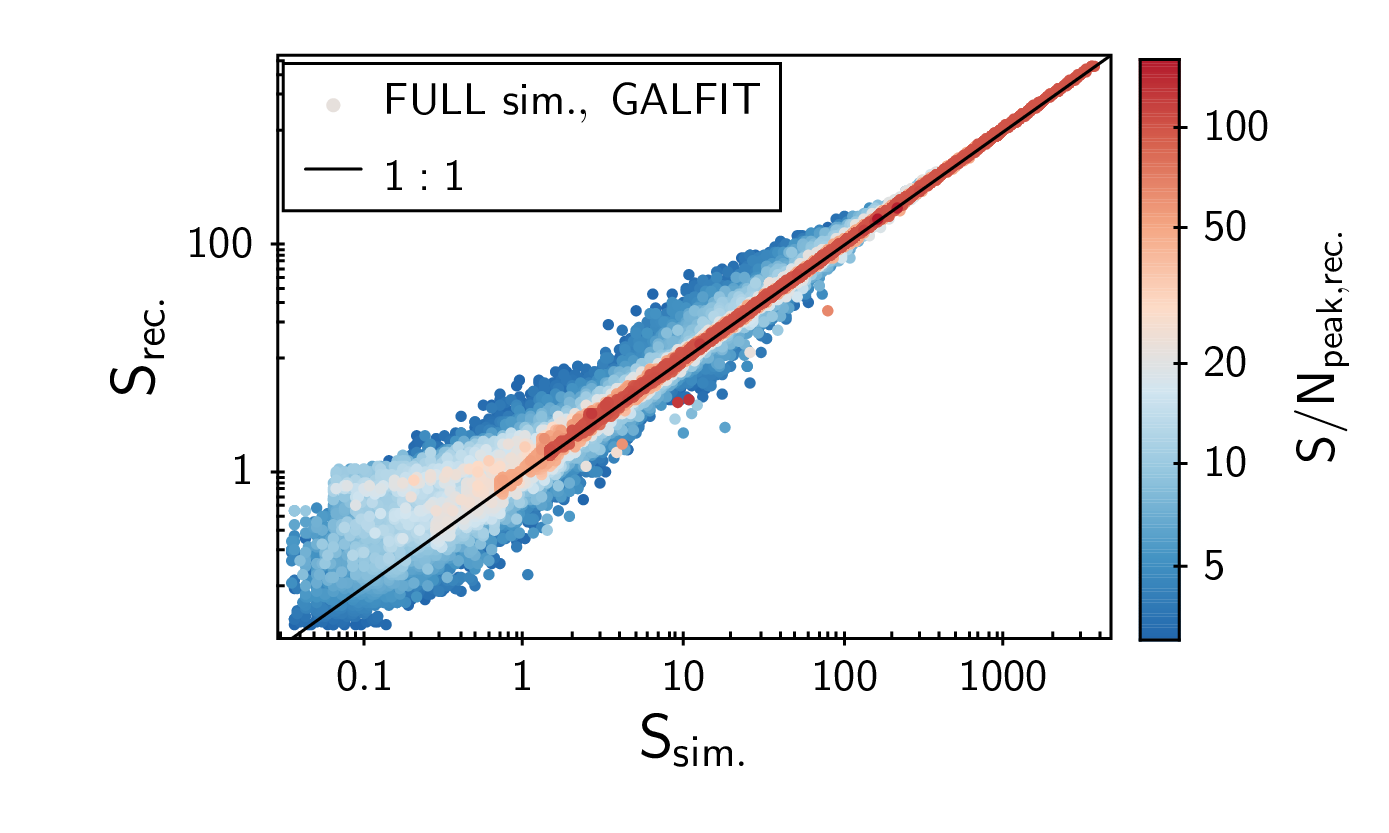}
\includegraphics[width=0.34\textwidth, trim=22mm 15mm 8mm 0]{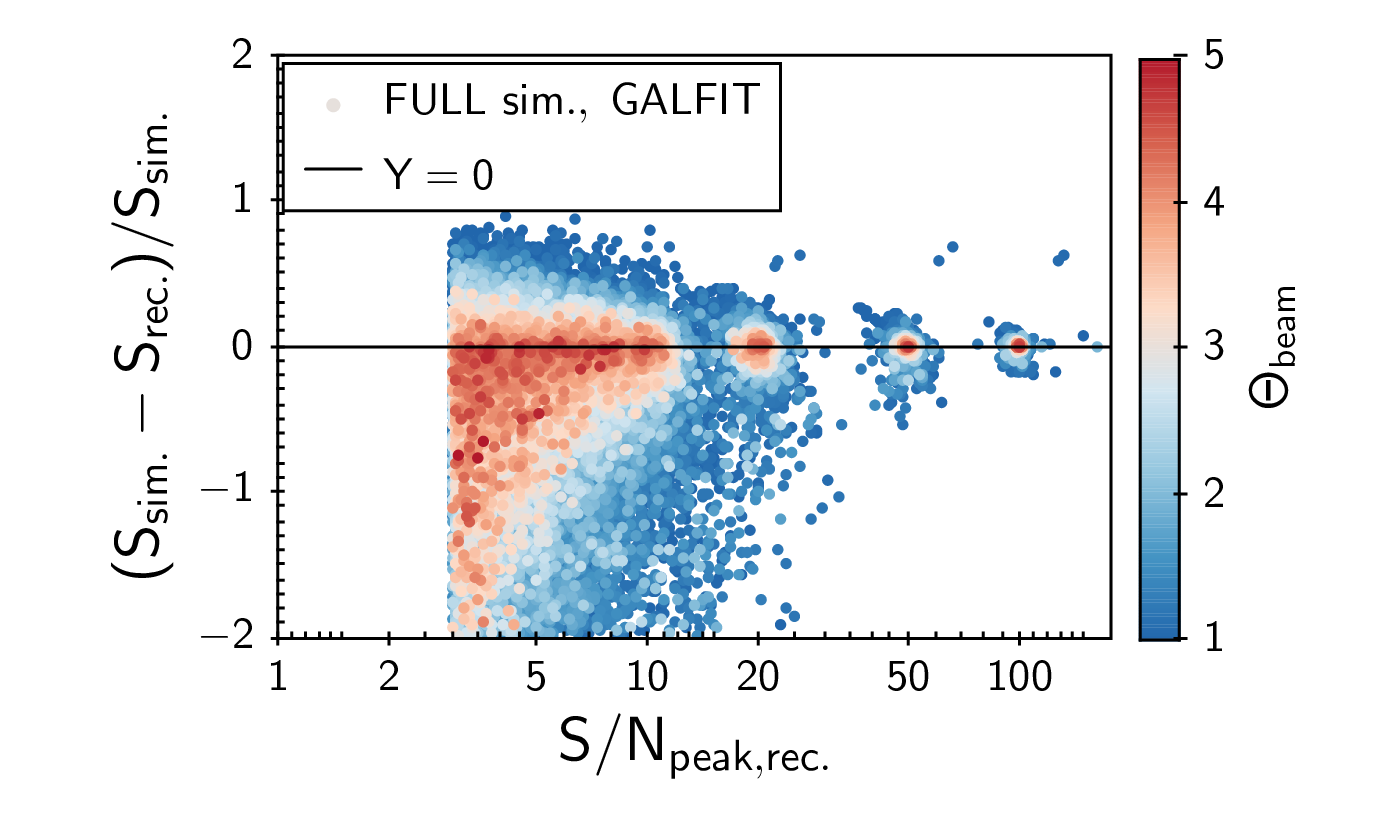}
\includegraphics[width=0.30\textwidth, trim=15mm 0 5mm 0]{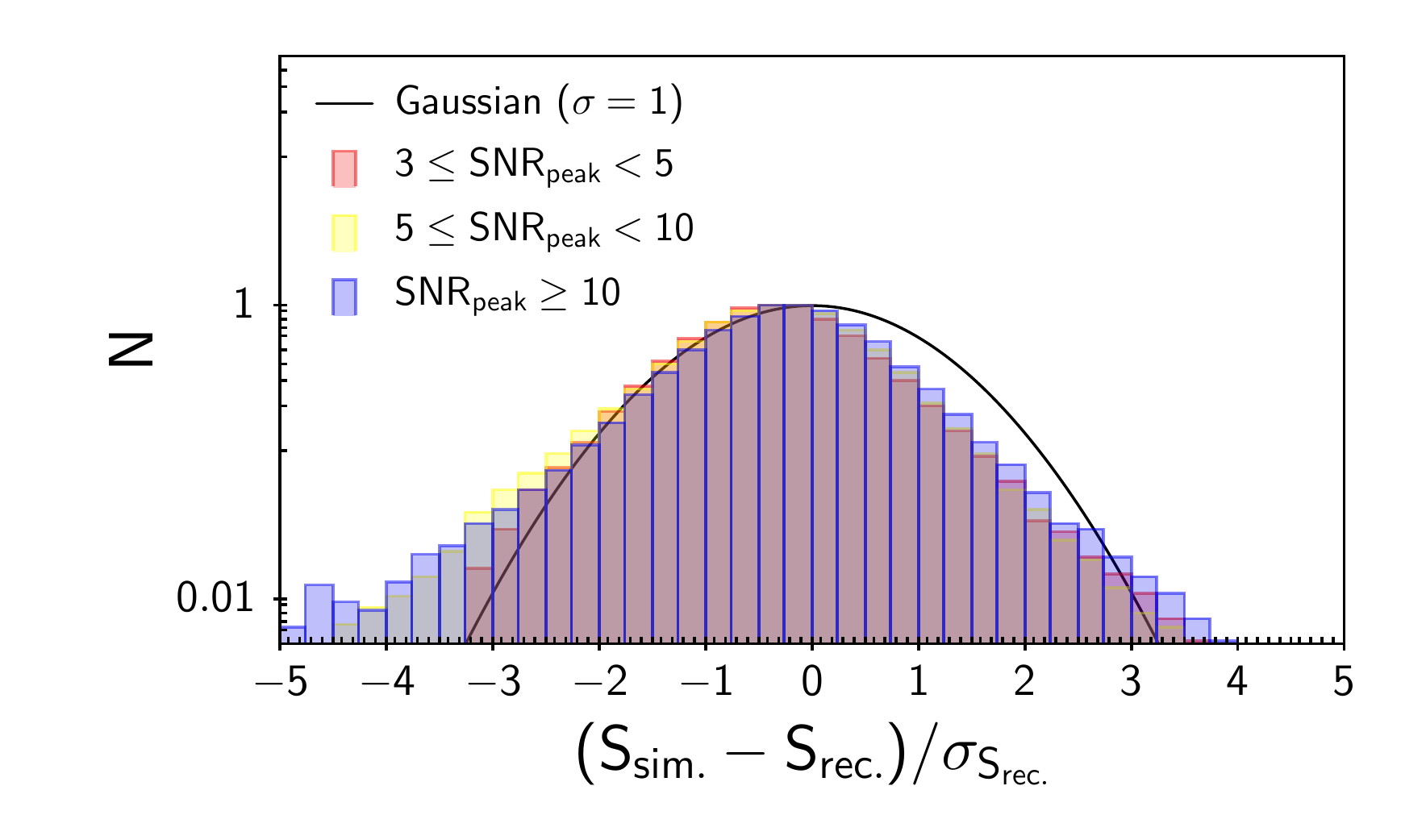}
\caption{%
Analysis of the \pybdsm{} ({\em top}) and \galfit{} ({\em bottom}) recovery of properties of the ``FULL'' simulation sources. {\bf Left panels} show the comparison of simulated versus recovered flux colored by $\SNRpeak$. {\bf Middle panels} present the relative flux difference ($(\Ssim - \Srec)/\Ssim$) versus $\SNRpeak$. {\bf Right panels} show the histogram of the normalized flux difference ($(\Ssim - \Srec)/\ESrec$), where $\ESrec$ is the \pybdsm{} output flux error following the \cite{Condon1997} equations. Compared to also our corrected histogram in Fig.~\ref{Plot_MC_sim_PHYS_PYBDSM_corrected}. 
\label{Plot_MC_sim_scatter_FULL}
}
\end{figure}

In Fig.~\ref{Plot_MC_sim_scatter_FULL} we show the comparisons of the simulated and recovered fluxes for \pybdsm{} ({\em top panels}) and \galfit{} ({\em bottom panels}). 
The left panels show the simulated versus recovered fluxes, colored by $\SNRpeak$. Here we only shows sources which have $\SNRpeak>3$, because lower $\SNR$ detections are mostly spurious, according to the spurious fraction analysis in Sect.~\ref{Section_Spurious_Fraction}, and eventually we select our ALMA detections with a much higher $\SNRpeak\sim5$ (Sect.~\ref{Section_Combining_two_photometry_catalogs}). 

The middle panels of Fig.~\ref{Plot_MC_sim_scatter_FULL} show the difference between the simulated and recovered fluxes $(\Ssim-\Srec)$ normalized by $\Ssim$ as a function of $\SNRpeak$ (source peak flux to rms noise ratio; Eq.~\ref{Equation_SNRpeak}), colored by $\Sbeam$ (source area to beam area ratio; Eq.~\ref{Equation_Sbeam}). In general, at a low $\SNRpeak$, $\Ssim$ is always smaller than $\Srec$, indicating that fluxes are boosted by noise. Such a flux-boosting is much smaller for a higher $\SNRpeak$. Therefore, based on these, we quantify the flux bias by the two parameters $\SNRpeak$ and $\Sbeam$ in the main text (Sect.~\ref{Section_MC_Sim_Statistics}). 
Meanwhile, the scatter of $(\Ssim-\Srec)$ reflects the uncertainty of the photometry, i.e., flux errors, which can also be quantified by the two parameters (Sect.~\ref{Section_MC_Sim_Flux_Error}). Note that the flux errors that came along with our two photometry pipelines are based on the equations in \cite{Condon1997}, where the author used about 3000 MC simulations to calibrate these equations. Our MC simulations offer the possibility for alternative assessments that show a broad consistency but also evidence for a second-order trend with $\SNRpeak$ (Sect.~\ref{Section_MC_Sim_Flux_Error}~and~\ref{Section_MC_Sim_Flux_Error_for_GALFIT}). 

The right panels of Fig.~\ref{Plot_MC_sim_scatter_FULL} show the histogram of $(\Ssim-\Srec)$ normalized by the flux errors $\ESrec$. Sources are grouped into subsamples according to their $\SNRpeak$. An 1D Gaussian fit ($e^{-(x-\mu)^2/2\,\sigma^2}$) to the histogram indicates whether the flux errors can statistically represent the uncertainty of the photometry. If the fitted Gaussian is too wide (i.e., $\sigma>1$), then the flux errors are underestimated, and vice versa. 
We overlay the $\sigma=1$, $\mu=0$ 1D Gaussian curve for comparison. Note that both flux bias and error affect these histograms. We demonstrate in Appx.~\ref{Section_Appendix_MC_Sim_Final_Correction} that after correcting flux biases and re-estimating flux errors, these histograms becomes much more close to $\sigma=1$, $\mu=0$ 1D Gaussian shapes, i.e., we can say that they follow a well-behaved Gaussian statistics.

\subsubsection{Limitations}
\label{Section_MC_Sim_FULL_Limitations}

We discuss several limitations related to the use of the ``FULL'' simulation to indicate flux bias, error and completeness in this section. 
First is the assumed source property distribution. Using uniform $\SNRpeak$ and $\Sbeam$ distributions is indeed a strong assumption, although it is perhaps the most commonly adopted way in IR/mm/radio photometry studies. We have to consider the following situation, which we refer to as the \textit{``resolution bias''}. A large, low $\SNR$ source can usually break up into several smaller clumps due to noise fluctuations. Our photometry code will then usually only detect a smaller, low $\SNR$ clump, therefore the source's total flux is only partially recovered. This acts opposite to the effect of flux-boosting, where our photometry code detects a low $\SNR$ source which is actually the peak of a noise fluctuation instead of a real galaxy. In reality, what we know about the detected sources are only the recovered fluxes and sizes, therefore we can not distinguish the two effects. As we parametrize the flux biases and errors by the recovered fluxes and sizes (as will be described in detail in Sect.~\ref{Section_MC_Sim_Statistics}), simulating more large size sources will lead to less flux-boosting (hence smaller flux biases), and larger flux errors that can be significant particularly at low $\SNR$. 

A second limitation is that sources are simulated and then recovered individually in our ``FULL'' simulation procedure. Thus there is no source blending or clustering effect. In reality, sources can be blended even at the arcsec resolution of the ALMA data, although this situation occurs at low probability, e.g., it depends on the galaxy merger fraction. This limitation could affect our estimation of the completeness of the \pybdsm{} photometry, because \pybdsm{} is a blind extraction tool and sometimes will extract two blended sources as a single source. The \galfit{} photometry should not be affected, if the prior source catalog has a high resolution (sub-arcsec) and is complete (not missing sources in ALMA bands). 

Another limitation is the input map we used for injecting source models. We use the \pybdsm{} residual maps for all of our analysis presented here, because the residual maps ideally should contain the exact noise as in the observations. However, imperfect source subtraction (by \pybdsm) could potentially increase the noise at certain positions. However, this is likely a very minor problem as we randomize the injection positions.
Injecting source models in the $uv$-plane (pure interferometry noise) instead of the image plane (\pybdsm{} residual image) could in principle help to assess the additional uncertainty introduced by the imaging/cleaning process. But the difference should be small because we have verified with our real data that the rms noise and source fluxes measured from the $uv$-plane and image plane are fully consistent.

\subsection{Physically-motivated (``PHYS'') MC Simulation}
\label{Section_Appendix_MC_Sim_PHYS}

\subsubsection{Source simulation}
\label{Section_Appendix_MC_Sim_PHYS_Source_Simulation}

In order to disentangle the major limitations from the ``FULL'' simulation, we have done another physically-motivated simulation (``PHYS'' simulation) where we try to reproduce the real physical properties of galaxies across cosmic time. 

In detail, we follow the two-star-formation-mode (2SFM) recipe \citep{Sargent2012_2SFM,Sargent2014,Bethermin2012_2SFM}, which assumes all star-forming galaxies are in two populations, with the population of starbursts being enhanced in their sSFRs by a range of factors under a normal distribution with a mean of 5 (here we adopt 5 because we find this better fits the millimeter number counts; \citealt{Sargent2014} suggest a value of 4; see their Fig.~10). We generate these star-forming galaxies within the cosmic volume of the 2 square degree COSMOS field with the following procedures: 
\begin{enumerate}[label=\arabic*), topsep=0pt, noitemsep]
    \item Defining 25 redshift bins from z=9.75 to 0; 
    \item Computing the number of star-forming galaxies using the stellar mass function (e.g., \citealt{Davidzon2017}) at each redshift and starting from $\rm \Mstar=10^{8.0} \; \Msun$; 
    \item Assuming a small fraction of these star-forming galaxies are in starburst (SB) mode while the rest are in MS mode. The fraction is set to be consistent with the merger fraction extrapolated from \cite{Conselice_2014_ARAA}; 
    \item Computing SFRs for MS and SB galaxies following the MS correlation (including the scatter) and SB boost function in \cite{Sargent2014}; 
    \item Then we generate an IR-to-radio SED for each model galaxy according to the redshift, stellar mass and SFR, following the SED modeling in \cite{Liudz2017}, which is based on \cite{Magdis2012SED}, assuming \cite{Draine2007SED} dust models and simplifying the dust SEDs by associating them only to redshift and the interstellar radiation field ($\left<U\right>$; see \citealt{Magdis2012SED}, \citealt{Bethermin2015} and \citealt{Liudz2017} for more details); 
    \item Estimating MS galaxies' sizes depending on their redshifts and stellar masses following (extrapolating from) \cite{vanderWel2014} as well as considering that dust sizes are a factor of about 2 smaller (\citealt{Fujimoto2017}); 
    \item Random minor/major axis ratio from 0.2 to 1; 
\end{enumerate}

In total about 0.7 million model galaxies are generated in 2 square degrees. As a validation, their number counts are also  estimated at each IR/mm/radio wavelength, these simulated counts are found to agree well with real (sub-)mm measurements at 500\,$\mu$m, 850\,$\mu$m, 1.1\,$\mu$m and 1.3\,$\mu$m (within the error bars), e.g., \cite{Bethermin2012}, \cite{Karim2013}, \cite{Carniani2015}, \cite{Hatsukade2016} and \cite{Geach_2016_SCUBA2}.

These model galaxies are then randomly injected into the 2 sq. deg. COSMOS field and recovered, which is described in the following.

\subsubsection{Source placement}
\label{Section_Appendix_MC_Sim_PHYS_Source_Placement}

We assign a random position within the 2 square degree COSMOS field for each mock galaxy. 
Then, we create artificial ALMA maps by inserting our mock galaxies into the ALMA residual images of the 150 representative programs we have selected in Section\ \ref{Section_MC_Sim_FULL}. For each of the residual maps, we create 273 of such artificial maps. Within each iteration, we select a mock galaxy at specific redshift and stellar mass out of the full mock galaxy catalog, and place it in the center of the residual map. We apply a small random offset (1--6$''$) to avoid imperfect source extraction at the center of the residual image. A subset of other remaining galaxies from the full mock catalog may fall within the same residual map according to their position within the full simulated 2 square degree map and are inserted as well.
In this way, we account for possible clustering of (sub-)mm sources in real observations. 
We loop for each simulated galaxy over a redshift grid ranging from 1.0 to 6.0 in steps of 0.25, with log stellar mass from $\rm log(\Mstar(M_{\odot}))=$9.0 to 12.0 in steps of 0.25, making a total of 273 iterations. Then, we extract all the sources simultaneously in the next section (unlike for the ``FULL'' simulation where we extract each simulated source individually, see Appx.~\ref{Section_MC_Sim_FULL}).

\subsubsection{Source recovery}
\label{Section_Appendix_MC_Sim_PHYS_Source_Recovery}

The simulated images are then treated by our \pybdsm{} and \galfit{} pipelines in the same manner as the original ALMA images. The simulated and original ALMA images have the same size, therefore the impact of simultaneous multi-source fitting is also considered. 
Moreover, for the prior fitting with \galfit{}, we use the catalog of simulated galaxies as the prior source list. 

\begin{figure}[htb]
\centering%
\includegraphics[width=0.34\textwidth, trim=30mm 15mm 0 0]{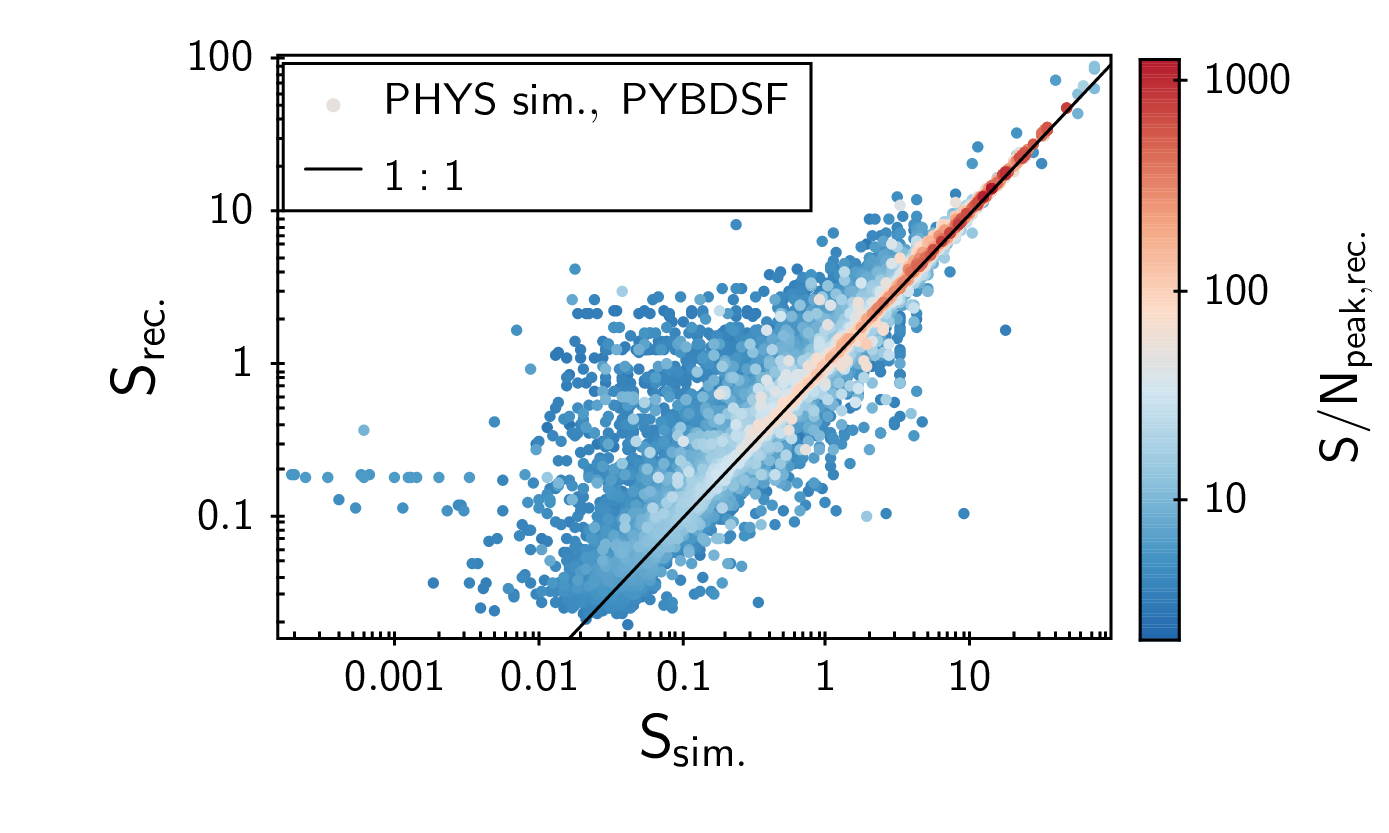}
\includegraphics[width=0.34\textwidth, trim=22mm 15mm 8mm 0]{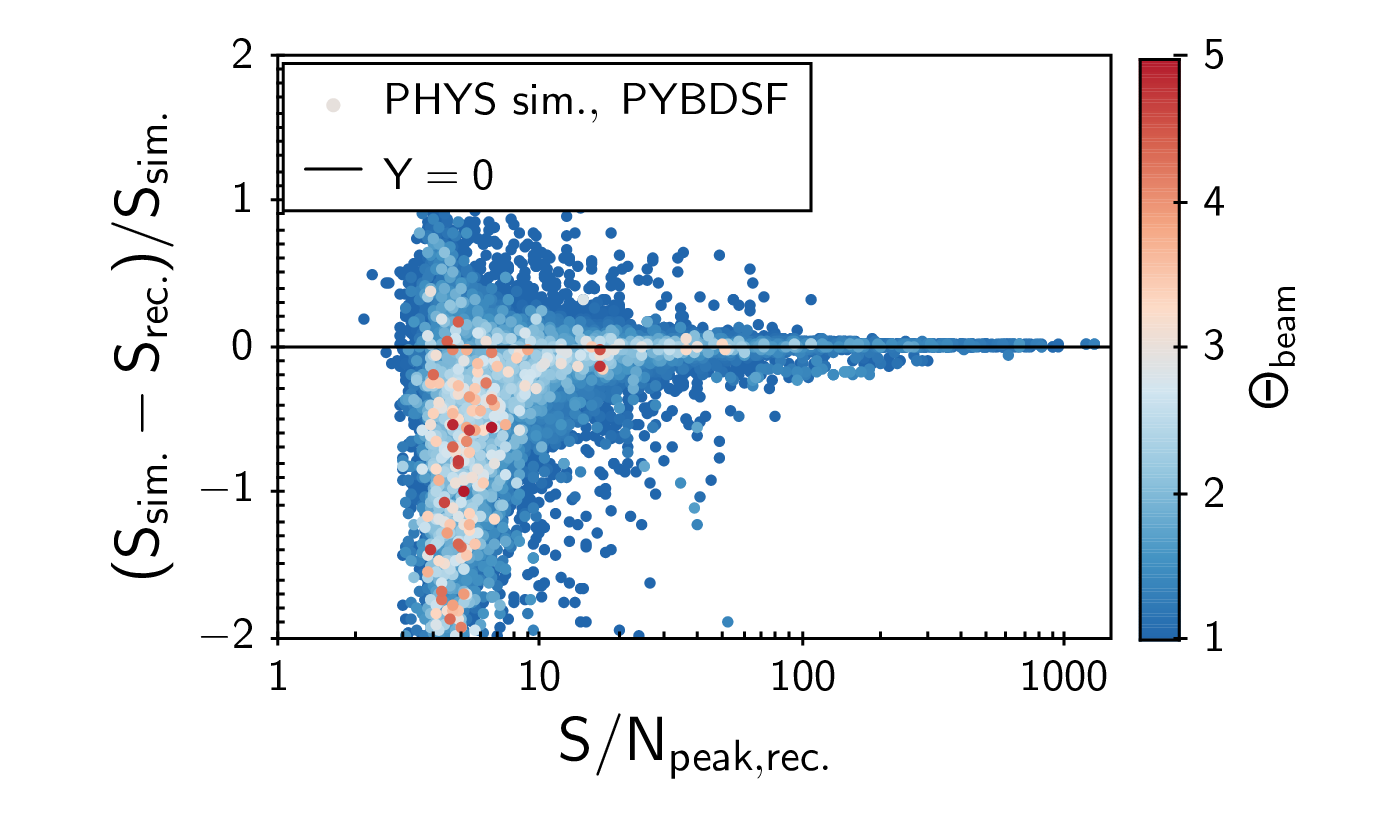}
\includegraphics[width=0.30\textwidth, trim=15mm 0 5mm 0]{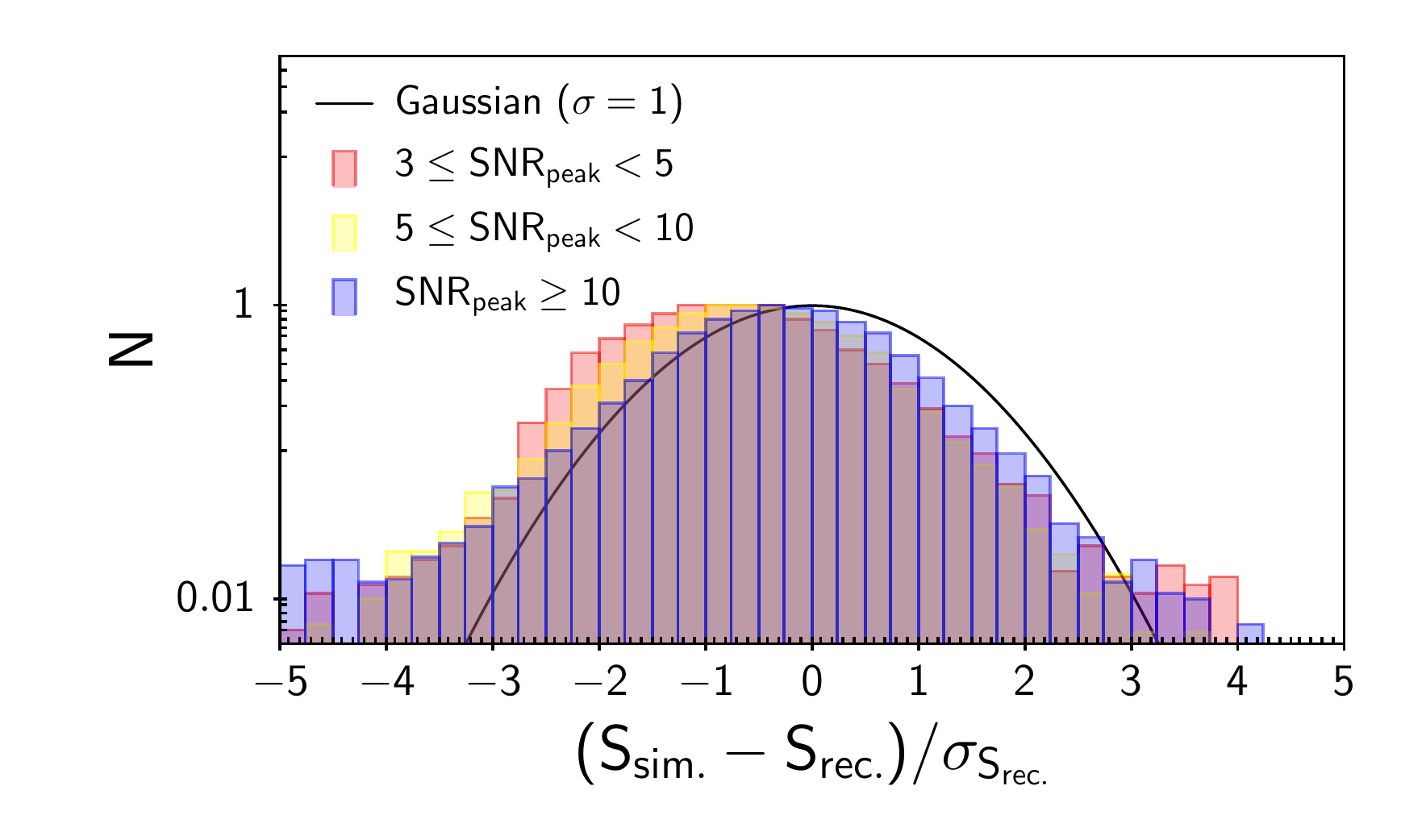}
\includegraphics[width=0.34\textwidth, trim=30mm 15mm 0 0]{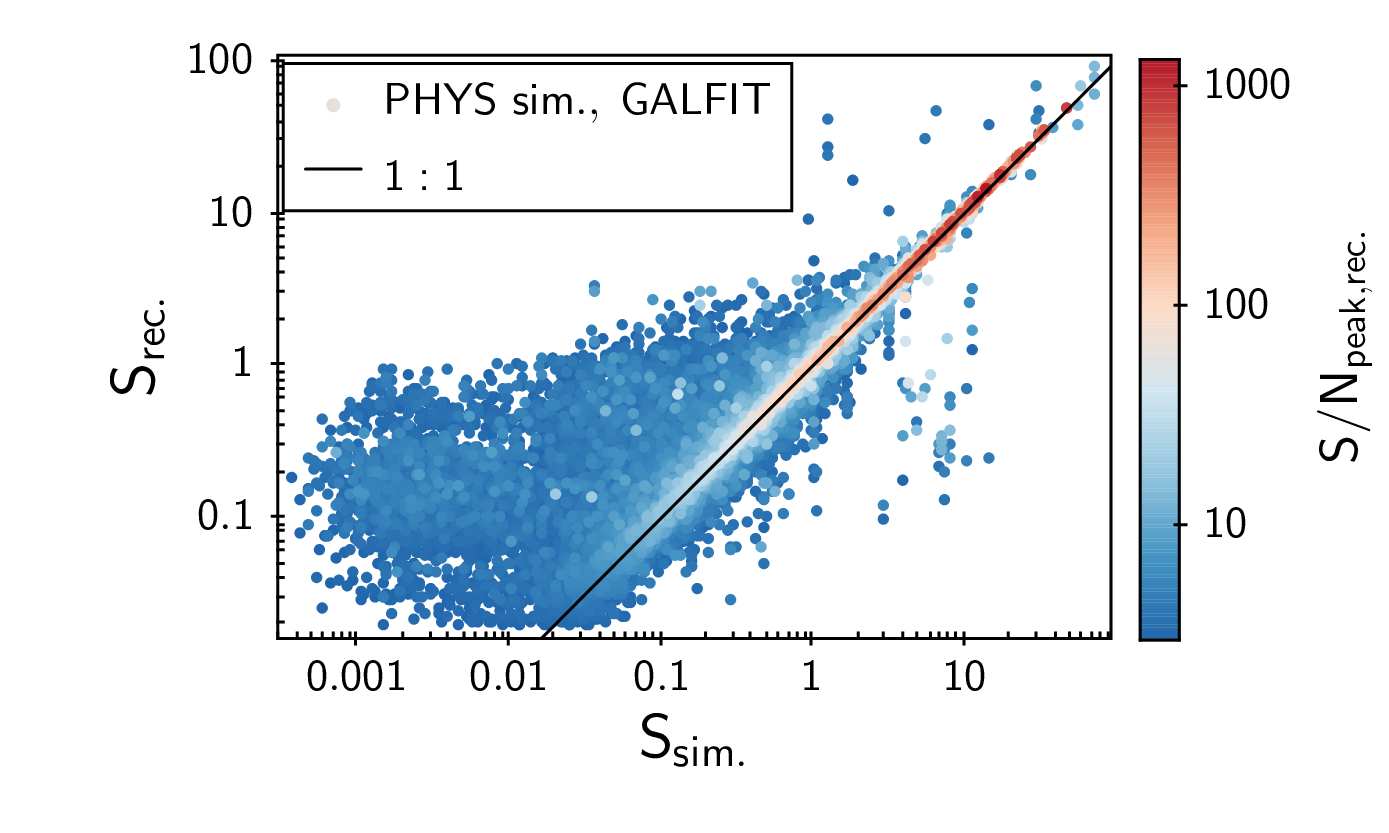}
\includegraphics[width=0.34\textwidth, trim=22mm 15mm 8mm 0]{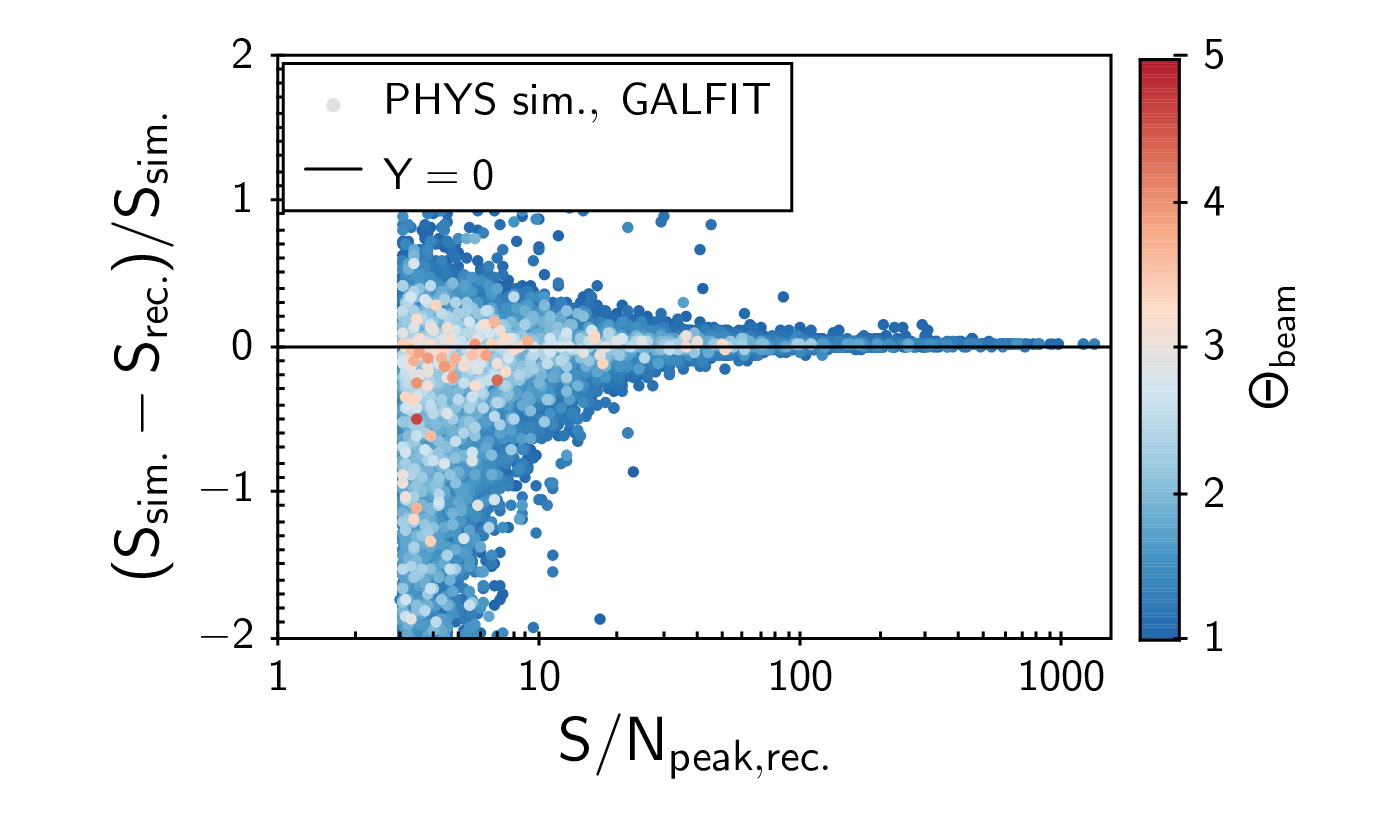}
\includegraphics[width=0.30\textwidth, trim=15mm 0 5mm 0]{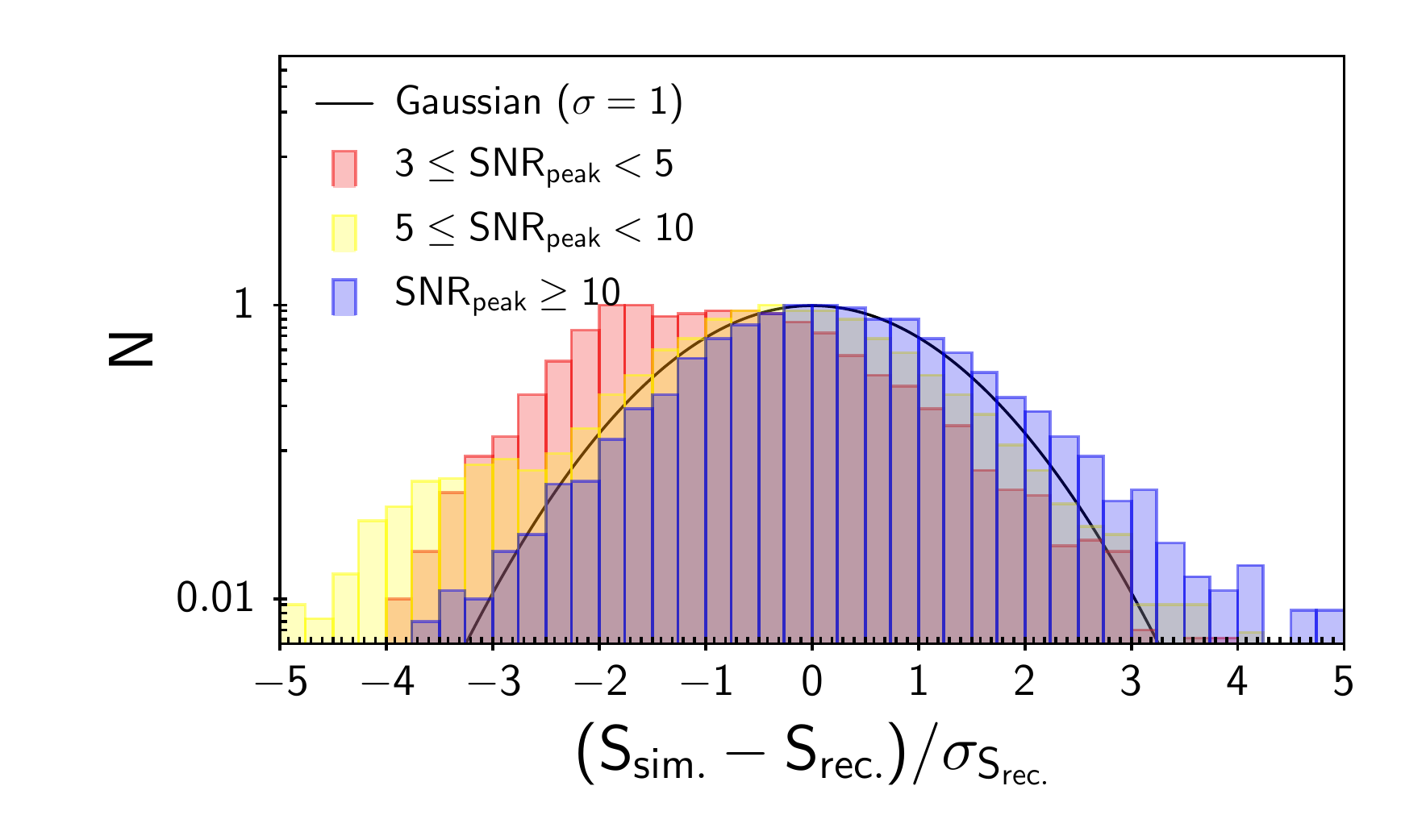}
\caption{%
Similar to Fig.~\ref{Plot_MC_sim_scatter_FULL} but for ``PHYS'' simulation. See Fig.~\ref{Plot_MC_sim_scatter_FULL} caption. 
\label{Plot_MC_sim_scatter_PHYS}
}
\end{figure}

Similar to Fig.~\ref{Plot_MC_sim_scatter_FULL}, we present the comparison of the simulated and recovered fluxes from the ``PHYS'' simulation in Fig.~\ref{Plot_MC_sim_scatter_PHYS}. The ``PHYS'' simulation contains many more faint sources (due to the realistic stellar mass function and main sequence correlation). The histograms in the right panels of Fig.~\ref{Plot_MC_sim_scatter_PHYS} are narrower than those of the ``FULL'' simulation as shown in Fig.~\ref{Plot_MC_sim_scatter_FULL}, especially for the low $\SNRpeak$ sources (shown as the red histograms in both figures). This means that the flux bias and error are different if we adopt different simulation methods (Fig.~\ref{Plot_MC_sim_fbias}~and~\ref{Plot_MC_sim_ecorr}), even when each simulation is repeated sufficiently to yield robust statistics. Here we emphasize that the prior information assumed in the simulations is important for analyzing the flux bias and error statistics, and making the simulation as close to real galaxy population distribution as possible will lead to more realistic results (Sect.~\ref{Section_MC_Sim_Statistics}). 

\subsubsection{Limitations}
\label{Section_Appendix_MC_Sim_PHYS_Limitations}

The physically-motivated simulation has the advantage of resembling closer the real situation for galaxy photometry in ALMA images. However, it also has limitations, both in the assumed galaxy evolution models and when comparing to the full-parameter-space simulation. 

Firstly, our model galaxies are built on the star-forming galaxy's stellar mass function at each redshift. These mass functions are not well constrained at redshift of $z\sim$4 and unconstrained at higher redshifts (e.g., \citealt{Grazian2015,Song2016,Davidzon2017}). Then, we assume a starburst fraction associated to the merger fraction, which is highly unconstrained at redshift $z\sim$2 and beyond. Moreover, because we aim to reproduce the majority of star-forming galaxies, we choose simplified galaxy SED models (\citealt{Magdis2012SED}; \citealt{Liudz2017}) which can represent the bulk of star-forming and starburst galaxies (see details in Appx.~\ref{Section_Appendix_MC_Sim_PHYS_Source_Simulation} and references there), but these SED models do not include extreme cases, e.g., galaxies with very high sSFR, very low or very high dust temperature, etc. 
The ``PHYS'' simulation also does not include a full implementation of the clustering effect as in \cite{Bethermin2017Model}. 
However, we emphasize that the aim of the ``PHYS'' simulation is to provide very different inputs from the ``FULL'' simulation, to see whether they can lead to different statistical results, and they do. Further the ``PHYS'' simulation we adopt here is sufficiently complex for testing our ALMA (sub-)mm photometry under all possible, physical situations that are not covered by the ``FULL'' simulation.

\subsection{Final statistical behaviour of corrected fluxes and errors}
\label{Section_Appendix_MC_Sim_Final_Correction}

\begin{figure}[htb]
\centering%
\setlength{\unitlength}{\textwidth}
\begin{picture}(0.49,0.46)
\put(0.02,0.25){%
\includegraphics[width=0.44\textwidth, trim=13mm 24mm 0 5mm, clip]{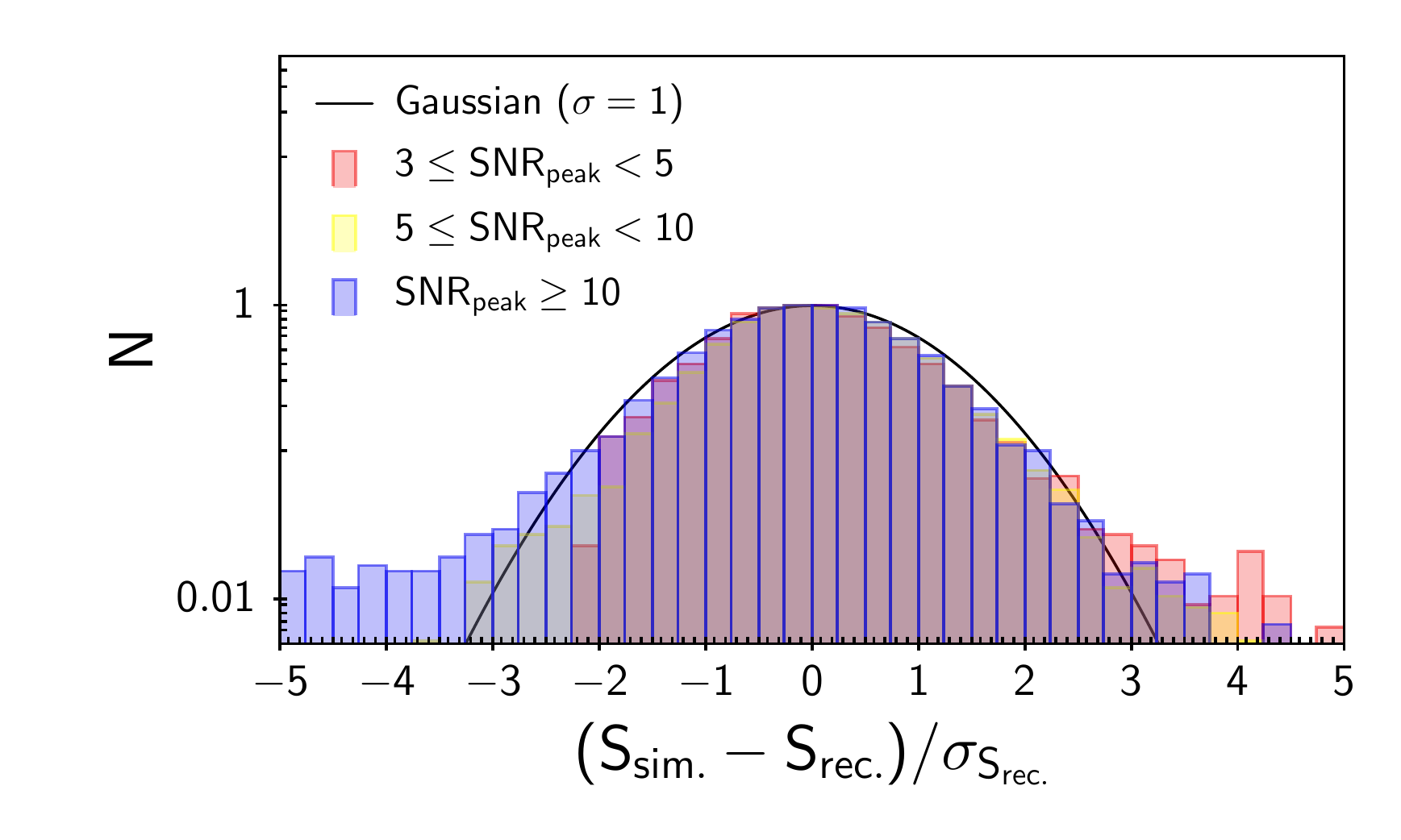}%
}
\put(0.02,0){%
\includegraphics[width=0.44\textwidth, trim=13mm 5mm 0 5mm, clip]{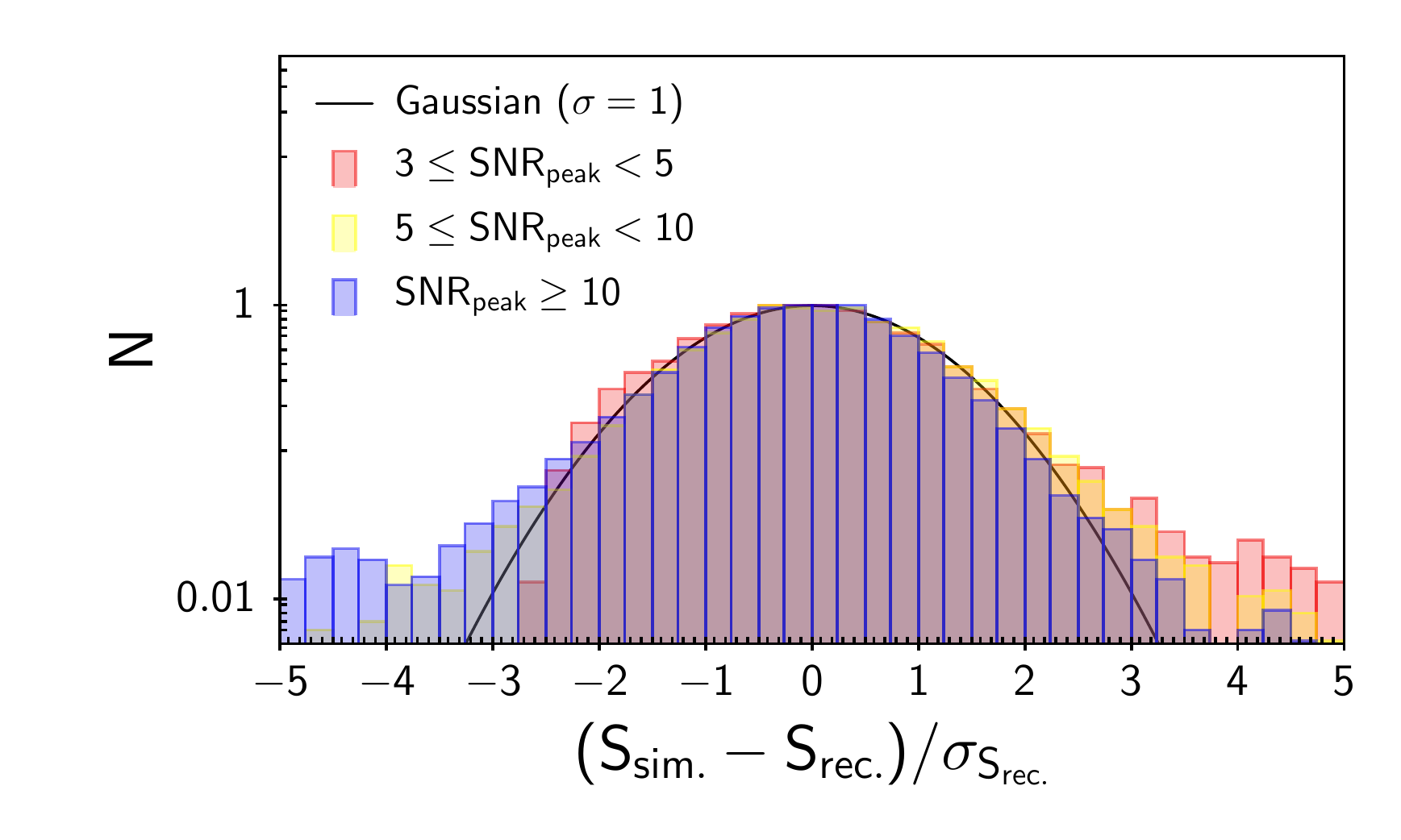}%
}
\put(0.293,0.425){\sffamily PHYS -- PYBDSF}
\put(0.261,0.405){\sffamily only flux bias corrected}
\put(0.293,0.225){\sffamily PHYS -- PYBDSF}
\put(0.225,0.205){\sffamily flux bias and error corrected}
\end{picture}
\begin{picture}(0.49,0.46)
\put(0.02,0.25){%
\includegraphics[width=0.44\textwidth, trim=13mm 24mm 0 5mm, clip]{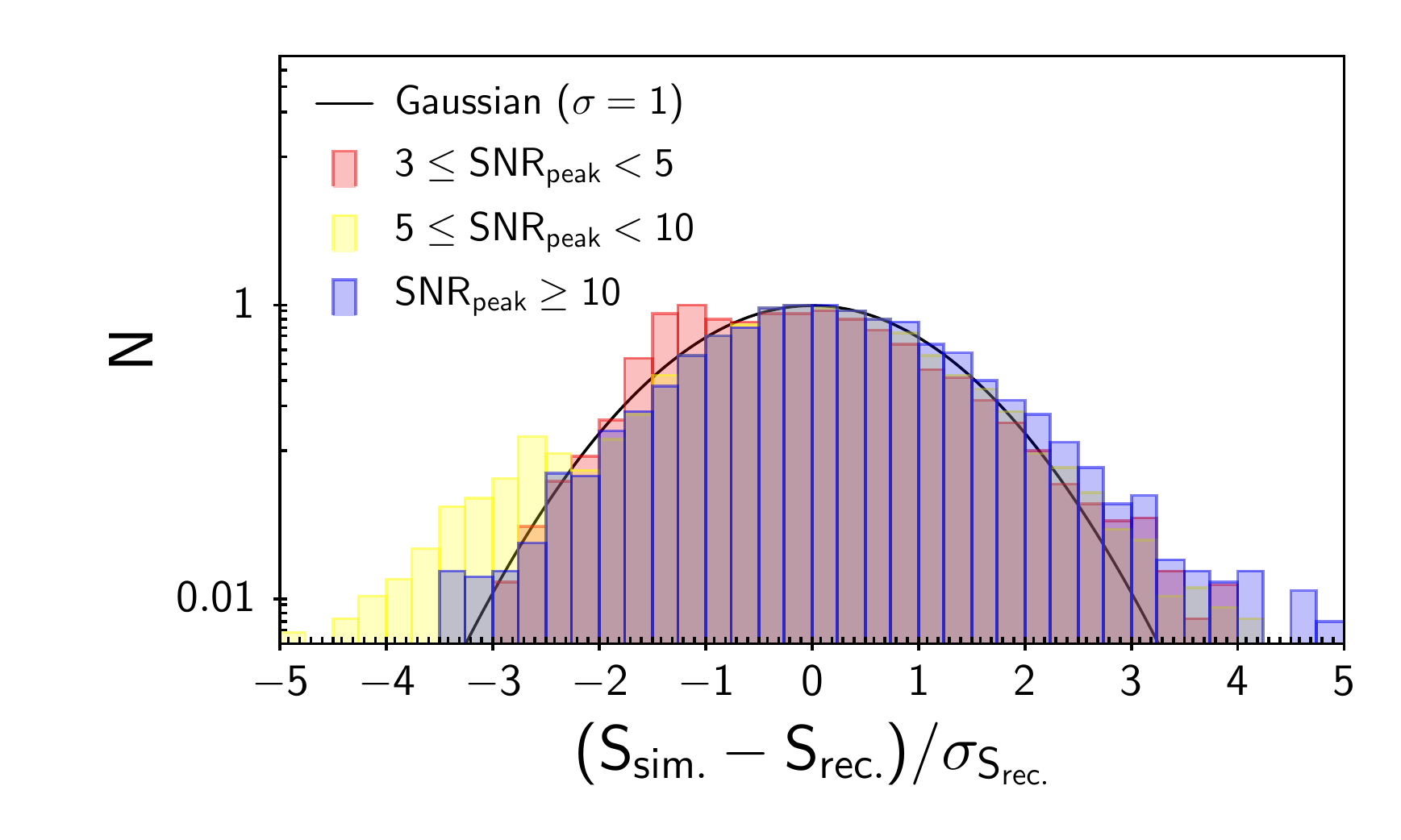}%
}
\put(0.02,0){%
\includegraphics[width=0.44\textwidth, trim=13mm 5mm 0 5mm, clip]{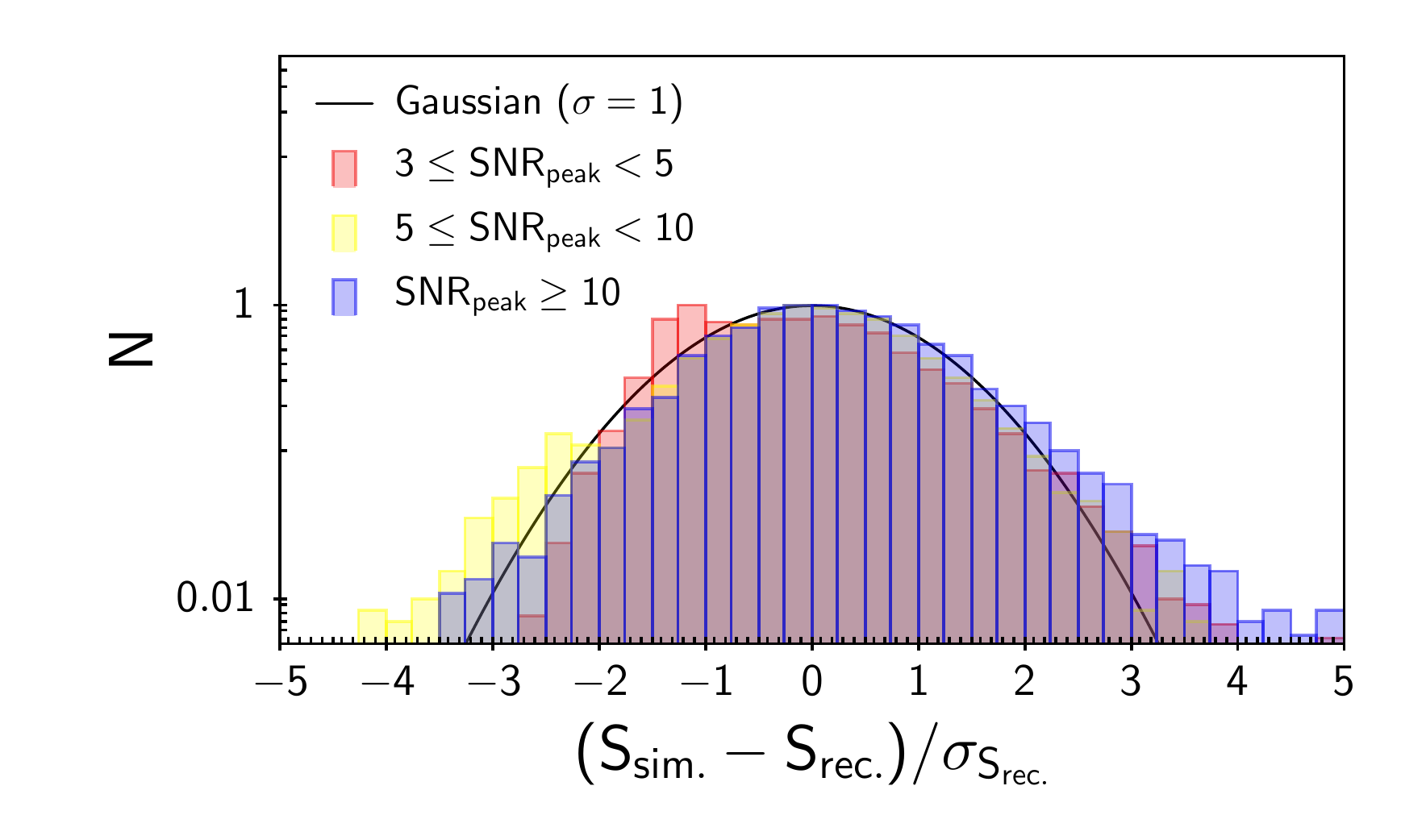}%
}
\put(0.305,0.425){\sffamily PHYS -- GALFIT}
\put(0.260,0.405){\sffamily only flux bias corrected}
\put(0.305,0.225){\sffamily PHYS -- GALFIT}
\put(0.225,0.205){\sffamily flux bias and error corrected}
\end{picture}\\
\vspace{-1.5ex}
\caption{%
\textbf{Left panels} are based on the ``PHYS'' simulation with \pybdsm{} photometry, and \textbf{right panels} are based on the ``PHYS'' simulation with \galfit{} photometry. In each left/right panel, the \textit{upper panel} is the histogram of $(\Ssim-\Scorr)/\sigma_{S_{\mathrm{rec.}},\,\mathrm{Condon1997}}$, where $\Scorr$ is the measured total flux after flux bias correction (Sect.~\ref{Section_MC_Sim_Flux_Bias}), and $\sigma_{S_{\mathrm{rec.}},\,\mathrm{Condon1997}}$ is the error in total flux shipped with \pybdsm{} based on \cite{Condon1997}. 
And the \textit{lower panel} is the histogram of $(\Ssim-\Scorr)/\EScorr$, where $\EScorr$ is the simulation-based error in total flux (Sect.~\ref{Section_MC_Sim_Flux_Error}).
An 1D Gaussian with mean~$=0$ and sigma~$=1$ is overlaid in each panel. \label{Plot_MC_sim_PHYS_PYBDSM_corrected}%
}
\end{figure}

We provide details on the statistical behaviour of flux errors here as an extension to the discussion in Appx.~\ref{Section_MC_Sim_Final_Correction}. 
We examined the histograms of $(\Ssim-\Scorr)/\EScorr$ as shown in Fig.~\ref{Plot_MC_sim_PHYS_PYBDSM_corrected} for final corrected \pybdsm{} and \galfit{} fluxes and errors. Such a histogram indicates how well our final flux errors $\EScorr$ can reflect the true scatter between $\Ssim$ and $\Scorr$. 
Ideally, if the flux error well represents the uncertainty in the photometry, the histogram should have the shape of an 1D Gaussian with mean~$=0$ and sigma~$=1$ (which is overlaid in the figure). Comparing these histograms to those before correction (Fig.~\ref{Plot_MC_sim_scatter_FULL}~and~\ref{Plot_MC_sim_scatter_PHYS}), we do find significant improvement in the shape of the histogram. In Fig.~\ref{Plot_MC_sim_PHYS_PYBDSM_corrected}, we show the histogram after each step of correction: 
(1) flux bias correction ({\em top row}) and
(2) both flux bias and error correction ({\em bottom row}). 
The final histograms nicely agree with the mean~$=0$, sigma~$=1$ 1D Gaussian. Although there are some outliers, but they only contribute a few percent in number. The outlier fraction is lower in the \galfit{} photometry compared to the \pybdsm{} photometry, probably because prior fitting uses known positional information and reduces the chance of recovering noise peaks as sources. 
Both of these effects are also related to the features in each photometry method. 
For example, at the high value-end of the histogram, 
outliers have highly underestimated fluxes likely due to the aforementioned resolution bias (large sources are break down and only partial flux are recovered). And at the low-value end, outliers have extra-boosted fluxes mostly because the recovered source sizes are significantly larger than their simulated sizes. 
For these outliers, we can tentatively identify them out by checking their \pybdsm{} multi-component flag (\incode{Flag_multi}), and also comparing their \pybdsm{} and \galfit{} fluxes and sizes. For example, there are 5\% of sources with \incode{Flag_multi=='M'} in our blind photometry catalog. Therefore, if excluding them, for the bulk of sources done with out photometry, the errors are quite well behaved in statistics.

\subsection{Discussion on the completeness of (sub-)mm/radio photometry}
\label{Section_Appendix_MC_Sim_Completeness} 

\begin{figure}[htb]
\includegraphics[width=0.48\textwidth]{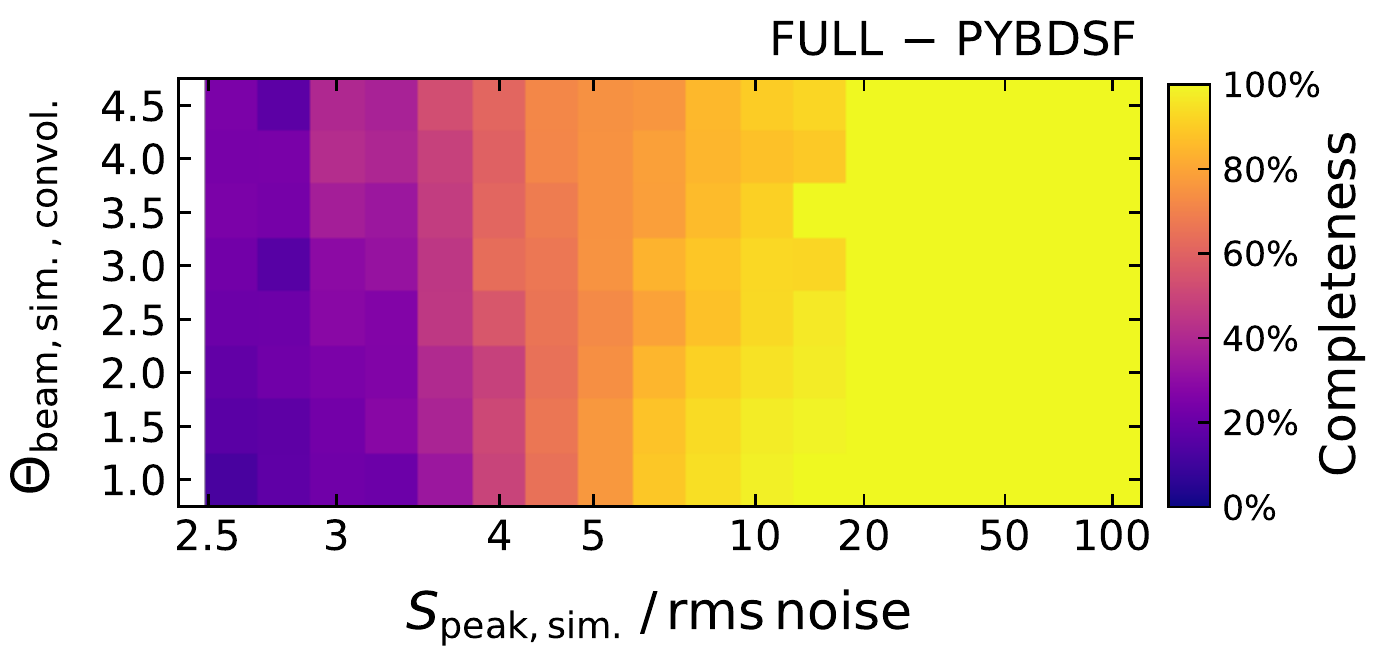}
\includegraphics[width=0.48\textwidth]{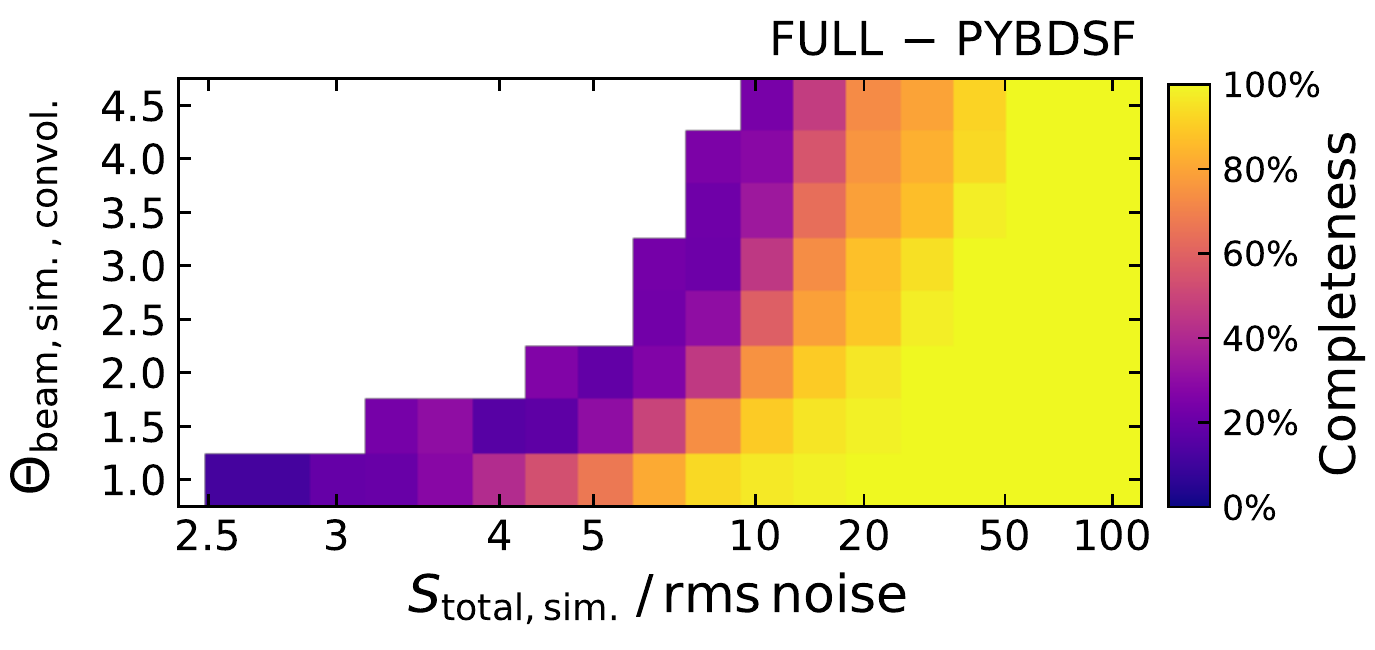}
\caption{%
Completeness of the \pybdsm{} source extraction as a 2D function of the simulated $\Speak/\noise$ (i.e., $\SNRpeak$; left panel) or $\Stotal/\noise$ (right panel) and $\Sbeam$, based on the ``FULL'' simulation (Sect.~\ref{Section_MC_Sim_FULL}). Color indicates the completeness fraction. 
\label{Plot_MC_sim_completeness_2D}}
\end{figure}

We present the 2D diagnostic diagram of completeness in Fig.~\ref{Plot_MC_sim_completeness_2D}. 
Such a diagram is also used in similar works (e.g., \citealt{JimenezAndrade2019}; \citealt{Franco2018}). 
The left panel shows the dependency on $\Speak$ and the right panel on $\Stotal$. Because $\Stotal \propto \Speak \times \ConvSbeam^2$, the completeness has a more complicated dependency on $\Stotal/\noise$ than $\SNRpeak$. Thus using $\SNRpeak$ to select the sample results in a more uniform completeness for various source sizes.

We compare our findings with other ALMA photometry work with completeness assessments in the literature: \cite{Hatsukade2011,Hatsukade2016},  \cite{Karim2013}, \cite{Ono2014}, \cite{Simpson2015b}, \cite{Aravena2016}, \cite{Umehata2017}, \cite{Dunlop2017} and \cite{Franco2018}. 
Our differential completeness at low $\SNRpeak$ is either lower than or consistent with the values in the literature. For example, \cite{Karim2013} estimated $\sim$70\% completeness at $\SNRpeak\sim3$ for their Gaussian-fitting photometry down to $2.5\,\sigma$ while we derive a value of $\sim$20\%. \cite{Aravena2016} estimated $\sim$50\% completeness at $\SNRpeak\sim3$ for their \textsc{SExtractor} \citep{Bertin1996} source extraction down to $3.5\,\sigma$. \cite{Hatsukade2016} found $\sim$25\% completeness at $\SNRpeak\sim3$ for their \textsc{AEGEAN} \citep{Hancock2012} Gaussian-fitting photometry down to $4\,\sigma$. And \cite{Franco2018} report $<$20\% completeness at the same $\SNRpeak$ for their \textsc{Blobcat} Gaussian-fitting photometry, which is consistent with ours. 
Note that all these studies mentioned are based on ALMA maps coming from a single program with similar beam size and rms noise. However, in our work these parameters vary significantly across the archival programs. Based on archival data, \cite{Ono2014} found a completeness of $\sim$85\% at $\SNRpeak\sim3$ for their \textsc{SExtractor} extraction down to as low as $1.8\,\sigma$. 

Our comparison above shows that both detection criterion and photometry method are important when deriving completeness fractions. A lower detection criterion leads to a higher completeness. However, we have also to consider the spurious detection fraction  (Fig.~\ref{Plot_spurious_fraction}) which rapidly increases by lowering the detection criterion. Our choice of \pybdsm{} parameters as described in Sect.~\ref{Section_Blind_Source_Extraction} is thus a compromise between the completeness and spurious fractions.

\vspace{1truecm}
\FloatBarrier

\section{An example of our automated counterpart association examination}
\label{Section_Counterpart_association_examples}

\begin{figure}[htb]
\centering
\includegraphics[width=0.49\textwidth, trim=31mm 20mm 10mm 0, clip]{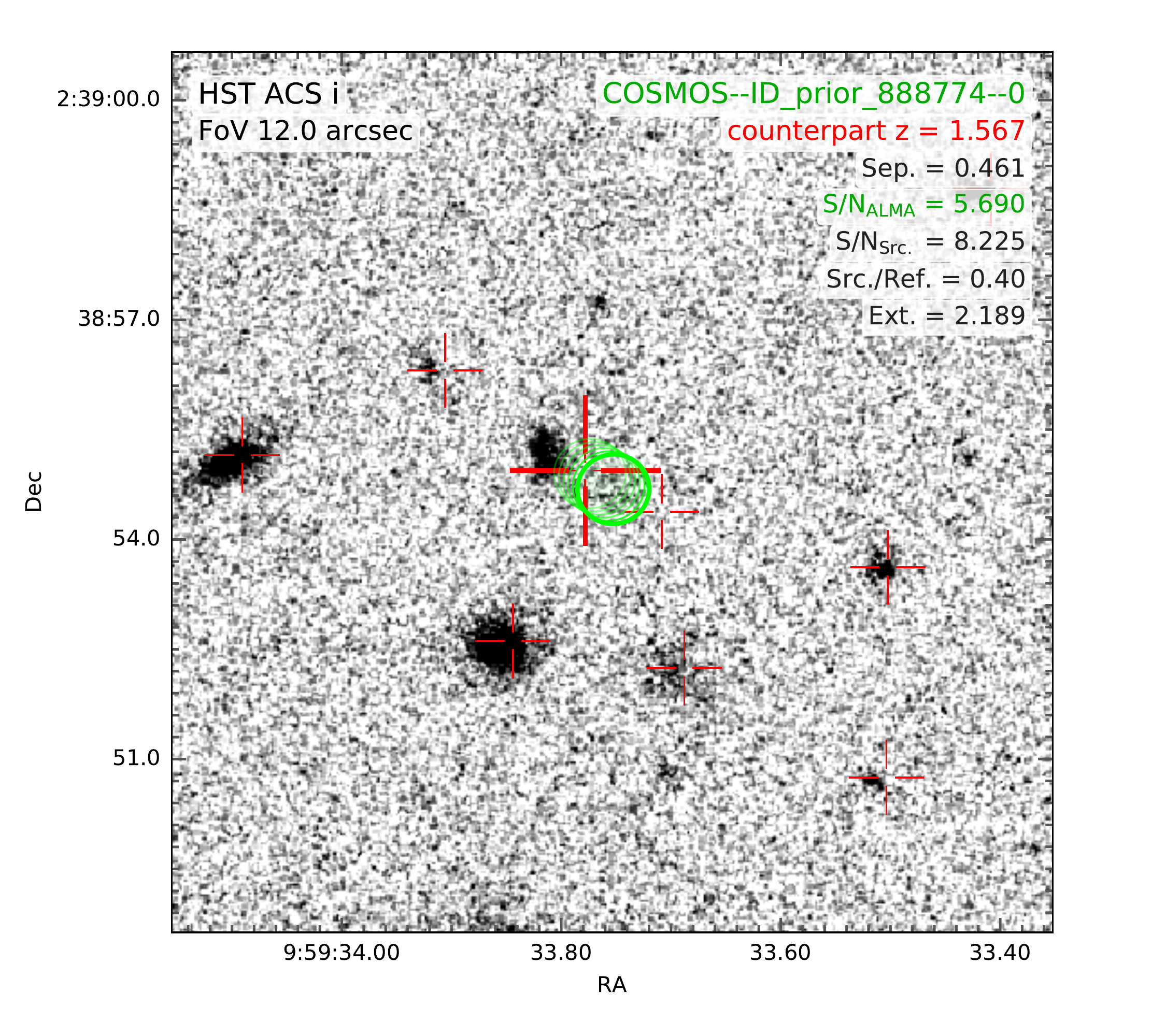}
\includegraphics[width=0.49\textwidth, trim=31mm 20mm 10mm 0, clip]{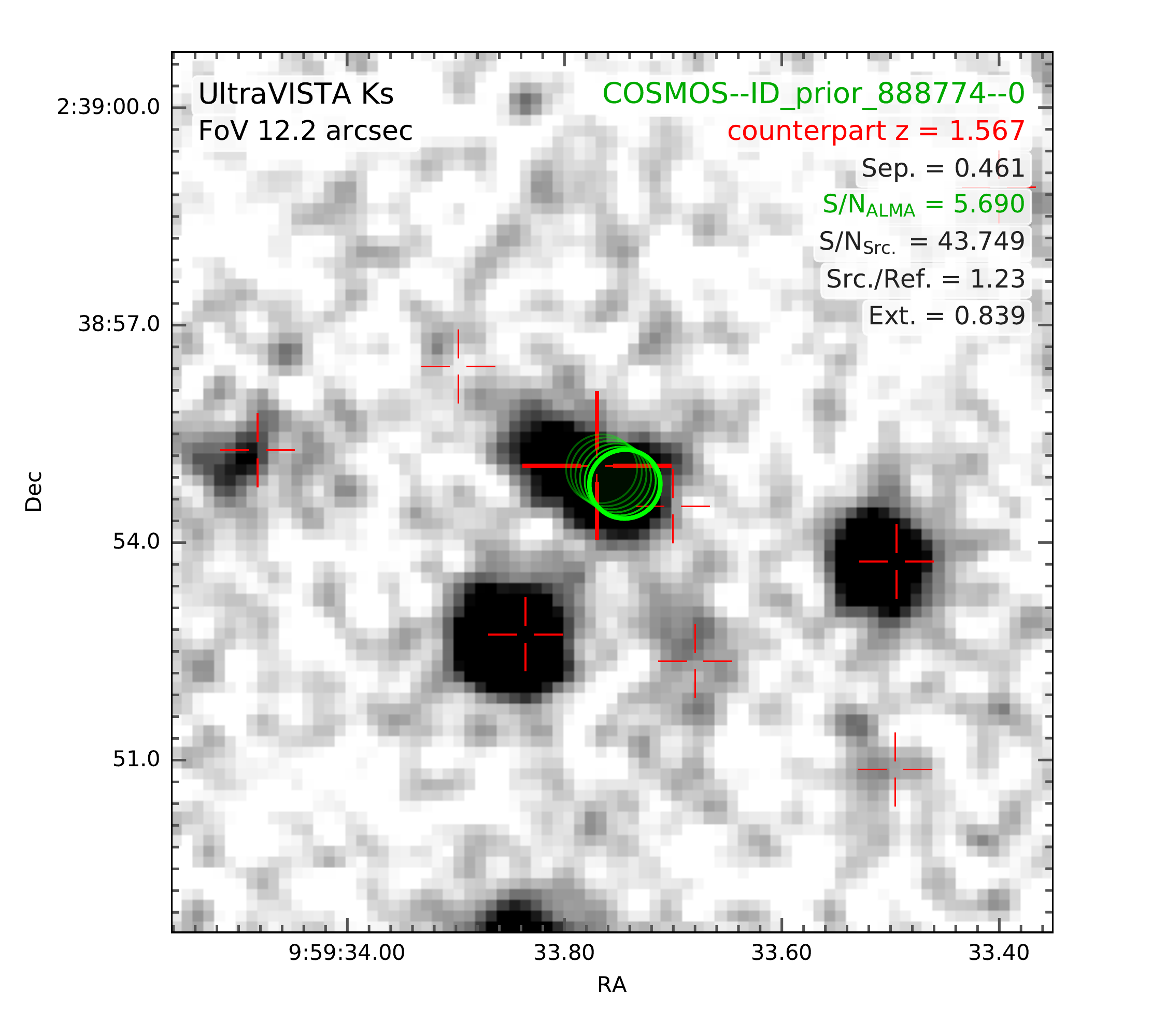}
\caption{%
    Example of our automated counterpart association (Sect.~\ref{Section_Examining_counterpart_association}) for the ALMA source with COSMOS2015 ID~888774 (\citealt{Laigle2016}). It is selected as an example because of its relatively large offset ($\mathrm{Sep.}=0.46$; see definition in Sect.~\ref{Section_Examining_counterpart_association}) between the ALMA and optical positions (corrected for astrometry). 
    The background image is the HST ACS $i$-band in the {\bf left} panel and the UltraVISTA $K_s$-band in the {\bf right} panel. Other symbols are the same in two panels: the thick green ellipse shows the position and size of the ALMA source, and the thin green ellipses show the aperture used for measuring the $\mathbf{Ext.}$ parameter as described in Sect.~\ref{Section_Examining_counterpart_association}. The largest red cross represents the counterpart position and smaller red crosses are other master catalog sources within this image. 
    This prior source is from the \cite{Laigle2016} catalog, therefore its prior position is based on an optical-to-near-IR combined SNR map and centered at the $K_s$ emission. There is a faint blue source with ID~1732293 in the \cite{Capak2007} $i$-band catalog (but not in the \cite{Laigle2016} catalog) at the northeast of the prior position with a small offset of $0.4''$. Current information (including astrometry) is not sufficient to distinguish whether this faint blue source is another galaxy or physically associated to the $K_s$ source. The latter situation (where UV stellar light offsets from dust emission) has been observed in many high-redshift dusty galaxies (e.g., \citealt{Hodge2016,Hodge2019}; \citealt{Rujopakarn2019}; \citealt{Lang2019}). 
    Due to the fact that the $i$-band source is quite faint, the $K_s$ galaxy's SED is mostly unaffected by the $i$-band and shorter wavelength photometry, and the ALMA photometry can be reasonably fitted for the $K_s$ galaxy at its redshift. 
    \label{Plot_counterpart_association_example_1}}
\end{figure}

In Fig.~\ref{Plot_counterpart_association_example_1} we show an example for our automated examination of the counterpart association between each ALMA source and its counterpart source in prior catalogs. For each counterpart image (HST ACS $i$-band, UltraVISTA $K_s$-band, SPLASH IRAC ch1, and VLA 3\,GHz), we measure several parameters as described in Sect.~\ref{Section_Examining_counterpart_association} and then link them to the likelihood of the counterpart association based on our visual inspection. For example, in Fig.~\ref{Plot_counterpart_association_example_1}, the bold green circle indicates the fitted ALMA source position and size (convolved with the ALMA beam), and the red crosses are sources in the prior catalog (Sect.~\ref{Section_Prior_Source_Catalogs}). The bold red cross 
indicates the counterpart of the ALMA source (or just the prior source used in prior fitting photometry).

\vspace{1truecm}
\FloatBarrier

\section{Problematic SED fitting cases}
\label{Section_Appendix_outliers_of_SED_fitting_AGNs}

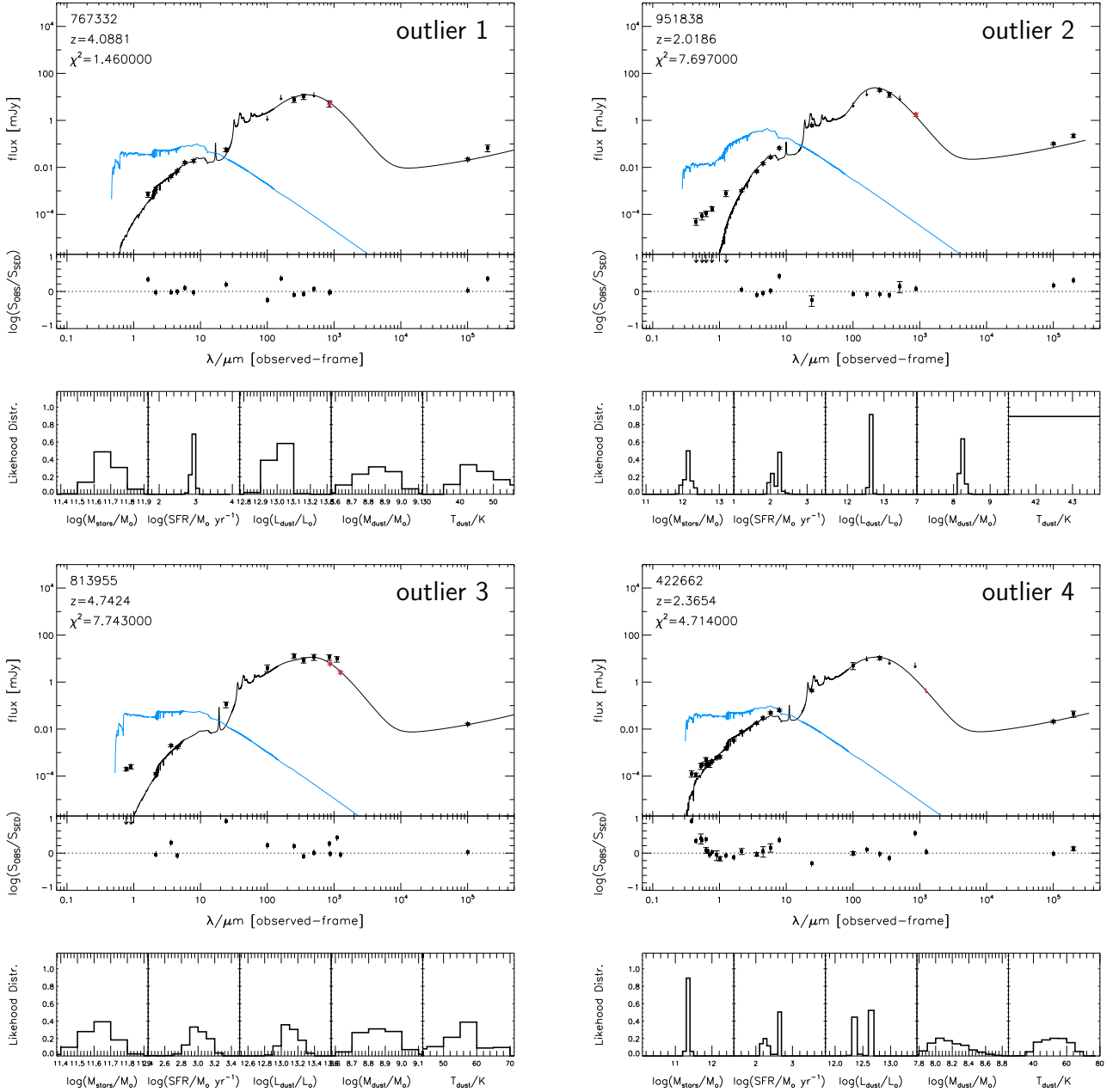
\begin{figure}[htb]
\centering%
\setlength{\unitlength}{\textwidth}
\begin{picture}(0.49,0.48)
\put(0,0){%
\includegraphics[width=0.49\textwidth]{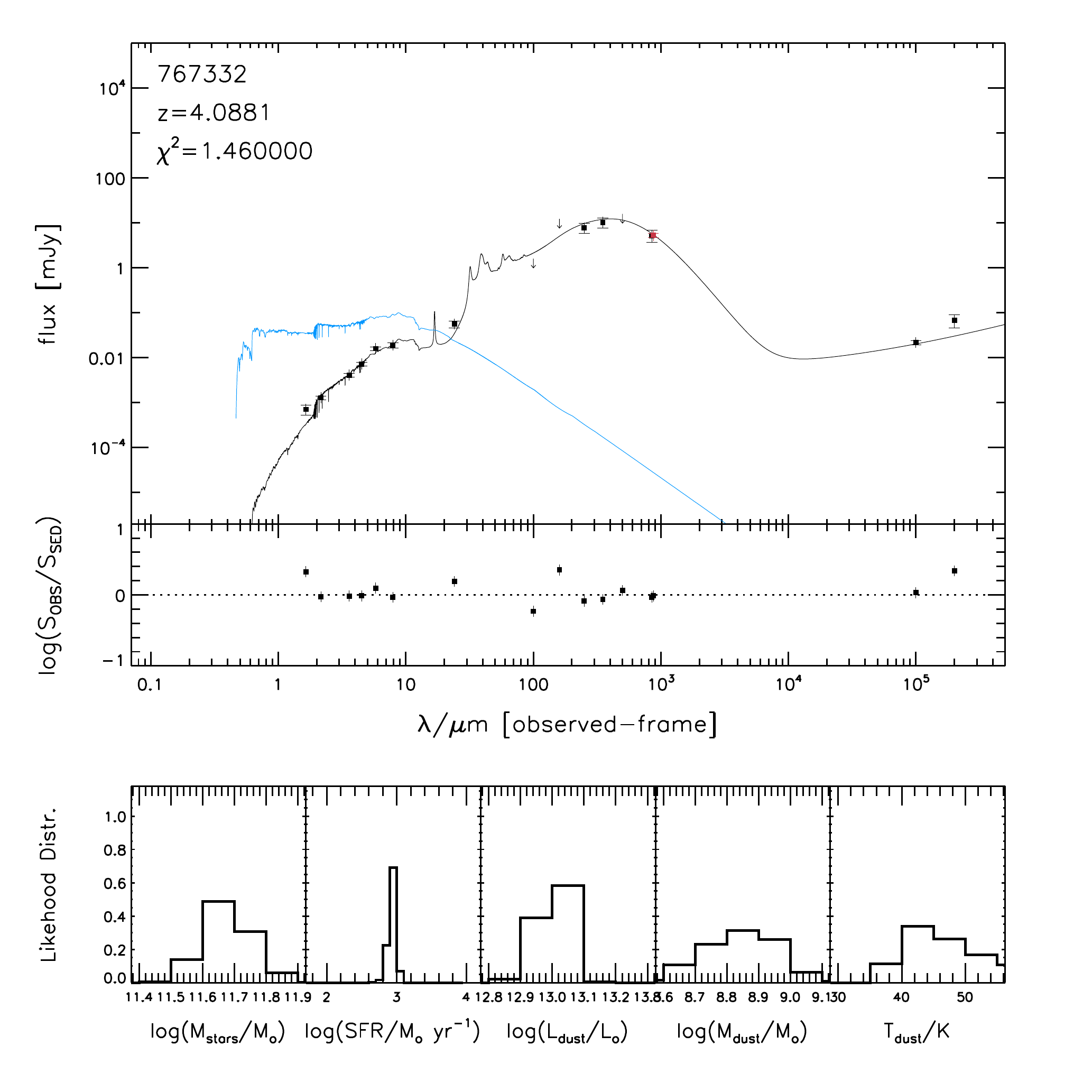}
}
\put(0.350,0.440){\large\sffamily outlier 1}
\end{picture}
\begin{picture}(0.49,0.48)
\put(0,0){%
\includegraphics[width=0.49\textwidth]{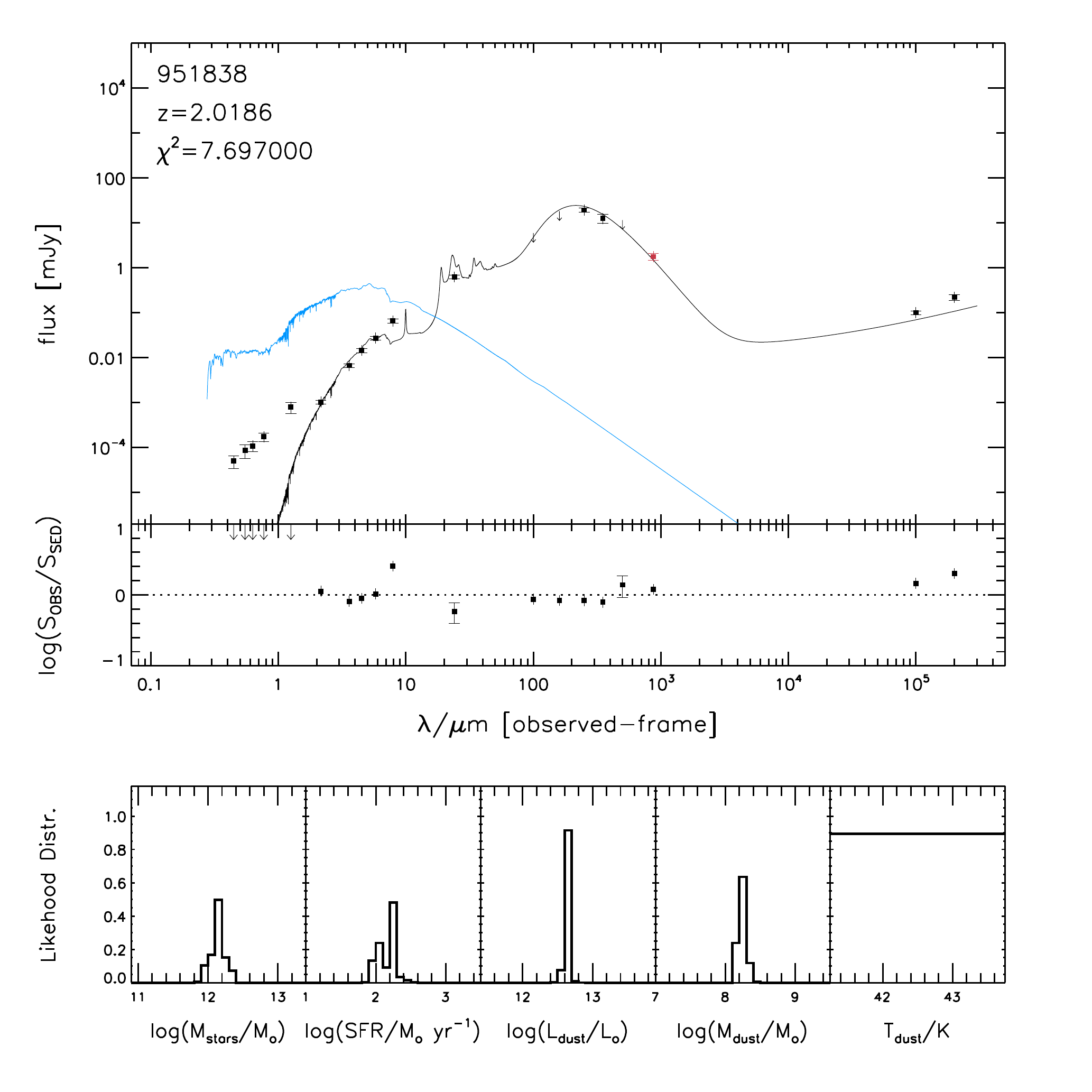}
}
\put(0.350,0.440){\large\sffamily outlier 2}
\end{picture}
\begin{picture}(0.49,0.48)
\put(0,0){%
\includegraphics[width=0.49\textwidth]{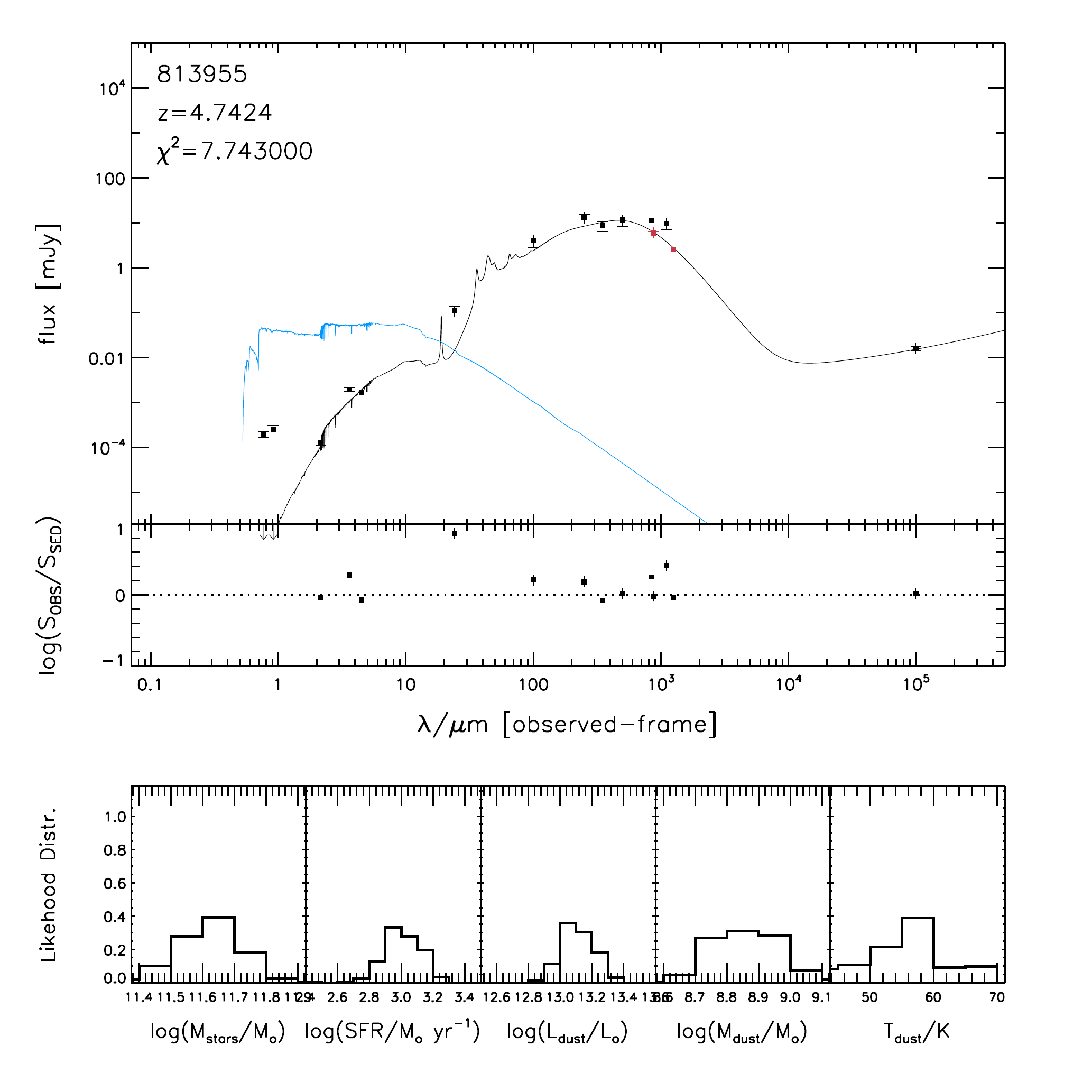}
}
\put(0.350,0.440){\large\sffamily outlier 3}
\end{picture}
\begin{picture}(0.49,0.48)
\put(0,0){%
\includegraphics[width=0.49\textwidth]{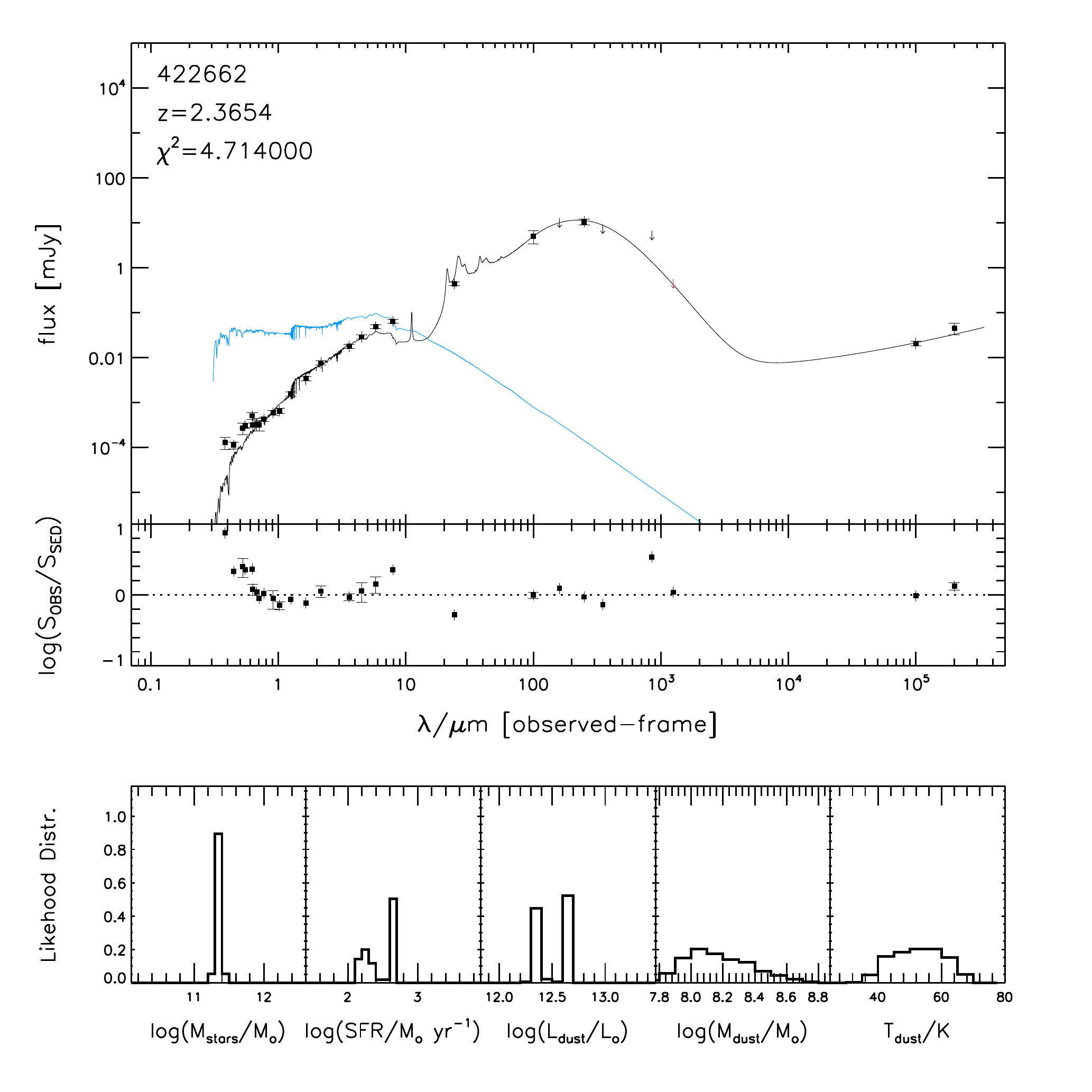}
}
\put(0.350,0.440){\large\sffamily outlier 4}
\end{picture}
\vspace{-6mm}
\caption{%
Example of SEDs for sources \incode{1}, \incode{2}, \incode{3} and \incode{4} listed in Fig.~\ref{Plot_comparison_with_Delvecchio2017_for_Mstar} which exhibit strong mid-IR AGN emission and disagreed in the stellar masses between this work and \cite{Delvecchio2017}. 
See discussion of each source in Sect.~\ref{Section_Obtaining_Galaxy_Properties}. 
For each source, the best-fit full SED (black line) and unattenuated stellar SED (cyan line) as well as the photometric data points are shown in the upper, large  sub-panel. The ALMA data point from this work is highlighted in red. Optical to near-IR $K$ band data are from the ``COSMOS2015'' catalog \citep{Laigle2016} with 3$''$ aperture photometry; near-IR IRAC to mm and radio data are from the ``super-deblended'' FIR/mm catalog \citep{Jin2018}. The SED fitting residuals at each band ($\log_{10} S_{\mathrm{OBS}}/S_{\mathrm{SED}}$) are shown in each middle sub-panel. The probability distribution histograms of each fitted physical parameter ($\log_{10}\Mstar$, $\log_{10} \mathrm{SFR}$, IR luminosity $\log_{10} L_{\mathrm{dust}}$, dust mass $\log_{10} M_{\mathrm{dust}}$ and dust temperature $T_{\mathrm{dust}}$) are shown in the sub-panels at the bottom. 
\label{Plot_SED_Magphys_813955}
}
\end{figure}

In Fig.~\ref{Plot_SED_Magphys_813955} we show the four problematic SED fitting cases which are labeled in Fig.~\ref{Plot_comparison_with_Delvecchio2017_for_Mstar} and discussed in Sect.~\ref{Section_Obtaining_Galaxy_Properties}. Their stellar masses derived from our optimized iterative \textsc{MAGPHYS} fitting (Sect.~\ref{Section_Running_SED_Fitting}) without an AGN component are higher by a factor of 10--100 compared to the masses reported by \cite{Delvecchio2017}, who used \textsc{SED3FIT} to account for an AGN SED component (and both redshifts are consistent). However, we have the following reasons to believe that these are just rare cases that do not indicate an obvious bias in our SED fitting affected by mid-IR AGN contamination. First, these sources are very rare, with only 4 out of a total of 396 sources cross-matched with the \cite{Delvecchio2017} catalog. All other AGNs have no such large difference between this work and \cite{Delvecchio2017}. Second, the advantage of this work is that we have the ALMA and far-IR/(sub-)mm constraint, and \textsc{MAGPHYS} has the energy balance assumption that links optical dust attenuation to dust luminosity in IR; therefore, the degeneracy between age and attenuation can be reduced (if the energy balance is valid, which could be true for our ALMA-selected dusty sample). Thirdly, the optical photometry itself has uncertainties from either noise or galaxy--galaxy blending. For example, for outlier \incode{3}, its HST $i$-band and Subaru Suprime-Cam $z'$-band data could not be fit well, but its IR SED looks reasonable and counterpart association shows no problem. It has no obvious blended optical source within 3$''$, but we could not rule out the chance of line-of-sight blending of two sources, i.e., like the case of the galaxy CRLE reported in \cite{Pavesi2018}. 

To verify the degeneracy between age and attenuation, we ran some additional SED fitting with our multi-component $\chi^2$ fitting code under development\,\footnote{It is still in the experimental phase but is available at \url{http://github.com/1054/Crab.Toolkit.michi2}.}, similar to \cite{Liudz2017}. We fit two components for test purposes: one uses \cite{BC03} stellar SED models at solar metallicity, with constant SFH, with various ages from 0.1 to 1\,Gyr and with a varied dust attenuation ($E(B-V)=0$ to 1.2) following the \cite{Calzetti2000} attenuation law; and the other uses the \cite{Mullaney2011} AGN SED models. We fit only the optical to mid-IR part ($\lambda<20\,\mathrm{\mu}m$) of the SED. We loop each combination of the two component models and find a best $\chi^2$ fit, and then we merge all combinations together to analyze the $\chi^2$ distributions for $\Mstar$ and age (to obtain a median value and error for each parameter, similar to \citealt{Liudz2017}). We find that such a fitting without constraints from far-IR dust emission leads to very large uncertainties in age and stellar mass. For example, for outlier \incode{2} (ID\,951838), the best-fit $\log_{10}(\Mstar/\Msun)$ dramatically varies from $9.5\pm0.5$ to $11.4\pm0.6$ with an age that varies from 200\,Myr to 1\,Gyr (with best-fit $E(B-V)=0.5$ and 0.7, respectively). For comparison, the \cite{Delvecchio2017} $\log_{10}\Mstar$ of 10.24 and our \textsc{MAGPHYS} value of 12.15 agreed within the errors.
For our outlier \incode{3} (ID\,813955), a fixed age of 200\,Myr fitting gives $E(B-V)=0.5$ and $\log_{10}\Mstar=11.0\pm0.1$, while a free-age fitting gives $E(B-V)=1.2$ and $\log_{10}\Mstar=12.6\pm0.3$ with an age of 450\,Myr. In comparison, \cite{Delvecchio2017} report a $\log_{10}\Mstar$ of 10.34, and our \textsc{MAGPHYS} value is 11.63, also in agreement within the range of uncertainties. Outlier \incode{4} (ID\,842140) presents a similar situation. While for outlier \incode{1} (ID\,422662), our experimental stellar+AGN SED fitting ($\log_{10}\Mstar\sim11.2-12.6\pm0.6$) cannot recover the low $\log_{10}\Mstar=9.45$ reported in the \cite{Delvecchio2017} catalog (their \incode{ID_VLA3}\,4136). Therefore, as mentioned in the text of Sect.~\ref{Section_Obtaining_Galaxy_Properties}, we think that its stellar mass is more reliable from the \textsc{MAGPHYS} fitting because of the inclusion of ALMA data here. Finally, we conclude that these sources are just low-probability outliers suffering from the large uncertainty in stellar mass estimation and possibly also optical line-of-sight blending. The latter needs further follow-up observations, which is beyond the scope of this paper.

\clearpage

\DeclareRobustCommand{\disambiguate}[3]{#1}

\nocite{}

\bibliography{Biblio}

\end{document}

%% file: Input_usepackage.tex
\usepackage{graphicx}
\usepackage{float}
\usepackage{amsfonts,amsmath,amssymb}
\usepackage{natbib}
\usepackage{url}
\usepackage{xcolor}
\usepackage{hyperref}
\usepackage[numbered]{bookmark} 
\usepackage{calculator}
\usepackage{placeins} 
\let\splitbox\relax
\usepackage{adjustbox} 
\usepackage{enumitem} 
\usepackage{xifthen}
\usepackage{enumitem} 
\usepackage{chngcntr}

\hypersetup{
    colorlinks,
    linkcolor={magenta!80!white},
    citecolor={blue!80!black},
    urlcolor={magenta!80!black}
}

\usepackage{listings} 
\usepackage{color}
\definecolor{lightgray}{gray}{0.9}

\usepackage{lmodern}
\usepackage{slantsc}

\usepackage{tabularx} 

\usepackage{xstring} 

\usepackage[nomessages]{fp}

%% file: Input_newcommand.tex
\setcitestyle{notesep={; }}

\DeclareRobustCommand{\disambiguate}[3]{#2~#3} 

\renewcommand{\figurename}{Fig.}

\newcommand{\TODO}[2][]{%
\ifthenelse{\isempty{#1}}%
{\textbf{\textcolor{red}{TODO: #2}}}%
{\textbf{\textcolor{red}{TODO (#1) #2}}}%
}

\usepackage{twoopt}
\usepackage[normalem]{ulem}

\def\FinalRevision{Yes}
\newcommandtwoopt{\REVISED}[3][][]{%
	\ifthenelse{\isempty{#3}}%
	{%
		\ifthenelse{\isempty{#2}}%
		{
			\ifthenelse{\equal{\FinalRevision}{Yes}}{%
			}{%
		        \textbf{\textcolor{blue}{#1}}%
			}%
		}%
		{
			\ifthenelse{\equal{\FinalRevision}{Yes}}{%
			}{%
			    \textcolor{gray}{\sout{#2}}%
			}%
		}%
	}%
	{%
		\ifthenelse{\isempty{#2}}%
		{
			\ifthenelse{\equal{\FinalRevision}{Yes}}{%
			    {#3}%
			}{%
			\textbf{\textcolor{blue}{#3}}%
			}%
		}%
		{
			\ifthenelse{\equal{\FinalRevision}{Yes}}{%
			    {#3}%
			}{%
			    \textcolor{gray}{\sout{#2}}%
			    \textbf{\textcolor{blue}{#3}}%
			}%
		}%
	}%
}

\def\FinalFinalRevision{Yes}
\newcommandtwoopt{\REREVISED}[3][][]{%
	\ifthenelse{\isempty{#3}}%
	{%
		\ifthenelse{\isempty{#2}}%
		{
			\ifthenelse{\equal{\FinalFinalRevision}{Yes}}{%
			}{%
			    \textbf{\textcolor{blue}{#1}}%
			}%
		}%
		{
			\ifthenelse{\equal{\FinalFinalRevision}{Yes}}{%
			}{%
			    \textcolor{gray}{\sout{#2}}%
			}%
		}%
	}%
	{%
		\ifthenelse{\isempty{#2}}%
		{
			\ifthenelse{\equal{\FinalFinalRevision}{Yes}}{%
			    {#3}%
			}{%
			    \textbf{\textcolor{blue}{#3}}%
			}%
		}%
		{
			\ifthenelse{\equal{\FinalFinalRevision}{Yes}}{%
			    {#3}%
			}{%
			    \textcolor{gray}{\sout{#2}}%
			    \textbf{\textcolor{blue}{#3}}%
			}%
		}%
	}%
}

\newcommand{\DONE}[2][]{%
\ifthenelse{\isempty{#1}}%
{%
\textit{\textcolor{green!55!black}{DONE: #2}}}%
{%
\IfSubStr{#1}{ES}{%
\textit{\textcolor{magenta!95!black}{DONE (#1) #2}}%
}{%
\IfSubStr{#1}{SKL}{%
\textit{\textcolor{orange!95!black}{DONE (#1) #2}}%
}{%
\IfSubStr{#1}{PL}{%
\textit{\textcolor{cyan!95!black}{DONE (#1) #2}}%
}{%
\textit{\textcolor{green!55!black}{DONE (#1) #2}}%
}%
}%
}%
}%
}

\def\gtsima{$\; \buildrel > \over \sim \;$}
\def\ltsima{$\; \buildrel < \over \sim \;$}
\def\prosima{$\; \buildrel \propto \over \sim \;$}
\def\gsim{\lower.5ex\hbox{\gtsima}}
\def\lsim{\lower.5ex\hbox{\ltsima}}
\def\simgt{\lower.5ex\hbox{\gtsima}}
\def\simlt{\lower.5ex\hbox{\ltsima}}
\def\simpr{\lower.5ex\hbox{\prosima}}

\def\h1{$h^{-1}$}
\def\beq{\begin{equation}}
\def\eeq{\end{equation}}
\def\8mu{8\,$\mu{\rm m}$}
\def\16mu{16\,$\mu{\rm m}$}
\def\24mu{24\,$\mu{\rm m}$}
\def\70mu{70\,$\mu{\rm m}$}

\def\SNR{\mathrm{S/N}}

\def\deltaGDR{\delta_{\mathrm{GDR}}}

\def\DeltaMS{\Delta{\mathrm{MS}}}

\def\nH2{n_{\mathrm{H}_2}}
\def\H2{\mathrm{H}_2}
\def\HI{\mathrm{H}\textnormal{\textup{\uppercase\expandafter{\romannumeral 1}}}}

\def\Msun{\mathrm{M}_{\odot}}
\def\Msyr{\mathrm{M}_{\odot}\,\mathrm{yr^{-1}}}
\def\Kkmspc2{\mathrm{K} \mathrm{km} \mathrm{s}^{-1} \mathrm{pc}^{2}}
\def\SFR{\mathrm{SFR}}
\def\Mstar{M_\mathrm{{*}}}

\def\R52{R_{52}}

\def\galfit{\textsc{galfit}}
\def\getpix{\textsc{getpix}}
\def\pybdsm{\textsc{PyBDSF}} 
\def\Speak{S_{\mathrm{peak}}}
\def\Stotal{S_{\mathrm{total}}}
\def\SNRpeak{\mathrm{S/N}_{\mathrm{peak}}}
\def\SNRtotal{\mathrm{S/N}_{\mathrm{total}}}
\def\noise{{\mathrm{rms\;noise}}}

\def\EStotal{\sigma_{S_{\mathrm{total}}}}
\def\Sbeam{\Theta_{\mathrm{beam}}}
\def\ConvSbeam{\Theta_{\mathrm{beam,\,convol.}}}
\def\ConvSimSbeam{\Theta_{\mathrm{beam,\,sim.\,convol.}}}
\def\Ssim{S_{\mathrm{sim.}}}
\def\Srec{S_{\mathrm{rec.}}}
\def\ESrec{\sigma_{S_{\mathrm{rec.}}}}
\def\Scorr{S_{\mathrm{rec.}}^{\mathrm{corr.}}}
\def\EScorr{\sigma_{S_{\mathrm{rec.}}^{\mathrm{corr.}}}}

\def\CII_{[C\,{\sc ii}]}
\def\NII_{[N\,{\sc ii}]}
\def\CI_{[C\,{\sc i}]}
\def\OIII_{[O\,{\sc iii}]}
\def\aaacosmosAlmaDataDate{Jan. 2nd, 2018}
\def\aaacosmosNextUpdateAlmaDataDate{Aug. 1st, 2018}
\def\aaacosmosAlmaPointingNumber{1534}
\def\aaacosmosAlmaPointingArea{164}
\def\aaacosmosNextUpdateAlmaPointingArea{280}
\def\aaacosmosAlmaDetectionNumber{1146}
\def\aaacosmosPriorDetectionNumber{1027}
\def\aaacosmosBlindDetectionNumber{939}
\FPeval{result}{clip(\aaacosmosAlmaDetectionNumber{}-\aaacosmosPriorDetectionNumber{})}
\edef\aaacosmosBlindOnlyNumber{\result} 
\FPeval{result}{clip(\aaacosmosAlmaDetectionNumber{}-\aaacosmosBlindDetectionNumber{})}
\edef\aaacosmosPriorOnlyNumber{\result} 
\FPeval{result}{clip(\aaacosmosAlmaDetectionNumber{}-\aaacosmosBlindOnlyNumber{}-\aaacosmosPriorOnlyNumber{})}
\edef\aaacosmosPriorAndBlindDetectionNumber{\result} 
\FPeval{result}{clip(\aaacosmosAlmaDetectionNumber{}-\aaacosmosPriorAndBlindDetectionNumber{})}
\edef\aaacosmosPriorXorBlindDetectionNumber{\result} 
\def\aaacosmosGalaxyNumber{823}
\FPeval{result}{clip(\aaacosmosPriorDetectionNumber{}-\aaacosmosGalaxyNumber{})}
\edef\aaacosmosAlmaDetectionNonUniqueNumber{\result} 
\FPeval{result}{round((\aaacosmosAlmaDetectionNonUniqueNumber{}/\aaacosmosGalaxyNumber{}*100):0)}

\FPeval{result}{clip(\aaacosmosPriorDetectionNumber{}-\aaacosmosGalaxyNumber{})}
\edef\aaacosmosGalaxyWithMultiplePhotometryNumber{\result} 
\FPeval{result}{round((\aaacosmosGalaxyWithMultiplePhotometryNumber{}/\aaacosmosGalaxyNumber{}*100):0)}
\edef\aaacosmosGalaxyWithMultiplePhotometryFraction{\result{}\%}
\def\aaacosmosNumberOfOutlierICF{4} 
\FPeval{result}{round((\aaacosmosNumberOfOutlierICF{}/\aaacosmosGalaxyNumber{}*100):1)}
\edef\aaacosmosFractionOfOutlierICF{\result{}\%}
\def\aaacosmosNumberOfOutlierCPA{36} 
\FPeval{result}{round((\aaacosmosNumberOfOutlierCPA/\aaacosmosGalaxyNumber{}*100):0)}
\edef\aaacosmosFractionOfOutlierCPA{\result{}\%}
\def\aaacosmosNumberOfOutlierSED{21} 
\FPeval{result}{round((\aaacosmosNumberOfOutlierSED/\aaacosmosGalaxyNumber{}*100):0)}
\edef\aaacosmosFractionOfOutlierSED{\result{}\%}
\def\aaacosmosNumberOfPriorGalaxiesHavingNoPriorz{43} 
\def\aaacosmosNumberOfPriorGalaxiesHavingOnlyJinPriorz{53} 
\FPeval{result}{clip(\aaacosmosNumberOfOutlierICF{}+\aaacosmosNumberOfOutlierCPA{}+\aaacosmosNumberOfOutlierSED{})}
\edef\aaacosmosNumberOfQADiscardedGalaxies{\result}

\FPeval{result}{clip(\aaacosmosNumberOfPriorGalaxiesHavingNoPriorz{}+\aaacosmosBlindOnlyNumber{}+\aaacosmosNumberOfPriorGalaxiesHavingOnlyJinPriorz{})}
\edef\aaacosmosNumberOfNoPriorOrPriorz{\result}
\FPeval{result}{round((\aaacosmosNumberOfNoPriorOrPriorz{}/\aaacosmosGalaxyNumber{}*100):0)}
\edef\aaacosmosFractionOfNoPriorz{\result{}\%} 
\FPeval{result}{clip(\aaacosmosGalaxyNumber{}-\aaacosmosNumberOfPriorGalaxiesHavingNoPriorz{}-\aaacosmosNumberOfPriorGalaxiesHavingOnlyJinPriorz{}{})}
\edef\aaacosmosNumberOfQAedGalaxies{\result}
\def\aaacosmosNumberOfDuplicatedOutliers{10} 
\FPeval{result}{clip(\aaacosmosNumberOfPriorGalaxiesHavingNoPriorz{}+\aaacosmosNumberOfPriorGalaxiesHavingOnlyJinPriorz{}+\aaacosmosNumberOfOutlierICF{}+\aaacosmosNumberOfOutlierCPA{}+\aaacosmosNumberOfOutlierSED{}-\aaacosmosNumberOfDuplicatedOutliers{})}
\edef\aaacosmosNumberOfAllDiscardedGalaxies{\result} 
\FPeval{result}{clip(\aaacosmosGalaxyNumber{}-\aaacosmosNumberOfAllDiscardedGalaxies{})}
\edef\aaacosmosGoodGalaxyNumber{\result} 
\FPeval{result}{round((\aaacosmosGoodGalaxyNumber{}/\aaacosmosGalaxyNumber{}*100):0)}
\def\CASA{\textsc{CASA}}

\newcommand{\rom}[1]{{{\uppercase\expandafter{\romannumeral #1}}}}
\newcommand{\romup}[1]{{\textup{\uppercase\expandafter{\romannumeral #1}}}}

\usepackage[T1]{fontenc}
\usepackage{textcomp}
\usepackage{listings}
\newcommand{\incode}[1]{\mbox{\raggedright\lstinline|#1|}}
\newcommand{\incodep}[1]{\mbox{\raggedright\lstinline|"#1"|}}

\newcolumntype{R}{>{\raggedleft\arraybackslash}X}

%% file: Input_flow_chart_style.tex
\makeatletter
\newcommand\HUGE{\@setfontsize\HUGE{32}{32}}
\makeatother

\definecolor{link(red)}{rgb}{1.0, 0.55, 0.12}
\definecolor{blue(ncs)}{rgb}{0.0, 0.53, 0.74}

\dimendef\prevdepth=0 

\tikzstyle{Project} = [rectangle, draw, color=white, fill=blue(ncs), text=white, text centered, inner sep=0.75cm, minimum width=10cm, line width=0.5mm, execute at begin node={\bf\HUGE}, rounded corners]

\tikzstyle{S02} = [rectangle, draw, color=white, fill=blue(ncs), text=white, text centered, inner sep=0.75cm, minimum width=11.2cm, line width=0.5mm, execute at begin node={\begin{varwidth}{11.2cm}\centering\bf\HUGE}, execute at end node={\end{varwidth}}, rounded corners]

\tikzstyle{S03} = [rectangle, draw, color=blue(ncs), fill=white, text=black, text centered, inner sep=0.50cm, minimum width=5.2cm, line width=0.5mm, execute at begin node={\begin{varwidth}{5.2cm}\centering\bf\HUGE}, execute at end node={\end{varwidth}}, rounded corners]

\tikzstyle{S03W7p5cm} = [rectangle, draw, color=blue(ncs), fill=white, text=black, text centered, inner sep=0.50cm, minimum width=5.2cm, line width=1.5mm, execute at begin node={\begin{varwidth}{7.5cm}\centering\bf\HUGE}, execute at end node={\end{varwidth}}, rounded corners]

\tikzstyle{S04} = [ellipse, draw, color=blue(ncs), fill=white, text=black, text centered, inner sep=0.25cm, line width=0.5mm, execute at begin node={\begin{varwidth}{4.8cm}\centering\bf\Huge}, execute at end node={\end{varwidth}}, rounded corners]

\tikzstyle{S04W7p5cm} = [ellipse, draw, color=blue(ncs), fill=white, text=black, text centered, inner sep=0.25cm, line width=0.5mm, execute at begin node={\begin{varwidth}{7.5cm}\centering\bf\Huge}, execute at end node={\end{varwidth}}, rounded corners]

\tikzstyle{S05} = [ellipse, draw, color=blue(ncs), fill=white, text=black, text centered, inner sep=0.09cm, line width=0.5mm, execute at begin node={\begin{varwidth}{3.8cm}\centering\bf\Huge}, execute at end node={\end{varwidth}}, rounded corners]

\tikzstyle{S06} = [ellipse, draw, color=blue(ncs), fill=white, text=black, text centered, inner sep=0.09cm, line width=0.5mm, execute at begin node={\begin{varwidth}{2.8cm}\centering\bf\Huge}, execute at end node={\end{varwidth}}, rounded corners]

\tikzstyle{emp} = [fill=white, inner sep=0cm, minimum width=0cm]

\tikzstyle{app} = [diamond, draw, fill=cyan!15, text centered, rounded corners, inner sep=0.09cm, line width=0.5mm, execute at begin node={\begin{varwidth}{1.8cm}\centering\bf\LARGE}, execute at end node={\end{varwidth}}]

\tikzstyle{app2} = [diamond, draw, fill=cyan!15, text centered, rounded corners, inner sep=0.09cm, line width=0.5mm, execute at begin node={\begin{varwidth}{2.3cm}\centering\bf\LARGE}, execute at end node={\end{varwidth}}]

\tikzstyle{app3} = [diamond, draw, fill=cyan!15, text centered, rounded corners, inner sep=0.09cm, minimum width=8.5cm, minimum height=8.5cm, line width=0.5mm, execute at begin node={\begin{varwidth}{6.2cm}\centering\bf\Huge}, execute at end node={\end{varwidth}}]

\tikzstyle{app4} = [diamond, draw, fill=cyan!15, text centered, rounded corners, inner sep=0.09cm, minimum width=7.5cm, minimum height=7.5cm, line width=0.5mm, execute at begin node={\begin{varwidth}{5.2cm}\centering\bf\Huge}, execute at end node={\end{varwidth}}]

\tikzstyle{line} = [draw, -{Latex[scale=3,length=5,width=3,angle'=25,open]}, line width=0.8pt, color=blue(ncs)]

\tikzstyle{line(thick)} = [draw, -{Latex[scale=3,length=5,width=3,angle'=25,open]}, line width=1.4pt, color=blue(ncs)]

\tikzstyle{line(thicker)} = [draw, -{Latex[scale=3,length=5,width=3,angle'=25,open]}, line width=1.8pt, color=blue(ncs)]

\tikzstyle{ldot} = [draw, dotted, -latex', line width=2.0pt, blue]

\tikzstyle{ldas} = [draw, dashed, -latex', line width=2.5pt, blue]

%% file: Input_galaxy_flow_chart_style.tex
\makeatletter
\newcommand\XHUGE{\@setfontsize\Huge{29}{29}}
\makeatother

\definecolor{link(red)}{rgb}{1.0, 0.55, 0.12}
\definecolor{blue(ncs)}{rgb}{0.0, 0.53, 0.74}

\dimendef\prevdepth=0 

\tikzstyle{StyleOfCatalog} = [rectangle, draw, color=black, fill=white, text=black, text centered, inner sep=0.5cm, line width=0.75mm, minimum width=8.0cm, execute at begin node={\begin{varwidth}{9.5cm}\centering\bf\XHUGE}, execute at end node={\end{varwidth}}, rounded corners]

\tikzstyle{StyleOfCatalog(OneLine)} = [rectangle, draw, color=black, fill=white, text=black, text centered, inner sep=0.45cm, line width=0.5mm, execute at begin node={\centering\bf\Huge}, execute at end node={}, rounded corners]

\tikzstyle{StyleOfEmptyNode} = [fill=white, inner sep=0cm, minimum width=0cm]

\tikzstyle{StyleOfProcess} = [diamond, draw, fill=cyan!15, text centered, rounded corners, inner sep=0.09cm, line width=0.5mm, execute at begin node={\begin{varwidth}{6.5cm}\centering\bf\Huge}, execute at end node={\end{varwidth}}]

\tikzstyle{StyleOfFlag} = [ellipse, draw, color=blue(ncs), fill=white, text=black, text centered, inner sep=0.09cm, line width=0.5mm, minimum width=4.4cm, execute at begin node={\begin{varwidth}{4.8cm}\centering\bf\Huge}, execute at end node={\end{varwidth}}, rounded corners]

\tikzstyle{StyleOfFlag(long)} = [ellipse, draw, color=blue(ncs), fill=white, text=black, text centered, inner sep=0.01cm, line width=0.5mm, minimum width=8.8cm, minimum height=1.8cm, execute at begin node={\begin{varwidth}{8.8cm}\centering\bf\Huge}, execute at end node={\end{varwidth}}, rounded corners]

\tikzstyle{StyleOfFinalProduct} = [rectangle, draw, color=blue(ncs), fill=white, text=black, text centered, inner sep=0.50cm, minimum width=16.5cm, line width=1.5mm, execute at begin node={\begin{varwidth}{18.5cm}\centering\bf\XHUGE}, execute at end node={\end{varwidth}}, rounded corners]

\tikzstyle{StyleOfLine} = [draw, -{Latex[scale=3,length=5,width=3,angle'=25,open]}, line width=0.8pt, color=blue(ncs)]

\tikzstyle{StyleOfLine(thick)} = [draw, -{Latex[scale=3,length=5,width=3,angle'=25,open]}, line width=1.4pt, color=blue(ncs)]

\tikzstyle{StyleOfLine(thicker)} = [draw, -{Latex[scale=3,length=9pt,width=15pt,angle'=25,open]}, line width=2.8pt, color=blue(ncs)]

\tikzstyle{StyleOfLine(dotted)} = [draw, dotted, -latex', line width=2.0pt, color=blue(ncs)]

\tikzstyle{StyleOfLine(dashed)} = [draw, dashed, -latex', line width=2.5pt, color=blue(ncs)]

%% file: Input_flow_chart.tikz

\def\hideLiteratureCatalogs{1}
\def\hideSpectralDataCubes{1}

\def\angleout{70}
\def\anglein{20}

\hypersetup{linkcolor=link(red)}

\begin{tikzpicture}[node distance = 1.5cm and 1.0cm, auto]

\node[Project] (Project) {\HUGE $\bm{A^3}$COSMOS};

\node[emp, right =of Project, xshift=+6cm] (Empty Node Next Project) {};

\node[S02, below =of Empty Node Next Project, yshift=-1cm] (ALMA continuum images) {ALMA Continuum Images 
\ifx\isStandalone\undefined%
\textcolor{white}{(}%
\textcolor{link(red)}{Sect.~\ref{Section_ALMA_Continuum_Images}}%
\textcolor{white}{)}%
\fi%
};

\draw[line(thicker)] 
(Project.east)
to 
(ALMA continuum images.north);

\node[S02, right =of ALMA continuum images, xshift=+2.5cm] (Catalogs) {Multi-wavelength Catalogs 
\ifx\isStandalone\undefined%
\textcolor{white}{(}%
\textcolor{link(red)}{Sect.~\ref{Section_Prior_Source_Catalogs}}%
\textcolor{white}{)}%
\fi%
};

\draw[line(thicker)] 
(Project.east)
to 
(Catalogs.north);

\node[S02, right =of Catalogs, xshift=+2.5cm] (Simulations) {Monte Carlo Simulations 
\ifx\isStandalone\undefined%
\textcolor{white}{(}%
\textcolor{link(red)}{Sect.~\ref{Section_Monte_Carlo_Simulation_and_Correction}}%
\textcolor{white}{)}%
\fi%
};

\draw[line(thicker)] 
(Project.east)
to 
(Simulations.north);

\ifx\hideSpectralDataCubes\undefined
\node[S02, right =of Simulations, xshift=+4.0cm] (ALMA spectral data cubes) {ALMA spectral data cubes};
\fi

\ifx\hideSpectralDataCubes\undefined
\draw[line(thicker)] 
(Project.south)
to [out={-90+\angleout+15},in={90+\anglein+50}]
(ALMA spectral data cubes.north);
\fi

\node[emp, below =of Catalogs] (Catalogs Downstream node) {};
\node[app2, below =of Catalogs Downstream node, yshift=+1.75cm] (Catalogs Combination) {combine};
\node[S04, below =of Catalogs Combination), yshift=+0.5cm] (COSMOS master catalog) {COSMOS master catalog
\ifx\isStandalone\undefined%
\textcolor{black}{(}%
\textcolor{link(red)}{Sect.~\ref{Section_Prior_Source_Catalogs}}%
\textcolor{black}{)}%
\fi%
};

\draw[line(thick)] (Catalogs.south) to (Catalogs Combination.north);
\draw[line(thick)] (Catalogs Combination.south) to [out=165, in=-15] (COSMOS master catalog.north);

\ifx\hideLiteratureCatalogs\undefined
\node[COSMOS literature catalogs, left =of Catalogs Downstream node), xshift=-5cm, yshift=-3cm] (COSMOS literature catalogs) {%
COSMOS literature catalogs:\\[6pt]
{McCracken priv.}, \\
	--- UltraVISTA DR4\\[6pt]
\href{http://adsabs.harvard.edu/abs/2016ApJS..224...24L}{Laigle et al. (2016)}
\href{ftp://ftp.iap.fr/pub/from_users/hjmcc/COSMOS2015/}{$\bm{p}$+$\bm{z}$-cat.}, \\
	--- UltraVISTA DR2\\[6pt]
\href{http://adsabs.harvard.edu/abs/2013A&A...556A..55I}{Ilbert et al. (2013)}
\href{http://cdsarc.u-strasbg.fr/viz-bin/qcat?J/A+A/556/A55}{$\bm{p}$+$\bm{z}$-cat.}, \\
	--- UltraVISTA DR1\\[6pt]
\href{http://adsabs.harvard.edu/abs/2013ApJS..206....8M}{Muzzin et al. (2013)}
\href{http://cdsarc.u-strasbg.fr/viz-bin/qcat?J/ApJS/206/8}{$\bm{p}$+$\bm{z}$-cat.}, \\
	--- UltraVISTA DR1\\[6pt]
\href{http://adsabs.harvard.edu/abs/2009ApJ...690.1236I}{Ilbert et al. (2009)}
\href{http://adsabs.harvard.edu/abs/2017yCat..16901236I}{$\bm{z}$-cat.}, \\
	--- Subaru/SupCAM $\bm{i^{+}_{\mathrm{AB}}<26.5}$ \\[6pt]
\href{http://adsabs.harvard.edu/abs/2009ApJ...703..222L}{Le Floc'h et al. (2009)}
\href{http://irsa.ipac.caltech.edu/data/COSMOS/tables/scosmos/}{$\bm{p}$-cat.}, \\
	--- S-COSMOS GO2+3 MIPS24\\[6pt]
\href{http://adsabs.harvard.edu/abs/2007ApJS..172...86S}{Sander et al. (2007)}
\href{http://irsa.ipac.caltech.edu/data/COSMOS/tables/scosmos/}{$\bm{p}$-cat.}, \\
	--- S-COSMOS GO2 IRAC+MIPS\\[6pt]
\href{http://adsabs.harvard.edu/abs/2007ApJS..172...99C}{Capak et al. (2007)}
\href{http://cdsarc.u-strasbg.fr/viz-bin/qcat?II/284}{$\bm{p}$-cat.}, \\
	--- Ground-based $\bm{i^{+}+i^{*}}$\\[6pt]
{Davidzon priv.}, \\
	--- SPLASH IRAC ch1,2 \\[6pt]
\href{https://adsabs.harvard.edu/abs/2017A&A...602A...1S}{Smolcic et al. (2017)}
\href{http://irsa.ipac.caltech.edu/data/COSMOS/tables/vla/}{$\bm{p}$-cat.}, \\
	--- VLA 3~GHz\\[6pt]
};
\fi

\node[emp, below =of ALMA continuum images, yshift=-3cm] (ALMA continuum images Downstream node) {};

\node[app3, left  =of ALMA continuum images Downstream node, xshift=+0.75cm, yshift=-2.25cm] (Blind extraction) {Blind source extraction (PyBDSF%
\ifx\isStandalone\undefined%
; \ \textcolor{link(red)}{Sect.~\ref{Section_Blind_Source_Extraction}}%
\fi%
)
};

\node[app3, right =of ALMA continuum images Downstream node, xshift=-0.75cm, yshift=-2.25cm] (Prior extraction) {Prior source fitting (getpix+galfit%
\ifx\isStandalone\undefined%
; \textcolor{link(red)}{Sect.~\ref{Section_Prior_Source_Fitting}}%
\fi%
)
};

\draw[line(thick)] 
(ALMA continuum images.south)
to [out={-90-\angleout},in={90-\anglein}]
(Blind extraction.north);

\draw[line(thick)] 
(ALMA continuum images.south)
to [out={-90+\angleout},in={90+\anglein}]
(Prior extraction.north);

\draw[line(thick)] (COSMOS master catalog.west) to[out=180,in=0] (Prior extraction.east);

\node[S04, left =of Blind extraction, minimum height=3cm] (Data release) {Meta data};

\draw[line(thick)] 
([yshift=-0.5cm] ALMA continuum images.west)
to [out=-145,in=75]
(Data release.north);

\node[S04, below =of Prior extraction] (Prior extraction catalog) {Prior extraction catalog};

\node[S04, below =of Blind extraction] (Blind extraction catalog) {Blind extraction catalog};

\node[app, below =of Prior extraction catalog] (Prior extraction catalog Sim corr) {apply sim. corr.};

\node[app, below =of Blind extraction catalog] (Blind extraction catalog Sim corr) {apply sim. corr.};

\node[S04, below =of Prior extraction catalog Sim corr] (Prior extraction catalog Corrected) {Final prior extraction catalog};

\node[S04, below =of Blind extraction catalog Sim corr] (Blind extraction catalog Corrected) {Final blind extraction catalog};

\node[emp, below =of Simulations] (S03c) {};

\node[app4, left  =of S03c, xshift=+0.42cm, yshift=-4.5cm] (Physic MC sim) {Physically-motivated MC simulation};
\node[app4, right =of S03c, xshift=-0.42cm, yshift=-4.5cm] (Param MC sim) {Full parameter-space MC simulation};

\draw[line(thick)] 
(Simulations.south)
to [out={-90-\angleout},in={90-\anglein}]
(Physic MC sim.north);

\draw[line(thick)] 
(Simulations.south)
to [out={-90+\angleout},in={90+\anglein}]
(Param MC sim.north);

\node[S04, below =of Physic MC sim, yshift=0.55cm] (Physic MC Simulated images) {Simulated images};
\node[S04, below =of Param MC sim, yshift=0.55cm] (Param MC Simulated images) {Simulated images};

\draw[line(thick)] 
(Physic MC sim.south)
to 
(Physic MC Simulated images.north);

\draw[line(thick)] 
(Param MC sim.south)
to 
(Param MC Simulated images.north);

\draw[line(thick)] 
([xshift=0.4cm, yshift=0.4cm] Blind extraction.south)
to 
([xshift=+1cm, yshift=+1.2cm] Physic MC Simulated images.west);

\draw[line(thick)] 
([xshift=0.5cm, yshift=0.5cm] Blind extraction.south)
to 
([xshift=+2cm, yshift=+1.6cm] Param MC Simulated images.west);

\node[emp, below =of Physic MC Simulated images] (S05s1) {};
\node[app2, left  =of S05s1, xshift=+1.05cm, yshift=-2.0cm] (Physic MC Simulated images Prior extraction) {Prior extraction};
\node[app2, right =of S05s1, xshift=-1.05cm, yshift=-2.0cm] (Physic MC Simulated images Blind extraction) {Blind extraction};

\node[emp, below =of Param MC Simulated images] (S05s2) {};
\node[app2, left  =of S05s2, xshift=+1.05cm, yshift=-2.0cm] (Param MC Simulated images Prior extraction) {Prior extraction};
\node[app2, right =of S05s2, xshift=-1.05cm, yshift=-2.0cm] (Param MC Simulated images Blind extraction) {Blind extraction};

\draw[line(thick)] 
(Physic MC Simulated images.south)
to [out={-90-\angleout},in={90-\anglein}]
(Physic MC Simulated images Prior extraction.north);

\draw[line(thick)] 
(Physic MC Simulated images.south)
to [out={-90+\angleout},in={90+\anglein}]
(Physic MC Simulated images Blind extraction.north);

\draw[line(thick)] 
(Param MC Simulated images.south)
to [out={-90-\angleout},in={90-\anglein}]
(Param MC Simulated images Prior extraction.north);

\draw[line(thick)] 
(Param MC Simulated images.south)
to [out={-90+\angleout},in={90+\anglein}]
(Param MC Simulated images Blind extraction.north);

\node[S06, below =of Physic MC Simulated images Prior extraction] (Physic MC Simulated images Prior extraction Sim stat) {sim. stat.};
\node[S06, below =of Physic MC Simulated images Blind extraction] (Physic MC Simulated images Blind extraction Sim stat) {sim. stat.};

\node[S06, below =of Param MC Simulated images Prior extraction] (Param MC Simulated images Prior extraction Sim stat) {sim. stat.};
\node[S06, below =of Param MC Simulated images Blind extraction] (Param MC Simulated images Blind extraction Sim stat) {sim. stat.};

\draw[line(thick)] 
(Physic MC Simulated images Prior extraction.south)
to 
(Physic MC Simulated images Prior extraction Sim stat.north);

\draw[line(thick)] 
(Physic MC Simulated images Blind extraction.south)
to 
(Physic MC Simulated images Blind extraction Sim stat.north);

\draw[line(thick)] 
(Param MC Simulated images Prior extraction.south)
to 
(Param MC Simulated images Prior extraction Sim stat.north);

\draw[line(thick)] 
(Param MC Simulated images Blind extraction.south)
to 
(Param MC Simulated images Blind extraction Sim stat.north);

\node[S04W7p5cm, below =of Physic MC Simulated images Prior extraction Sim stat, xshift=6.5cm, yshift=-0.5cm] (sim corr) {simulation-based correction tables 
\ifx\isStandalone\undefined%
(Sect.~\ref{Section_MC_Sim_Final_Correction})%
\fi%
};

\draw[line(thick)] (Physic MC Simulated images Prior extraction Sim stat.south) to[out=-90+25,in=165] ([xshift=-1cm] sim corr.north);

\draw[line(thick)] (Physic MC Simulated images Blind extraction Sim stat.south) to[out=-90,in=135] ([xshift=-0.2cm] sim corr.north);

\draw[line(thick)] (Param MC Simulated images Prior extraction Sim stat.south) to[out=-90,in=45] ([xshift=+0.2cm] sim corr.north);

\draw[line(thick)] (Param MC Simulated images Blind extraction Sim stat.south) to[out=-90-25,in=25] ([xshift=+1cm] sim corr.north);

\draw[line(thick)] 
(sim corr.west)
to [out=168,in=-20] 
(Prior extraction catalog Sim corr.east);

\draw[line(thick)] 
(sim corr.west)
to [out=172,in=-15] 
(Blind extraction catalog Sim corr.east);

\ifx\hideSpectralDataCubes\undefined

\node[S03, below =of ALMA spectral data cubes] (ALMA spectral data cubes Spectral line finder) {Spectral line finder};

\node[S04, below =of ALMA spectral data cubes Spectral line finder] (ALMA spectral data cubes Spectral line identifier) {Spectral line identifier (ML)};

\node[emp, below =of ALMA spectral data cubes Spectral line identifier] (ALMA spectral data cubes Spectral line identifier Downstrem node) {};

\node[S05, left  =of ALMA spectral data cubes Spectral line identifier Downstrem node, xshift=+0.5cm] (ALMA spectral data cubes Spectral line flux) {Spectral line flux};

\node[S05, right =of ALMA spectral data cubes Spectral line identifier Downstrem node, xshift=-0.5cm] (ALMA spectral data cubes Spectral line redshift) {Spectroscopic redshift};

\fi






\draw[line(thick)] 
(Prior extraction.south)
to 
(Prior extraction catalog.north);

\draw[line(thick)] 
(Blind extraction.south)
to 
(Blind extraction catalog.north);

\draw[line(thick)] 
(Prior extraction catalog.south)
to 
(Prior extraction catalog Sim corr.north);

\draw[line(thick)] 
(Blind extraction catalog.south)
to 
(Blind extraction catalog Sim corr.north);

\draw[line(thick)] 
(Prior extraction catalog.south)
to 
(Prior extraction catalog Sim corr.north);

\draw[line(thick)] 
(Blind extraction catalog.south)
to 
(Blind extraction catalog Sim corr.north);

\draw[line(thick)] 
(Prior extraction catalog Sim corr.south)
to 
(Prior extraction catalog Corrected.north);

\draw[line(thick)] 
(Blind extraction catalog Sim corr.south)
to 
(Blind extraction catalog Corrected.north);





%

%
%
%
%


%
%

\node[S02, below =of COSMOS master catalog, xshift=+0.25cm, yshift=-15.75cm, minimum width=12.25cm, minimum height=3.75cm] (SED Fitting) {Source selection, SED fitting, and galaxy catalogs 
\ifx\isStandalone\undefined%
\textcolor{white}{(}%
\textcolor{link(red)}{Sects.~\ref{Section_Galaxy_Sample_and_Properties}--\ref{Section_Data_Delivery}}%
\textcolor{white}{)}%
\fi%
};


\draw[line(thick)] (Blind extraction catalog Corrected.south) to[out=-30,in=-170] ([yshift=-0.5cm] SED Fitting.west);

\draw[line(thick)] (Prior extraction catalog Corrected.south) to[out=-35,in=-175] ([yshift=-0.4cm] SED Fitting.west);

\ifdefined\isStandalone
\node[emp, below =of SED Fitting, yshift=-1cm] (Empty Node Below SED Fitting) {};
\fi


\ifx\hideSpectralDataCubes\undefined

\draw[line(thick)] 
(ALMA spectral data cubes.south)
to 
(ALMA spectral data cubes Spectral line finder.north);

\draw[line(thick)] 
(ALMA spectral data cubes Spectral line finder.south)
to 
(ALMA spectral data cubes Spectral line identifier.north);

\draw[line(thick)] 
(ALMA spectral data cubes Spectral line identifier.south)
to [out={-90-70},in={90-70}]
(ALMA spectral data cubes Spectral line flux.north);

\draw[line(thick)] 
(ALMA spectral data cubes Spectral line identifier.south)
to [out={-90+70},in={90+70}]
(ALMA spectral data cubes Spectral line redshift.north);

\fi


%
%
%
%
%
%

\end{tikzpicture}

\hypersetup{linkcolor=magenta!80!white}

%% file: Table_per_band_info_table.tex

\begin{table*}[htbp]
\caption{%
    Information per ALMA Band
    \label{Table_per_band_info_table}
}
\begin{tabularx}{1.0\textwidth}{l *{7}{R}}
\hline
\hline
Info Type                      &       Band 3 &       Band 4 &       Band 5 &       Band 6 &       Band 7 &       Band 8 &       Band 9 \\
\hline
Number of Images                   &           34 &            6 &            2 &          633 &          857 &            1 &            1 \\
Sum Beam Area [arcmin$^2$]\,$^{a}$ &       26.639 &        2.294 &        0.329 &       79.511 &       54.729 &        0.044 &        0.016 \\
Mean Beam Size [arcsec]            &        2.164 &        1.098 &        1.548 &        1.202 &        0.772 &        0.526 &        0.305 \\
Mean RMS Noise [mJy/beam]          &        0.039 &        0.025 &        0.090 &        0.077 &        0.160 &        0.034 &        1.757 \\
PYBDSF $\SNRpeak>5.40$             &           24 &            5 &            3 &          371 &          524 &            1 &            2 \\
GALFIT $\SNRpeak>4.35$\,$^{b}$     &       20 (7) &       10 (7) &        2 (2) &    452 (342) &    553 (461) &        1 (1) &        1 (1) \\
\hline
%
\end{tabularx}
$^{a}$ The areas are the sum of primary beam circular area only. \\
$^{b}$ The number in parentheses corresponds to the sources that passed our quality assessments from Sects.~\ref{Section_Combining_two_photometry_catalogs}~to~\ref{Section_Running_SED_Fitting}. 
Based on the spurious fraction analysis in Sect.~\ref{Section_Spurious_Fraction}, we expect about 8\% spurious sources in total for the PYBDSF selection and $\sim$12\% for the GALFIT selection. \\
\end{table*}

%% file: Table_prior_catalogs_tabular.tex
\newlength{\mycolaa}
\newlength{\mycolab}
\newlength{\mycolac}
\newlength{\mycolad}
\newlength{\mycolae}
\newlength{\mycolaf}
\newlength{\mycolag}
\newlength{\mycolah}
\newlength{\mycolai}
\newlength{\myrowsep}
\setlength{\mycolaa}{3.4cm}
\setlength{\mycolab}{1.0cm}
\setlength{\mycolac}{1.1cm}
\setlength{\mycolad}{1.1cm}
\setlength{\mycolae}{1.1cm}
\setlength{\mycolaf}{2.8cm}
\setlength{\mycolag}{2.5cm}
\setlength{\mycolah}{2.0cm}
\setlength{\mycolai}{3.5cm}
\setlength{\myrowsep}{0.35cm}


\begin{table*}[htpb]
\caption{%
	Prior Catalogs Used for Constructing the COSMOS Master Catalog\,$^{\dagger}$
	\label{Table_prior_catalogs}
}
\begin{tabularx}{1.0\textwidth}{X X X X X X X X}
\hline
\hline
\multicolumn{1}{p{\mycolaa}}{\raggedright 
	{Catalog Name (and Reference)}
} & 
\multicolumn{1}{p{\mycolab}}{\raggedright 
	{Area (deg$^2$)}
} & 
\multicolumn{1}{p{\mycolac}}{\raggedright 
	{$N_{\mathrm{Catalog}}\,^{a}$}
} &
\multicolumn{1}{p{\mycolad}}{\raggedright 
	{$N_{\mathrm{Master}}\,^{b}$}
} &
\multicolumn{1}{p{\mycolad}}{\raggedright 
	{$N_{\mathrm{Unique}}\,^{c}$}
} &
\multicolumn{1}{p{\mycolae}}{\raggedright 
	{Detection Map}
} & 
\multicolumn{1}{p{\mycolaf}}{\raggedright 
	{Depth\,$^{d}$}
} & 
\multicolumn{1}{p{\mycolag}}{\raggedright 
	{Res.\,$^{e}$}
} 
\\
\hline
\multicolumn{1}{p{\mycolaa}}{\raggedright 
	COSMOS2015 catalog \quad {\citep{Laigle2016}}
} & 
\multicolumn{1}{p{\mycolab}}{\raggedright 
	$1.8$
} & 
\multicolumn{1}{p{\mycolac}}{\raggedright 
	1182108
} & 
\multicolumn{1}{p{\mycolad}}{\raggedright 
	1182108
} & 
\multicolumn{1}{p{\mycolae}}{\raggedright 
	738420
} & 
\multicolumn{1}{p{\mycolaf}}{\raggedright 
	VISTA $z \, Y J H K_s$
} & 
\multicolumn{1}{p{\mycolag}}{\raggedright 
	24.0 ($3\,\sigma$, 3$''$, $K_s$)
} & 
\multicolumn{1}{p{\mycolah}}{\raggedright 
	$\sim1''$
} 
\\[\myrowsep]
\multicolumn{1}{p{\mycolaa}}{\raggedright 
	$K_s$-band catalog \quad {\citep{Muzzin2013}}\,$^{f}$
} & 
\multicolumn{1}{p{\mycolab}}{\raggedright 
	$1.6\,^{f}$
} & 
\multicolumn{1}{p{\mycolac}}{\raggedright 
	263229
} & 
\multicolumn{1}{p{\mycolad}}{\raggedright 
	18536
} & 
\multicolumn{1}{p{\mycolae}}{\raggedright 
	10799
} & 
\multicolumn{1}{p{\mycolaf}}{\raggedright 
	VISTA $K_s$
} & 
\multicolumn{1}{p{\mycolag}}{\raggedright 
	24.35 ($3\,\sigma$, 2$\farcs$1)
} & 
\multicolumn{1}{p{\mycolah}}{\raggedright 
	$\sim1''$
} 
\\[\myrowsep]
\multicolumn{1}{p{\mycolaa}}{\raggedright 
	$i$-band catalog \quad\quad\quad {\citep{Capak2007}}
} & 
\multicolumn{1}{p{\mycolab}}{\raggedright 
	$1.8$
} & 
\multicolumn{1}{p{\mycolac}}{\raggedright 
	386125
} & 
\multicolumn{1}{p{\mycolad}}{\raggedright 
	31159
} & 
\multicolumn{1}{p{\mycolae}}{\raggedright 
	30146
} & 
\multicolumn{1}{p{\mycolaf}}{\raggedright 
	CFHT $i^\star$ $+$ \quad \textit{Subaru} $i^+$
} & 
\multicolumn{1}{p{\mycolag}}{\raggedright 
	26.2 ($5\,\sigma$, 3$''$)
} & 
\multicolumn{1}{p{\mycolah}}{\raggedright 
	$\sim0\farcs5$ 
} 
%
\\[\myrowsep]
\multicolumn{1}{p{\mycolaa}}{\raggedright 
	SPLASH IRAC supplementary catalog\,$^{g}$
} & 
\multicolumn{1}{p{\mycolab}}{\raggedright 
	$1.6$ 
} & 
\multicolumn{1}{p{\mycolac}}{\raggedright 
	5390
} & 
\multicolumn{1}{p{\mycolad}}{\raggedright 
	4685
} & 
\multicolumn{1}{p{\mycolae}}{\raggedright 
	3690
} & 
\multicolumn{1}{p{\mycolaf}}{\raggedright 
	\textit{Spitzer}/IRAC 3.6$\,+\,$4.5$\,\mu$m
} & 
\multicolumn{1}{p{\mycolag}}{\raggedright 
	25.5 ($3\,\sigma$)
} & 
\multicolumn{1}{p{\mycolah}}{\raggedright 
	$\sim1\farcs6$
} 
%
\\[\myrowsep]
\multicolumn{1}{p{\mycolaa}}{\raggedright 
	VLA catalog \quad\quad\quad {\citep{Smolcic2017a}}
} & 
\multicolumn{1}{p{\mycolab}}{\raggedright 
	$2$
} & 
\multicolumn{1}{p{\mycolac}}{\raggedright 
	10922
} & 
\multicolumn{1}{p{\mycolad}}{\raggedright 
	1042
} & 
\multicolumn{1}{p{\mycolae}}{\raggedright 
	644
} & 
\multicolumn{1}{p{\mycolaf}}{\raggedright 
	VLA 3\,GHz
} & 
\multicolumn{1}{p{\mycolag}}{\raggedright 
	$2.3\,\mu$Jy ($1\,\sigma$)
} & 
\multicolumn{1}{p{\mycolah}}{\raggedright 
	$\sim0\farcs75$
} 
\\[\myrowsep]
\multicolumn{1}{p{\mycolaa}}{\raggedright 
	IRAC catalog \quad\quad\quad {\citep{Sanders2007}}
} & 
\multicolumn{1}{p{\mycolab}}{\raggedright 
	$2.3$
} & 
\multicolumn{1}{p{\mycolac}}{\raggedright 
	347332
} & 
\multicolumn{1}{p{\mycolad}}{\raggedright 
	55346
} & 
\multicolumn{1}{p{\mycolae}}{\raggedright 
	54642
} & 
\multicolumn{1}{p{\mycolaf}}{\raggedright 
	\textit{Spitzer}/IRAC $3.6-8.0\,\mu$m
} & 
\multicolumn{1}{p{\mycolag}}{\raggedright 
	24.01 ($5\,\sigma$, 3.6\,$\mu$m)
} & 
\multicolumn{1}{p{\mycolah}}{\raggedright 
	$\sim1\farcs6$
} 
\\
\hline
\end{tabularx}
$^{a}$ Total number of sources in each prior catalog. 
$^{b}$ The number of sources in each prior catalog that are not in higher order catalogs (the order is as listed from top to bottom). Which is, the number of sources in our ``COSMOS master catalog'' that are originated from the current prior catalog. 
$^{c}$ The number of unique sources in each prior catalog, which means these sources have no counterpart in any other prior catalog. 
$^{d}$ The depth is in AB magnitude, and is in an aperture if indicated in brackets. 
$^{e}$ Spatial resolution, or point spread function size of the detection map. 
$^{f}$ The \cite{Muzzin2013} catalog contains sources in the masked area of the COSMOS2015 catalog which are close to bright, saturated stars. 
$^{g}$ Based on the source extraction in the Spitzer Large Area Survey with Hyper-Suprime-Cam (SPLASH; PI: P. Capak) survey data after fitting and removing all COSMOS2015 catalog sources (I. Davidzon; priv. comm.)
$^{\dagger}$ The master catalog used in this paper's work has a version code of 20170426. 
%
%
%
\end{table*}

%% file: Input_galaxy_flow_chart.tikz


\def\hideLiteratureCatalogs{1}
\def\hideSpectralDataCubes{1}

\def\angleout{70}
\def\anglein{20}

\hypersetup{linkcolor=red!50!white}

\begin{tikzpicture}[node distance = 2.5cm and 1.0cm, auto]

\node[StyleOfEmptyNode] (NodeOfEmpty) {};


\node[StyleOfCatalog, left=of NodeOfEmpty, yshift=+3cm] (NodeOfBlindCat) {Blind Extraction Photometry Catalog 
\ifx\isStandalone\undefined%
(%
\textcolor{red!50!white}{Sect.~\ref{Section_Blind_Source_Extraction}}%
)%
\fi%
};


\node[StyleOfCatalog, right=of NodeOfEmpty, yshift=+3cm] (NodeOfPriorCat) {Prior Fitting Photometry Catalog 
\ifx\isStandalone\undefined%
(%
\textcolor{red!50!white}{Sect.~\ref{Section_Prior_Source_Fitting}}%
)%
\fi%
};


\node[StyleOfProcess, below=of NodeOfBlindCat, yshift=+1cm](NodeOfBlindCatSNRCut){$\mathbf{S/N_{{peak}}}$ Cut 
\ifx\isStandalone\undefined%
(%
\textcolor{red!50!white}{Sect.~\ref{Section_Combining_two_photometry_catalogs}}%
)%
\fi%
};

\draw[StyleOfLine(thicker)] 
(NodeOfBlindCat.south)
to
(NodeOfBlindCatSNRCut.north);


\node[StyleOfProcess, below=of NodeOfPriorCat, yshift=+1cm](NodeOfPriorCatSNRCut){$\mathbf{S/N_{{peak}}}$ Cut 
\ifx\isStandalone\undefined%
(%
\textcolor{red!50!white}{Sect.~\ref{Section_Combining_two_photometry_catalogs}}%
)%
\fi%
};

\draw[StyleOfLine(thicker)] 
(NodeOfPriorCat.south)
to
(NodeOfPriorCatSNRCut.north);


\node[StyleOfProcess, below=of NodeOfEmpty, yshift=-4cm](NodeOfFluxComparison){Flux \ \ \ \ \ \ Comparison 
\ifx\isStandalone\undefined%
(%
\textcolor{red!50!white}{Sect.~\ref{Section_Combining_two_photometry_catalogs}}%
)%
\fi%
};

\draw[StyleOfLine(thicker)] 
(NodeOfBlindCatSNRCut.south)
to[out=-75,in=135]
(NodeOfFluxComparison.west);

\draw[StyleOfLine(thicker)] 
(NodeOfPriorCatSNRCut.south)
to[out=-105,in=45]
(NodeOfFluxComparison.east);


\node[StyleOfProcess, below=of NodeOfFluxComparison, xshift=+7.5cm, yshift=+5.5cm](NodeOfAutoCounterpart){Examining Counterpart Association 
\ifx\isStandalone\undefined%
(%
\textcolor{red!50!white}{Sect.~\ref{Section_Examining_counterpart_association}}%
)%
\fi%
};

\draw[StyleOfLine(thicker)] 
(NodeOfFluxComparison.south)
to[out=-45,in=180]
(NodeOfAutoCounterpart.west);


\node[StyleOfCatalog, below=of NodeOfAutoCounterpart, xshift=2.5cm, yshift=+1.0cm] (NodeOfMasterCat) {COSMOS Master Catalog 
\ifx\isStandalone\undefined%
(%
\textcolor{red!50!white}{Sect.~\ref{Section_Prior_Source_Catalogs}}%
)%
\fi%
};

\draw[StyleOfLine(thicker)] 
(NodeOfMasterCat.north)
to[out=135,in=-45]
(NodeOfAutoCounterpart.south);


\node[StyleOfFlag(long), below=of NodeOfFluxComparison, xshift=-6.5cm, yshift=+2.8cm] (NodeOfWrongFluxComparison) {Inconsistent-flux outliers (0.5\%)};

\draw[StyleOfLine(dashed)] 
([xshift=-2cm,yshift=2cm]NodeOfFluxComparison.south)
to
([xshift=+2cm]NodeOfWrongFluxComparison.north);


\node[StyleOfFlag(long), below=of NodeOfAutoCounterpart, xshift=-8.0cm, yshift=+3.8cm] (NodeOfWrongCounterpart) {Unreliable-counterpart outliers (4\%)};

\draw[StyleOfLine(dashed)] 
([xshift=-2cm,yshift=2cm]NodeOfAutoCounterpart.south)
to
([xshift=+2cm]NodeOfWrongCounterpart.north);


\node[StyleOfCatalog, right=of NodeOfAutoCounterpart, xshift=+0.5cm, yshift=+9.0cm](NodeOfGalaxyCat){Galaxy Multi-wavelength SEDs and prior-redshifts
\ifx\isStandalone\undefined%
(%
\textcolor{red!50!white}{Sect.~\ref{Section_Combining_prior_redshifts}}%
)%
\fi%
};

\draw[StyleOfLine(thicker)] 
(NodeOfAutoCounterpart.east)
to[out=45,in=-135]
(NodeOfGalaxyCat.west);


\node[StyleOfFlag(long), above=of NodeOfGalaxyCat, xshift=+2cm, yshift=-1.35cm] (NodeOfMultiALMADataGalaxies) {Galaxies having more than one ALMA data (25\%)};

\draw[StyleOfLine(dashed)] 
(NodeOfGalaxyCat.north)
to
(NodeOfMultiALMADataGalaxies.south);


\node[StyleOfFlag(long), below=of NodeOfGalaxyCat, xshift=+2cm, yshift=+1.35cm] (NodeOfNoPriorRedshift) {Galaxies having no optical/near-IR counterpart/prior-redshift (26\%)};

\draw[StyleOfLine(dashed)] 
(NodeOfGalaxyCat.south)
to
(NodeOfNoPriorRedshift.north);

\node[StyleOfProcess, right=of NodeOfGalaxyCat, xshift=+1.0cm](NodeOfSedFitting){MAGPHYS SED Fitting 
\ifx\isStandalone\undefined%
(%
\textcolor{red!50!white}{Sect.~\ref{Section_Running_SED_Fitting}}%
)%
\fi%
};

\draw[StyleOfLine(thicker)] 
(NodeOfGalaxyCat.east)
to
(NodeOfSedFitting.west);

\node[StyleOfCatalog, right=of NodeOfSedFitting, xshift=+0.5cm] (NodeOfFirstGalaxyProperties) {First Galaxy Properties: $\boldsymbol{z}$, $\boldsymbol{M_{\star}}$, SFR
\ifx\isStandalone\undefined%
(%
\textcolor{red!50!white}{Sect.~\ref{Section_Obtaining_Galaxy_Properties}}%
)%
\fi%
};

\draw[StyleOfLine(thicker)] 
(NodeOfSedFitting.east)
to
(NodeOfFirstGalaxyProperties.west);


\node[StyleOfProcess, above=of NodeOfFirstGalaxyProperties, xshift=-1.5cm, yshift=-1.5cm](NodeOfLineDecontamination){Predicting and Subtracting mm Line Fluxes
\ifx\isStandalone\undefined%
(%
\textcolor{red!50!white}{Sect.~\ref{Section_Subtracting_Strong_Emission_Lines}}%
)%
\fi%
};

\draw[StyleOfLine(thicker)] 
(NodeOfFirstGalaxyProperties.east)
to[out=+15,in=-15]
(NodeOfLineDecontamination.east);


\draw[StyleOfLine(thicker)] 
(NodeOfLineDecontamination.west)
to[out=+180,in=+90]
(NodeOfSedFitting.north);


\node[StyleOfProcess, below=of NodeOfSedFitting, yshift=+1.0cm](NodeOfCheckResidual){Examining ALMA-band $\boldsymbol{\chi^2}$
\ifx\isStandalone\undefined%
(%
\textcolor{red!50!white}{Sect.~\ref{Section_Running_SED_Fitting}}%
)%
\fi%
};

\draw[StyleOfLine(thicker)] 
(NodeOfSedFitting.south)
to
(NodeOfCheckResidual.north);

\node[StyleOfFlag(long), below=of NodeOfCheckResidual, xshift=-6.5cm, yshift=+2.8cm] (NodeOfWrongSedResidual) {SED-excess outliers (3\%)};

\draw[StyleOfLine(dashed)] 
([xshift=-2cm,yshift=2cm]NodeOfCheckResidual.south)
to
([xshift=+2cm]NodeOfWrongSedResidual.north);


\node[StyleOfFinalProduct, below=of NodeOfCheckResidual, xshift=9.5cm, yshift=+2.0cm](NodeOfFinalProduct){Final Galaxy Properties: z, $\boldsymbol{M_{\star}}$, SFR, sSFR, rest-frame mm fluxes, etc.
\ifx\isStandalone\undefined%
(%
\textcolor{red!50!white}{Sect.~\ref{Section_Obtaining_Galaxy_Properties}},~\textcolor{red!50!white}{\ref{Section_Data_Delivery}}%
)%
\fi%
};

\draw[StyleOfLine(thicker)] 
([xshift=2cm,yshift=2cm]NodeOfCheckResidual.south)
to[out=-45,in=125]
(NodeOfFinalProduct.north);

\ifdefined\isStandalone
\node[StyleOfEmptyNode, below =of NodeOfFinalProduct, yshift=0cm] (EmptyNodeAtBottom) {};
\fi

\end{tikzpicture}

\hypersetup{linkcolor=magenta!80!white}

%% file: Table_number_of_sources_excluded.tex

\begin{table}[htpb]
\caption{%
	Number of Sources in A$^3$COSMOS Catalogs and Excluded at Each Step in Sect.~\ref{Section_Galaxy_Sample_and_Properties} (Version \incode{20180201}; See Also Workflow in Fig.~\ref{Figure_galaxy_flow_chart})
	\label{Table_number_of_sources_excluded}
}
%
%
\begin{tabularx}{1.0\linewidth}{X@{\extracolsep{\fill}} r r}
\hline
\hline
\multicolumn{1}{p{0.65\linewidth}}{\raggedright
Catalog/Step
} & Number & Fraction \\
\hline
\multicolumn{1}{p{0.65\linewidth}}{\raggedright
Prior-photometry catalog 
} & \aaacosmosPriorDetectionNumber{} & \nodata \\
\multicolumn{1}{p{0.65\linewidth}}{\raggedright
Blind-photometry catalog 
} & \aaacosmosBlindDetectionNumber{} & \nodata \\
\multicolumn{1}{p{0.65\linewidth}}{\raggedright
Combined ALMA detections
} & \aaacosmosAlmaDetectionNumber{} & \nodata \\
\hline
%
%
%
%
\multicolumn{1}{p{0.65\linewidth}}{\raggedright
Galaxies having more than one ALMA data points (Sect.~\ref{Section_Combining_two_photometry_catalogs})\,$^{a}$
} & \aaacosmosGalaxyWithMultiplePhotometryNumber{} & \aaacosmosGalaxyWithMultiplePhotometryFraction{} \\[0.735ex]
\multicolumn{1}{p{0.65\linewidth}}{\raggedright
Galaxies having no optical/near-IR counterpart/prior-redshift (Sects.~\ref{Section_Combining_two_photometry_catalogs},~\ref{Section_Examining_counterpart_association}~and~\ref{Section_Combining_prior_redshifts})\,$^{b,\,c}$
} & \aaacosmosNumberOfNoPriorOrPriorz{} & \aaacosmosFractionOfNoPriorz{} \\[0.735ex]
\multicolumn{1}{p{0.65\linewidth}}{\raggedright
Inconsistent flux outliers (Sect.~\ref{Section_Photometry_Quality_Check_1}; \incode{Flag_inconsistent_flux})\,$^{c}$
} & \aaacosmosNumberOfOutlierICF{} & \aaacosmosFractionOfOutlierICF{} \\[0.735ex]
\multicolumn{1}{p{0.65\linewidth}}{\raggedright
Unreliable counterpart outliers (Sect.~\ref{Section_Examining_counterpart_association}; \incode{Flag_outlier_CPA})\,$^{c}$
} & \aaacosmosNumberOfOutlierCPA{} & \aaacosmosFractionOfOutlierCPA{} \\[0.735ex]
\multicolumn{1}{p{0.65\linewidth}}{\raggedright 
SED excess outliers (Sect.~\ref{Section_Running_SED_Fitting}; \incode{Flag_outlier_SED})\,$^{c}$
} & \aaacosmosNumberOfOutlierSED{} & \aaacosmosFractionOfOutlierSED{} \\
\hline
\multicolumn{1}{p{0.65\linewidth}}{\raggedright
Final galaxy catalog (Sect.~\ref{Section_Data_Delivery})\,$^{d}$
} & \aaacosmosGoodGalaxyNumber{} & \nodata \\
\hline
\end{tabularx}
$^{a}$ In this step, \REREVISED[RefereeReport2.9][]{we sorted \aaacosmosPriorDetectionNumber{} ALMA prior detections into \aaacosmosGalaxyNumber{} unique galaxies, while discarded \aaacosmosBlindOnlyNumber{} blind-only sources (see discussion in Sect.~\ref{Section_Discareded_Sources}). The fractions in the third column are of the \aaacosmosGalaxyNumber{} unique galaxies.} \\
%
$^{b}$ This includes the \aaacosmosBlindOnlyNumber{} blind-only sources, \aaacosmosNumberOfPriorGalaxiesHavingNoPriorz{} galaxies which have no redshift from literature as prior information and \aaacosmosNumberOfPriorGalaxiesHavingOnlyJinPriorz{} galaxies which only have a far-IR/mm photo-$z$ from \cite{Jin2018}. They are excluded \REREVISED[RefereeReport2.9][]{from the further quality assessments} due to too poor constraints on galaxy properties. \\
$^{c}$ 10 sources are duplicated among these flags. \\
$^{d}$ Our approach aims at keeping only galaxies with most reliable properties (redshift, stellar mass and dust-obscured SFR), therefore the number of galaxies is significantly reduced compared to the number of ALMA detections. The exclusion of galaxies does not mean they are all not real, but just their properties could not be reliably estimated with current data. Future follow-ups will be needed to explore their properties. \\
\end{table}

%% file: Table_photometry_catalog_columns.tex

\begin{table*}[hb]
\caption{%
	Columns in the two photometry catalogs
	\label{Table_photometry_catalog_columns}
}
\begin{tabularx}{1.0\textwidth}{l l l X}
\hline
\hline

Column Name & File & Units & Description \\

\hline

\incode{ID} & Prior & --- & A3COSMOS master catalog ID (version 20170426), which equals \cite{Laigle2016} COSMOS2015 catalog ID when ID~$\le 1182108$. \\

\incode{ID_PriorCat} & Prior & --- & The original ID in the \incode{Ref_ID_PriorCat}-th prior catalog. \\

\incode{Ref_ID_PriorCat} & Prior & --- & The reference number of the prior catalog in which the source is first included (see Table~\ref{Table_prior_catalogs}). \\

\incode{RA} & Blind+Prior & degree & The fitted R.A. coordinate of the ALMA emission with Gaussian source models, in equatorial coordinate in the epoch of J2000. \\

\incode{Dec} & Blind+Prior & degree & Same as above but is Dec. coordinate, in equatorial coordinate in the epoch of J2000. \\

\incode{Total_flux_pbcor} & Blind+Prior & mJy & The fitted total flux with Gaussian source models, corrected for flux bias and primary beam attenuation. \\

\incode{E_Total_flux_pbcor} & Blind+Prior & mJy & Error in \incode{Total_flux_pbcor}, provided by photometry pipelines based on \cite{Condon1997} simulation statistics and equations. \\

\incode{E_Total_flux_sim_pbcor} & Blind+Prior & mJy & Error in \incode{Total_flux_pbcor}, but estimated from our own simulation statistics. \\

\incode{Pbcor} & Blind+Prior & --- & Primary beam attenuation factor. \\






\incode{Primary_beam} & Blind+Prior & arcsec & ALMA 12 meter antenna's primary beam FWHM size at the observing frequency. \\

\incode{Peak_flux} & Blind+Prior & $\mathrm{mJy/beam}$ & Fitted ALMA continuum emission's peak flux, uncorrected for primary beam attenuation. \\ 

\incode{RMS_noise} & Blind+Prior & $\mathrm{mJy/beam}$ & Pixel rms noise in the continuum image. \\

\incode{Obs_frequency} & Blind+Prior & GHz & Observing frequency, i.e., the center frequency of all collapsed spectral windows. \\

\incode{Obs_wavelength} & Blind+Prior & $\mu$m & Observing wavelength, $=(2.99792458\times10^{5})/$~\incode{Obs_frequency}. \\

\incode{Maj_beam} & Blind+Prior & arcsec & Synthesized beam's major axis FWHM size. \\

\incode{Min_beam} & Blind+Prior & arcsec & Synthesized beam's minor axis FWHM size. \\

\incode{PA_beam} & Blind+Prior & degree & Synthesized beam's position angle, zero means to the North. \\

\incode{Image_file} & Blind+Prior & --- & Image file name. \\

\incode{Flag_multi} & Blind & --- & \incode{Flag}~$=$~\incode{S} (or \incode{M}) means the source is fitted with single (or multiple) Gaussian component(s). \\

\incode{Galfit_reduced_chi_square} & Prior & --- & The reduced $\chi^2$ of galfit prior source fitting, measured from the residual image for each source with an aperture of 1.0 arcsec in diameter. \\

\incode{Flag_size_upper_boundary} & Prior & --- & \incode{Flag}~$=1$ means the fitted source major axis FWHM size reaches the upper boundary of 3.0 arcsec and should be used with caution. \\

\incode{Flag_inconsistent_flux} & Blind+Prior & --- & \incode{Flag}~$=1$ means the source has $>5\,\sigma$ inconsistent total fluxes from our prior and blind photometry. \\

\incode{Flag_outlier_CPA} & Prior & --- & \incode{Flag}~$=1$ means the source is flagged as an outlier in our counterpart association analysis (Sect.~\ref{Section_Examining_counterpart_association}). \\

\incode{Flag_outlier_SED} & Prior & --- & \incode{Flag}~$=1$ means the source is flagged as an outlier in our SED fitting analysis (Sect.~\ref{Section_Running_SED_Fitting}). \\

\hline
\end{tabularx}
%
%
%
\end{table*}

%% file: Table_galaxy_catalog_columns.tex

\begin{table*}[hb]
\caption{%
	Columns in the final galaxy property catalog
	\label{Table_galaxy_catalog_columns}
}
\begin{tabularx}{1.0\textwidth}{l l X}
\hline
\hline

Column Name & Units & Description \\

\hline

\incode{ID} & --- & A3COSMOS master catalog ID (version 20170426), which equals \cite{Laigle2016} COSMOS2015 catalog ID when ID~$\le 1182108$. \\

\incode{RA} & degree & Fitted ALMA continuum emission's R.A. with Gaussian source models, in the equatorial coordinate in the epoch of J2000. \\

\incode{Dec} & degree & Same as above but is Dec., in the equatorial coordinate in the epoch of J2000. \\

\incode{z} & --- & SED best-fit redshift from the list of prior redshifts in \incode{z_prior}. \\

\incode{z_prior} & --- & Prior redshifts (prior-$z$) in the literature, multiple values are separated by white spaces. \\

\incode{Ref_z_prior} & --- & References of \incode{z_prior}~$^{a}$. \\

\incode{M_star} & $\mathrm{M_{\odot}}$ & Stellar mass from our SED fitting at redshift \incode{z}. Assumed \cite{Chabrier2003} initial mass function (IMF). \\

\incode{L_dust} & $\mathrm{L_{\odot}}$ & Infrared 8--1000~$\mu$m luminosity from dust from the same SED fitting as above. \\

\incode{SFR} & $\mathrm{M_{\odot}\,yr^{-1}}$ & Star formation rate integrated from star formation history from the same SED fitting as above. Same IMF as above. \\ 

\incode{sSFR} & $\mathrm{Gyr^{-1}}$ & Specific SFR from star formation history, $=$~\incode{SFR}~$/$~\incode{M_star}~$\times10^{9}$. \\

\hline
\end{tabularx}
$^{a}$ 
    \incode{A3COSMOS_specz} means the source has the spectroscopic redshift (spec-$z$) confirmed in our A3COSMOS data cube analysis with at least one $\SNR>6$ spectral line (Sect.~\ref{Section_Subtracting_Strong_Emission_Lines}; Liu et al. in prep). 
    \incode{Salvato2017_specz} means the source has spec-$z$ in the COSMOS spec-$z$ catalog compiled by M. Salvato et al. (available in the COSMOS collaboration; version 07SEP2017 with 103,964 rows). 
    \incode{Salvato2011_Chandra_photoz} means the source has photometric redshift (photo-$z$) (optimized for AGNs) in \cite{Salvato2011} \textit{Chandra} source catalog and the photo-$z$ is inconsistent with any previous redshift (by $>0.15\times(1+z)$ difference, same condition afterwards). 
    \incode{Salvato2011_XMM_photoz} means the source has photo-$z$ (optimized for AGNs) in \cite{Salvato2011} \textit{XMM-Newton} source catalog and the photo-$z$ is inconsistent with any previous redshift. 
    \incode{Laigle2016_photoz} means the source has photo-$z$ in the COSMOS2015 catalog provided by \cite{Laigle2016} and the photo-$z$ is inconsistent with any previous redshift. 
    \incode{Davidzon2017_photoz} means the source has photo-$z$ (optimized for $z>2.5$ sources) in \cite{Davidzon2017} catalog and the photo-$z$ is inconsistent with any previous redshift. 
    \incode{Delvecchio2017_photoz} means the source has photo-$z$ (considered mid-IR AGN component) in \cite{Delvecchio2017} catalog and the photo-$z$ is inconsistent with any previous redshift. 
    \incode{Jin2018_photoz} means the source has photo-$z$ (with far-IR/mm photometry) in \cite{Jin2018} catalog and the photo-$z$ is inconsistent with any previous redshift. 
\end{table*}